\documentclass[11pt,english,aps,manuscript,aps,superscriptaddress,eqsecnum]{revtex4}
\usepackage[T1]{fontenc}
\usepackage[latin9]{inputenc}
\usepackage{babel}
\usepackage{verbatim}
\usepackage{amsmath}
\usepackage{amssymb}
\usepackage{graphicx}
\usepackage[unicode=true,pdfusetitle,
 bookmarks=true,bookmarksnumbered=false,bookmarksopen=false,
 breaklinks=false,pdfborder={0 0 1},backref=false,colorlinks=false]
 {hyperref}

\makeatletter

\providecommand{\tabularnewline}{\\}

\@ifundefined{textcolor}{}
{%
 \definecolor{BLACK}{gray}{0}
 \definecolor{WHITE}{gray}{1}
 \definecolor{RED}{rgb}{1,0,0}
 \definecolor{GREEN}{rgb}{0,1,0}
 \definecolor{BLUE}{rgb}{0,0,1}
 \definecolor{CYAN}{cmyk}{1,0,0,0}
 \definecolor{MAGENTA}{cmyk}{0,1,0,0}
 \definecolor{YELLOW}{cmyk}{0,0,1,0}
}

\makeatother

\begin{document}

\pagenumbering{Roman}
\setcounter{page}{0}
\title{Wigner function for spin-1/2 fermions in electromagnetic fields}

\author{Xin-li Sheng}
\affiliation{Institute for Theoretical Physics, Goethe University,
Max-von-Laue-Str.\ 1, D-60438 Frankfurt am Main, Germany}
\affiliation{Interdisciplinary Center for Theoretical Study and Department of
Modern Physics, University of Science and Technology of China, Hefei,
Anhui 230026, China}

\date{07/24/19}

\hspace{-1.5cm}\includegraphics[width=18cm]{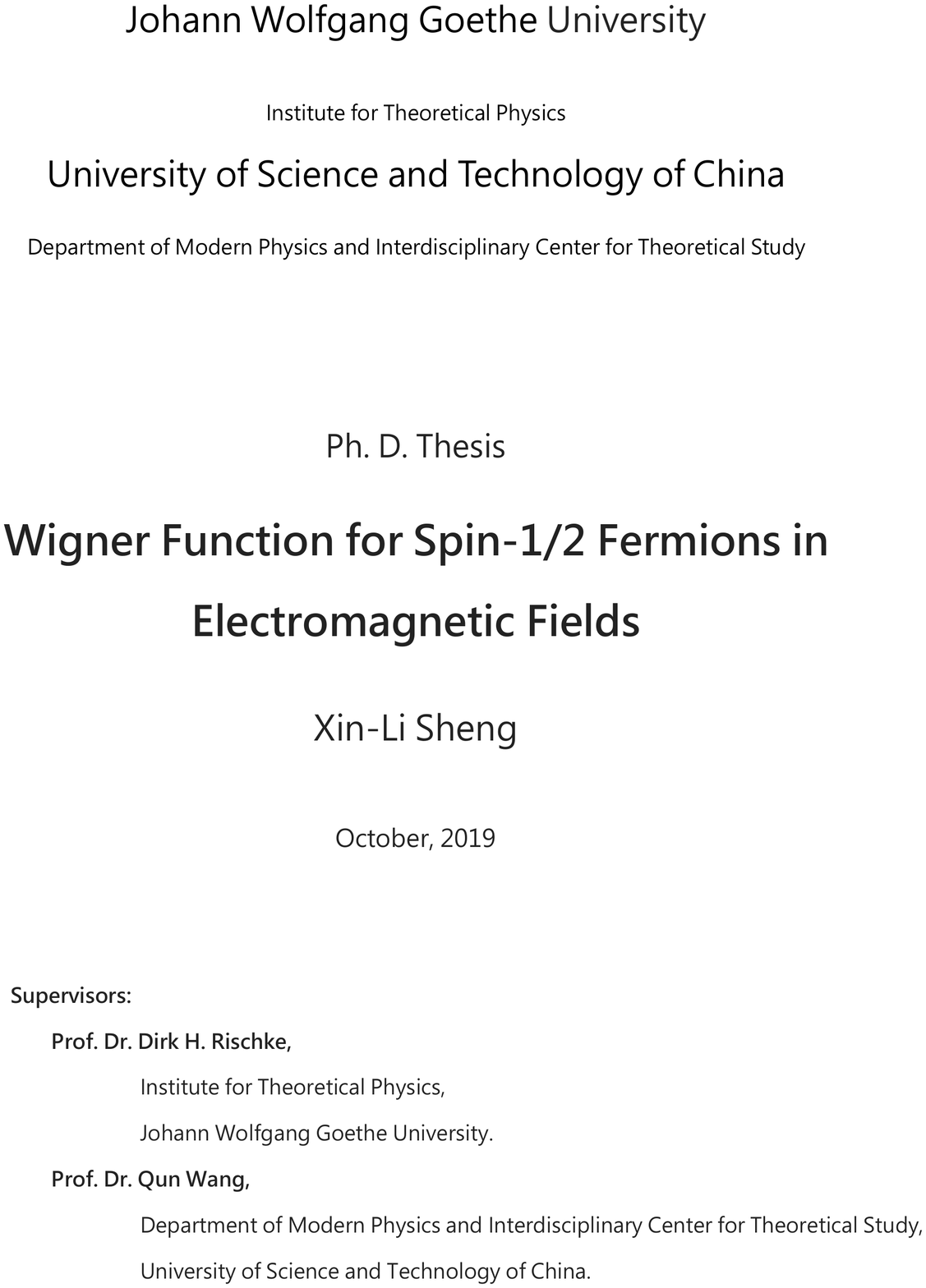}

\begin{abstract}
We study the Wigner function for massive spin-1/2 fermions in electromagnetic
fields. The covariant Wigner function is a four by four matrix function
in 8-dimensional phase space $\{x^{\mu},p^{\mu}\}$, whose components
give various physical quantities such as the particle distribution,
the current density, and the spin distribution, etc.. The kinetic
equations for the Wigner function are obtained from the Dirac equation.
We derive the Dirac form equations with first order differential operators,
as well as the Klein-Gordon form equations with second order differential
operators, both are matrix equations in Dirac space. We prove that
some component equations are automatically satisfied if the rest are
fulfilled, which means both the Dirac form and the Klein-Gordon form
equations have redundancy. In this thesis two methods are proposed
for calculating the Wigner function, which are proved to be equivalent.
In addition to the covariant Wigner function, the equal-time Wigner
function will also be introduced. The equal-time one is a function
of time and 6-dimensional phase space variables $\{\mathbf{x},\mathbf{p}\}$,
which can be derived from the covariant one by taking an integration
over energy $p^{0}$. The equal-time Wigner function is not Lorentz-covariant
but it is a powerful tool to deal with dynamical problems. In this
thesis, it is used to study the Schwinger pair-production in the presence
of an electric field.

The Wigner function can be analytically calculated following the standard
second-quantization procedure. We consider three cases: free fermions
with or without chiral imbalance, and fermions in constant magnetic
field with chiral imbalance. The computations are achieved via firstly
deriving a set of orthonormal single-particle wavefunctions from the
Dirac equation, then constructing the quantized field operator, and
finally inserting the field operator into the Wigner function and
determine the expectation values of operators under the wave-packet
description. The Wigner functions are computed to leading order in
spatial gradients. In Strong electric field the vacuum can decay into
a pair of particle and anti-particle. The pair-production process
is studied using the equal-time Wigner function. General solutions
are obtained for pure constant electric fields and for constant parallel
electromagnetic fields. We also solve the case of a Sauter-type electric
field numerically.

For an arbitrary space-time dependent electromagnetic field, the Dirac
equation does not have an analytical solution and neither has the
Wigner function. A semi-classical expansion with respect to the reduced
Planck's constant $\hbar$ are performed for the Wigner function as
well as the kinetic equations. We calculate the Wigner function (and
all of its components equivalently) to leading order in $\hbar$,
in which order the spin component start playing a role. Up to this
order, the Wigner function contains four independent degrees of freedoms,
three of which describe the polarization density and the remaining
one describes the net particle number density. A generalized Bargmann-Michel-Telegdi
(BMT) equation and a generalized Boltzmann equation are obtained for
these undetermined parts, which can be used to construct spin-hydrodynamics
in the future.

Using analytical results and semi-classical solutions, we compute
physical quantities in thermal equilibrium. In semi-classical expansion,
we introduce the chiral chemical potential $\mu_{5}$ in the thermal
distribution. This naive treatment is straightforward extension of
the massless case but provides a good estimate of physical quantities
when $\mu_{5}$ is comparable or smaller than the typical energy scale,
i.e., the temperature in a thermal system. Meanwhile, by making comparison
of the results of the semi-classical expansion and the ones in a constant
magnetic field, we find that the semi-classical method works well
for the chiral effects, including the Chiral Magnetic Effect, the
Chiral Separation Effect, as well as the energy flux along the direction
of the magnetic field. But when the mass and chemical potentials are
much larger than the temperature, the semi-classical results over
estimate these chiral effects. The magnetic field strength dependence
of physical quantities is discussed. If we fix the thermodynamical
variables, the net fermion number density, energy density, and the
longitudinal pressure are proportional to the field strength, while
the axial-charge density and the transverse pressure are inversely
proportional to it.

Schwinger pair-production rates in a thermal background are computed
for a Sauter-type electric field and a constant parallel electromagnetic
field, respectively. For the Sauter-type field, the total number of
newly generated pairs is proportional to the field strength and the
life time of the field. On the other hand, a parallel magnetic field
will enhance the pair-production rate. Due to Pauli's exclusion principle,
the creation of pairs is forbidden for particles already exsiting
in the same quantum state. Thus in both cases, the pair-production
rate is proved to be inversely proportional to the chemical potential
and temperature.

$\textbf{Keywords:}$ Wigner function, electromagnetic fields, chiral
effect, pair-production.
\end{abstract}
\maketitle

\begin{center}
\Large{Zusammenfassung}
\end{center}

In dieser Arbeit haben wir uns mit dem Wignerfunktionsansatz f\"ur Spin-1/2-Teilchen beschäftigt und diese Herangehensweise verwendet, um die chiralen Effekte und die Paarbildung in Gegenwart eines elektromagnetischen Feldes zu untersuchen. Die Wignerfunktion ist als Quasiverteilungsfunktion im Phasenraum definiert. Die Wignerfunktion ist eine komplexe $4 \times 4$-Matrix, die in die Generatoren der Clifford-Algebra $\Gamma_{i}$ zerlegt werden kann. Die Zerlegungskoeffizienten werden gem\"a{\ss} ihrer Transformationseigenschaften unter Lorentz-Transformationen und Parit\"atsinversion jeweils als Skalar, Pseudoskalar, Vektor, Axialvektor und Tensor identifiziert. Sie k\"onnen nach Integration \"uber den Impuls mit verschiedenen Arten von makroskopischen physikalischen Gr\"o{\ss}en wie dem Fermionstrom, der Spinpolarisation und dem magnetischen Dipolmoment in Beziehung gesetzt werden. Sie werden also als die Dichten im Phasenraum interpretiert.

Da die Wignerfunktion mit Hilfe des Dirac-Feldes konstruiert wird, haben wir die kinetischen Gleichungen f\"ur die Wignerfunktion aus der Dirac-Gleichung erhalten. In dieser Arbeit haben wir die Dirac-Form-Gleichung abgeleitet, die im Differentialoperator linear ist. Daneben haben wir auch die Klein-Gordon-Form-Gleichung erhalten, die die Operatoren zur zweiten Ordnung enthält. Diese Gleichungen werden dann wie auch die Wignerfunktion selbst in $\Gamma_{i}$ zerlegt, sodass sie ein System mehrerer partieller Differentialgleichungen (PDG) liefern. Gl\"ucklicherweise sind die zerlegten Gleichungen nicht unabh\"angig voneinander. Durch Eliminieren der redundanten Gleichungen erhielten wir zwei M\"oglichkeiten, die L\"osung für die Wignerfunktion im massiven Fall zu bestimmen. Diese Redundanz beruht auf der Tatsache, dass die Vektor- und Axialvektorkomponenten $\mathcal{V}^{\mu}$, $\mathcal{A}^{\mu}$ der Wignerfunktion in Form der Skalar-, Pseudoskalar- und Tensorkomponenten $\mathcal{F}$, $\mathcal{P}$, $\mathcal{S}^{\mu \nu}$ ausgedrückt werden können oder umgekehrt. Ein Ansatz zur L\"osung des PDG-Systems besteht daher darin, $\mathcal{V}^{\mu}$, $\mathcal{A}^{\mu}$ als Basisfunktionen zu verwenden und sich auf ihre Massenschalenbedingungen zu konzentrieren. Daneben besteht der andere Ansatz darin, $\mathcal{F}$, $\mathcal{P}$, $\mathcal{S}^{\mu \nu}$ als Basisfunktionen zu verwenden. Durch eine Entwicklung in $\hbar$, die als semiklassische Entwicklung bezeichnet wird, haben wir die allgemeine L\"osung der Wignerfunktion bis zur ersten Ordnung in $\hbar$ erhalten. Es hat sich gezeigt, dass die beiden oben genannten Ans\"atze zu dem gleichen Ergebnis führen und somit \"aquivalent sind. Die endg\"ultige L\"osung hat nur vier unabh\"angige Freiheitsgrade, was durch eine Eigenwertanalyse bewiesen wird. Zur Ordnung $\hbar$ wird die \"ubliche Massenschale $p^{2}-m^{2}=0$ durch die spinmagnetische Kopplung verschoben.

Wir haben weiterhin die Wignerfunktion f\"ur den masselosen Fall durch eine semiklassische Entwicklung reproduziert. Im masselosen Fall k\"onnen die Fermionen nach ihrer Chiralit\"at in zwei Gruppen eingeteilt werden. Unter Verwendung von $\mathcal{V}^{\mu}$ und $\mathcal{A}^{\mu}$ konstruierten wir die linkshändigen und rechtshändigen Str\"ome, die zur Ordnung $\hbar$ bestimmt werden. Die \"ubrigen Komponenten $\mathcal{F}$, $\mathcal{P}$, $\mathcal{S}^{\mu \nu}$ sind proportional zur Teilchenmasse und verschwinden somit im masselosen Limes. In dieser Arbeit haben wir eine direkte Beziehung zwischen den masselosen und massiven Strömen $\mathcal{V}^{\mu}$, $\mathcal{A}^{\mu}$ gefunden. Dies k\"onnte darauf hinweisen, dass unsere massiven Ergebnisse allgemeiner sind als die chirale kinetische Theorie.

In dieser Arbeit haben wir mehrere analytisch l\"osbare F\"alle diskutiert. In den folgenden drei F\"allen konnten aus der Dirac-Gleichung analytisch Einteilchenwellenfunktionen bestimmt werden, aus denen die Wignerfunktion abgeleitet wird. Wir haben nur den Beitrag zur f\"uhrenden Ordnung in r\"aumlichen Gradienten der Wignerfunktionen aufgelistet, aber auch einen potentiellen Ansatz gefunden, um Beitr\"age h\"oherer Ordnung abzuleiten.

1. Quantisierung der ebenen Wellen: In diesem Fall enth\"alt die Dirac-Gleichung keine \"au{\ss}ere Wechselwirkung und wird somit durch freie ebene Wellen gel\"ost. Die Ergebnisse dieses Ansatzes bilden den Grundstein f\"ur die Methode der semiklassischen Entwicklung: Sie dienen als L\"osungen nullter Ordnung in $\hbar$, w\"ahrend solche h\"oherer Ordnung automatisch Ordnung für Ordnung erscheinen.

2. Chirale Quantisierung: In diesem Fall haben wir $\mu$ und $\mu_{5}$ als konstante Variablen f\"ur die Selbstenergie eingef\"uhrt. Diese Variablen tragen in Form von $\mu \hat{\mathbb{N}} + \mu_{5}\hat{\mathbb{N}}_{5}$ zum gesamten Hamiltonian bei, wobei $\hat{\mathbb{N}}$ und $\hat{\mathbb{N}}_{5}$ Operatoren f\"ur Teilchenzahl und Axialladungszahl sind. Im masselosen Limes konnten wir $\mu$ als das chemische Vektorpotential und $\mu_{5}$ als das chirale chemische Potential identifizieren. Wir betonen, dass das chirale chemische Potenzial im massiven Fall nicht wohldefiniert ist, da die konjugierte Grö{\ss}e, die Axialladung, nicht erhalten ist. Im massiven Fall ist $\mu_{5}$ also nur eine Variable, die das Spin-Ungleichgewicht beschreibt. Der modifizierte Hamilton-Operator f\"uhrt zu einer neuen Dirac-Gleichung, die gel\"ost werden k\"onnte, wenn wir annehmen, dass $\mu$ und $\mu_{5}$ Konstanten sind. Die Wignerfunktion wird dann durch Einteilchenwellenfunktionen konstruiert. Da jedoch das Vorhandensein von $\mu$ und $\mu_{5}$ die Dirac-Gleichung \"andert, m\"ussen auch die kinetischen Gleichungen f\"ur die Wignerfunktion weiter modifiziert werden. Dar\"uber hinaus k\"onnen wir die Einteilchenl\"osung f\"ur allgemeine raum- / zeitabh\"angige $\mu$ und $\mu_{5}$ nicht erhalten. Die Methode der chiralen Quantisierung dient somit nur als Gegenprobe f\"ur die Methode der semiklassischen Entwicklung. Hier sind die elektromagnetischen Felder noch nicht enthalten.

3. Landau-Quantisierung: Basierend auf dem Fall 2 f\"uhren wir weiterhin ein konstantes Magnetfeld ein. Die Energieniveaus werden dann durch die Landau-Niveaus mit Modifikation von den chemischen Potentialen $\mu$ und $\mu_{5}$ beschrieben. In diesem Fall k\"onnen wir die Ph\"anomene im Magnetfeld wie CME, CSE und anomalen Energiefluss explizit untersuchen. Da das Feld das Energiespektrum \"andert, haben wir festgestellt, dass die Gesamtfermionzahldichte, die Energiedichte und der Druck von der St\"arke des Magnetfelds abh\"angen.

Dar\"uber hinaus haben wir basierend auf der Quantisierung der ebenen Wellen eine semiklassische Entwicklung in der reduzierten Plancksche Konstante $\hbar$ durchgef\"uhrt. Die Wignerfunktion wird dann bis zur Ordnung $\hbar$ gel\"ost. Man beachte, dass die Methode der semiklassischen Entwicklung f\"ur ein elektromagnetisches Feld mit beliebiger Raum-/Zeitabhängigkeit verwendet werden kann. Bei dieser Methode setzen wir $\mu$ und $\mu_{5}$ in die thermischen Gleichgewichtsverteilungen anstatt in die Hamilton-Verteilung ein und machen Gebrauch von der spezifischen Annahme, dass alle Fermionen in longitudinaler Richtung polarisiert sind. Dieses Verfahren stellt eine naive Erweiterungen des Verfahrens f ür den masselosen Fall dar. Numerische Berechnungen zeigen, dass die auf diese Weise erhaltene Gesamtfermionzahldichte und Axialladungsdichte mit denjenigen aus der chiralen Quantisierung \"ubereinstimmen, wenn $\mu_{5}$ und Masse $m$ vergleichbar mit der oder kleiner als die Temperatur sind. Gleichzeitig zeigen die Energiedichten und Dr\"ucke bei diesen beiden Methoden ebenfalls \"Ubereinstimmungen.

Neben den obigen drei analytisch l\"osbaren F\"allen haben wir auch die Wignerfunktion im elektrischen Feld diskutiert. Basierend auf den Ergebnissen der Quantisierung der ebenen Wellen und der Landauquantisierung erhalten wir durch dynamische Betrachtungen jeweils Wignerfunktionen in Gegenwart eines konstanten elektrischen Feldes. Anschlie{\ss}end werden Paarproduktionen berechnet, die, wie bewiesen wird, durch ein paralleles Magnetfeld verst\"arkt,  durch Temperatur und chemisches Potential dagegen unterdr\"uckt werden. Die Unterdr\"uckung der Paarproduktion im thermischen System wird auf das Pauli'sch Ausschlussprinzip zur\"uckgef\"uhrt.

Die Methode der semiklassischen Entwicklung bietet eine allgemeine M\"oglichkeit, die Spinkorrektur zu berechnen. Zu nullter Ordnung in $\hbar$ haben wir die klassische spinlose Boltzmann-Gleichung reproduziert. Zur Ordnung $\hbar$ treten automatisch Spinkorrekturen wie die Energieverschiebung durch spin-magnetische Kopplung auf. In dieser Arbeit haben wir eine allgemeine Boltzmann-Gleichung und eine allgemeine BMT-Gleichung erhalten, die jeweils die Entwicklungen der Teilchenverteilung und der Spinpolarisationsdichte bestimmen. Kollisionen zwischen Teilchen sind jedoch noch nicht enthalten. Nach der Methode der Momente k\"onnten wir die semiklassischen Ergebnisse zu einer hydrodynamischen Beschreibung ausweiten, was unsere zuk\"unftige Arbeit w\"are. Allerdings erscheinen bei der Methode der semiklassischen Entwicklung die elektromagnetischen Felder zur Ordnung $\hbar$, was f\"ur den Limes schwacher Feldstärke gilt. Im Anfangsstadium von Schwerionenkollisionen ist die Magnetfeldst\"arke jedoch vergleichbar mit $m_{\pi}^{2}$. Auch in sp\"ateren Stadien k\"onnen die Schwankungen der elektromagnetischen Felder $m_{\pi}^{2}$ erreichen. In der starken Laserphysik sind die elektromagnetischen Felder von erheblicher St\"arke, aber es gibt fast keine Teilchen. Ob die semiklassische Entwicklung in diesen F\"allen eingesetzt werden kann oder nicht, bedarf einer genaueren Diskussion. Die Untersuchungen des konstanten Magnetfeldes in dieser Arbeit k\"onnen als Ausgangspunkt der kinetischen Theorie in einem starken Hintergrundfeld dienen.

Eine weitere m\"ogliche Erweiterung dieser Arbeit ist die Axialladungserzeugung. Bei Vorhandensein paralleler elektrischer und magnetischer Felder erzeugt das elektrische Feld Teilchenpaare aus dem Vakuum und die neu erzeugten Paare werden durch das Magnetfeld polarisiert. Infolgedessen tr\"agt die Paarproduktion auf dem niedrigsten Landau-Niveau zur axialen Ladungsdichte bei. Die Realzeitaxialladungserzeugung von massiven Teilchen im thermischen Hintergrund wurde noch nicht gel\"ost. Und der Wignerfunktionsansatz in dieser Arbeit k\"onnte einen m\"oglichen Zugang zu diesem Ziel liefern.

\newpage{}

\tableofcontents{}

\newpage{}

$\ $

\newpage{}

\pagenumbering{arabic}
\setcounter{page}{1}

\section{Introduction}

\subsection{Chiral effects and pair-production}

The quark-gluon plasma (QGP) is a new state of matter created in relativistic
heavy-ion collisions. It is the hot and dense matter of strong interaction
governed by quantum chromodynamics (QCD). The universe is in the QGP
phase in its early stage. Thus creating and studying the QGP helps
us to better understand both the properties of QCD and the evolution
of the universe. There are two big collider experiments for heavy-ion
collisions that are running in the world: the Large Hadron Collider
(LHC) at CERN \cite{LHC:website} and the Relativistic Heavy Ion Collider
(RHIC) at BNL \cite{RHIC:website}. There are also other colliders
under construction: the Facility for Antiproton and Ion Research (FAIR)
at GSI \cite{FAIR:website}, Nuclotron-based Ion Collider fAcility
(NICA) at Dubna, etc..

The evolution of the QGP is dominated by the strong interaction. The
interaction rate is sufficiently large such that the plasma reaches
hydrodynamization rapidly after its generation \cite{Busza:2018rrf}.
Here, hydrodynamization means the QGP can be accurately described
by relativistic hydrodynamics. In non-central collisions, a strong
magnetic field is generated by the fast-moving protons. The field
strength depends on the type of colliding nuclei and the center-of-mass
collision energy. For example, in Au+Au collisions at RHIC with collision
energy $\sqrt{s}=200\mathrm{GeV}$ per nucleon, the magnetic field
can reach several $m_{\pi}^{2}\sim10^{18}\mathrm{G}$ \cite{Skokov:2009qp,Bzdak:2011yy,Voronyuk:2011jd,Deng:2012pc},
which is the strongest field that humans have ever made. The magnetic
field decays quickly because it is mainly generated by the spectators
which move far away from the interaction region soon after the collision
moment. Most simulations use the Lienard-Wiechert potential, developed
by A. M Lienard in 1898 and E. Wiechert in 1900, to describe the electromagnetic
field of a moving point charge in vacuum. However, the QGP is a conducting
medium, with the conductivity having been calculated via lattice QCD
\cite{Ding:2010ga,Aarts:2014nba} and holographic models \cite{Finazzo:2013efa}.
A non-vanishing electrical conductivity will significantly extend
the life-time of the magnetic field \cite{McLerran:2013hla,Tuchin:2013apa}.
Recently, an analytical formula has been derived for the electromagnetic
field generated by a moving point charge in a medium with constant
electrical conductivity $\sigma$ and chiral magnetic conductivity
$\sigma_{\chi}$ \cite{Tuchin:2014iua,Li:2016tel}, which can serve
for numerical simulations in the future.

For massless particles, the chirality operator is commutable with
the Hamiltonian. One can then separate the massless particles into
the right-handed (RH) ones and left-handed (LH) ones according to
their chirality. In the QGP, an imbalance between RH and LH particles
can be generated by topological fluctuations of the gluonic sector,
fluctuations of the quark sector, or glasma flux tubes \cite{Kharzeev:2001ev,Kharzeev:2004ey,Kharzeev:2007tn,Kharzeev:2007jp,Fukushima:2010vw}.
The corresponding thermodynamic states can be specified by a chiral
chemical potential $\mu_{5}$, which is defined as the parameter conjugate
to the topological charge. A nonzero topological charge breaks the
charge-parity symmetry locally and induces a charge current along
the direction of the magnetic field, i.e. Chiral Magnetic Effect (CME)
\cite{Vilenkin:1980fu,Kharzeev:2007jp,Fukushima:2008xe}. The CME
can also be understood through Landau quantization: In the presence
of a magnetic field, charged particles will occupy energy states with
specific spin and orbital angular momentum, which are called Landau
levels. The ground state, i.e., the lowest Landau level is occupied
by positively charged particles whose spins are parallel to the magnetic
field, or negatively charged particles whose spins are anti-parallel
to the magnetic field. Such a spin configuration is required by the
principle of minimum energy. Because of the nonzero topological charge,
the momentum of positively (or negatively) charged particle has a
preference direction with respect to its spin and thus generates a
collective current. The CME is proportional to the magnetic field
and the chiral chemical potential $\mu_{5}$ \cite{Fukushima:2008xe,Landsteiner:2012kd},
\begin{equation}
\mathbf{J}=\frac{\mu_{5}}{2\pi^{2}}q\mathbf{B},
\end{equation}
where $q$ is the electric charge of particles. For systems with multiple
species of particles, we need to take a sum over all species.

In heavy-ion collision experiments, the chiral imbalance can be spontaneously
generated in the initial stage of the collision \cite{Kharzeev:2001ev,Kharzeev:2004ey,Kharzeev:2007tn,Kharzeev:2007jp,Fukushima:2010vw,Hirono:2014oda}.
Thus, the CME is expected to be observed in non-central collisions
through the azimuthal distribution of charge \cite{Voloshin:2004vk},
\begin{equation}
\frac{dN}{d\phi}\propto1+2\sum_{n}\left\{ v_{n}\cos\left[n\left(\phi-\Psi_{\text{RP}}\right)\right]+a_{n}\sin\left[n\left(\phi-\Psi_{\text{RP}}\right)\right]\right\} ,
\end{equation}
where $v_{n}$ and $a_{n}$ denote the parity-even and parity-odd
Fourier coefficients and $\phi-\Psi_{\text{RP}}$ is the azimuthal
angle with respect to the reaction plane. In experiments, the reaction
plane cannot be detected directly, thus in practice we use the event
plane $\Psi_{\text{EP}}$ as an approximation. Here the event plane
is determined by the beam direction and the direction of maximal energy
density. The CME was first expected to be observed through the charge
correlation \cite{Voloshin:2004vk,Abelev:2009ac},
\begin{equation}
\gamma_{\alpha\beta}=\left\langle \cos\left(\phi_{\alpha}+\phi_{\beta}-2\Psi_{\text{EP}}\right)\right\rangle ,
\end{equation}
where $\alpha,\ \beta$ denote particles with the same or opposite
charge sign and $\left\langle \cdots\right\rangle $ means average
over all the particles. Moreover, the determination of the event plane
is not necessary: $\Psi_{\text{EP}}$ can be replaced by a third particle,
which gives the three-particle correlation \cite{Voloshin:2004vk,Abelev:2009ad},
\begin{equation}
\gamma_{\alpha\beta}=\frac{1}{v_{2,c}}\left\langle \cos\left(\phi_{\alpha}+\phi_{\beta}-2\phi_{c}\right)\right\rangle .
\end{equation}
In the CME, the correlation for the same charge sign is observed to
be positive while that for the opposite sign is negative, at RHIC
\cite{Abelev:2009ad,Abelev:2009ac,Adamczyk:2014mzf} and at LHC \cite{Abelev:2012pa,Khachatryan:2016got,Acharya:2017fau}.
However, these correlation functions have significant background contributions
from the cluster particle correlations \cite{Wang:2009kd} and the
coupling between local charge conservation and $v_{2}$ of the QGP
\cite{Schlichting:2010qia}. Meanwhile, the difference between same-charge-sign
correlations and opposite-charge-sign correlations, $\Delta\gamma\equiv\gamma_{SS}-\gamma_{OS}$,
are of the same magnitude in Pb+Pb and p+Pb collisions at LHC \cite{Khachatryan:2016got}.
This is a challenge for the CME interpretation of the charge correlation
because the magnetic field in p+Pb collisions is expected to be much
smaller than in Pb+Pb collisions. On the other hand, the direction
of magnetic field is random with respect to the reaction plane in
p+Pb collisions according to the Glauber Monte Carlo simulation \cite{Alver:2008aq,Khachatryan:2016got},
thus the event-by-event average of the CME contribution is expected
to be small, which indicates that the large part of observables measured
in Pb+Pb collisions may come from the background instead of the CME.
Various new methods are proposed to isolate the CME from the background
\cite{Skokov:2016yrj,Zhao:2017nfq,Magdy:2017yje,Xu:2017qfs}. Isobaric
collisions have been proposed at RHIC for this purpose \cite{Skokov:2016yrj}.
The isobars are chosen to be $\,_{44}^{96}\text{Ru}$+$\,_{44}^{96}\text{Ru}$
and $\,_{40}^{96}\text{Zr}$+$\,_{40}^{96}\text{Zr}$ since they have
the same nucleon number but different proton number. Due to different
proton numbers in collisions of two isobars, the magnetic field would
be $10\%$ different and so is the CME signal, while the backgrounds
are expected to be of the same magnitude because they are dominated
by the strong interaction. Thus, the isobaric collisions would provide
controlled experiment for the CME \cite{Skokov:2016yrj}.

In non-central collisions, the colliding nuclei carry large orbital
angular momentum. For example, the total angular momentum is about
$10^{6}\hbar$ for Au+Au collisions at $\sqrt{s_{NN}}=200$ GeV and
$b=10$ fm \cite{Becattini:2007sr,Deng:2016gyh}. Most of the total
orbital momentum will be taken away by spectators while about 10\%
is left in the QGP \cite{Deng:2016gyh}. The rotation of QGP can be
described by a kinematic vorticity $\boldsymbol{\omega}=\frac{1}{2}\boldsymbol{\nabla}\times\mathbf{v}$,
where $\mathbf{v}$ is the fluid velocity. Analogous to the CME, the
vorticity can also induce an electrical current along its direction
because of the spin-orbit coupling, which is known as the Chiral Vortical
Effect (CVE) \cite{Banerjee:2008th,Son:2009tf}. In the massless case,
$2/3$ of the total CVE is attributed to the magnetization current
and the remaining $1/3$ is attributed to the modified particle distribution
because of the spin-vorticity coupling \cite{Becattini:2013fla,Chen:2014cla,Gao:2018jsi}.
Since the CVE is blind to the charge, it can induce a separation of
baryons. Thus, the CVE is expected to be detected through baryon-baryon
correlations \cite{Kharzeev:2010gr,Zhao:2014aja}.

Because of the spin-magnetic-field and spin-vorticity couplings, both
the magnetic field and vorticity can polarize particles, known as
the Chiral Separation Effect (CSE) \cite{Son:2004tq,Metlitski:2005pr}
and the Axial Chiral Vortical Effect (ACVE) \cite{Banerjee:2008th,Erdmenger:2008rm,Son:2009tf}.
Note that these two effects exist even if the chiral imbalance vanishes,
i.e. $\mu_{5}=0$. The ACVE can induce a global polarization of hyperons,
which has been observed through the polarization of $\Lambda$ hyperon
at STAR \cite{STAR:2017ckg,Adam:2018ivw}. Here the $\Lambda$ decays
into proton and $\pi^{-}$ through weak interaction, which breaks
the parity symmetry. The spin of the $\Lambda$ can then be detected
by the azimuthal distribution of daughter protons.

Since the CME depends on the axial-charge imbalance and induces an
electrical current, while the CSE depends on the electrical charge
imbalance and induces an axial current, the interplay between the
CME and the CSE will generate a propagating wave along the direction
of the magnetic field, the so-called Chiral Magnetic Wave \cite{Kharzeev:2010gd}.
Analogously, the interplay between the CVE and the ACVE excites collective
flow along the vorticity called the Chiral Vortical Wave \cite{Jiang:2015cva}.

In the past few years, a lot of progress has been made in the chiral
effects in heavy-ion collisions, see e.g. Ref. \cite{Kharzeev:2015znc,Huang:2015oca}
for a recent review. Note that the QGP is a complicated many-particle
system, in which the chiral conductivities receive many corrections
\cite{Jensen:2013vta,Gorbar:2013upa}. Thus a self-consistent kinetic
theory is needed for numerical simulations. Recent works along this
line include the kinetic theory with Berry curvature \cite{Son:2012zy,Son:2012bg,Son:2012wh,Chen:2012ca,Chen:2013iga},
the Chiral Kinetic Theory \cite{Stephanov:2012ki,Chen:2015gta,Hidaka:2016yjf,Hidaka:2017auj,Gao:2017gfq,Huang:2018wdl},
and Anomalous Hydrodynamics \cite{Son:2009tf,Neiman:2010zi,Kharzeev:2011ds,Hirono:2014oda,Hidaka:2017auj,Gorbar:2017toh,Florkowski:2017ruc}
but most of these works are for massless particles. Even through the
$u,d$ quarks are almost massless compared with the typical temperature
of the QGP, the $s$ quarks are quite massive. The Wigner-function
method in this thesis provides a possible way to develop the kinetic
theory for massive spin-1/2 particles \cite{Gao:2019znl,Weickgenannt:2019dks,Hattori:2019ahi,Wang:2019moi}.

In addition to the physics related to the magnetic field and vorticity,
we also study the effects of the strong electric field in the QGP.
This electric field is induced by the fast decreasing magnetic field
according to the Maxwell's equation. It is of the same magnitude as
the magnetic field and both of them are sufficiently large in the
initial stage of heavy-ion collisions, see Refs. \cite{Deng:2012pc,Li:2016tel}
for some numerical simulations. In a strong electric field, the QED
vacuum becomes unstable due to fermion/anti-fermion pair-production,
the so-called Schwinger process \cite{Schwinger:1951nm}. The pair-production
rate was first derived by Julian Schwinger in 1951 via quantum field
theory \cite{Schwinger:1951nm} and then reproduced through various
kinds of methods, such as the WKB method \cite{Brezin:1970xf,Popov:1971iga,Popov:1973az},
instanton method \cite{Affleck:1981bma,Kim:2000un,Dunne:2006ur},
holographic method \cite{Ambjorn:2011wz,Sato:2013pxa,Sato:2013dwa},
and the Wigner-function method \cite{BialynickiBirula:1991tx,Hebenstreit:2010vz,Kohlfurst:2015niu,Kohlfurst:2017git,Sheng:2018jwf}.
The pair-production process can be analytically solved for a constant
electric field $\mathbf{E}(t)=\mathbf{E}_{0}$ or a Sauter-type field
$\mathbf{E}(t)=\mathbf{E}_{0}\text{sech}^{2}(t/\tau)$ using the quantum
kinetic theory \cite{Smolyansky:1997fc,Kluger:1998bm}. In heavy-ion
collisions, the electric field strongly depends on the space and time.
One may estimate the pair-production rate via firstly dividing the
whole space into small cells, and then applying the pair-production
rate for constant electric field in each cell. However, in the instanton
method \cite{Dunne:2005sx}, one can show that spatial inhomogeneities
tend to suppress the pair-production while the temporal ones tend
to enhance it. Thus it is difficult to judge whether the constant-field
approximation overestimates or underestimates the total pair-production
rate. Some methods such as the Wigner-function method \cite{Hebenstreit:2010vz,BialynickiBirula:2011uh,Kohlfurst:2015niu,Kohlfurst:2017git,Sheng:2018jwf}
can deal with space- and time-dependent electric fields, but one has
to solve a system of non-linear partial differential equations. Thanks
to the development of computing power, the pair-production for more
general field configurations becomes numerically solvable \cite{Blinne:2013via,Berenyi:2013eia,Kasper:2014uaa,Buyens:2016hhu}.

Moreover, the thermal background and strong magnetic field makes the
calculation of pair-production more challenging in a medium than in
the vacuum. In a thermal system, the existing particles prohibit the
creation of new particles with the same quantum number because of
the Pauli exclusion principle. Thus the pair-production will be suppressed
in thermal systems \cite{Gies:1999vb,Kim:2007ra,Kim:2008em,Gould:2017fve,Sheng:2018jwf}.
Meanwhile, the magnetic field can increase the pair-production rate
if it is parallel to the electric field. This enhancement has been
analytically shown many years ago through the proper-time method \cite{Nikishov:1969tt,Bunkin:1970iz,Popov:1971iga}
and recently been reproduced in string theory \cite{Lu:2017tnm,Lu:2018nsc},
holographic theory \cite{Cai:2016jgr}, and the Wigner function approach
\cite{Sheng:2018jwf}. On the other hand, due to the fact that particles
in the lowest Landau level behave like chiral fermions, the pair-production
in parallel electromagnetic fields is related to the production of
axial charge \cite{Fukushima:2010vw,Cai:2016jgr,Copinger:2018ftr}
and to the pseudoscalar condensation \cite{Cao:2015cka,Fang:2016uds}.

Non-central heavy-ion collisions always lead to strong electromagnetic
fields. The study of fermions (quarks) in an electromagnetic field
will help us to better understand the early-stage evolution of the
QGP. In central collisions, the event-by-event average of the field
is zero but its fluctuation is sufficiently large \cite{Deng:2012pc}.
Although the chiral effects and the pair-production, have been extensively
studied for many years, they have not yet been fully understood in
terms of experimental observables. For example, how to extract the
very weak signal out of large backgrounds is still unsolved \cite{Skokov:2016yrj,Zhao:2017nfq,Magdy:2017yje,Xu:2017qfs}.
As another example, the $\Lambda$ polarization in the longitudinal
and the transverse directions: the results of hydrodynamical calculation
\cite{Becattini:2017gcx,Wei:2018zfb,Xia:2018tes} have opposite sign
with respect to the experiment data \cite{Adam:2018ivw,Adam:2019srw}.
Thus deeper and more extensive studies about fermions in electromagnetic
and vorticity fields are necessary in the frontier of high-energy
physics.

\subsection{Wigner-function method}

In classical statistical theory, a multi-particle system is described
by a classical particle distribution $f(t,\mathbf{x},\mathbf{p})$,
as a function of the time $t$, the spatial coordinates $\mathbf{x}$,
and the 3-momentum $\mathbf{p}$ of the particle. This description
is valid because the spatial position and 3-momentum of a classical
particle can be determined simultaneously to arbitrary precision.
However, in quantum mechanics, the classical distribution $f(t,\mathbf{x},\mathbf{p})$
is not well-defined because Heisenberg's uncertainty principle states
that the more precisely the momentum of one particle is determined,
the more uncertainty is its position, and vice versa. This principle
was firstly proposed by Werner Heisenberg \cite{aHeisenberg:1927zz}
in 1927 and then mathematically derived by Earle Kennard \cite{kennard1927quantenmechanik}
and Hermann Weyl \cite{weyl1928gruppentheorie},
\begin{equation}
\sigma_{x}\sigma_{p}\geq\frac{\hbar}{2},\label{eq:Heisenberg uncertainty principle}
\end{equation}
where $\sigma_{x(p)}$ is the standard deviation when measuring the
position $x$ or the momentum $p$, and $\hbar$ is the reduced Planck's
constant. In 1932, Eugene Wigner introduced a quasi-probability distribution
to study quantum statistical mechanics \cite{Wigner:1932eb}, which
is now called the Wigner function (or Wigner quasi-probability distribution).
The Wigner function is derived from a two-point correlation function
by taking the Fourier transform with respect to the distance between
the two points, so the Wigner function is a function in phase space
$\left\{ x^{\mu},p^{\mu}\right\} $. The spatial densities of physical
observables can be derived form the Wigner function by integrating
over the 4-momentum $p^{\mu}$\cite{Vasak:1987um}. A more detailed
discussion of the Wigner function will be presented in Sec. \ref{subsec:Definition-of-Wigner}.
For spin-1/2 particles, the Wigner function is defined using the Dirac
field operator, thus the kinetic equations for the Wigner function
can be derived from the Dirac equation without loss of generality
\cite{Vasak:1987um}.

The Wigner function can be analytically computed only in very limited
cases, such as in constant electromagnetic fields \cite{Hebenstreit:2010vz,Gorbar:2017awz,Sheng:2017lfu,Sheng:2018jwf}.
Meanwhile, the numerical calculation is challenging because the parameter
space is $8$-dimensional, which is too large for finite-difference
methods. A general way to deal with the Wigner function is to treat
the space-time derivative and electromagnetic field as small quantities
and expand all the Wigner function as well as all the operators in
terms of $\hbar$. This method is known as the semi-classical expansion
\cite{Vasak:1987um,Zhuang:1995pd}. Since $\hbar$ is the unit of
the angular momentum, the expansion in $\hbar$ is also an expansion
in spin. Up to $\mathcal{O}(\hbar)$, general solutions for the Wigner
function have been obtained, in both the massless \cite{Gao:2012ix,Chen:2012ca,Hidaka:2016yjf,Huang:2018wdl}
and the massive \cite{Gao:2019znl,Weickgenannt:2019dks,Hattori:2019ahi,Wang:2019moi}
case. In the massless case, the Chiral Kinetic Theory can be obtained
from the Wigner-function approach \cite{Hidaka:2016yjf,Hidaka:2017auj,Gao:2017gfq,Huang:2018wdl}.
The chiral effects are successfully reproduced at the order $\mathcal{O}(\hbar)$
\cite{Gao:2012ix,Chen:2012ca,Hidaka:2016yjf,Gao:2017gfq,Huang:2018wdl}.
In the massive case, kinetic equations are obtained which agree with
the relativistic Boltzmann-Vlasov equation and the Bargmann-Michel-Telegdi
(BMT) equation in the classical limit \cite{Gao:2019znl,Weickgenannt:2019dks,Hattori:2019ahi,Wang:2019moi}
and recover the Chiral Kinetic Theory in the massless limit \cite{Weickgenannt:2019dks,Hattori:2019ahi,Wang:2019moi}.

On the other hand, the equal-time Wigner function can be derived from
the covariant one by integrating out the energy $p^{0}$ \cite{BialynickiBirula:1991tx,Zhuang:1998bqx,Gorbar:2017awz}.
The equal-time formula only depends on $\left\{ t,\mathbf{x},\mathbf{p}\right\} $,
thus is suitable for time-dependent problems, such as out-of-equilibrium
physics \cite{Guo:2017dzf} and pair-production \cite{Hebenstreit:2010vz,BialynickiBirula:2011uh,Kohlfurst:2015niu,Kohlfurst:2017git,Sheng:2018jwf}.

\subsection{System of units, notations and conventions\label{subsec:System-of-Units,}}

In this subsection, we declare the system of units, notations and
conventions we will use throughout this thesis. In the International
System of Units (SI), the reduced Planck's constant, the speed of
light, and the electron volt are
\begin{equation}
[\hbar]=[kg\cdot m^{2}\cdot s^{-1}],\ \ [c]=[m\cdot s^{-1}],\ \ [\text{eV}]=[kg\cdot m^{2}\cdot s^{-2}].\label{eq:SI units}
\end{equation}
Here the square bracket $[\cdots]$ represents the unit or the dimension,
and $m$, $kg$, $s$ are meter, kilogram and second, the unit of
mass, length, and time, respectively. On the other hand, from Eq.
(\ref{eq:SI units}), we can express the units of mass, length and
time in terms of $\hbar$, $c$ and eV,
\begin{equation}
[kg]=[c^{-2}\cdot\text{eV}],\ \ [m]=[\hbar\cdot c\cdot\text{eV}^{-1}],\ \ [s]=[\hbar^{2}\cdot c\cdot\text{eV}^{-2}].
\end{equation}
Natural units are convenient to us, in which the units of physical
quantities are selected as physical constants. For example, the speed
of light is the natural unit of speed. In natural units, the values
of the reduced Planck's constant $\hbar$ and the speed of light $c$
are set to $1$, while the unit of energy is set to eV. In SI units,
the unit of any physical quantity can be expressed as $m^{a}\cdot s^{b}\cdot kg^{c}$,
with some rational numbers $a,\ b,\ c$. Then it can be rewritten
as $\mathrm{eV}^{c-a-2b}$ in natural units, where $\hbar$ and $c$
are hidden because they are set to $1$. If we consider, for example,
a charged particle in a static electric field, the unit of electric
force is given by
\begin{equation}
[qE]=[kg\cdot m\cdot s^{-2}]=[\hbar^{-1}\cdot c^{-1}\cdot\text{eV}^{2}].
\end{equation}
Thus in natural units, $qE$ has the unit of energy squared, $\text{eV}^{2}$.
Here $q$ denotes the charge of the particles considered. As a convention,
$q$ always comes in front of the gauge potential $\mathbb{A}^{\mu}$,
electric and magnetic fields $E^{\mu}$, $B^{\mu}$ and the field
strength tensor $F^{\mu\nu}$. In the thesis, we only consider charged
fermions of one species, where the charge of fermions is $+1$ (and
$-1$ for anti-fermions). Thus we can absorb the charge $q$ in the
definition of the electromagnetic field tensor $F^{\mu\nu}$ and the
gauge potential $\mathbb{A}^{\mu}$.

Since the spin has the unit of $\hbar$, we will use $\hbar$ as a
parameter to label its quantum nature. In the calculation of the Wigner
function, we will recover $\hbar$ in Sec. \ref{sec:Overview-of-Wigner}
and treat $\hbar$ as an expansion parameter in Sec. \ref{sec:Semi-classical-expansion}.
This method is already known as the semi-classical expansion \cite{Vasak:1987um,Zhuang:1995pd},
which at leading order in $\hbar$ can reproduce the classical results.

Throughout this thesis, we assume the mass $m$ of a particle is constant.
We use natural units $\hbar=c=k_{B}=1$ but show $\hbar$ explicitly
in Sec. \ref{sec:Semi-classical-expansion} since $\hbar$ is used
as a parameter for power-counting in that section. We work in Minkowski
space with the metric tensor $g^{\mu\nu}=\text{diag}(1,-1,-1,-1)$
and the Levi-Civita symbol $\epsilon^{0123}=-\epsilon_{0123}=1$.
We use bold symbols such as $\mathbf{p}$ to represent 3-dimensional
vectors. The electromagnetic potential is denoted by $\mathbb{A}^{\mu}$
in order to distinguish it from the axial-vector component of the
Wigner function. The electric charge $q$ is set to $+1$ for fermions
and $-1$ for anti-fermions and thus $q$ will be hidden in this thesis.
The operators in the Dirac theory, such as the Dirac field operator,
the Hamiltonian operator, etc., are denoted with hat. Meanwhile, operators
constructed by the space-time derivative $\partial_{x}^{\mu}$ and
the momentum derivative $\partial_{p}^{\mu}$ are denoted without
hat. The components of a Lorentz vector are labeled by $\left\{ 0,x,y,z\right\} $,
for example, the 4-momentum is denoted as $p^{\mu}=(p^{0},p^{x},p^{y},p^{z})^{T}$.
Sometime we use the transpose operation to change a 4-momentum in
the line vector into a column vector. In this thesis we also used
$M^{-1}$ to denote the inverse of the matrix $M$. The unit matrix
is denoted by $\mathbb{I}_{n}$, with $n$ being the dimension of
the matrix.

\subsection{Outline}

In this thesis we will first give an overview of the Wigner-function
method in Sec. \ref{sec:Overview-of-Wigner}. This section includes
the definition of the covariant Wigner function and its kinetic equations.
Two kinds of kinetic equations are obtained, one of which is the analog
of the Dirac equation and the other one is the analog of the Klein-Gordon
equation. These equations are differential equations of first order
and second order in time, respectively. The Wigner function, as well
as its kinetic equations, are then decomposed in terms of the generators
of the Clifford algebra, i.e., the gamma matrices $\{\mathbb{I}_{4},\ i\gamma^{5},\ \gamma^{\mu},\ \gamma^{5}\gamma^{\mu},\ \frac{1}{2}\sigma^{\mu\nu}\}$.
The equal-time Wigner function is also introduced in Sec. \ref{sec:Overview-of-Wigner}.
In Sec. \ref{sec:Analytically-solvable-cases} we focus on several
analytically solvable cases:
\begin{enumerate}
\item Free fermions, without electromagnetic field and without chiral imbalance.
\item Without electromagnetic field, but a chiral imbalance is introduced
in the Dirac equation as a self-energy correction.
\item A constant magnetic field, otherwise as in case 2.
\item A constant electric field, otherwise as in case 1.
\item Constant electromagnetic fields added to case 1, where the electric
and magnetic fields are assumed to be parallel to each other.
\end{enumerate}
In the cases 1, 2, and 3, the Dirac equation has analytical single-particle
solutions, which are then used to derive the Wigner functions. Note
that the magnetic field and the additional self-energy term break
the Lorentz symmetry. For example, if we take a Lorentz boost along
the direction perpendicular to the magnetic field, we find that an
electric field automatically appears in the new frame. That is, the
magnetic field itself is not Lorentz covariant. Meanwhile, chemical
potentials also breaks the Lorentz covariance. So for cases 2 and
3, we are working in specific frames in which the chemical potential
$\mu$ is the conjugate parameter for the net particle number and
$\mu_{5}$ is the one for the axial charge. In the presence of an
electric field, i.e., cases 4 and 5, the existing particles will be
accelerated and new fermion/anti-fermion pairs will be excited from
the vacuum. Thus systems in an electric field are evolving over time
and the equal-time Wigner function is used for these systems. In Sec.
\ref{sec:Analytically-solvable-cases} we analytically solve the Wigner
function for the case of a constant electric field. Meanwhile in subsection
\ref{subsec:Solutions-for-Sauter-type} we numerically calculate the
solution for a Sauter-type electric field $E(t)=E_{0}\text{sech}^{2}(t/\tau)$.
A more general field configuration is considered in Sec. \ref{sec:Semi-classical-expansion}
using the method of semi-classical expansion. Solutions are obtained
up to order $\hbar$ for both massless and massive particles. In Sec.
\ref{sec:Physical quantities} we relate the Wigner function with
several physical quantities such as the net fermion current, spin
polarization, energy-momentum tensor, etc.. The analytical results
from Sec. \ref{sec:Analytically-solvable-cases} are used under a
thermal-equilibrium assumption. The physical quantities show a dependence
with respect to the thermodynamical variables and the magnetic field.
Different methods are used and compared with each other, which show
both coincidences and differences. pair-production is also discussed
in Sec. \ref{sec:Physical quantities}. The results show that the
magnetic field enhances the pair-production rate, while the thermal
background suppresses it. A summary and outlook of this thesis are
given in Sec. \ref{sec:Summary}. In App. \ref{sec:Gamma-matrices}
we listed the gamma matrices and their properties. Other useful auxiliary
functions are discussed in App. \ref{sec:Auxiliary-functions}, which
appear when dealing with the Wigner function in constant electromagnetic
fields. In App. \ref{sec:Wave-packet-description} we present the
standard wave-packet description for a quantum particle, which will
be used when solving the Wigner function. The relation between the
pair-production rate and the Wigner function is derived from a quantum
field description in App. \ref{sec:Pair-production-in}.

\newpage{}

$\ $

\newpage{}

\section{Overview of Wigner function\label{sec:Overview-of-Wigner}}

\subsection{Definition of Wigner function \label{subsec:Definition-of-Wigner}}

In quantum mechanics, the space-time position $x^{\mu}$ and the 4-momentum
$p^{\mu}$ cannot be specified simultaneously for a single particle,
which is a straightforward consequence of the Heisenberg's uncertainty
principle \cite{aHeisenberg:1927zz,kennard1927quantenmechanik,weyl1928gruppentheorie}.
Thus, the classical particle distribution function $f(t,\mathbf{x},\mathbf{p})$
is not well-defined in the quantum case. In order to find a proper
way to describe quantum kinetics, we first consider a system of two
particles, whose space-time coordinate operators are $\hat{x}_{1}^{\mu}$
and $\hat{x}_{2}^{\mu}$ and 4-momentum operators are $\hat{p}_{1}^{\mu}$
and $\hat{p}_{2}^{\mu}$, respectively. The uncertainty principle
gives the following relations
\begin{equation}
\left[\hat{x}_{a}^{\mu},\,\hat{p}_{b}^{\nu}\right]=-i\hbar g^{\mu\nu}\delta_{ab},\ \ \left[\hat{x}_{a}^{\mu},\,\hat{x}_{b}^{\nu}\right]=0,\ \ \left[\hat{p}_{a}^{\mu},\,\hat{p}_{b}^{\nu}\right]=0.\label{eq:commutators of two particles}
\end{equation}
where $a,b=1,2$. We now define the center position and the relative
momentum as
\begin{equation}
\hat{x}^{\mu}\equiv\frac{1}{2}\left(\hat{x}_{1}^{\mu}+\hat{x}_{2}^{\mu}\right),\ \ \hat{p}^{\mu}\equiv\frac{1}{2}\left(\hat{p}_{1}^{\mu}-\hat{p}_{2}^{\mu}\right).
\end{equation}
Then using the commutators in Eq. (\ref{eq:commutators of two particles})
we can check that these two quantities are commutable with each other
\begin{equation}
\left[\hat{x}^{\mu},\,\hat{p}^{\nu}\right]=0,\ \ \left[\hat{x}^{\mu},\,\hat{x}^{\nu}\right]=0,\ \ \left[\hat{p}^{\mu},\,\hat{p}^{\nu}\right]=0.
\end{equation}
Thus according to the uncertainty principle, even if we can not determine
the position and momentum simultaneously for each particle, the center
position and relative momentum can specified simultaneously. Meanwhile,
the relative position and the total momentum are defined as
\begin{equation}
\hat{y}^{\mu}\equiv\hat{x}_{1}^{\mu}-\hat{x}_{2}^{\mu},\ \ \hat{q}^{\mu}\equiv\hat{p}_{1}^{\mu}+\hat{p}_{2}^{\mu}.
\end{equation}
These operators also commute with each other
\begin{equation}
\left[\hat{y}^{\mu},\,\hat{q}^{\nu}\right]=0,\ \ \left[\hat{y}^{\mu},\,\hat{y}^{\nu}\right]=0,\ \ \left[\hat{q}^{\mu},\,\hat{q}^{\nu}\right]=0,
\end{equation}
and thus they are not constrained by the uncertainty principle. On
the other hand, we have the following commutators
\begin{equation}
\left[\hat{y}^{\mu},\,\hat{p}^{\nu}\right]=\left[\hat{x}^{\mu},\,\hat{q}^{\nu}\right]=-i\hbar g^{\mu\nu}\delta_{ab},
\end{equation}
which indicates that the relative momentum is the conjugate variable
of the relative position, while the total momentum is the conjugate
variable of the center position.

The Wigner operator for a free Dirac field is defined from the two-point
correlation function \cite{Wigner:1932eb},
\begin{equation}
\hat{\mathcal{W}}_{\mathrm{free}}(x,p)\equiv\int\frac{d^{4}y}{(2\pi)^{4}}\exp\left(-iy^{\mu}p_{\mu}\right)\hat{\bar{\psi}}\left(x+\frac{y}{2}\right)\otimes\hat{\psi}\left(x-\frac{y}{2}\right),\label{def:free Wigner operator}
\end{equation}
where the operator $\otimes$ represents the tensor product and $\hat{\psi}$
is the Dirac field operator. In this definition, the two field operators
are defined at two different space-time points, $x^{\mu}\pm\frac{y^{\mu}}{2}$,
where $x^{\mu}$ the center position and $y^{\mu}$ the relative position.
A Fourier transform is taken with respect to $y^{\mu}$, whose conjugate
momentum $p_{\mu}$ can be identified as the relative momentum of
two fields in classical mechanics. According to discussion in previous
paragraph, $p^{\mu}$ and $x^{\mu}$ can be determined simultaneously.
Thus the Wigner operator is a well-defined quasi-distribution in phase-space
$\{x^{\mu},p^{\mu}\}$. We will show in Sec. \ref{sec:Analytically-solvable-cases}
that the Wigner function is related to the classical distribution
$f(t,\mathbf{x},\mathbf{p})$ at leading order in spatial gradients.

Note that the Wigner operator defined in Eq. (\ref{def:free Wigner operator})
is not gauge-invariant. Under a local gauge transformation $\theta(x)$,
the field operators transform as follows,
\begin{equation}
\hat{\psi}\left(x-\frac{y}{2}\right)\rightarrow e^{i\theta(x-y/2)}\hat{\psi}\left(x-\frac{y}{2}\right),\ \ \hat{\bar{\psi}}\left(x+\frac{y}{2}\right)\rightarrow e^{-i\theta(x+y/2)}\hat{\bar{\psi}}\left(x+\frac{y}{2}\right),
\end{equation}
and the Wigner operator transforms as
\begin{equation}
\hat{\mathcal{W}}_{\mathrm{free}}(x,p)\rightarrow\exp\left[i\theta(x-y/2)-i\theta(x+y/2)\right]\hat{\mathcal{W}}_{\mathrm{free}}(x,p).
\end{equation}
The exponential factor is not $1$ for general local transformation
with $\theta(x-y/2)\neq\theta(x+y/2)$. Thus the Wigner operator in
Eq. (\ref{def:free Wigner operator}) is not gauge-invariant. In order
to define a gauge-invariant quantity, we first express the Dirac field
at the position $x-\frac{y}{2}$ in a Taylor expansion as follows,
\begin{equation}
\hat{\psi}\left(x-\frac{y}{2}\right)=\sum_{n=0}^{\infty}\frac{1}{n!}\left(-\frac{1}{2}y_{\mu}\partial_{x}^{\mu}\right)^{n}\hat{\psi}(x)=\exp\left(-\frac{1}{2}y_{\mu}\partial_{x}^{\mu}\right)\hat{\psi}(x).
\end{equation}
In the presence of an electromagnetic field, we replace the ordinary
derivative $\partial_{x}^{\mu}$ by the covariant one $D_{x}^{\mu}=\partial_{x}^{\mu}+i\mathbb{A}^{\mu}$
so that the gauge invariance is automatically ensured. Here $\mathbb{A}^{\mu}$
is the four-vector potential of the electromagnetic field. Inserting
the field operator into Eq. (\ref{def:free Wigner operator}), we
define the following Wigner operator,
\begin{equation}
\hat{\mathcal{W}}(x,p)\equiv\int\frac{d^{4}x}{(2\pi)^{4}}\exp\left(-iy^{\mu}p_{\mu}\right)\left[\exp\left(\frac{1}{2}y_{\mu}D_{x}^{\mu}\right)\hat{\psi}(x)\right]^{\dagger}\gamma^{0}\otimes\left[\exp\left(-\frac{1}{2}y_{\mu}D_{x}^{\mu}\right)\hat{\psi}(x)\right],\label{def:covariant Wigner function}
\end{equation}
which is covariant and gauge-invariant.

Now we define the gauge link between two space-time points
\begin{equation}
U(x_{2},x_{1})=\exp\left[-i\int_{x_{1}}^{x_{2}}dx^{\mu}\mathbb{A}_{\mu}(x)\right],
\end{equation}
where the path of integration is taken as a straight line between
two points. Then we obtain the gauge link between $x^{\mu}-\frac{1}{2}y^{\mu}$
and $x^{\mu}+\frac{1}{2}y^{\mu}$,
\begin{equation}
U\left(x+\frac{y}{2},\,x-\frac{y}{2}\right)=\exp\left[-iy^{\mu}\int_{-1/2}^{1/2}ds\mathbb{A}_{\mu}(x+sy)\right],\label{eq:gauge link}
\end{equation}
We can prove
\begin{equation}
e^{\frac{1}{2}y_{\mu}D_{x}^{\mu}}=U\left(x,\,x+\frac{y}{2}\right)e^{\frac{1}{2}y_{\mu}\partial_{x}^{\mu}},\label{eq:gauge link relation}
\end{equation}
with the help of two auxiliary functions
\begin{eqnarray}
f(s) & \equiv & e^{sy_{\mu}D_{x}^{\mu}},\nonumber \\
g(s) & \equiv & U\left(x,\,x+sy\right)e^{sy_{\mu}\partial_{x}^{\mu}}.\label{def:auxiliary functions for gauge link}
\end{eqnarray}
Their derivatives with respect to the parameter $s$ are%
\begin{eqnarray}
\frac{d}{ds}f(s) & = & f(s)y_{\nu}D_{x}^{\nu},\nonumber \\
\frac{d}{ds}g(s) & = & U\left(x,\,x+sy\right)e^{sy_{\mu}\partial_{x}^{\mu}}y_{\nu}\partial_{x}^{\nu}+\left[\frac{d}{ds}U\left(x,\,x+sy\right)\right]e^{sy_{\mu}\partial_{x}^{\mu}}\nonumber \\
 & = & U\left(x,\,x+sy\right)\left\{ e^{sy_{\mu}\partial_{x}^{\mu}}y_{\nu}\partial_{x}^{\nu}+\left[iy^{\nu}\mathbb{A}_{\nu}(x+sy)\right]e^{sy_{\mu}\partial_{x}^{\mu}}\right\} \nonumber \\
 & = & g(s)y_{\nu}D_{x}^{\nu},
\end{eqnarray}
which means the two functions defined in Eq. (\ref{def:auxiliary functions for gauge link})
satisfy the same differential equation. Furthermore they also share
the same value at point $s=0$, we conclude that they are equivalent
for arbitrary $s$. Equation (\ref{eq:gauge link relation}) is then
proved by taking $s=\frac{1}{2}$. Substituting Eq. (\ref{eq:gauge link relation})
into Eq. (\ref{def:covariant Wigner function}), we obtain another
form of the Wigner operator
\begin{equation}
\hat{\mathcal{W}}(x,p)=\int\frac{d^{4}y}{(2\pi)^{4}}\exp\left(-iy^{\mu}p_{\mu}\right)U\left(x+\frac{y}{2},\,x-\frac{y}{2}\right)\hat{\bar{\psi}}\left(x+\frac{y}{2}\right)\otimes\hat{\psi}\left(x-\frac{y}{2}\right).
\end{equation}
The Wigner function is then derived by taking the expectation value
of the Wigner operator on the physical state of the system $|\Omega\rangle$
\begin{equation}
W(x,p)=\int\frac{d^{4}y}{(2\pi)^{4}}\exp\left(-iy^{\mu}p_{\mu}\right)U\left(x+\frac{y}{2},\,x-\frac{y}{2}\right)\left\langle \Omega\left|\hat{\bar{\psi}}\left(x+\frac{y}{2}\right)\otimes\hat{\psi}\left(x-\frac{y}{2}\right)\right|\Omega\right\rangle .\label{def:Wigner function}
\end{equation}
Note that here we have taken the gauge link out of the expectation
value. We treat the gauge field (i.e., the electromagnetic field in
this thesis) as a classical C-number field, while the fermionic field
as a quantum field. This is known as the Hartree approximation, which
is valid when higher-loop corrections are negligible or the field
is large enough.

The Wigner function in Eq. (\ref{def:Wigner function}) is not Hermitian
but it transforms as
\begin{equation}
W^{\dagger}=\gamma^{0}W\gamma^{0},\label{prop:Hermitian property of Wigner}
\end{equation}
which is the same as the property (\ref{eq:Cliddord algebra}) of
the generators of the Clifford algebra $\Gamma_{i}=\{\mathbb{I}_{4},\ i\gamma^{5},\ \gamma^{\mu},\ \gamma^{5}\gamma^{\mu},\ \frac{1}{2}\sigma^{\mu\nu}\}$,
where $\sigma^{\mu\nu}=\frac{i}{2}\left[\gamma^{\mu},\gamma^{\nu}\right]$
and $\gamma^{5}=i\gamma^{0}\gamma^{1}\gamma^{2}\gamma^{3}$. Thus
the Wigner function can be expanded in terms of $\Gamma_{i}$,
\begin{equation}
W(x,p)=\frac{1}{4}\left(\mathbb{I}_{4}\mathcal{F}+i\gamma^{5}\mathcal{P}+\gamma^{\mu}\mathcal{V}_{\mu}+\gamma^{5}\gamma^{\mu}\mathcal{A}_{\mu}+\frac{1}{2}\sigma^{\mu\nu}\mathcal{S}_{\mu\nu}\right).\label{def:Wigner function decomposition}
\end{equation}
The Wigner function has 16 independent components because it is a
complex $4\times4$ matrix and Eq. (\ref{prop:Hermitian property of Wigner})
provides 16 constraints, which correspond to $\Gamma_{i}$. Inserting
the decomposition (\ref{def:Wigner function decomposition}) into
Eq. (\ref{prop:Hermitian property of Wigner}) and using the property
(\ref{eq:Cliddord algebra}) of $\Gamma_{i}$, one can prove that
all the coefficients in Eq. (\ref{def:Wigner function decomposition})
are real functions. %
{} The expansion coefficients can be derived from the Wigner function
by multiplying the corresponding generators and then taking the trace,
\begin{eqnarray}
\mathcal{F} & = & \mathrm{Tr}(W),\nonumber \\
\mathcal{P} & = & -\mathrm{Tr}(i\gamma^{5}W),\nonumber \\
\mathcal{V}^{\mu} & = & \mathrm{Tr}(\gamma^{\mu}W),\nonumber \\
\mathcal{A}^{\mu} & = & \mathrm{Tr}(\gamma^{\mu}\gamma^{5}W),\nonumber \\
\mathcal{S}^{\mu\nu} & = & \mathrm{Tr}(\sigma^{\mu\nu}W).\label{eq:reproduce components of Wigner funtion}
\end{eqnarray}
The tensor component is anti-symmetric and has 6 independent members.
We can equivalently introduce two vector functions
\begin{equation}
\mathcal{\boldsymbol{T}}=\frac{1}{2}\mathbf{e}_{i}\left(\mathcal{S}^{i0}-\mathcal{S}^{0i}\right),\ \ \mathcal{\boldsymbol{S}}=\frac{1}{2}\epsilon_{ijk}\mathbf{e}_{i}\mathcal{S}_{jk}.
\end{equation}
By Lorentz and parity transformations, $\mathcal{F}$, $\mathcal{P}$,
$\mathcal{V}^{\mu}$, $\mathcal{A}^{\mu}$, $\mathcal{S}^{\mu\nu}$
are the scalar, pseudo-scalar, vector, axial-vector, and tensor, respectively.
The properties under charge conjugation, parity and time reversal
are shown in Tab. \ref{tab:Transformation-properties}. It has been
shown in Ref. \cite{Vasak:1987um} that some components in Eq. (\ref{def:Wigner function decomposition})
have obvious physical meaning. For example, the vector component is
the fermion number current density and the axial-vector component
is the polarization density. The physical meanings are listed in Tab.
\ref{tab:Physical meanings}. We will have a more detailed discussion
in Sec. (\ref{sec:Physical quantities}).

\begin{table}
\begin{tabular}{|c|c|c|c|c|c|}
\hline
 & $\mathcal{F}(t,\mathbf{x})$ & $\mathcal{P}(t,\mathbf{x})$ & $\mathcal{V}_{\mu}(t,\mathbf{x})$ & $\mathcal{A}_{\mu}(t,\mathbf{x})$ & $\mathcal{S}_{\mu\nu}(t,\mathbf{x})$\tabularnewline
\hline
\hline
C & $\mathcal{F}(t,\mathbf{x})$ & $\mathcal{P}(t,\mathbf{x})$ & $-\mathcal{V}_{\mu}(t,\mathbf{x})$ & $\mathcal{A}_{\mu}(t,\mathbf{x})$ & $-\mathcal{S}_{\mu\nu}(t,\mathbf{x})$\tabularnewline
\hline
P & $\mathcal{F}(t,-\mathbf{x})$ & $-\mathcal{P}(t,-\mathbf{x})$ & $\mathcal{V}_{\mu}(t,-\mathbf{x})$ & $-\mathcal{A}_{\mu}(t,-\mathbf{x})$ & $\mathcal{S}_{\mu\nu}(t,-\mathbf{x})$\tabularnewline
\hline
T & $\mathcal{F}(-t,\mathbf{x})$ & $-\mathcal{P}(-t,\mathbf{x})$ & $\mathcal{V}_{\mu}(-t,\mathbf{x})$ & $\mathcal{A}_{\mu}(-t,\mathbf{x})$ & $-\mathcal{S}_{\mu\nu}(-t,\mathbf{x})$\tabularnewline
\hline
CPT & $\mathcal{F}(-t,-\mathbf{x})$ & $\mathcal{P}(-t,-\mathbf{x})$ & $-\mathcal{V}_{\mu}(-t,-\mathbf{x})$ & $-\mathcal{A}_{\mu}(-t,-\mathbf{x})$ & $\mathcal{S}_{\mu\nu}(-t,-\mathbf{x})$\tabularnewline
\hline
\end{tabular}

\caption{Transformation properties of the components of the Wigner function
under charge conjugation (C), parity (P), and time reversal (T). The
dependence on the momentum $p^{\mu}$ is suppressed here.}

\label{tab:Transformation-properties}
\end{table}

\begin{table}
\begin{tabular}{|c|c|}
\hline
Component & Physical meaning (distribution in phase space)\tabularnewline
\hline
\hline
$\mathcal{F}$ & Mass\tabularnewline
\hline
$\mathcal{P}$ & Pesudoscalar condensate\tabularnewline
\hline
$\mathcal{V}^{\mu}$ & Net fermion current\tabularnewline
\hline
$\mathcal{A}^{\mu}$ & Polarization (or spin current, or axial-charge current)\tabularnewline
\hline
$\mathcal{\boldsymbol{T}}$ & Electric dipole-moment\tabularnewline
\hline
$\mathcal{\boldsymbol{S}}$ & Magnetic dipole-moment\tabularnewline
\hline
\end{tabular}

\caption{Physical meaning of the components of the Wigner function.}

\label{tab:Physical meanings}
\end{table}

\subsection{Equations for the Wigner function }

In this subsection we will derive kinetic equations for the Wigner
function. The Wigner function is defined in Eq. (\ref{def:Wigner function})
for spin-1/2 fermions, whose kinetic equation will be derived from
the Dirac equation and its conjugate, %
\begin{eqnarray}
[i\gamma^{\mu}(\overrightarrow{\partial}_{x\mu}+i\mathbb{A}_{\mu})-m\mathbb{I}_{4}]\psi & = & 0,\nonumber \\
\bar{\psi}[i\gamma^{\mu}(\overleftarrow{\partial}{}_{x\mu}-i\mathbb{A}_{\mu})+m\mathbb{I}_{4}] & = & 0.\label{eq:Dirac equation and conjugate}
\end{eqnarray}
Note that we have adopted the Hartree approximation, in which the
electromagnetic field is assumed to be a classical field instead of
a quantum one. In these equations, $\psi$ and $\bar{\psi}$ represent
either the field or the field operators after second quantization.

\subsubsection{Dirac form}

In order to derive a Dirac form kinetic equation with the first order
in the time derivative, we first act with $i\gamma^{\sigma}\partial_{x\sigma}$
on Eq. (\ref{def:Wigner function}),
\begin{eqnarray}
 &  & i\gamma^{\sigma}\partial_{x\sigma}W(x,p)\nonumber \\
 & = & i\gamma^{\sigma}\int\frac{d^{4}y}{(2\pi)^{4}}\exp\left(-iy^{\mu}p_{\mu}\right)\left[\partial_{x\sigma}U\left(x+\frac{y}{2},\,x-\frac{y}{2}\right)\right]\left\langle \Omega\left|\hat{\bar{\psi}}\left(x+\frac{y}{2}\right)\otimes\hat{\psi}\left(x-\frac{y}{2}\right)\right|\Omega\right\rangle \nonumber \\
 &  & +\int\frac{d^{4}y}{(2\pi)^{4}}\exp\left(-iy^{\mu}p_{\mu}\right)U\left(x+\frac{y}{2},\,x-\frac{y}{2}\right)\left\langle \Omega\left|\left[\partial_{x\sigma}\hat{\bar{\psi}}\left(x+\frac{y}{2}\right)\right]\otimes i\gamma^{\sigma}\hat{\psi}\left(x-\frac{y}{2}\right)\right|\Omega\right\rangle \nonumber \\
 &  & +\int\frac{d^{4}y}{(2\pi)^{4}}\exp\left(-iy^{\mu}p_{\mu}\right)U\left(x+\frac{y}{2},\,x-\frac{y}{2}\right)\left\langle \Omega\left|\hat{\bar{\psi}}\left(x+\frac{y}{2}\right)\otimes\left[i\gamma^{\sigma}\partial_{x\sigma}\hat{\psi}\left(x-\frac{y}{2}\right)\right]\right|\Omega\right\rangle .\nonumber \\
\label{eq:i=00005Cgamma=00005CpartialW}
\end{eqnarray}
In the second and third lines we have used the following property
of the tensor product to put the gamma matrix into the expectation
value,
\begin{equation}
A(B\otimes C)=B\otimes(AC).
\end{equation}
With this property, the operator $i\gamma^{\sigma}\partial_{x\sigma}$
in the last line of Eq. (\ref{eq:i=00005Cgamma=00005CpartialW}) directly
acts on $\psi\left(x-\frac{y}{2}\right)$ and thus the Dirac equation
can be used for further simplification. On the other hand, the conjugate
of the Dirac equation in Eq. (\ref{eq:Dirac equation and conjugate})
cannot be directly used to simplify the second line of Eq. (\ref{eq:i=00005Cgamma=00005CpartialW})
because $\gamma^{\sigma}$ does not come with $\partial_{x\sigma}\bar{\psi}\left(x+\frac{y}{2}\right)$.
However, since the field operator $\hat{\bar{\psi}}\left(x+\frac{y}{2}\right)$
depends on $x^{\mu}+\frac{1}{2}y^{\mu}$, we can replace the derivative
with respect to $x^{\mu}$ by that with respect to $y^{\mu}$,
\begin{equation}
\partial_{x\sigma}\hat{\bar{\psi}}\left(x+\frac{y}{2}\right)=2\partial_{y\sigma}\hat{\bar{\psi}}\left(x+\frac{y}{2}\right).
\end{equation}
Inserting this into Eq. (\ref{eq:i=00005Cgamma=00005CpartialW}) and
integrating by parts, we obtain %
\begin{eqnarray}
i\gamma^{\sigma}\partial_{x\sigma}W(x,p) & = & -2\gamma^{\sigma}p_{\sigma}W(x,p)\nonumber \\
 &  & +i\gamma^{\sigma}\int\frac{d^{4}y}{(2\pi)^{4}}\exp\left(-iy^{\mu}p_{\mu}\right)\left[\left(\partial_{x\sigma}-2\partial_{y\sigma}\right)U\left(x+\frac{y}{2},\,x-\frac{y}{2}\right)\right]\nonumber \\
 &  & \quad\times\left\langle \Omega\left|\hat{\bar{\psi}}\left(x+\frac{y}{2}\right)\otimes\hat{\psi}\left(x-\frac{y}{2}\right)\right|\Omega\right\rangle \nonumber \\
 &  & +2\int\frac{d^{4}y}{(2\pi)^{4}}\exp\left(-iy^{\mu}p_{\mu}\right)U\left(x+\frac{y}{2},\,x-\frac{y}{2}\right)\nonumber \\
 &  & \quad\times\left\langle \Omega\left|\hat{\bar{\psi}}\left(x+\frac{y}{2}\right)\otimes\left[i\gamma^{\sigma}\partial_{x\sigma}\hat{\psi}\left(x-\frac{y}{2}\right)\right]\right|\Omega\right\rangle .\label{eq:after integrating by parts}
\end{eqnarray}
Here we have dropped the boundary term which is assumed to vanish
if we take an infinitely large volume. Now the Dirac equation (\ref{eq:Dirac equation and conjugate})
can be used to further simplify the last term in Eq. (\ref{eq:after integrating by parts}),
\begin{eqnarray}
i\gamma^{\sigma}\partial_{x\sigma}W(x,p) & = & -2\gamma^{\sigma}p_{\sigma}W(x,p)+2m\mathbb{I}_{4}W(x,p)\nonumber \\
 &  & +\gamma^{\sigma}\int\frac{d^{4}y}{(2\pi)^{4}}\exp\left(-iy^{\mu}p_{\mu}\right)\left\langle \Omega\left|\hat{\bar{\psi}}\left(x+\frac{y}{2}\right)\otimes\hat{\psi}\left(x-\frac{y}{2}\right)\right|\Omega\right\rangle \nonumber \\
 &  & \quad\times\left[i\left(\partial_{x\sigma}-2\partial_{y\sigma}\right)+2\mathbb{A}_{\sigma}\left(x-\frac{y}{2}\right)\right]U\left(x+\frac{y}{2},\,x-\frac{y}{2}\right).\label{eq:kin equation 0}
\end{eqnarray}
Using the definition of the gauge link in Eq. (\ref{eq:gauge link}),
we can explicitly calculate the derivative of the gauge link,%
\begin{eqnarray}
 &  & \left[i\left(\partial_{x\sigma}-2\partial_{y\sigma}\right)+2\mathbb{A}_{\sigma}\left(x-\frac{y}{2}\right)\right]U\left(x+\frac{y}{2},\,x-\frac{y}{2}\right)\nonumber \\
 & = & U\left(x+\frac{y}{2},\,x-\frac{y}{2}\right)\left[y^{\mu}\int_{-1/2}^{1/2}ds(1-2s)F_{\sigma\mu}(x+sy)\right].\label{eq:derivative of gauge link}
\end{eqnarray}
Inserting Eq. (\ref{eq:derivative of gauge link}) into Eq. (\ref{eq:kin equation 0})
we can obtain
\begin{eqnarray}
i\gamma^{\sigma}\partial_{x\sigma}W(x,p) & = & -2(\gamma^{\sigma}p_{\sigma}-m\mathbb{I}_{4})W(x,p)\nonumber \\
 &  & +\gamma^{\sigma}\int\frac{d^{4}y}{(2\pi)^{4}}\exp\left(-iy^{\mu}p_{\mu}\right)y^{\rho}\int_{-1/2}^{1/2}ds(1-2s)F_{\sigma\rho}(x+sy)\nonumber \\
 &  & \quad\times U\left(x+\frac{y}{2},\,x-\frac{y}{2}\right)\left\langle \Omega\left|\hat{\bar{\psi}}\left(x+\frac{y}{2}\right)\otimes\hat{\psi}\left(x-\frac{y}{2}\right)\right|\Omega\right\rangle .
\end{eqnarray}
Due to the phase factor $\exp\left(-iy^{\mu}p_{\mu}\right)$, the
relative coordinate $y^{\rho}$ can be replaced by the momentum derivative
$i\partial_{p}^{\rho}$. Furthermore, the integral over the field
tensor can be calculated using a Taylor expansion
\begin{eqnarray}
\int_{-1/2}^{1/2}ds(1-2s)F_{\sigma\rho}(x+sy) & = & \sum_{n=0}^{\infty}\frac{1}{n!}(y^{\nu}\partial_{x\nu})^{n}F_{\sigma\rho}(x)\int_{-1/2}^{1/2}ds(1-2s)s^{n}\nonumber \\
 & = & \sum_{n=0}^{\infty}\frac{1+(-1)^{n}(3+2n)}{(n+2)!2^{n+1}}(y^{\nu}\partial_{x\nu})^{n}F_{\sigma\mu}(x)\nonumber \\
 & = & \sum_{n=0}^{\infty}\frac{1+(-1)^{n}(3+2n)}{(n+2)!2^{n+1}}(i\partial_{p}^{\nu}\partial_{x\nu})^{n}F_{\sigma\rho}(x).
\end{eqnarray}
We now separate the even and odd terms in the series expansion, so
the above formula can be written in a more concise form,%
\begin{eqnarray}
 &  & \sum_{n=0}^{\infty}\frac{1+(-1)^{n}(3+2n)}{(n+2)!2^{n+1}}(i\partial_{p}^{\nu}\partial_{x\nu})^{n}\nonumber \\
 &  & \qquad\qquad=\sum_{n=0}^{\infty}\frac{(-1)^{n}}{(2n+1)!2^{2n}}(\partial_{p}^{\nu}\partial_{x\nu})^{2n}-i\sum_{n=0}^{\infty}\frac{(-1)^{n}}{(2n+3)(2n+1)!2^{2n+1}}(\partial_{p}^{\nu}\partial_{x\nu})^{2n+1}\nonumber \\
 &  & \qquad\qquad=j_{0}\left(\frac{1}{2}\partial_{p}^{\nu}\partial_{x\nu}\right)-i\,j_{1}\left(\frac{1}{2}\partial_{p}^{\nu}\partial_{x\nu}\right),
\end{eqnarray}
where the following spherical Bessel functions were used
\begin{equation}
j_{0}(x)=\frac{\sin x}{x},\ \ j_{1}(x)=\frac{\sin x-x\cos x}{x^{2}}.
\end{equation}
Thus we finally obtain the following kinetic equation
\begin{eqnarray}
i\gamma^{\sigma}\partial_{x\sigma}W(x,p) & = & -2(\gamma^{\sigma}p_{\sigma}-m\mathbb{I}_{4})W(x,p)\nonumber \\
 &  & +i\gamma^{\sigma}\left[j_{0}\left(\frac{1}{2}\partial_{p}^{\nu}\partial_{x\nu}\right)F_{\sigma\rho}(x)-i\,j_{1}\left(\frac{1}{2}\partial_{p}^{\nu}\partial_{x\nu}\right)F_{\sigma\rho}(x)\right]\partial_{p}^{\rho}W(x,p).
\end{eqnarray}
Defining the following operators
\begin{eqnarray}
K^{\mu} & \equiv & \Pi^{\mu}+\frac{i}{2}\nabla^{\mu},\nonumber \\
\Pi^{\mu} & \equiv & p^{\mu}-\frac{1}{2}j_{1}(\Delta)F^{\mu\nu}(x)\partial_{p\nu},\nonumber \\
\nabla^{\mu} & \equiv & \partial_{x}^{\mu}-j_{0}(\Delta)F^{\mu\nu}(x)\partial_{p\nu},\label{def:operators K^=00005Cmu}
\end{eqnarray}
with $\Delta\equiv\frac{1}{2}\partial_{p}^{\nu}\partial_{x\nu}$,
the kinetic equation for the Wigner function can be written in a compact
form,
\begin{equation}
(\gamma^{\mu}K_{\mu}-m\mathbb{I}_{4})W(x,p)=0,\label{eq:Dirac equation for Wigner}
\end{equation}
We note that the derivative $\partial_{x\nu}$ in the operator $\Delta$
only acts on $F^{\mu\nu}(x)$ but not on the Wigner function. The
operators $\Pi^{\mu}$ and $\nabla^{\mu}$ are generalized 4-momentum
and spatial-derivative operators, respectively, which can be reduced
to the ordinary ones without the electromagnetic field. In the limit
of vanishing electromagnetic field, the equation for the Wigner function
in Eq. (\ref{eq:Dirac equation for Wigner}) takes the same form as
the Dirac equation. Note that Eq. (\ref{eq:Dirac equation for Wigner})
is first order in space-time derivatives. Thus, in the remainder part
of the thesis, we call Eq. (\ref{eq:Dirac equation for Wigner}) the
Dirac-form kinetic equation for the Wigner function.

Note that in Sec. \ref{sec:Semi-classical-expansion} we will expand
the Wigner function in powers of the Planck's constant. To this end,
we will show $\hbar$ explicitly. Recalling the discussions about
natural units in Sec. \ref{subsec:System-of-Units,}, the product
of $p^{\mu}$ and $x_{\mu}$ has the unit of $\hbar$. The field strength
$F^{\mu\nu}$ has the unit $[kg\cdot m\cdot s^{-2}]$ in the SI units,
thus $F^{\mu\nu}(x)\partial_{p\nu}$ has the unit $[s^{-1}]$. In
order to make sure $K^{\mu}$ has the unit of momentum and $\Delta$
is unit-less, we recover $\hbar$ as follows,
\begin{eqnarray}
K^{\mu} & \equiv & \Pi^{\mu}+\frac{i\hbar}{2}\nabla^{\mu},\nonumber \\
\Pi^{\mu} & \equiv & p^{\mu}-\frac{\hbar}{2}j_{1}(\Delta)F^{\mu\nu}(x)\partial_{p\nu},\nonumber \\
\nabla^{\mu} & \equiv & \partial_{x}^{\mu}-j_{0}(\Delta)F^{\mu\nu}(x)\partial_{p\nu},\label{def:operators Kmu with hbar}
\end{eqnarray}
with the operator $\Delta\equiv\frac{\hbar}{2}\partial_{p}^{\nu}\partial_{x\nu}$.

\subsubsection{Klein-Gordon form}

In the previous part of this subsection, we have derived the Dirac-form
kinetic equation (\ref{eq:Dirac equation for Wigner}), which is of
first order in space-time derivatives. A second-order equation can
be obtained by multiplying the Dirac-form equation (\ref{eq:Dirac equation for Wigner})
with $\gamma^{\mu}K_{\mu}+m$ and using the following relation,
\begin{equation}
\gamma^{\mu}\gamma^{\nu}=\frac{1}{2}\left\{ \gamma^{\mu},\gamma^{\nu}\right\} +\frac{1}{2}\left[\gamma^{\mu},\gamma^{\nu}\right]=g^{\mu\nu}-i\sigma^{\mu\nu}.
\end{equation}
Then the new kinetic equation, which is called the Klein-Gordon-form
kinetic equation, reads, %
\begin{equation}
\left(K^{\mu}K_{\mu}-\frac{i}{2}\sigma^{\mu\nu}\left[K_{\mu},K_{\nu}\right]-m^{2}\right)W(x,p)=0,\label{eq:K-G equation for Wigner}
\end{equation}
where $K^{\mu}$ is defined in Eq. (\ref{def:operators K^=00005Cmu}).
Since both the Wigner function and the operators are complex matrices,
taking the Hermitian conjugate of Eq. (\ref{eq:K-G equation for Wigner})
and using the property (\ref{prop:Hermitian property of Wigner})
of the Wigner function, we obtain the conjugate equation %
\begin{equation}
\left[\left(K^{\mu}K_{\mu}\right)^{\ast}-m^{2}\right]W(x,p)+\frac{i}{2}\left[K_{\mu},K_{\nu}\right]^{\ast}W(x,p)\sigma^{\mu\nu}=0.\label{eq:Conjugate K-G equation}
\end{equation}
Since $K^{\mu}$ is complex, one can separate the real and imaginary
parts of $K^{\mu}K_{\mu}$ and $\left[K_{\mu},K_{\nu}\right]$ as
$K^{\mu}K_{\mu}=\Re K^{2}+i\Im K^{2}$ and $[K_{\mu},K_{\nu}]=\Re K_{\mu\nu}+i\Im K_{\mu\nu}$.
We will give the explicit expressions for these operators later. The
Klein-Gordon-form equation (\ref{eq:K-G equation for Wigner}) and
its conjugate (\ref{eq:Conjugate K-G equation}) now become
\begin{eqnarray}
\left(\Re K^{2}-m^{2}\right)W+i\Im K^{2}W-\frac{i}{2}\Re K_{\mu\nu}\sigma^{\mu\nu}W+\frac{1}{2}\Im K_{\mu\nu}\sigma^{\mu\nu}W & = & 0,\nonumber \\
\left(\Re K^{2}-m^{2}\right)W-i\Im K^{2}W+\frac{i}{2}\Re K_{\mu\nu}W\sigma^{\mu\nu}+\frac{1}{2}\Im K_{\mu\nu}W\sigma^{\mu\nu} & = & 0.
\end{eqnarray}
Note that the above two equations should be satisfied simultaneously.
Thus we can form linear combinations by taking the sum and the difference,
\begin{eqnarray}
\left(\Re K^{2}-m^{2}\right)W-\frac{i}{4}\Re K_{\mu\nu}\left[\sigma^{\mu\nu},W\right]+\frac{1}{4}\Im K_{\mu\nu}\left\{ \sigma^{\mu\nu},W\right\}  & = & 0,\nonumber \\
\Im K^{2}W-\frac{1}{4}\Re K_{\mu\nu}\left\{ \sigma^{\mu\nu},W\right\} -\frac{i}{4}\Im K_{\mu\nu}\left[\sigma^{\mu\nu},W\right] & = & 0.\label{eq:on-shell and Vlasov equations}
\end{eqnarray}
The first equation obviously depends on the mass while the second
one does not. These equations are the generalized on-shell condition
and the Vlasov equation, respectively.

The operator $K^{\mu}$ is the linear combination of the generalized
momentum operator $\Pi^{\mu}$ and the generalized space-time derivative
operator $\nabla^{\mu}$, as defined in Eq. (\ref{def:operators Kmu with hbar}).
Using the operators $\Pi^{\mu}$ and $\nabla^{\mu}$, we obtain %
\begin{eqnarray}
K^{\mu}K_{\mu} & = & \Pi^{\mu}\Pi_{\mu}-\frac{\hbar^{2}}{4}\nabla^{\mu}\nabla_{\mu}+\frac{i\hbar}{2}\left\{ \nabla^{\mu},\Pi_{\mu}\right\} ,\nonumber \\
\left[K_{\mu},K_{\nu}\right] & = & -\frac{\hbar^{2}}{4}\left[\nabla_{\mu},\nabla_{\nu}\right]+\left[\Pi_{\mu},\Pi_{\nu}\right]+\frac{i\hbar}{2}\left(\left[\Pi_{\mu},\nabla_{\nu}\right]-\left[\Pi_{\nu},\nabla_{\mu}\right]\right).
\end{eqnarray}
Since both $\Pi^{\mu}$ and $\nabla^{\mu}$ are real-defined operators,
one can read off the real and imaginary parts
\begin{eqnarray}
\Re K^{2} & \equiv & \Pi^{\mu}\Pi_{\mu}-\frac{\hbar^{2}}{4}\nabla^{\mu}\nabla_{\mu},\nonumber \\
\Im K^{2} & \equiv & \frac{\hbar}{2}\left\{ \nabla^{\mu},\Pi_{\mu}\right\} ,\nonumber \\
\Re K_{\mu\nu} & \equiv & -\frac{\hbar^{2}}{4}\left[\nabla_{\mu},\nabla_{\nu}\right]+\left[\Pi_{\mu},\Pi_{\nu}\right],\nonumber \\
\Im K_{\mu\nu} & \equiv & \frac{\hbar}{2}\left(\left[\Pi_{\mu},\nabla_{\nu}\right]-\left[\Pi_{\nu},\nabla_{\mu}\right]\right).\label{eq:real and imaginary parts of second order operators}
\end{eqnarray}
More detailed calculations give %
\begin{eqnarray}
\Re K^{2} & = & p_{\mu}p^{\mu}-\frac{\hbar^{2}}{4}\partial_{x\mu}\partial_{x}^{\mu}-\hbar p^{\mu}\left[j_{1}(\triangle)F_{\mu\nu}\right]\partial_{p}^{\nu}\nonumber \\
 &  & +\frac{\hbar^{2}}{2}\left[j_{0}(\triangle)F_{\mu\nu}\right]\partial_{p}^{\nu}\left\{ \partial_{x}^{\mu}-\frac{1}{2}\left[j_{0}(\triangle)F^{\mu\alpha}\right]\partial_{p,\alpha}\right\} \nonumber \\
 &  & -\frac{\hbar^{2}}{4}\left\{ \left[j_{1}^{\prime}(\triangle)-j_{0}(\triangle)\right]\partial_{x}^{\mu}F_{\mu\nu}\right\} \partial_{p}^{\nu}\nonumber \\
 &  & +\frac{\hbar^{2}}{4}[j_{1}(\triangle)F_{\mu\nu}][j_{1}(\triangle)F^{\mu\alpha}]\partial_{p}^{\nu}\partial_{p,\alpha},\nonumber \\
\Im K^{2} & = & \hbar p_{\mu}\left\{ \partial_{x}^{\mu}-\left[j_{0}(\triangle)F^{\mu\alpha}\right]\partial_{p,\alpha}\right\} \nonumber \\
 &  & -\frac{\hbar^{2}}{2}\left\{ \partial_{x}^{\mu}-\left[j_{0}(\triangle)F^{\mu\alpha}\right]\partial_{p,\alpha}\right\} \left[j_{1}(\triangle)F_{\mu\nu}\right]\partial_{p}^{\nu},\nonumber \\
\Re K_{\mu\nu} & = & -\hbar\Delta j_{0}(\Delta)F_{\mu\nu}(x),\nonumber \\
\Im K_{\mu\nu} & = & -\hbar j_{0}(\triangle)F_{\mu\nu}+\hbar\triangle j_{1}(\triangle)F_{\mu\nu}.
\end{eqnarray}
These expressions seem to be complicated but if we truncate $\mathcal{O}(\hbar^{2})$
and higher order terms, these operators become
\begin{eqnarray}
\Re K^{2} & = & p^{2}+\mathcal{O}(\hbar^{2}),\nonumber \\
\Im K^{2} & = & \hbar p_{\mu}\left(\partial_{x}^{\mu}-F^{\mu\alpha}\partial_{p,\alpha}\right)+\mathcal{O}(\hbar^{2}),\nonumber \\
\Re K_{\mu\nu} & = & \mathcal{O}(\hbar^{2}),\nonumber \\
\Im K_{\mu\nu} & = & -\hbar F_{\mu\nu}+\mathcal{O}(\hbar^{2}),
\end{eqnarray}
which are quite concise and will be useful in semi-classical expansion.
Inserting the truncated operators into Eq. (\ref{eq:on-shell and Vlasov equations}),
we obtain
\begin{eqnarray}
\left(p^{2}-m^{2}\right)W-\frac{\hbar}{4}F_{\mu\nu}\left\{ \sigma^{\mu\nu},W\right\}  & = & 0,\nonumber \\
p_{\mu}\left(\partial_{x}^{\mu}-F^{\mu\alpha}\partial_{p,\alpha}\right)W+\frac{i}{4}F_{\mu\nu}\left[\sigma^{\mu\nu},W\right] & = & 0.\label{eq:on-shell and Vlasov equations at order =00005Chbar}
\end{eqnarray}
The first equation coincides with the on-shell condition, because
if the electromagnetic field vanishes, the non-trivial solution of
$W(x,p)$ should ensure $p^{2}=m^{2}$. Here the term $-\frac{\hbar}{4}F_{\mu\nu}\left\{ \sigma^{\mu\nu},W\right\} $
in the first equation plays the role of a coupling between the electromagnetic
field and the dipole-moment. The second equation is the Vlasov equation.
Note that the first equation does not contain any information on the
dynamical evolution. Up to order $\hbar$, the evolution of the Wigner
function is determined by the second equation (\ref{eq:on-shell and Vlasov equations at order =00005Chbar})
while the first equation just provides a constraint.

\subsection{Component equations}

In the previous subsection we have derived the Dirac form and Klein-Gordon
form of the kinetic equation. In this subsection, we decompose the
Wigner function into the scalar, pseudoscalar, vector, axial-vector,
and tensor parts, as shown in Eq. (\ref{def:Wigner function decomposition})
and derive the equations for all 16 independent components. Inserting
the decomposition (\ref{def:Wigner function decomposition}) into
the Dirac-form equation (\ref{eq:Dirac equation for Wigner}) and
extracting the coefficients of different matrices $\Gamma_{i}$, we
find following complex-valued equations
\begin{eqnarray}
K^{\mu}\mathcal{V}_{\mu}-m\mathcal{F} & = & 0,\nonumber \\
K^{\mu}\mathcal{A}_{\mu}+im\mathcal{P} & = & 0,\nonumber \\
K_{\mu}\mathcal{F}+iK^{\nu}\mathcal{S}_{\nu\mu}-m\mathcal{V}_{\mu} & = & 0,\nonumber \\
iK_{\mu}\mathcal{P}+\frac{1}{2}\epsilon_{\mu\nu\alpha\beta}K^{\nu}\mathcal{S}^{\alpha\beta}+m\mathcal{A}_{\mu} & = & 0,\nonumber \\
-iK_{[\mu}\mathcal{V}_{\nu]}-\epsilon_{\mu\nu\alpha\beta}K^{\alpha}\mathcal{A}^{\beta}-m\mathcal{S}_{\mu\nu} & = & 0,\label{eq:kinetic equations for components}
\end{eqnarray}
where $A_{[\mu}B_{\nu]}\equiv A_{\mu}B_{\nu}-A_{\nu}B_{\mu}$. Since
all components $\left\{ \mathcal{F},\mathcal{P},\mathcal{V}^{\mu},\mathcal{A}^{\mu},\mathcal{S}^{\mu\nu}\right\} $
are real functions and the operator $K^{\mu}$ is given by Eq. (\ref{def:operators Kmu with hbar}),
the real and imaginary parts of the above equations can be easily
separated and the real parts read
\begin{eqnarray}
\Pi^{\mu}\mathcal{V}_{\mu}-m\mathcal{F} & = & 0,\nonumber \\
\frac{\hbar}{2}\nabla^{\mu}\mathcal{A}_{\mu}+m\mathcal{P} & = & 0,\nonumber \\
\Pi_{\mu}\mathcal{F}-\frac{\hbar}{2}\nabla^{\nu}\mathcal{S}_{\nu\mu}-m\mathcal{V}_{\mu} & = & 0,\nonumber \\
-\frac{\hbar}{2}\nabla_{\mu}\mathcal{P}+\frac{1}{2}\epsilon_{\mu\nu\alpha\beta}\Pi^{\nu}\mathcal{S}^{\alpha\beta}+m\mathcal{A}_{\mu} & = & 0,\nonumber \\
\frac{\hbar}{2}\nabla_{[\mu}\mathcal{V}_{\nu]}-\epsilon_{\mu\nu\alpha\beta}\Pi^{\alpha}\mathcal{A}^{\beta}-m\mathcal{S}_{\mu\nu} & = & 0,\label{eq:real parts of kinetic equation}
\end{eqnarray}
while the imaginary parts are
\begin{eqnarray}
\hbar\nabla^{\mu}\mathcal{V}_{\mu} & = & 0,\nonumber \\
\Pi^{\mu}\mathcal{A}_{\mu} & = & 0,\nonumber \\
\frac{\hbar}{2}\nabla_{\mu}\mathcal{F}+\Pi^{\nu}\mathcal{S}_{\nu\mu} & = & 0,\nonumber \\
\Pi_{\mu}\mathcal{P}+\frac{\hbar}{4}\epsilon_{\mu\nu\alpha\beta}\nabla^{\nu}\mathcal{S}^{\alpha\beta} & = & 0,\nonumber \\
\Pi_{[\mu}\mathcal{V}_{\nu]}+\frac{\hbar}{2}\epsilon_{\mu\nu\alpha\beta}\nabla^{\alpha}\mathcal{A}^{\beta} & = & 0.\label{eq:imaginary parts of kinetic equation}
\end{eqnarray}
Note that the real parts of these equations explicitly depend on the
particle mass while the imaginary parts do not. Equations (\ref{eq:real parts of kinetic equation})
and (\ref{eq:imaginary parts of kinetic equation}) contains $32$
component equations in total, but they can be simplified in massless
case. If the mass is zero, then the terms proportional to the mass
in Eq. (\ref{eq:real parts of kinetic equation}) vanish, while the
imaginary parts (\ref{eq:imaginary parts of kinetic equation}) do
not change. Then the vector and axial-vector components decouple from
the other components, the corresponding equations read,
\begin{eqnarray}
\hbar\nabla^{\mu}\mathcal{V}_{\mu}=0, &  & \hbar\nabla^{\mu}\mathcal{A}_{\mu}=0,\nonumber \\
\Pi^{\mu}\mathcal{V}_{\mu}=0, &  & \Pi^{\mu}\mathcal{A}_{\mu}=0,\nonumber \\
\Pi_{[\mu}\mathcal{V}_{\nu]}+\frac{\hbar}{2}\epsilon_{\mu\nu\alpha\beta}\nabla^{\alpha}\mathcal{A}^{\beta}=0, &  & \Pi_{[\mu}\mathcal{A}_{\nu]}+\frac{\hbar}{2}\epsilon_{\mu\nu\alpha\beta}\nabla^{\alpha}\mathcal{V}^{\beta}=0.\label{eq:massless equations}
\end{eqnarray}
We observe that these equations are symmetric with respect to $\mathcal{V}^{\mu}\rightleftarrows\mathcal{A}^{\mu}$.
This can be understood from another point of view: in the massless
limit the chiral symmetry is restored, thus the net fermion number
current $\mathcal{\ensuremath{V}^{\mu}}$ and the axial current $\mathcal{A}^{\mu}$
are related by chiral symmetry. The remaining equations are for the
scalar, pseudoscalar and tensor parts
\begin{eqnarray}
\Pi_{\mu}\mathcal{F}-\frac{\hbar}{2}\nabla^{\nu}\mathcal{S}_{\nu\mu}=0, &  & \frac{\hbar}{2}\nabla_{\mu}\mathcal{F}+\Pi^{\nu}\mathcal{S}_{\nu\mu}=0,\nonumber \\
\Pi_{\mu}\mathcal{P}+\frac{\hbar}{4}\epsilon_{\mu\nu\alpha\beta}\nabla^{\nu}\mathcal{S}^{\alpha\beta}=0, &  & -\frac{\hbar}{2}\nabla_{\mu}\mathcal{P}+\frac{1}{2}\epsilon_{\mu\nu\alpha\beta}\Pi^{\nu}\mathcal{S}^{\alpha\beta}=0.
\end{eqnarray}

On the other hand, we can derive the on-shell conditions and Vlasov
equations from Eq. (\ref{eq:on-shell and Vlasov equations}). Combining
commutators and anti-commutators between gamma matrices and $\sigma^{\mu\nu}$,
which is listed in Eq. (\ref{eq:Commutators and anticommutators}),
with the decomposition (\ref{def:Wigner function decomposition}),
we have %
\begin{eqnarray}
[\sigma^{\mu\nu},W(x,p)] & = & \frac{i}{2}\left\{ \gamma^{[\mu}\mathcal{V}^{\nu]}+\gamma^{5}\gamma^{[\mu}\mathcal{A}^{\nu]}+\frac{1}{2}\left(g^{\rho[\mu}\sigma^{\nu]\sigma}-g^{\sigma[\mu}\sigma^{\nu]\rho}\right)\mathcal{S}_{\sigma\rho}\right\} ,\nonumber \\
\left\{ \sigma^{\mu\nu},W(x,p)\right\}  & = & \frac{1}{2}\biggl\{\sigma^{\mu\nu}\mathcal{F}-\frac{1}{2}\epsilon^{\mu\nu\alpha\beta}\sigma_{\alpha\beta}\mathcal{P}+\epsilon^{\mu\nu\alpha\beta}\gamma^{5}\gamma_{\beta}\mathcal{V}_{\alpha}+\epsilon^{\mu\nu\alpha\beta}\gamma_{\beta}\mathcal{A}_{\alpha}\nonumber \\
 &  & \quad+\frac{1}{2}\left(g^{\mu[\sigma}g^{\rho]\nu}+i\epsilon^{\mu\nu\sigma\rho}\gamma^{5}\right)\mathcal{S}_{\sigma\rho}\biggr\}.
\end{eqnarray}
Inserting this into Eq. (\ref{eq:on-shell and Vlasov equations})
and separating different coefficients of the matrices, we obtain the
on-shell conditions%
\begin{eqnarray}
\left(\Re K^{2}-m^{2}\right)\mathcal{F}+\frac{1}{2}\Im K_{\mu\nu}\mathcal{S}^{\mu\nu} & = & 0,\nonumber \\
\left(\Re K^{2}-m^{2}\right)\mathcal{P}+\frac{1}{4}\epsilon_{\mu\nu\alpha\beta}\Im K^{\mu\nu}\mathcal{S}^{\alpha\beta} & = & 0,\nonumber \\
\left(\Re K^{2}-m^{2}\right)\mathcal{V}_{\mu}+\Re K_{\mu\nu}\mathcal{V}^{\nu}-\frac{1}{2}\epsilon_{\mu\nu\alpha\beta}\Im K^{\alpha\beta}\mathcal{A}^{\nu} & = & 0,\nonumber \\
\left(\Re K^{2}-m^{2}\right)\mathcal{A}_{\mu}+\Re K_{\mu\nu}\mathcal{A}^{\nu}-\frac{1}{2}\epsilon_{\mu\nu\alpha\beta}\Im K^{\alpha\beta}\mathcal{V}^{\nu} & = & 0,\nonumber \\
\left(\Re K^{2}-m^{2}\right)\mathcal{S}_{\mu\nu}+\Re K_{\ [\mu}^{\alpha}\mathcal{S}_{\nu]\alpha}+\Im K_{\mu\nu}\mathcal{F}-\frac{1}{2}\epsilon_{\mu\nu\alpha\beta}\Im K^{\alpha\beta}\mathcal{P} & = & 0,\label{eq:decomposed on-shell}
\end{eqnarray}
and the Vlasov equations
\begin{eqnarray}
\Im K^{2}\mathcal{F}-\frac{1}{2}\Re K_{\mu\nu}\mathcal{S}^{\mu\nu} & = & 0,\nonumber \\
\Im K^{2}\mathcal{P}-\frac{1}{4}\epsilon_{\mu\nu\alpha\beta}\Re K^{\mu\nu}\mathcal{S}^{\alpha\beta} & = & 0,\nonumber \\
\Im K^{2}\mathcal{V}_{\mu}+\Im K_{\mu\nu}\mathcal{V}^{\nu}+\frac{1}{2}\epsilon_{\mu\nu\alpha\beta}\Re K^{\alpha\beta}\mathcal{A}^{\nu} & = & 0,\nonumber \\
\Im K^{2}\mathcal{A}_{\mu}+\Im K_{\mu\nu}\mathcal{A}^{\nu}+\frac{1}{2}\epsilon_{\mu\nu\alpha\beta}\Re K^{\alpha\beta}\mathcal{V}^{\nu} & = & 0,\nonumber \\
\Im K^{2}\mathcal{S}_{\mu\nu}+\Im K_{\ [\mu}^{\alpha}\mathcal{S}_{\nu]\alpha}-\Re K_{\mu\nu}\mathcal{F}+\frac{1}{2}\epsilon_{\mu\nu\alpha\beta}\Re K^{\alpha\beta}\mathcal{P} & = & 0.\label{eq:decomposed Vlasov}
\end{eqnarray}
Here the operators $\Re K^{2}$, $\Im K^{2}$, $\Re K_{\mu\nu}$,
$\Im K_{\mu\nu}$ are given in Eq. (\ref{eq:real and imaginary parts of second order operators}).
We observe that in these equations, the vector and axial-vector components
$\mathcal{V}_{\mu}$, $\mathcal{A}_{\mu}$ decouple from all the other
components $\mathcal{F}$, $\mathcal{P}$, $\mathcal{S}_{\mu\nu}$,
in both the massive and the massless cases. We will show in the next
subsection that, the scalar, pseudoscalar, and tensor components $\mathcal{F}$,
$\mathcal{P}$, $\mathcal{S}_{\mu\nu}$ can be derived form the other
components $\mathcal{V}_{\mu}$, $\mathcal{A}_{\mu}$, or vice versa.
Thus the on-shell conditions and Vlasov equations for $\mathcal{F}$,
$\mathcal{P}$, $\mathcal{S}_{\mu\nu}$ in Eqs. (\ref{eq:decomposed on-shell}),
(\ref{eq:decomposed Vlasov}) can be obtained using the equations
for components $\mathcal{V}_{\mu}$, $\mathcal{A}_{\mu}$ (and vice
versa). This means that equations (\ref{eq:real parts of kinetic equation}),
(\ref{eq:imaginary parts of kinetic equation}), (\ref{eq:decomposed on-shell}),
and (\ref{eq:decomposed Vlasov}) are reducible: when solving for
the Wigner function, it is not necessary to check that all these equations
are fulfilled. More detailed arguments will be given in the next subsection.

Both the decomposed equations (\ref{eq:real parts of kinetic equation}),
(\ref{eq:imaginary parts of kinetic equation}) and the above on-shell
conditions (\ref{eq:decomposed on-shell}) and Vlasov equations (\ref{eq:decomposed Vlasov})
are derived from the Dirac-form equation for the Wigner function (\ref{eq:Dirac equation for Wigner}).
The difference is that Eqs. (\ref{eq:real parts of kinetic equation}),
(\ref{eq:imaginary parts of kinetic equation}) are first-order equations
with respect to the operators $\Pi^{\mu}$, $\nabla^{\mu}$ while
Eqs. (\ref{eq:decomposed on-shell}), (\ref{eq:decomposed Vlasov})
are second-order ones. Analogous to the fact that the Klein-Gordon
equation for fermions can be derived from the Dirac equation, one
can reproduce Eqs. (\ref{eq:decomposed on-shell}), (\ref{eq:decomposed Vlasov})
from Eqs. (\ref{eq:real parts of kinetic equation}), (\ref{eq:imaginary parts of kinetic equation})
by multiplying with the appropriate  operators and then taking linear
combinations. Taking the on-shell and Vlasov equations for the scalar
component $\mathcal{F}$ as an example, from the third line of Eq.
(\ref{eq:real parts of kinetic equation}) we express $\mathcal{V}_{\mu}$
in terms of $\mathcal{F}$ and $\mathcal{S}_{\nu\mu}$
\begin{equation}
\mathcal{V}_{\mu}=\frac{1}{m}\Pi_{\mu}\mathcal{F}-\frac{\hbar}{2m}\nabla^{\nu}\mathcal{S}_{\nu\mu}.\label{eq:derive V from F and S}
\end{equation}
Multiplying the first line of Eq. (\ref{eq:real parts of kinetic equation})
by the mass $m$ and using the above relation, an on-shell condition
for the scalar component $\mathcal{F}$ is obtained,
\begin{equation}
(\Pi^{\mu}\Pi_{\mu}-m^{2})\mathcal{F}-\frac{\hbar}{2}\Pi^{\mu}\nabla^{\nu}\mathcal{S}_{\nu\mu}=0.
\end{equation}
Using the anti-symmetric property of $\mathcal{S}_{\mu\nu}$, the
second term can be simplified as
\begin{eqnarray}
-\frac{\hbar}{2}\Pi^{\mu}\nabla^{\nu}\mathcal{S}_{\nu\mu} & = & \frac{\hbar}{2}\left[\Pi^{\mu},\nabla^{\nu}\right]\mathcal{S}_{\mu\nu}+\frac{\hbar}{2}\nabla^{\nu}\Pi^{\mu}\mathcal{S}_{\mu\nu}\nonumber \\
 & = & \frac{\hbar}{4}\left(\left[\Pi^{\mu},\nabla^{\nu}\right]-\left[\Pi^{\nu},\nabla^{\mu}\right]\right)\mathcal{S}_{\mu\nu}+\frac{\hbar^{2}}{4}\nabla^{\nu}\nabla_{\nu}\mathcal{F},
\end{eqnarray}
where we have used the third line of Eq. (\ref{eq:imaginary parts of kinetic equation})
in the last step. Comparing with the operators listed in Eq. (\ref{eq:real and imaginary parts of second order operators})
we directly reproduce the first line in Eq. (\ref{eq:decomposed on-shell}).
On the other hand, inserting the relation (\ref{eq:derive V from F and S})
into the first line of Eq. (\ref{eq:imaginary parts of kinetic equation})
we obtain
\begin{equation}
\frac{\hbar}{2}\nabla^{\mu}\Pi_{\mu}\mathcal{F}-\frac{\hbar^{2}}{4}\nabla^{\mu}\nabla^{\nu}\mathcal{S}_{\nu\mu}=0.\label{eq:temp equation-1}
\end{equation}
Then multiplying the third line of Eq. (\ref{eq:imaginary parts of kinetic equation})
with operator $\Pi^{\mu}$ gives
\begin{equation}
\frac{\hbar}{2}\Pi^{\mu}\nabla_{\mu}\mathcal{F}+\Pi^{\mu}\Pi^{\nu}\mathcal{S}_{\nu\mu}=0.\label{eq:temp equation-2}
\end{equation}
Taking the sum of Eqs. (\ref{eq:temp equation-1}) and (\ref{eq:temp equation-2})
and comparing with the operators in Eq. (\ref{eq:real and imaginary parts of second order operators}),
we reproduce the Vlasov equation for the scalar component $\mathcal{F}$,
which is the first line of Eq. (\ref{eq:decomposed Vlasov}). Similar
procedures can be performed for all the other components of the Wigner
function, which proves that the on-shell and Vlasov equations in (\ref{eq:decomposed on-shell}),
(\ref{eq:decomposed Vlasov}) can be derived from Eqs. (\ref{eq:real parts of kinetic equation}),
(\ref{eq:imaginary parts of kinetic equation}) without any additional
assumptions.

\subsection{Redundancy of equations}

In this section we will prove that Eqs. (\ref{eq:real parts of kinetic equation}),
(\ref{eq:imaginary parts of kinetic equation}) are reducible: the
third and fourth lines of Eq. (\ref{eq:imaginary parts of kinetic equation})
can be derived from the others and $\mathcal{F}$, $\mathcal{P}$,
and $\mathcal{S}_{\mu\nu}$ can be expressed by $\mathcal{V}_{\mu}$
and $\mathcal{A}_{\mu}$, or vice versa. Thus the kinetic equations
can be further simplified. Two approaches are proposed for solving
the Wigner function: one approach is based on $\mathcal{V}_{\mu}$
and $\mathcal{A}_{\mu}$, and the other approach is based on $\mathcal{F}$,
$\mathcal{P}$, $\mathcal{S}_{\mu\nu}$. In Sec. \ref{sec:Semi-classical-expansion}
both of these two methods are used and obtain the same results.

Now we prove that Eqs. (\ref{eq:real parts of kinetic equation}),
(\ref{eq:imaginary parts of kinetic equation}) are not independent
from each other for the massive case. We make the following combination
using the first and last lines in Eq. (\ref{eq:real parts of kinetic equation})
and Eq. (\ref{eq:imaginary parts of kinetic equation})
\begin{eqnarray}
0 & = & \frac{\hbar}{2m}\nabla_{\mu}\left(\Pi^{\nu}\mathcal{V}_{\nu}-m\mathcal{F}\right)-\frac{1}{2m}\Pi_{\mu}\left(\hbar\nabla^{\nu}\mathcal{V}_{\nu}\right)\nonumber \\
 &  & -\frac{1}{m}\Pi^{\nu}\left(\frac{\hbar}{2}\nabla_{[\mu}\mathcal{V}_{\nu]}-\epsilon_{\mu\nu\alpha\beta}\Pi^{\alpha}\mathcal{A}^{\beta}-m\mathcal{S}_{\mu\nu}\right)\nonumber \\
 &  & +\frac{\hbar}{2m}\nabla^{\nu}\left(\Pi_{[\mu}\mathcal{V}_{\nu]}+\frac{\hbar}{2}\epsilon_{\mu\nu\alpha\beta}\nabla^{\alpha}\mathcal{A}^{\beta}\right).
\end{eqnarray}
The right-hand-side vanishes because all the terms inside parentheses
vanish. After some calculations we obtain
\begin{eqnarray}
\frac{\hbar}{2}\nabla_{\mu}\mathcal{F}+\Pi^{\nu}\mathcal{S}_{\nu\mu} & = & -\frac{\hbar}{2m}\left(\left[\Pi_{\mu},\nabla_{\nu}\right]+\left[\Pi_{\nu},\nabla_{\mu}\right]\right)\mathcal{V}^{\nu}+\frac{\hbar}{2m}\left[\Pi_{\nu},\nabla^{\nu}\right]\mathcal{V}_{\mu}\nonumber \\
 &  & +\frac{1}{2m}\epsilon_{\mu\nu\alpha\beta}\left(\left[\Pi^{\nu},\Pi^{\alpha}\right]+\frac{\hbar^{2}}{4}\left[\nabla^{\nu},\nabla^{\alpha}\right]\right)\mathcal{A}^{\beta}.\label{eq:reducible equation1}
\end{eqnarray}
Using Eq. (\ref{def:operators Kmu with hbar}) one can calculate the
commutators
\begin{eqnarray}
\left[\Pi_{\mu},\Pi_{\nu}\right] & = & -\hbar\left[j_{1}(\Delta)+\frac{1}{2}\Delta j_{1}^{\prime}(\Delta)\right]F_{\mu\nu},\nonumber \\
\left[\Pi_{\mu},\nabla_{\nu}\right] & = & \left[\Delta j_{1}(\Delta)-j_{0}(\Delta)\right]F_{\mu\nu},\nonumber \\
\hbar^{2}\left[\nabla_{\mu},\nabla_{\nu}\right] & = & 2\hbar\Delta j_{0}(\Delta)F_{\mu\nu},
\end{eqnarray}
where $j_{1}^{\prime}(x)\equiv\frac{d}{dx}j_{1}(x)$. Thus we can
find the following relations
\begin{eqnarray}
\left[\Pi_{\mu},\nabla_{\nu}\right]+\left[\Pi_{\nu},\nabla_{\mu}\right] & = & \left[\Delta j_{1}(\Delta)-j_{0}(\Delta)\right]\left(F_{\mu\nu}+F_{\nu\mu}\right)=0,\nonumber \\
\left[\Pi_{\nu},\nabla^{\nu}\right] & = & \left[\Delta j_{1}(\Delta)-j_{0}(\Delta)\right]F_{\nu}^{\ \nu}=0,\nonumber \\
\left[\Pi_{\mu},\Pi_{\nu}\right]+\frac{\hbar^{2}}{4}\left[\nabla_{\mu},\nabla_{\nu}\right] & = & \frac{\hbar}{2}\left[\Delta j_{0}(\Delta)-2j_{1}(\Delta)-\Delta j_{1}^{\prime}(\Delta)\right]F_{\mu\nu}=0,\label{eq:commutators of operators Pi and Nabla}
\end{eqnarray}
where we have used the anti-symmetry of $F_{\mu\nu}$ and the following
relation for spherical Bessel functions
\begin{equation}
xj_{0}(x)-2j_{1}(x)-xj_{1}^{\prime}(x)=0.
\end{equation}
Inserting the commutators in Eq. (\ref{eq:commutators of operators Pi and Nabla})
into Eq. (\ref{eq:reducible equation1}), we confirm that the right-hand
side vanishes and we obtain the third line of Eq. (\ref{eq:imaginary parts of kinetic equation}).
Analogously, we construct the following equation from the second and
the last lines of Eqs. (\ref{eq:real parts of kinetic equation}),
(\ref{eq:imaginary parts of kinetic equation}),
\begin{eqnarray}
0 & = & \frac{1}{m}\Pi_{\mu}\left(\frac{\hbar}{2}\nabla^{\nu}\mathcal{A}_{\nu}+m\mathcal{P}\right)-\frac{\hbar}{2m}\nabla_{\mu}\left(\Pi^{\nu}\mathcal{A}_{\nu}\right)\nonumber \\
 &  & -\frac{\hbar}{4m}\epsilon_{\mu\nu\alpha\beta}\nabla^{\nu}\left(\frac{\hbar}{2}\nabla^{[\alpha}\mathcal{V}^{\beta]}-\epsilon^{\alpha\beta\rho\sigma}\Pi_{\rho}\mathcal{A}_{\sigma}-m\mathcal{S}^{\alpha\beta}\right)\nonumber \\
 &  & -\frac{1}{2m}\epsilon_{\mu\nu\alpha\beta}\Pi^{\nu}\left(\Pi^{[\alpha}\mathcal{V}^{\beta]}+\frac{\hbar}{2}\epsilon^{\alpha\beta\rho\sigma}\nabla_{\rho}\mathcal{A}_{\sigma}\right).
\end{eqnarray}
After some calculations we obtain
\begin{eqnarray}
\Pi_{\mu}\mathcal{P}+\frac{\hbar}{4}\epsilon_{\mu\nu\alpha\beta}\nabla^{\nu}\mathcal{S}^{\alpha\beta} & = & -\frac{\hbar}{2m}\left(\left[\Pi_{\mu},\nabla_{\nu}\right]+\left[\Pi_{\nu},\nabla_{\mu}\right]\right)\mathcal{A}^{\nu}+\frac{\hbar}{2m}\left[\Pi_{\nu},\nabla^{\nu}\right]\mathcal{A}_{\mu}\nonumber \\
 &  & +\frac{1}{4m}\epsilon_{\mu\nu\alpha\beta}\left(\left[\Pi^{\nu},\Pi^{\alpha}\right]+\frac{\hbar^{2}}{4}\left[\nabla^{\nu},\nabla^{\alpha}\right]\right)\mathcal{V}^{\beta}.
\end{eqnarray}
where the right-hand side vanishes according to (\ref{eq:commutators of operators Pi and Nabla}).
In this way we recover the fourth line in Eq. (\ref{eq:imaginary parts of kinetic equation}).
Thus according to the above discussions, the third and fourth lines
in Eq. (\ref{eq:imaginary parts of kinetic equation}) can be obtained
from the other lines in Eqs. (\ref{eq:real parts of kinetic equation})
and (\ref{eq:imaginary parts of kinetic equation}).

Now we will construct a proper way for computing the Wigner function.
As discussed in the previous subsection, the dynamical evolution and
constraints of the Wigner function are determined by Eqs. (\ref{eq:real parts of kinetic equation})
and (\ref{eq:imaginary parts of kinetic equation}), or equivalently
by Eqs. (\ref{eq:decomposed on-shell}) and (\ref{eq:decomposed Vlasov}).
However, we note that according to the first, second, and last lines
of Eq. (\ref{eq:real parts of kinetic equation}), we can express
the scalar, pseudo-scalar, and tensor components in terms of $\mathcal{V}^{\mu}$,
$\mathcal{A}^{\mu}$,
\begin{eqnarray}
\mathcal{F} & = & \frac{1}{m}\Pi^{\mu}\mathcal{V}_{\mu},\nonumber \\
\mathcal{P} & = & -\frac{\hbar}{2m}\nabla^{\mu}\mathcal{A}_{\mu},\nonumber \\
\mathcal{S}_{\mu\nu} & = & \frac{\hbar}{2m}\nabla_{[\mu}\mathcal{V}_{\nu]}-\frac{1}{m}\epsilon_{\mu\nu\alpha\beta}\Pi^{\alpha}\mathcal{A}^{\beta},\label{eq:FPS from VA}
\end{eqnarray}
Substituting $\mathcal{F}$, $\mathcal{P}$, $\mathcal{S}_{\mu\nu}$
into the third and fourth lines of Eq. (\ref{eq:real parts of kinetic equation})
by Eq. (\ref{eq:FPS from VA}), one obtains
\begin{eqnarray}
\left(\Re K^{2}-m^{2}\right)\mathcal{V}_{\mu}+\Re K_{\mu\nu}\mathcal{V}^{\nu}-\frac{1}{2}\epsilon_{\mu\nu\alpha\beta}\Im K^{\alpha\beta}\mathcal{A}^{\nu} & = & 0,\nonumber \\
\left(\Re K^{2}-m^{2}\right)\mathcal{A}_{\mu}+\Re K_{\mu\nu}\mathcal{A}^{\nu}-\frac{1}{2}\epsilon_{\mu\nu\alpha\beta}\Im K^{\alpha\beta}\mathcal{V}^{\nu} & = & 0,\label{eq:on-shell of VA}
\end{eqnarray}
where the operators $\Re K^{2}$, $\Im K^{2}$, $\Re K_{\mu\nu}$,
$\Im K_{\mu\nu}$ are defined in Eq. (\ref{eq:real and imaginary parts of second order operators}).
These equations are nothing new but the vector and axial-vector components
of the on-shell conditions in Eq. (\ref{eq:decomposed on-shell}).
The functions $\mathcal{V}^{\mu}$, $\mathcal{A}^{\mu}$ should satisfy
the equations listed in Eq. (\ref{eq:imaginary parts of kinetic equation}),
\begin{eqnarray}
\hbar\nabla^{\mu}\mathcal{V}_{\mu} & = & 0,\nonumber \\
\Pi^{\mu}\mathcal{A}_{\mu} & = & 0,\nonumber \\
\Pi_{[\mu}\mathcal{V}_{\nu]}+\frac{\hbar}{2}\epsilon_{\mu\nu\alpha\beta}\nabla^{\alpha}\mathcal{A}^{\beta} & = & 0,\label{eq:constraints for VA}
\end{eqnarray}
while the remaining two equations, i.e., the third and fourth lines
of Eq. (\ref{eq:real parts of kinetic equation}), are satisfied automatically
according to the previous discussion. In the massless limit, we have
chiral fermion whose spin is quantized along its momentum. We define
spin-up and spin-down currents as
\begin{equation}
\mathcal{J}_{\chi}^{\mu}\equiv\frac{1}{2}\left(\mathcal{V}^{\mu}+\chi\mathcal{A}^{\mu}\right),\label{def:left- and right-handed currents}
\end{equation}
where $\chi=\pm$ labels chirality. Analogous to the massless case,
we adopt the same definition (\ref{def:left- and right-handed currents})
in the massive case. The corresponding on-shell conditions for $\mathcal{J}_{\chi}^{\mu}$
are derived from Eq. (\ref{eq:on-shell of VA}),
\begin{equation}
\left(\Re K^{2}-m^{2}\right)\mathcal{J}_{\chi}^{\mu}+\Re K^{\mu\nu}\mathcal{J}_{\chi\nu}-\frac{\chi}{2}\epsilon^{\mu\nu\alpha\beta}\Im K_{\alpha\beta}\mathcal{J}_{\chi\nu}=0.\label{eq:on-shell equation for f_pm}
\end{equation}
We conclude that one method for computing the Wigner function is firstly
solving $\mathcal{V}_{\mu}$, $\mathcal{A}_{\mu}$ from the on-shell
equations (\ref{eq:on-shell equation for f_pm}) together with Eq.
(\ref{eq:constraints for VA}). Then the remaining components $\mathcal{F}$,
$\mathcal{P}$, and $\mathcal{S}_{\mu\nu}$ are given by Eq. (\ref{eq:FPS from VA}).

On the other hand, according to Eq. (\ref{eq:real parts of kinetic equation}),
we can prove that $\mathcal{V}^{\mu}$ and $\mathcal{A}^{\mu}$ can
also be expressed by $\mathcal{F}$, $\mathcal{P}$, and $\mathcal{S}^{\mu\nu}$.
This can be done by
\begin{eqnarray}
\mathcal{V}_{\mu} & = & \frac{1}{m}\Pi_{\mu}\mathcal{F}+\frac{\hbar}{2m}\nabla^{\nu}\mathcal{S}_{\mu\nu},\nonumber \\
\mathcal{A}_{\mu} & = & \frac{\hbar}{2m}\nabla_{\mu}\mathcal{P}-\frac{1}{2m}\epsilon_{\mu\nu\alpha\beta}\Pi^{\nu}\mathcal{S}^{\alpha\beta}.\label{eq:VA from FPS}
\end{eqnarray}
The functions $\mathcal{F}$, $\mathcal{P}$, and $\mathcal{S}^{\mu\nu}$
satisfy Eq. (\ref{eq:imaginary parts of kinetic equation}), which
gives the following constraints
\begin{eqnarray}
\frac{\hbar}{2}\nabla_{\mu}\mathcal{F}+\Pi^{\nu}\mathcal{S}_{\nu\mu} & = & 0,\nonumber \\
\Pi_{\mu}\mathcal{P}+\frac{\hbar}{4}\epsilon_{\mu\nu\alpha\beta}\nabla^{\nu}\mathcal{S}^{\alpha\beta} & = & 0.\label{eq:constraints of FPS}
\end{eqnarray}
The other equations, i.e., the first, second, and last lines of Eq.
(\ref{eq:imaginary parts of kinetic equation}) are automatically
fulfilled. In order to prove this, we form the combinations,
\begin{eqnarray}
0 & = & \frac{1}{m}\Pi^{\mu}\left(-\frac{\hbar}{2}\nabla_{\mu}\mathcal{P}+\frac{1}{2}\epsilon_{\mu\nu\alpha\beta}\Pi^{\nu}\mathcal{S}^{\alpha\beta}+m\mathcal{A}_{\mu}\right)\nonumber \\
 &  & +\frac{\hbar}{2m}\nabla^{\mu}\left(\Pi_{\mu}\mathcal{P}+\frac{\hbar}{4}\epsilon_{\mu\nu\alpha\beta}\nabla^{\nu}\mathcal{S}^{\alpha\beta}\right),
\end{eqnarray}
and
\begin{eqnarray}
0 & = & -\frac{\hbar}{m}\nabla^{\mu}\left(\Pi_{\mu}\mathcal{F}-\frac{\hbar}{2}\nabla^{\nu}\mathcal{S}_{\nu\mu}-m\mathcal{V}_{\mu}\right)\nonumber \\
 &  & +\frac{2}{m}\Pi^{\mu}\left(\frac{\hbar}{2}\nabla_{\mu}\mathcal{F}+\Pi^{\nu}\mathcal{S}_{\nu\mu}\right),
\end{eqnarray}
together with
\begin{eqnarray}
0 & = & -\frac{1}{m}\Pi_{[\mu}\left(\Pi_{\nu]}\mathcal{F}+\frac{\hbar}{2}\nabla^{\alpha}\mathcal{S}_{\nu]\alpha}-m\mathcal{V}_{\nu]}\right)\nonumber \\
 &  & -\frac{\hbar}{2m}\nabla_{[\mu}\left(\frac{\hbar}{2}\nabla_{\nu]}\mathcal{F}-\Pi^{\alpha}\mathcal{S}_{\nu]\alpha}\right)\nonumber \\
 &  & -\frac{1}{m}\epsilon_{\mu\nu\alpha\beta}\Pi^{\alpha}\left(\Pi^{\beta}\mathcal{P}+\frac{\hbar}{4}\epsilon^{\beta\gamma\rho\sigma}\nabla_{\gamma}\mathcal{S}_{\rho\sigma}\right)\nonumber \\
 &  & +\frac{\hbar}{2m}\epsilon_{\mu\nu\alpha\beta}\nabla^{\alpha}\left(-\frac{\hbar}{2}\nabla^{\beta}\mathcal{P}+\frac{1}{2}\epsilon^{\beta\gamma\rho\sigma}\Pi_{\gamma}\mathcal{S}_{\rho\sigma}+m\mathcal{A}^{\beta}\right).
\end{eqnarray}
These equations are satisfied because the terms inside the parentheses
are zero according to Eqs. (\ref{eq:VA from FPS}), (\ref{eq:constraints of FPS}).
{} After complicated but straightforward calculations and with the help
of Eq. (\ref{eq:commutators of operators Pi and Nabla}), we reproduce
the first, second, and last lines of Eq. (\ref{eq:imaginary parts of kinetic equation}).
Meanwhile, substituting Eq. (\ref{eq:VA from FPS}) into Eq. (\ref{eq:real parts of kinetic equation}),
one obtains the following on-shell conditions
\begin{eqnarray}
\left(\Re K^{2}-m^{2}\right)\mathcal{F}+\frac{1}{2}\Im K_{\mu\nu}\mathcal{S}^{\mu\nu} & = & 0,\nonumber \\
\left(\Re K^{2}-m^{2}\right)\mathcal{P}+\frac{1}{4}\epsilon_{\mu\nu\alpha\beta}\Im K^{\mu\nu}\mathcal{S}^{\alpha\beta} & = & 0,\nonumber \\
\left(\Re K^{2}-m^{2}\right)\mathcal{S}_{\mu\nu}+\Re K_{\ [\mu}^{\alpha}\mathcal{S}_{\nu]\alpha}+\Im K_{\mu\nu}\mathcal{F}-\frac{1}{2}\epsilon_{\mu\nu\alpha\beta}\Im K^{\alpha\beta}\mathcal{P} & = & 0.\label{eq:on-shell of FPS}
\end{eqnarray}
So we conclude that another approach for computing the Wigner function
is: first obtain a solution for $\mathcal{F}$, $\mathcal{P}$, and
$\mathcal{S}^{\mu\nu}$ which satisfies Eqs. (\ref{eq:constraints of FPS})
and (\ref{eq:on-shell of FPS}) and then derive $\mathcal{V}^{\mu}$
and $\mathcal{A}^{\mu}$ using Eq. (\ref{eq:VA from FPS}).

Note that in the massless case, the above discussion seem to be useless
because the mass appears in the denominators in Eqs. (\ref{eq:FPS from VA}),
(\ref{eq:VA from FPS}) and $1/m$ will be divergent when $m\rightarrow0$.
However, detailed calculations in the next section show that the numerators
are also proportional to the mass, which leads to a finite quotient.
In this way, the results in the massive case are expected to smoothly
converge to the results in the massless case. For massless particles,
the vector and axial-vector components $\mathcal{V}^{\mu}$ and $\mathcal{A}^{\mu}$
decouple from the other components, as given in (\ref{eq:massless equations}).
Adopting the definition (\ref{def:left- and right-handed currents})
of spin-up/spin-down currents, the equations can be rewritten in a
compact form,
\begin{eqnarray}
 &  & \hbar\nabla_{\mu}\mathcal{J}_{\chi}^{\mu}=0,\ \ \Pi_{\mu}\mathcal{J}_{\chi}^{\mu}=0,\nonumber \\
 &  & \Pi_{[\mu}\mathcal{J}_{\nu]}^{\chi}+\frac{\chi\hbar}{2}\epsilon_{\mu\nu\alpha\beta}\nabla^{\alpha}\mathcal{J}_{\chi}^{\beta}=0.\label{eq:equations of currents massless}
\end{eqnarray}
Properly taking combinations of the above equations, one can derive
the on-shell equations, which agree with Eq. (\ref{eq:on-shell equation for f_pm})
by putting $m=0$.

As a brief summary of this subsection, we list once more the approaches
for computing the Wigner function:
\begin{enumerate}
\item For the massless case, the vector and axial-vector components $\mathcal{V}^{\mu}$
and $\mathcal{A}^{\mu}$ can be written in terms of the spin-up and
spin-down currents as shown in Eq. (\ref{def:left- and right-handed currents}).
These currents should satisfy Eq. (\ref{eq:equations of currents massless}).
\item For massive particles, one can take the vector and axial-vector components
$\mathcal{V}^{\mu}$ and $\mathcal{A}^{\mu}$ as basic quantities,
which satisfy the on-shell conditions (\ref{eq:on-shell of VA}) and
Eq. (\ref{eq:constraints for VA}). The other components, $\mathcal{F}$,
$\mathcal{P}$, and $\mathcal{S}^{\mu\nu}$, are then derived from
$\mathcal{V}^{\mu}$, $\mathcal{A}^{\mu}$ using relation (\ref{eq:FPS from VA}).
\item For massive particles, another possible method is to take the scalar,
pseudoscalar, and tensor components $\mathcal{F}$, $\mathcal{P}$,
and $\mathcal{S}^{\mu\nu}$ as basic quantities, which satisfy Eq.
(\ref{eq:constraints of FPS}) and the on-shell conditions (\ref{eq:on-shell of FPS}).
Equation (\ref{eq:VA from FPS}) shows how to derive the other components
$\mathcal{V}^{\mu}$ and $\mathcal{A}^{\mu}$ from $\mathcal{F}$,
$\mathcal{P}$, and $\mathcal{S}^{\mu\nu}$.
\end{enumerate}
The more detailed semi-classical calculations in Sec. \ref{sec:Semi-classical-expansion}
show that the approach 2 is equivalent with 3.

\subsection{Equal-time Wigner function\label{subsec:Equal-time-formula}}

In some dynamical problems, it appears to be more convenient to use
the equal-time Wigner function, which was first proposed in Refs.
\cite{BialynickiBirula:1991tx,Zhuang:1998bqx}. In this thesis, we
define the equal-time Wigner function as follows
\begin{equation}
W(t,\mathbf{x},\mathbf{p})=\int dp^{0}\,W(x,p),
\end{equation}
which is derived from the covariant Wigner function by integrating
over energy $p^{0}$. Obviously the equal-time Wigner function is
not Lorentz covariant because the observer's frame has been fixed.
From Eq. (\ref{def:Wigner function}), we can finish the integration
over energy $p^{0}$ and obtain
\begin{equation}
W(t,\mathbf{x},\mathbf{p})\equiv\int\frac{d^{3}\mathbf{y}}{(2\pi)^{3}}\exp\left(i\mathbf{x}\cdot\mathbf{p}\right)U\left(t,\mathbf{x}+\frac{\mathbf{y}}{2},\mathbf{x}-\frac{\mathbf{y}}{2}\right)\left\langle \Omega\left|\hat{\bar{\psi}}\left(t,\mathbf{x}+\frac{\mathbf{y}}{2}\right)\otimes\hat{\psi}\left(t,\mathbf{x}-\frac{\mathbf{y}}{2}\right)\right|\Omega\right\rangle .\label{def:equal-time Wigner function}
\end{equation}
Here the two field operators are defined at the same time $t$ but
at different spatial points. A 3-dimensional Fourier transform is
made with respect to the relative coordinate $\mathbf{y}$, which
gives the dependence on the kinetic 3-momentum $\mathbf{p}$. Similar
to the covariant form, the gauge field (i.e., the electromagnetic
field) is assumed to be a classical $C$-number and thus the gauge
link is taken out of the quantum expectation value. Meanwhile, the
covariant Wigner function can be described by its energy moments $\int dp^{0}\ (p^{0})^{n}W(x,p)$
and the equal-time Wigner function is just the zeroth order moment.
Thus, from the covariant Wigner function one can derive the equal-time
Wigner function, but from the equal-time one we cannot reproduce the
covariant one because the higher-order energy moments, $\int dp^{0}\ (p^{0})^{n}W(x,p)$
for $n>0$, are unknown. But if particles are on the usual mass-shell
$p^{2}=m^{2}$, the covariant Wigner function and the equal-time one
are equivalent to each other.

The equation of motion for the equal-time Wigner function can be obtained
from the Dirac equation, or equivalently from the equation of motion
for the covariant Wigner function via taking an energy integral. From
Eq. (\ref{eq:Dirac equation for Wigner}), we can obtain the Dirac-form
equation for the equal-time Wigner function by integrating over $p^{0}$
and dropping boundary terms such as $\int dp^{0}\ \partial_{p^{0}}W(x,p)$,
\begin{equation}
\gamma^{0}\int dp^{0}\ p^{0}W(x,p)+\gamma^{0}\Pi^{0}W(x,p)+\left(\frac{i\hbar}{2}\gamma^{0}D_{t}-\boldsymbol{\gamma}\cdot\mathbf{K}-m\right)W(t,\mathbf{x},\mathbf{p})=0,\label{eq:Dirac like equation of equal-time formula}
\end{equation}
where the operator is defined as
\begin{equation}
\mathbf{K}\equiv\boldsymbol{\Pi}-\frac{i\hbar}{2}\mathbf{D}_{\mathbf{x}}.
\end{equation}
Here the generalized time derivative operator $D_{t}$, the spatial
derivative operator $\mathbf{D}_{\mathbf{x}}$, the energy shift $\Pi^{0}$,
and the momentum operator $\boldsymbol{\Pi}$ are given by
\begin{eqnarray}
D_{t} & \equiv & \partial_{t}+j_{0}(\Delta)\mathbf{E}(x)\cdot\boldsymbol{\nabla}_{\mathbf{p}},\nonumber \\
\mathbf{D}_{\mathbf{x}} & \equiv & \boldsymbol{\nabla}_{\mathbf{x}}+j_{0}(\Delta)\mathbf{B}(x)\times\boldsymbol{\nabla}_{\mathbf{p}},\\
\Pi^{0} & = & \frac{\hbar}{2}j_{1}(\Delta)\mathbf{E}(x)\cdot\boldsymbol{\nabla}_{\mathbf{p}},\nonumber \\
\boldsymbol{\Pi} & \equiv & \mathbf{p}-\frac{\hbar}{2}j_{1}(\Delta)\mathbf{B}(x)\times\boldsymbol{\nabla}_{\mathbf{p}},\label{eq:generalized operators for equal-t formula}
\end{eqnarray}
with $\Delta\equiv-\frac{\hbar}{2}\boldsymbol{\nabla}_{\mathbf{p}}\cdot\boldsymbol{\nabla}_{\mathbf{x}}$
where $\boldsymbol{\nabla}_{\mathbf{x}}$ only acts on the electromagnetic
fields. These generalized operators $D_{t}$, $\mathbf{D}_{\mathbf{x}}$,
and $\boldsymbol{\Pi}$ are reduced to the normal time derivative,
spatial derivative, and 3-momentum when the electromagnetic fields
vanish. They are real operators, thus the Hermitian conjugate of Eq.
(\ref{eq:Dirac like equation of equal-time formula}) reads,
\begin{equation}
\gamma^{0}\int dp^{0}\ p^{0}W(x,p)+\gamma^{0}\Pi^{0}W(x,p)+\gamma^{0}W(t,\mathbf{x},\mathbf{p})\left[-\frac{i\hbar}{2}\left(\gamma^{0}D_{t}+\boldsymbol{\gamma}\cdot\mathbf{D}_{\mathbf{x}}\right)-\boldsymbol{\gamma}\cdot\boldsymbol{\Pi}-m\right]\gamma^{0}=0,\label{eq:Conjugate equation of equal-time}
\end{equation}
where we have used the property $W^{\dagger}=\gamma^{0}W\gamma^{0}$.
Multiplying Eqs. (\ref{eq:Dirac like equation of equal-time formula})
and (\ref{eq:Conjugate equation of equal-time}) with $\gamma^{0}$
from the left, we obtain
\begin{eqnarray}
\int dp^{0}\ p^{0}W(x,p)+\Pi^{0}W(x,p)+\left[\frac{i\hbar}{2}\left(D_{t}+\gamma^{0}\boldsymbol{\gamma}\cdot\mathbf{D}_{\mathbf{x}}\right)-\gamma^{0}\boldsymbol{\gamma}\cdot\boldsymbol{\Pi}-m\gamma^{0}\right]W(t,\mathbf{x},\mathbf{p}) & = & 0,\nonumber \\
\int dp^{0}\ p^{0}W(x,p)+\Pi^{0}W(x,p)+W(t,\mathbf{x},\mathbf{p})\left[-\frac{i\hbar}{2}\left(D_{t}-\gamma^{0}\boldsymbol{\gamma}\cdot\mathbf{D}_{\mathbf{x}}\right)+\gamma^{0}\boldsymbol{\gamma}\cdot\boldsymbol{\Pi}-m\gamma^{0}\right] & = & 0.\nonumber \\
\end{eqnarray}
Taking the difference of these two equations we obtain the equation
of motion for the equal-time Wigner function
\begin{equation}
i\hbar D_{t}W(t,\mathbf{x},\mathbf{p})+\frac{i\hbar}{2}\mathbf{D}_{\mathbf{x}}\cdot\left[\gamma^{0}\boldsymbol{\gamma},W(t,\mathbf{x},\mathbf{p})\right]-\boldsymbol{\Pi}\cdot\left\{ \gamma^{0}\boldsymbol{\gamma},W(t,\mathbf{x},\mathbf{p})\right\} -m\left[\gamma^{0},W(t,\mathbf{x},\mathbf{p})\right]=0,\label{eq:Equation of motion of equal-time}
\end{equation}
while the sum gives
\begin{eqnarray}
 &  & \int dp^{0}\ p^{0}W(x,p)\nonumber \\
 &  & \qquad=-\frac{i\hbar}{2}\mathbf{D}_{\mathbf{x}}\cdot\left\{ \gamma^{0}\boldsymbol{\gamma},W(t,\mathbf{x},\mathbf{p})\right\} -\Pi^{0}W(x,p)+\boldsymbol{\Pi}\cdot\left[\gamma^{0}\boldsymbol{\gamma},W(t,\mathbf{x},\mathbf{p})\right]+m\left\{ \gamma^{0},W(t,\mathbf{x},\mathbf{p})\right\} .\nonumber \\
\label{eq:constraint equation for equal-time}
\end{eqnarray}
We note that the time-evolution of the equal-time Wigner function
is determined by Eq. (\ref{eq:Equation of motion of equal-time})
while Eq. (\ref{eq:constraint equation for equal-time}) provides
the relation between the first-order energy moment $\int dp^{0}\ p^{0}W(x,p)$
and the equal-time Wigner function.

Analogously to the covariant Wigner function, the equal-time Wigner
function can be decomposed in 16 independent generators of the Clifford
algebra, $\Gamma_{i}=\{1,\ i\gamma^{5},\ \gamma^{\mu},\ \gamma^{\mu}\gamma^{5},\ \frac{1}{2}\sigma^{\mu\nu}\}$,
as shown in Eq. (\ref{def:Wigner function decomposition}). Here the
coefficients are now functions of $\{t,\mathbf{x},\mathbf{p}\}$.
Inserting the decomposed Wigner function into Eq. (\ref{eq:Equation of motion of equal-time})
and taking the trace over $\Gamma_{i}$ we obtain the following equations
of motion,
\begin{eqnarray}
\hbar D_{t}\mathcal{F} & = & 2\boldsymbol{\Pi}\cdot\boldsymbol{\mathcal{T}},\nonumber \\
\hbar D_{t}\mathcal{P} & = & -2\boldsymbol{\Pi}\cdot\boldsymbol{\mathcal{S}}+2m\mathcal{A}^{0},\nonumber \\
\hbar D_{t}\mathcal{V}^{0} & = & -\hbar\mathbf{D}_{\mathbf{x}}\cdot\boldsymbol{\mathcal{V}},\nonumber \\
\hbar D_{t}\boldsymbol{\mathcal{V}} & = & -\hbar\mathbf{D}_{\mathbf{x}}\mathcal{V}^{0}+2\boldsymbol{\Pi}\times\boldsymbol{\mathcal{A}}-2m\boldsymbol{\mathcal{T}},\nonumber \\
\hbar D_{t}\mathcal{A}^{0} & = & -\hbar\mathbf{D}_{\mathbf{x}}\cdot\boldsymbol{\mathcal{A}}-2m\mathcal{P},\nonumber \\
\hbar D_{t}\boldsymbol{\mathcal{A}} & = & -\hbar\mathbf{D}_{\mathbf{x}}\mathcal{A}^{0}+2\boldsymbol{\Pi}\times\boldsymbol{\mathcal{V}},\nonumber \\
\hbar D_{t}\boldsymbol{\mathcal{T}} & = & -\hbar\mathbf{D}_{\mathbf{x}}\times\boldsymbol{\mathcal{S}}-2\boldsymbol{\Pi}\mathcal{F}+2m\boldsymbol{\mathcal{V}},\nonumber \\
\hbar D_{t}\boldsymbol{\mathcal{S}} & = & \hbar\mathbf{D}_{\mathbf{x}}\times\boldsymbol{\mathcal{T}}+2\boldsymbol{\Pi}\mathcal{P},\label{eq:equation of motion equal-time component}
\end{eqnarray}
where we have suppressed the dependence on $\{t,\mathbf{x},\mathbf{p}\}$
for all component functions. These equations describe how these component
functions evolve with time. On the other hand, decomposing Eq. (\ref{eq:constraint equation for equal-time})
we derive the first-order energy moments,
\begin{eqnarray}
\int dp^{0}\ p^{0}\mathcal{F}(x,p) & = & \frac{\hbar}{2}\mathbf{D}_{\mathbf{x}}\cdot\boldsymbol{\mathcal{T}}+m\mathcal{V}^{0}-\Pi^{0}\mathcal{F},\nonumber \\
\int dp^{0}\ p^{0}\mathcal{P}(x,p) & = & -\frac{\hbar}{2}\mathbf{D}_{\mathbf{x}}\cdot\boldsymbol{\mathcal{S}}-\Pi^{0}\mathcal{P},\nonumber \\
\int dp^{0}\ p^{0}\mathcal{V}^{0}(x,p) & = & \boldsymbol{\Pi}\cdot\boldsymbol{\mathcal{V}}+m\mathcal{F}-\Pi^{0}\mathcal{V}^{0},\nonumber \\
\int dp^{0}\ p^{0}\boldsymbol{\mathcal{V}}(x,p) & = & \frac{\hbar}{2}\mathbf{D}_{\mathbf{x}}\times\boldsymbol{\mathcal{A}}+\boldsymbol{\Pi}\mathcal{V}^{0}-\Pi^{0}\boldsymbol{\mathcal{V}},\nonumber \\
\int dp^{0}\ p^{0}\mathcal{A}^{0}(x,p) & = & \boldsymbol{\Pi}\cdot\boldsymbol{\mathcal{A}}-\Pi^{0}\mathcal{A}^{0},\nonumber \\
\int dp^{0}\ p^{0}\boldsymbol{\mathcal{A}}(x,p) & = & \frac{\hbar}{2}\mathbf{D}_{\mathbf{x}}\times\boldsymbol{\mathcal{V}}+\boldsymbol{\Pi}\mathcal{A}^{0}+m\boldsymbol{\mathcal{S}}-\Pi^{0}\boldsymbol{\mathcal{A}},\nonumber \\
\int dp^{0}\ p^{0}\boldsymbol{\mathcal{T}}(x,p) & = & -\frac{\hbar}{2}\mathbf{D}_{\mathbf{x}}\mathcal{F}+\boldsymbol{\Pi}\times\boldsymbol{\mathcal{S}}-\Pi^{0}\boldsymbol{\mathcal{T}},\nonumber \\
\int dp^{0}\ p^{0}\boldsymbol{\mathcal{S}}(x,p) & = & \frac{\hbar}{2}\mathbf{D}_{\mathbf{x}}\mathcal{P}-\boldsymbol{\Pi}\times\boldsymbol{\mathcal{T}}+m\boldsymbol{\mathcal{A}}-\Pi^{0}\boldsymbol{\mathcal{S}},
\end{eqnarray}
where the functions on the right-hand side are equal-time ones, while
the functions on the left-hand side are covariant ones.

Now we divide the 16 functions into four groups, each group having
four functions,
\begin{eqnarray}
\mathcal{G}_{1}=\left(\begin{array}{c}
\mathcal{F}\\
\boldsymbol{\mathcal{S}}
\end{array}\right), &  & \mathcal{G}_{2}=\left(\begin{array}{c}
\mathcal{V}^{0}\\
\boldsymbol{\mathcal{A}}
\end{array}\right),\nonumber \\
\mathcal{G}_{3}=\left(\begin{array}{c}
\mathcal{A}^{0}\\
\boldsymbol{\mathcal{V}}
\end{array}\right), &  & \mathcal{G}_{4}=\left(\begin{array}{c}
\mathcal{P}\\
\boldsymbol{\mathcal{T}}
\end{array}\right).\label{def:definition of G_i}
\end{eqnarray}
The introduction of these four groups proves to be useful for dealing
with the Wigner function when the observer's frame is fixed \cite{Sheng:2017lfu,Sheng:2018jwf}.
This form will be used in Sec. \ref{sec:Analytically-solvable-cases}
for the case of constant electromagnetic fields. Using Eq. (\ref{def:definition of G_i}),
Eq. (\ref{eq:equation of motion equal-time component}) takes a matrix
form,
\begin{equation}
\hbar D_{t}\left(\begin{array}{c}
\mathcal{G}_{1}\{t,\mathbf{x},\mathbf{p}\}\\
\mathcal{G}_{2}\{t,\mathbf{x},\mathbf{p}\}\\
\mathcal{G}_{3}\{t,\mathbf{x},\mathbf{p}\}\\
\mathcal{G}_{4}\{t,\mathbf{x},\mathbf{p}\}
\end{array}\right)=\left(\begin{array}{cccc}
0 & 0 & 0 & M_{1}\\
0 & 0 & -M_{2} & 0\\
0 & -M_{2} & 0 & -2m\mathbb{I}_{4}\\
M_{1} & 0 & 2m\mathbb{I}_{4} & 0
\end{array}\right)\left(\begin{array}{c}
\mathcal{G}_{1}\{t,\mathbf{x},\mathbf{p}\}\\
\mathcal{G}_{2}\{t,\mathbf{x},\mathbf{p}\}\\
\mathcal{G}_{3}\{t,\mathbf{x},\mathbf{p}\}\\
\mathcal{G}_{4}\{t,\mathbf{x},\mathbf{p}\}
\end{array}\right),\label{eq:EOM of 4 groups}
\end{equation}
while the constraint equation reads %
\begin{equation}
\int dp^{0}\ p^{0}\left(\begin{array}{c}
\mathcal{G}_{1}(x,p)\\
\mathcal{G}_{2}(x,p)\\
\mathcal{G}_{3}(x,p)\\
\mathcal{G}_{4}(x,p)
\end{array}\right)=\frac{1}{2}\left(\begin{array}{cccc}
0 & 2m\mathbb{I}_{4} & 0 & M_{2}\\
2m\mathbb{I}_{4} & 0 & M_{1} & 0\\
0 & M_{1} & 0 & 0\\
-M_{2} & 0 & 0 & 0
\end{array}\right)\left(\begin{array}{c}
\mathcal{G}_{1}\{t,\mathbf{x},\mathbf{p}\}\\
\mathcal{G}_{2}\{t,\mathbf{x},\mathbf{p}\}\\
\mathcal{G}_{3}\{t,\mathbf{x},\mathbf{p}\}\\
\mathcal{G}_{4}\{t,\mathbf{x},\mathbf{p}\}
\end{array}\right)-\Pi^{0}\left(\begin{array}{c}
\mathcal{G}_{1}\{t,\mathbf{x},\mathbf{p}\}\\
\mathcal{G}_{2}\{t,\mathbf{x},\mathbf{p}\}\\
\mathcal{G}_{3}\{t,\mathbf{x},\mathbf{p}\}\\
\mathcal{G}_{4}\{t,\mathbf{x},\mathbf{p}\}
\end{array}\right).
\end{equation}
Here we define two matrices which are constructed from $\mathbf{D}_{\mathbf{x}}$,
and $\boldsymbol{\Pi}$,
\begin{equation}
M_{1}=\left(\begin{array}{cc}
0 & 2\boldsymbol{\Pi}^{T}\\
2\boldsymbol{\Pi} & \hbar\mathbf{D}_{\mathbf{x}}^{\times}
\end{array}\right),\ M_{2}=\left(\begin{array}{cc}
0 & \hbar\mathbf{D}_{\mathbf{x}}^{T}\\
\hbar\mathbf{D}_{\mathbf{x}} & -2\boldsymbol{\Pi}^{\times}
\end{array}\right).\label{def:matrices M1M2}
\end{equation}
For any 3-dimensional column vector, for example, the momentum operator
$\boldsymbol{\Pi}$, we use $\boldsymbol{\Pi}^{T}$ for its transpose,
a row vector. In Eq. (\ref{def:matrices M1M2}), $\boldsymbol{\Pi}^{\times}$
represents an anti-symmetric matrix whose elements are $(\boldsymbol{\Pi}^{\times})^{ij}=-\epsilon^{ijk}\Pi^{k}$,
\begin{equation}
\boldsymbol{\Pi}^{\times}=\left(\begin{array}{ccc}
0 & -\Pi^{z} & \Pi^{y}\\
\Pi^{z} & 0 & -\Pi^{x}\\
-\Pi^{y} & \Pi^{x} & 0
\end{array}\right).
\end{equation}
When acting with the matrix $\boldsymbol{\Pi}^{\times}$ onto another
column vector $\mathbf{V}$, we obtain the cross product of two vectors,
\begin{equation}
\boldsymbol{\Pi}^{\times}\mathbf{V}=\boldsymbol{\Pi}\times\mathbf{V}.
\end{equation}
The operators defined in Eq. (\ref{eq:generalized operators for equal-t formula})
coincide with the ones used in Refs. \cite{BialynickiBirula:1991tx,Zhuang:1998bqx,Wang:2019moi}
because we have the following relations %
\begin{eqnarray}
j_{0}(\Delta)\mathbf{E}(x) & = & \int_{-1/2}^{1/2}ds\mathbf{E}(\mathbf{x}+is\hbar\boldsymbol{\nabla}_{\mathbf{p}}),\nonumber \\
-\frac{i}{2}j_{1}(\Delta)\mathbf{E}(x) & = & \int_{-1/2}^{1/2}dss\mathbf{E}(\mathbf{x}+is\hbar\boldsymbol{\nabla}_{\mathbf{p}}).
\end{eqnarray}
With the help of these relations, the operators in Eq. (\ref{eq:generalized operators for equal-t formula})
can be written in another form,
\begin{eqnarray}
D_{t} & = & \partial_{t}+\int_{-1/2}^{1/2}ds\mathbf{E}(\mathbf{x}+is\hbar\boldsymbol{\nabla}_{\mathbf{p}})\cdot\boldsymbol{\nabla}_{\mathbf{p}},\nonumber \\
\mathbf{D}_{\mathbf{x}} & = & \boldsymbol{\nabla}_{\mathbf{x}}+\int_{-1/2}^{1/2}ds\mathbf{B}(\mathbf{x}+is\hbar\boldsymbol{\nabla}_{\mathbf{p}})\times\boldsymbol{\nabla}_{\mathbf{p}},\nonumber \\
\Pi^{0} & = & i\hbar\int_{-1/2}^{1/2}dss\mathbf{E}(\mathbf{x}+is\hbar\boldsymbol{\nabla}_{\mathbf{p}})\cdot\boldsymbol{\nabla}_{\mathbf{p}},\nonumber \\
\boldsymbol{\Pi} & = & \mathbf{p}-i\hbar\int_{-1/2}^{1/2}dss\mathbf{B}(\mathbf{x}+is\hbar\boldsymbol{\nabla}_{\mathbf{p}})\times\boldsymbol{\nabla}_{\mathbf{p}},
\end{eqnarray}
which are used in Refs. \cite{BialynickiBirula:1991tx,Zhuang:1998bqx,Wang:2019moi}.

\newpage{}

$\ $

\newpage{}

\section{Analytical solutions\label{sec:Analytically-solvable-cases}}

In the previous section we have introduced the definition of the covariant
Wigner function in Eq. (\ref{def:Wigner function}) and its equal-time
formula in Eq. (\ref{def:equal-time Wigner function}). Kinetic equations
are also derived but we still need the initial conditions for numerically
solving the equations. In this section we will give several analytically
solvable cases. The results of this section can serve as initial conditions
for numerical calculations. In the following three cases, the Dirac
equation has analytical solutions
\begin{enumerate}
\item A system consisting of fermions without any interaction.
\item Fermions with chiral imbalance. The chemical potential $\mu$ and
the chiral chemical potential $\mu_{5}$ are included in the Dirac
equation but still without the electromagnetic field.
\item Fermions in a constant magnetic field. As in case 2, $\mu$ and $\mu_{5}$
are included in the Dirac equation.
\end{enumerate}
In all three cases, the Dirac equation can be analytically solved
and we derive the eigenenergies and corresponding eigenwavefunctions.
Then the field operator is derived following the standard procedure
of second quantization. The Wigner function is then solved up to zeroth
order in the spatial derivative. The chiral chemical potential $\mu_{5}$
is included for the further study of chiral effects. In this section,
two dynamical problems will also be considered,
\begin{enumerate}
\item Fermions in an electric field. The existence of the electric field
leads to the decay of the vacuum into fermion/anti-fermion pairs.
At the same time, charged particles in the system will be accelerated.
\item Fermions in constant electromagnetic fields. The magnetic field is
assumed to be parallel to the electric field.
\end{enumerate}
We use the equal-time Wigner function for these dynamical problems.
These discussions show that the Wigner function approach can also
be used for the study of pair-production. Furthermore, in parallel
electric and magnetic field, the existence of the magnetic field will
enhance the pair-production rate since it changes the structure of
energy levels. Meanwhile, at the lowest Landau level, spins of positive
charged particles are locked to the direction of the magnetic field,
and the newly generated positively charged particles move along the
electric field. Thus these particles have RH chirality. Similar arguements
show that the negatively charged particles (anti-fermions) have RH
chirality, too. This gives rise to interesting effects, such as axial-charge
production and axial-current production \cite{Copinger:2018ftr}.
In this section we display the analytical procedure for deriving the
Wigner function in the above five cases, while in Sec. \ref{sec:Physical quantities}
we will numerically calculate physical quantities. Throughout this
section we will suppress $\hbar$ but it can be recovered by carefully
counting the units.

\subsection{Free fermions\label{subsec:Free-fermions}}

\subsubsection{Plane-wave solutions}

In this subsection we will focus on free fermions with spin-$\frac{1}{2}$
in the absence of electromagnetic fields. Interactions among particles
are also neglected. In this case, fermions satisfy the free Dirac
equation (\ref{eq:Dirac equation and conjugate}) with vanishing gauge
potential $\mathbb{A}_{\mu}=0$,
\begin{equation}
\left(i\gamma^{\mu}\partial_{x\mu}-m\mathbb{I}_{4}\right)\psi(x)=0.
\end{equation}
The Dirac equation can be rewritten in the form of a Schr\"{o}dinger
equation %
\begin{equation}
i\frac{\partial}{\partial t}\psi=(-i\gamma^{0}\boldsymbol{\gamma}\cdot\boldsymbol{\partial}_{\mathbf{x}}+m\gamma^{0})\psi.\label{eq:free Schroedinger equation}
\end{equation}
Note that the spatial derivative operator $\boldsymbol{\partial}_{\mathbf{x}}$
commutes with the Hamilton operator $\hat{H}=-i\gamma^{0}\boldsymbol{\gamma}\cdot\boldsymbol{\partial}_{\mathbf{x}}+m\gamma^{0}$,
so we can introduce a kinetic 3-momentum $\mathbf{p}$ by making a
Fourier expansion for the field $\psi(x)$,
\begin{equation}
\psi(x)=\int\frac{d^{4}p}{(2\pi)^{4}}e^{-ip^{\mu}x_{\mu}}\psi(p).
\end{equation}
Applying this into the Dirac equation we obtain
\begin{equation}
p^{0}\psi(p)=(\gamma^{0}\boldsymbol{\gamma}\cdot\mathbf{p}+m\gamma^{0})\psi(p).\label{eq:free Dirac eq in momentum space}
\end{equation}
The on-shell condition can be obtained by acting with $(\gamma^{0}\boldsymbol{\gamma}\cdot\mathbf{p}+m\gamma^{0})$
onto Eq. (\ref{eq:free Dirac eq in momentum space}), %
\begin{equation}
(p^{0})^{2}\psi(p)=(m^{2}+\mathbf{p}^{2})\psi(p).
\end{equation}
Solving the on-shell condition, we obtain positive-energy states with
$p^{0}>0$, and negative-energy states with $p^{0}<0$.
\begin{equation}
p^{0}=\pm E_{\mathbf{p}}=\pm\sqrt{m^{2}+\mathbf{p}^{2}}.
\end{equation}
The positive-energy states will be identified as fermions, while the
negative-energy states are identified as anti-fermions.

In order to obtain the corresponding eigenwavefunctions, we perform
a Lorentz transformation  and work in the particle's rest frame. We
parameterize the Lorentz transformation using $\omega_{\mu\nu}$,
which is anti-symmetric with respect to $\mu\leftrightarrow\nu$.
The transformation matrix for a Lorentz vector is
\begin{equation}
\Lambda_{\ \nu}^{\mu}=\exp\left[-\frac{i}{2}\omega_{\alpha\beta}(\mathcal{J}^{\alpha\beta})_{\ \nu}^{\mu}\right],
\end{equation}
where $(\mathcal{J}^{\alpha\beta})_{\ \nu}^{\mu}$ is the generator
of the Lorentz algebra. In the coordinate representation, this generator
is given by
\begin{equation}
(\mathcal{J}^{\alpha\beta})_{\mu\nu}=i(\delta_{\mu}^{\alpha}\delta_{\nu}^{\beta}-\delta_{\nu}^{\alpha}\delta_{\mu}^{\beta}).
\end{equation}
Inserting $(\mathcal{J}^{\alpha\beta})_{\mu\nu}$ into the transformation
matrix $\Lambda_{\ \nu}^{\mu}$, we obtain
\begin{equation}
\Lambda_{\ \nu}^{\mu}=\exp\left(\omega_{\ \nu}^{\mu}\right).
\end{equation}
Any vector, for example the 4-momentum, transforms as
\begin{equation}
p^{\mu}\rightarrow\Lambda_{\ \nu}^{\mu}p^{\nu}.
\end{equation}
Meanwhile, the Dirac-spinor field $\psi(x)$ transforms as
\begin{equation}
\psi(x)\rightarrow\Lambda_{\frac{1}{2}}\psi(\Lambda^{-1}x),
\end{equation}
where the spinor representation of the Lorentz transformation is given
by
\begin{equation}
\Lambda_{\frac{1}{2}}=\exp\left(-\frac{i}{4}\omega_{\mu\nu}\sigma^{\mu\nu}\right),
\end{equation}
with $\sigma^{\mu\nu}\equiv\frac{i}{2}\left[\gamma^{\mu},\,\gamma^{\nu}\right]$.
Now we consider the transformation from the particle's rest frame
to the lab frame. For one particle which has 4-momentum $(E_{\mathbf{p}},\,\mathbf{p})$
in the lab frame, its 3-velocity is
\begin{equation}
\boldsymbol{\beta}=\frac{\mathbf{p}}{E_{\mathbf{p}}},
\end{equation}
and we define the rapidity vector as
\begin{equation}
\boldsymbol{\zeta}=\frac{\boldsymbol{\beta}}{\beta}\tanh^{-1}\beta,\label{def:rapidity vector}
\end{equation}
with $\beta\equiv\left|\boldsymbol{\beta}\right|$. Then we define
the parameters for the Lorentz transformation from the particle's
rest frame to the lab frame
\begin{equation}
\omega^{0i}=-\zeta^{i},\ \ \omega^{ij}=0,
\end{equation}
which leads to the following transformation matrix
\begin{equation}
\Lambda_{\ \nu}^{\mu}=\exp\left[\left(\begin{array}{cccc}
0 & \beta^{x} & \beta^{y} & \beta^{z}\\
\beta^{x} & 0 & 0 & 0\\
\beta^{y} & 0 & 0 & 0\\
\beta^{z} & 0 & 0 & 0
\end{array}\right)\frac{1}{\beta}\tanh^{-1}\beta\right].
\end{equation}
This matrix can be calculated using the Taylor expansion, %
\begin{equation}
\Lambda_{\ \nu}^{\mu}=\left(\begin{array}{cccc}
\gamma & \gamma\beta^{x} & \gamma\beta^{y} & \gamma\beta^{z}\\
\gamma\beta^{x} & 1+(\gamma-1)(\hat{\beta}^{x})^{2} & (\gamma-1)\hat{\beta}^{x}\hat{\beta}^{y} & (\gamma-1)\hat{\beta}^{x}\hat{\beta}^{z}\\
\gamma\beta^{y} & (\gamma-1)\hat{\beta}^{x}\hat{\beta}^{y} & 1+(\gamma-1)(\hat{\beta}^{y})^{2} & (\gamma-1)\hat{\beta}^{y}\hat{\beta}^{z}\\
\gamma\beta^{z} & (\gamma-1)\hat{\beta}^{x}\hat{\beta}^{z} & (\gamma-1)\hat{\beta}^{y}\hat{\beta}^{z} & 1+(\gamma-1)(\hat{\beta}^{z})^{2}
\end{array}\right),
\end{equation}
where $\beta^{x,y,z}$ are the three components of the 3-velocity
$\boldsymbol{\beta}$, $\hat{\beta}^{x,y,z}$ are the components of
the velocity direction $\boldsymbol{\beta}/\beta$, and the Lorentz
factor $\gamma=1/\sqrt{1-\beta^{2}}=E_{\mathbf{p}}/m$. The spinor
representation of this transformation is
\begin{equation}
\Lambda_{\frac{1}{2}}=\exp\left(\frac{1}{2}\gamma^{0}\boldsymbol{\gamma}\cdot\boldsymbol{\zeta}\right).\label{def:Lambda1/2}
\end{equation}
 Thus the Dirac field $\psi(p)$ can be written in terms of the field
in the particle's rest frame
\begin{equation}
\psi(p)=\Lambda_{\frac{1}{2}}\psi_{\text{rf}}.
\end{equation}

In the particle's rest frame, where the 3-momentum vanishes $\mathbf{p}=0$
and $E_{\mathbf{p}}=m$, the Dirac equation (\ref{eq:free Dirac eq in momentum space})
reads
\begin{equation}
\pm m\psi_{\text{rf}}=m\gamma^{0}\psi_{\text{rf}}.\label{eq:free Dirac eq in rest frame}
\end{equation}
Here we adopt the Weyl basis for the gamma matrices in Eq. (\ref{eq:gamma matrices-1}).
Then the wavefunctions for positive- and negative-energy states are
given by
\begin{equation}
\psi_{\text{rf},s}^{(+)}=\sqrt{m}\left(\begin{array}{c}
\xi_{s}\\
\xi_{s}
\end{array}\right),\ \ \psi_{\text{rf},s}^{(-)}=\sqrt{m}\left(\begin{array}{c}
\xi_{s}\\
-\xi_{s}
\end{array}\right),\label{sol:free wave function in rest frame}
\end{equation}
where $\xi_{s}$ are two-component spinors which satisfy the orthonormality
relation $\xi_{r}^{\dagger}\xi_{s}=\delta_{rs}$. We have introduced
a factor $\sqrt{m}$ in these solutions for convenience. The spinor
$\xi_{s}$ define the spin direction in the rest frame. For example,
$\xi=(1,0)^{T}$ corresponds to a spin-up state in the z-direction
and $\xi=\frac{1}{\sqrt{2}}(1,1)^{T}$ corresponds to a spin-up state
in the x-direction. Note that $\xi=\frac{1}{\sqrt{2}}(1,1)^{T}$ is
a superposition of $(1,0)^{T}$ and $(0,1)^{T}$, which respectively
represent the spin-up and spin-down states in the z-direction. Thus
we can choose $\xi_{+}=(1,0)^{T}$ and $\xi_{-}=(0,1)^{T}$ without
loss of generality and all possible spin configuration can be written
as a superposition of $\xi_{\pm}$. Generally, the spinors $\xi_{s}$
can be choosen as the eigenvectors of an arbitrary linear combination
of Pauli matrices. If we choose $\xi_{s}$ as the eigenvectors of
the $2\times2$ matrix $\mathbf{n}\cdot\boldsymbol{\sigma}$, then
$\psi_{\text{rf},s}^{(+)}$ represent fermions with spin parallel
($s=+$) or anti-parallel ($s=-$) to the vector $\mathbf{n}$ in
their rest frame, while $\psi_{\text{rf},s}^{(-)}$ represent anti-fermions
with spin parallel ($s=-$) or anti-parallel ($s=+$) to $\mathbf{n}$.

Then we boost from the particle's rest frame to the lab frame. Inserting
the gamma matrices (\ref{eq:gamma matrices-1}) and the rapidity vector
(\ref{def:rapidity vector}) into the definition of $\Lambda_{\frac{1}{2}}$
in Eq. (\ref{def:Lambda1/2}), we obtain
\begin{equation}
\Lambda_{\frac{1}{2}}=\exp\left[-\frac{1}{2\left|\mathbf{p}\right|}\left(\begin{array}{cc}
\boldsymbol{\sigma}\cdot\mathbf{p} & 0\\
0 & -\boldsymbol{\sigma}\cdot\mathbf{p}
\end{array}\right)\tanh^{-1}\frac{\left|\mathbf{p}\right|}{E_{\mathbf{p}}}\right],\label{eq:Lambda1/2}
\end{equation}
where $\boldsymbol{\sigma}$ are the Pauli matrices. Note that an
exponential of a matrix is defined as the Taylor expansion
\begin{equation}
\Lambda_{\frac{1}{2}}=\sum_{n=0}^{\infty}\frac{1}{n!}\left(-\frac{1}{2\left|\mathbf{p}\right|}\tanh^{-1}\frac{\left|\mathbf{p}\right|}{E_{\mathbf{p}}}\right)^{n}\left(\begin{array}{cc}
(\boldsymbol{\sigma}\cdot\mathbf{p})^{n} & 0\\
0 & (-\boldsymbol{\sigma}\cdot\mathbf{p})^{n}
\end{array}\right).\label{eq:another form of lambda1/2}
\end{equation}
In order to calculate $\Lambda_{\frac{1}{2}}$, we first focus on
the 2-dimensional matrix $\boldsymbol{\sigma}\cdot\mathbf{p}$. Note
that $\boldsymbol{\sigma}\cdot\mathbf{p}$ is Hermitian, which means
that it can be diagonalized. The normalized eigenstates of $\boldsymbol{\sigma}\cdot\mathbf{p}$
are given by
\begin{equation}
\frac{1}{\sqrt{2\left|\mathbf{p}\right|(p^{z}+\left|\mathbf{p}\right|)}}\left(\begin{array}{c}
p^{z}+\left|\mathbf{p}\right|\\
p^{x}+ip^{y}
\end{array}\right),\ \ \frac{1}{\sqrt{2\left|\mathbf{p}\right|(\left|\mathbf{p}\right|-p^{z})}}\left(\begin{array}{c}
p^{z}-\left|\mathbf{p}\right|\\
p^{x}+ip^{y}
\end{array}\right),
\end{equation}
which correspond to the eigenvalues $\pm\left|\mathbf{p}\right|$,
respectively. With these eigenvectors, one can define the following
transformation matrix $S_{\mathbf{p}}$ and its Hermitian conjugate,
\begin{equation}
S_{\mathbf{p}}=\left(\begin{array}{cc}
\frac{p^{z}+\left|\mathbf{p}\right|}{\sqrt{2\left|\mathbf{p}\right|(p^{z}+\left|\mathbf{p}\right|)}} & \frac{p^{z}-\left|\mathbf{p}\right|}{\sqrt{2\left|\mathbf{p}\right|(\left|\mathbf{p}\right|-p^{z})}}\\
\frac{p^{x}+ip^{y}}{\sqrt{2\left|\mathbf{p}\right|(p^{z}+\left|\mathbf{p}\right|)}} & \frac{p^{x}+ip^{y}}{\sqrt{2\left|\mathbf{p}\right|(\left|\mathbf{p}\right|-p^{z})}}
\end{array}\right),\ \ S_{\mathbf{p}}^{\dagger}=\left(\begin{array}{cc}
\frac{p^{z}+\left|\mathbf{p}\right|}{\sqrt{2\left|\mathbf{p}\right|(p^{z}+\left|\mathbf{p}\right|)}} & \frac{p^{x}-ip^{y}}{\sqrt{2\left|\mathbf{p}\right|(p^{z}+\left|\mathbf{p}\right|)}}\\
\frac{p^{z}-\left|\mathbf{p}\right|}{\sqrt{2\left|\mathbf{p}\right|(\left|\mathbf{p}\right|-p^{z})}} & \frac{p^{x}-ip^{y}}{\sqrt{2\left|\mathbf{p}\right|(\left|\mathbf{p}\right|-p^{z})}}
\end{array}\right).\label{def:transformation matrix Sp}
\end{equation}
The matrix $S_{\mathbf{p}}$ is a unitary matrix, i.e., its inverse
is equivalent to its Hermitian conjugate, $S_{\mathbf{p}}^{\dagger}S_{\mathbf{p}}=S_{\mathbf{p}}S_{\mathbf{p}}^{\dagger}=\mathbb{I}_{2}$.
The matrix $\boldsymbol{\sigma}\cdot\mathbf{p}$ is then diagonalized
as
\begin{equation}
\boldsymbol{\sigma}\cdot\mathbf{p}=\left|\mathbf{p}\right|S_{\mathbf{p}}\left(\begin{array}{cc}
1 & 0\\
0 & -1
\end{array}\right)S_{\mathbf{p}}^{\dagger}.\label{eq:diagonalization of sigma cdot p}
\end{equation}
With the help of Eq. (\ref{eq:diagonalization of sigma cdot p}),
we can calculate the $n$-th power of $\boldsymbol{\sigma}\cdot\mathbf{p}$,
\begin{equation}
\left(\boldsymbol{\sigma}\cdot\mathbf{p}\right)^{n}=\left|\mathbf{p}\right|^{n}S_{\mathbf{p}}\left(\begin{array}{cc}
1 & 0\\
0 & -1
\end{array}\right)^{n}S_{\mathbf{p}}^{\dagger}=\left|\mathbf{p}\right|^{n}S_{\mathbf{p}}\left(\begin{array}{cc}
1 & 0\\
0 & (-1)^{n}
\end{array}\right)S_{\mathbf{p}}^{\dagger}.
\end{equation}
We can calculate the terms in Eq. (\ref{eq:another form of lambda1/2})
\begin{eqnarray}
\sum_{n=0}^{\infty}\frac{1}{n!}\left(-\frac{1}{2\left|\mathbf{p}\right|}\tanh^{-1}\frac{\left|\mathbf{p}\right|}{E_{\mathbf{p}}}\right)^{n}(\boldsymbol{\sigma}\cdot\mathbf{p})^{n} & = & \frac{1}{\sqrt{m}}\sqrt{p_{\mu}\sigma^{\mu}},\nonumber \\
\sum_{n=0}^{\infty}\frac{1}{n!}\left(-\frac{1}{2\left|\mathbf{p}\right|}\tanh^{-1}\frac{\left|\mathbf{p}\right|}{E_{\mathbf{p}}}\right)^{n}(-\boldsymbol{\sigma}\cdot\mathbf{p})^{n} & = & \frac{1}{\sqrt{m}}\sqrt{p_{\mu}\bar{\sigma}^{\mu}},\label{eq:calculations of series}
\end{eqnarray}
where we have used
\begin{equation}
\sum_{n=0}^{\infty}\frac{1}{n!}\left(\pm\frac{1}{2}\tanh^{-1}\frac{\left|\mathbf{p}\right|}{E_{\mathbf{p}}}\right)^{n}=\exp\left[\pm\frac{1}{2}\tanh^{-1}\frac{\left|\mathbf{p}\right|}{E_{\mathbf{p}}}\right]=\frac{1}{\sqrt{m}}\sqrt{E_{\mathbf{p}}\pm\left|\mathbf{p}\right|},
\end{equation}
and introduced the following short-hand notations
\begin{equation}
\sqrt{p_{\mu}\sigma^{\mu}}=S_{\mathbf{p}}\left(\begin{array}{cc}
\sqrt{E_{\mathbf{p}}-\left|\mathbf{p}\right|} & 0\\
0 & \sqrt{E_{\mathbf{p}}+\left|\mathbf{p}\right|}
\end{array}\right)S_{\mathbf{p}}^{\dagger},\ \ \sqrt{p_{\mu}\bar{\sigma}^{\mu}}=S_{\mathbf{p}}\left(\begin{array}{cc}
\sqrt{E_{\mathbf{p}}+\left|\mathbf{p}\right|} & 0\\
0 & \sqrt{E_{\mathbf{p}}-\left|\mathbf{p}\right|}
\end{array}\right)S_{\mathbf{p}}^{\dagger}.\label{def:short notations}
\end{equation}
These matrices are real,
\begin{equation}
\left(\sqrt{p_{\mu}\sigma^{\mu}}\right)^{\dagger}=\sqrt{p_{\mu}\sigma^{\mu}},\ \ \left(\sqrt{p_{\mu}\bar{\sigma}^{\mu}}\right)^{\dagger}=\sqrt{p_{\mu}\bar{\sigma}^{\mu}},
\end{equation}
and satisfy following relations,
\begin{eqnarray}
 &  & \left(\sqrt{p_{\mu}\sigma^{\mu}}\right)^{2}=E_{\mathbf{p}}-\boldsymbol{\sigma}\cdot\mathbf{p},\ \ \left(\sqrt{p_{\mu}\bar{\sigma}^{\mu}}\right)^{2}=E_{\mathbf{p}}+\boldsymbol{\sigma}\cdot\mathbf{p},\nonumber \\
 &  & \sqrt{p_{\mu}\sigma^{\mu}}\sqrt{p_{\nu}\bar{\sigma}^{\nu}}=\sqrt{p_{\nu}\bar{\sigma}^{\nu}}\sqrt{p_{\mu}\sigma^{\mu}}=m,\label{eq:product of Sqrt p=00005Csigma}
\end{eqnarray}
which will be useful in checking the normalization relation of the
wavefunctions and calculating the Wigner function. Substituting the
Taylor series in Eq. (\ref{eq:another form of lambda1/2}) into Eq.
(\ref{eq:calculations of series}), the transformation matrix $\Lambda_{\frac{1}{2}}$
has the form
\begin{equation}
\Lambda_{\frac{1}{2}}=\frac{1}{\sqrt{m}}\left(\begin{array}{cc}
\sqrt{p_{\mu}\sigma^{\mu}} & 0\\
0 & \sqrt{p_{\mu}\bar{\sigma}^{\mu}}
\end{array}\right),\label{eq:lorentz trans lambda1/2}
\end{equation}
where $\sqrt{p_{\mu}\sigma^{\mu}}$ and $\sqrt{p_{\mu}\bar{\sigma}^{\mu}}$
are defined in Eq. (\ref{def:short notations}). Now we act with the
transformation matrix $\Lambda_{\frac{1}{2}}$ in Eq. (\ref{eq:lorentz trans lambda1/2})
onto the wavefunctions in the particle's rest frame to obtain the
wavefunctions in the lab frame %
\begin{equation}
\psi_{s}^{(+)}(\mathbf{p})=\left(\begin{array}{c}
\sqrt{p_{\mu}\sigma^{\mu}}\xi_{s}\\
\sqrt{p_{\mu}\bar{\sigma}^{\mu}}\xi_{s}
\end{array}\right),\ \ \psi_{s}^{(-)}(\mathbf{p})=\left(\begin{array}{c}
\sqrt{p_{\mu}\sigma^{\mu}}\xi_{s}\\
-\sqrt{p_{\mu}\bar{\sigma}^{\mu}}\xi_{s}
\end{array}\right).\label{sol:free wave functions}
\end{equation}
They are properly normalized, %
\begin{eqnarray}
\psi_{s}^{(+)\dagger}(\mathbf{p})\psi_{s^{\prime}}^{(+)}(\mathbf{p}) & = & 2E_{\mathbf{p}}\delta_{ss^{\prime}},\nonumber \\
\psi_{s}^{(-)\dagger}(\mathbf{p})\psi_{s^{\prime}}^{(-)}(\mathbf{p}) & = & 2E_{\mathbf{p}}\delta_{ss^{\prime}},\label{eq:normalization of psi pm}
\end{eqnarray}
and the positive-energy states are orthogonal to the negative-energy
ones,
\begin{equation}
\psi_{s}^{(+)\dagger}(\mathbf{p})\psi_{s^{\prime}}^{(-)}(-\mathbf{p})=\psi_{s}^{(-)\dagger}(\mathbf{p})\psi_{s^{\prime}}^{(+)}(-\mathbf{p})=0.\label{eq:orthogonality of psi pm}
\end{equation}
Although these solutions have already been obtained in many textbooks,
we have repeated the details in this thesis because we want to clarify
how to calculate the square root of a matrix, i.e., the terms $\sqrt{p_{\mu}\sigma^{\mu}}$
and $\sqrt{p_{\mu}\bar{\sigma}^{\mu}}$ in the solutions (\ref{sol:free wave functions}).
These details will help us in calculating the Wigner function in the
latter part of this subsection.

\subsubsection{Plane-wave quantization}

Using the single-particle wavefunction in Eq. (\ref{sol:free wave functions}),
the Dirac-field operator can be quantized as
\begin{equation}
\hat{\psi}(x)=\sum_{s}\int\frac{d^{3}\mathbf{p}}{(2\pi)^{3}\sqrt{2E_{\mathbf{p}}}}\left[e^{-iE_{\mathbf{p}}t+i\mathbf{p}\cdot\mathbf{x}}\psi_{s}^{(+)}(\mathbf{p})\hat{a}_{\mathbf{p},s}+e^{iE_{\mathbf{p}}t-i\mathbf{p}\cdot\mathbf{x}}\psi_{s}^{(-)}(\mathbf{p})\hat{b}_{\mathbf{p},s}^{\dagger}\right],\label{def:quantized free field}
\end{equation}
where $\hat{a}_{\mathbf{p},s}$ represents the annihilation operator
for a fermion with momentum $\mathbf{p}$ and spin $s$, and $\hat{b}_{\mathbf{p},s}^{\dagger}$
is the creation operator for an anti-fermion with the same quantum
numbers $\left\{ \mathbf{p},s\right\} $. Here the particle energy
is on the mass-shell $p^{2}=m^{2}$, thus we can rewrite the integration
over the 3-momentum $\mathbf{p}$ as a 4-dimensional covariant integration
over the 4-momentum $p^{\mu}$,
\begin{equation}
\hat{\psi}(x)=\sum_{s}\int\frac{d^{4}p}{(2\pi)^{3}}e^{-ip^{\mu}x_{\mu}}\delta(p^{2}-m^{2})\sqrt{2E_{\mathbf{p}}}\left[\theta(p^{0})\psi_{s}^{(+)}(\mathbf{p})\hat{a}_{\mathbf{p},s}+\theta(-p^{0})\psi_{s}^{(-)}(-\mathbf{p})\hat{b}_{-\mathbf{p},s}^{\dagger}\right],\label{def:covariant quantized free field}
\end{equation}
where we have used the following property of the delta-function,
\begin{equation}
\theta(\pm p^{0})\delta(p^{2}-m^{2})=\frac{1}{2E_{\mathbf{p}}}\delta(p^{0}\mp E_{\mathbf{p}}).
\end{equation}
We demand that the creation and annihilation operators satisfy the
following anti-commutation relations
\begin{equation}
\left\{ \hat{a}_{\mathbf{p},s},\,\hat{a}_{\mathbf{p}^{\prime},s^{\prime}}^{\dagger}\right\} =\left\{ \hat{b}_{\mathbf{p},s},\,\hat{b}_{\mathbf{p}^{\prime},s^{\prime}}^{\dagger}\right\} =(2\pi)^{3}\delta^{(3)}(\mathbf{p}-\mathbf{p}^{\prime})\delta_{ss^{\prime}},
\end{equation}
with all other anti-commutators vanishing,
\begin{equation}
\left\{ \hat{a}_{\mathbf{p},s},\,\hat{a}_{\mathbf{p}^{\prime},s^{\prime}}\right\} =\left\{ \hat{b}_{\mathbf{p},s},\,\hat{b}_{\mathbf{p}^{\prime},s^{\prime}}\right\} =\left\{ \hat{a}_{\mathbf{p},s},\,\hat{b}_{\mathbf{p}^{\prime},s^{\prime}}^{\dagger}\right\} =\left\{ \hat{b}_{\mathbf{p},s},\,\hat{a}_{\mathbf{p}^{\prime},s^{\prime}}^{\dagger}\right\} =0.
\end{equation}
Then it is easy to verify the equal-time anti-commutation relation
for the field operator %
\begin{equation}
\left\{ \hat{\psi}_{a}(t,\mathbf{x}),\hat{\psi}_{b}^{\dagger}(t,\mathbf{x}^{\prime})\right\} =\sum_{s}\int\frac{d^{3}\mathbf{p}}{(2\pi)^{3}2E_{\mathbf{p}}}e^{i\mathbf{p}\cdot(\mathbf{x}-\mathbf{x}^{\prime})}\left[u_{s,a}(\mathbf{p})u_{s,b}^{\dagger}(\mathbf{p})+v_{s,a}(\mathbf{p})v_{s,b}^{\dagger}(\mathbf{p})\right],
\end{equation}
where $a,b=1,2,3,4$ label components of $\hat{\psi}$ or $\hat{\psi}^{\dagger}$.
Inserting the explicit expressions for $\psi_{s}^{(+)}(\mathbf{p})$
and $\psi_{s}^{(-)}(\mathbf{p})$ in Eq. (\ref{sol:free wave functions})
into the above equation, one obtains %
\begin{equation}
\left\{ \hat{\psi}_{a}(t,\mathbf{x}),\hat{\psi}_{b}^{\dagger}(t,\mathbf{x}^{\prime})\right\} =\delta^{(3)}(\mathbf{x}-\mathbf{x}^{\prime})\delta_{ab},
\end{equation}
while the other anti-commutators are zero,
\begin{equation}
\left\{ \hat{\psi}_{a}(t,\mathbf{x}),\hat{\psi}_{b}(t,\mathbf{x}^{\prime})\right\} =\left\{ \hat{\psi}_{a}^{\dagger}(t,\mathbf{x}),\hat{\psi}_{b}^{\dagger}(t,\mathbf{x}^{\prime})\right\} =0.
\end{equation}
By checking these equal-time anti-commutators, we confirm that the
quantized field operator in Eq. (\ref{def:quantized free field})
has the correct property. A further calculation gives the Hamiltonian
operator,
\begin{equation}
\hat{H}=\int\frac{d^{3}\mathbf{p}}{(2\pi)^{3}}E_{\mathbf{p}}\sum_{s=\pm}\left(\hat{a}_{\mathbf{p},s}^{\dagger}\hat{a}_{\mathbf{p},s}+\hat{b}_{-\mathbf{p},s}^{\dagger}\hat{b}_{-\mathbf{p},s}-1\right),
\end{equation}
while the momentum operator is given by %
\begin{equation}
\hat{\mathbf{P}}=\int d^{3}\mathbf{x}\,\hat{\psi}^{\dagger}(-i\boldsymbol{\nabla}_{\mathbf{x}})\hat{\psi}=\sum_{s}\int\frac{d^{3}\mathbf{p}}{(2\pi)^{3}}\mathbf{p}\left(\hat{a}_{\mathbf{p},s}^{\dagger}\hat{a}_{\mathbf{p},s}-\hat{b}_{-\mathbf{p},s}^{\dagger}\hat{b}_{-\mathbf{p},s}\right).
\end{equation}
In deriving the Hamiltonian and momentum operators, we have used the
orthonormality relations in Eqs. (\ref{eq:normalization of psi pm})
and (\ref{eq:orthogonality of psi pm}). In quantum electrodynamics,
the Dirac spinor $\psi$ is used to describe spin-1/2 particles, such
as electrons and positrons. Here the spin is quantized along a given
direction, which is determined by the choice of Pauli spinors $\xi_{s}$
in (\ref{sol:free wave functions}). If we adopt the quantization
procedure in this subsection, the operator $\hat{a}_{\mathbf{p},s}^{\dagger}$
creates an electron with momentum $\mathbf{p}$ and spin parallel
($s=+$) or anti-parallel ($s=-$) to the spin quantization direction.
On the other hand, the operator $\hat{b}_{\mathbf{p},s}^{\dagger}$
creates a positron with momentum $\mathbf{p}$ and spin parallel ($s=-$)
or anti-parallel ($s=+$) to the spin quantization direction. The
interpretation of spin can be obtained via computing the spin angular
momentum operator.

\subsubsection{Wigner function}

In the previous parts of this subsection we have derived the plane-wave
solutions and quantized the Dirac-field in Eq. (\ref{def:covariant quantized free field}).
Inserting the field operator into the definition of the Wigner function
(\ref{def:Wigner function}), one obtains %
{}
\begin{eqnarray}
W(x,p) & = & \int\frac{d^{4}qd^{4}q^{\prime}}{(2\pi)^{6}}\sum_{ss^{\prime}}\exp\left[i(q^{\mu}-q^{\prime\mu})x_{\mu}\right]\sqrt{2E_{\mathbf{q}}}\sqrt{2E_{\mathbf{q}^{\prime}}}\nonumber \\
 &  & \times\delta^{(4)}\left(p_{\mu}-\frac{q_{\mu}+q_{\mu}^{\prime}}{2}\right)\delta(q^{\mu}q_{\mu}-m^{2})\delta(q^{\prime\mu}q_{\mu}^{\prime}-m^{2})\nonumber \\
 &  & \times\left\langle \Omega\left|\left[\theta(q^{0})\bar{\psi}_{s}^{(+)}(\mathbf{q})\hat{a}_{\mathbf{q},s}^{\dagger}+\theta(-q^{0})\bar{\psi}_{s}^{(-)}(-\mathbf{q})\hat{b}_{-\mathbf{q},s}\right]\right.\right.\nonumber \\
 &  & \left.\left.\otimes\left[\theta(q^{\prime0})\psi_{s^{\prime}}^{(+)}(\mathbf{q}^{\prime})\hat{a}_{\mathbf{q}^{\prime},s^{\prime}}+\theta(-q^{\prime0})\psi_{s^{\prime}}^{(-)}(-\mathbf{q}^{\prime})\hat{b}_{-\mathbf{q}^{\prime},s^{\prime}}^{\dagger}\right]\right|\Omega\right\rangle .\label{eq:free fermion Wigner function}
\end{eqnarray}
In the Wigner function two field operators are defined at different
space-time points, thus here after the Fourier transformations, we
have two momentum variables $q^{\mu}$ and $q^{\prime\mu}$. Then
we define the average and relative momentum as follows
\begin{equation}
k^{\mu}=\frac{1}{2}(q^{\mu}+q^{\prime\mu}),\ \ u^{\mu}=q^{\mu}-q^{\prime\mu},
\end{equation}
in terms of which can we express $q^{\mu}$ and $q^{\prime\mu}$,
\begin{equation}
q^{\mu}=k^{\mu}+\frac{1}{2}u^{\mu},\ \ q^{\prime\mu}=k^{\mu}-\frac{1}{2}u^{\mu}.
\end{equation}
Since the Jacobian for this substitution equals $1$, we have
\begin{equation}
d^{4}qd^{4}q^{\prime}=d^{4}kd^{4}u.
\end{equation}
Using the new variables $k^{\mu}$ and $u^{\mu}$, the delta functions
in Eq. (\ref{eq:free fermion Wigner function}) can be simplified
as %
\begin{eqnarray}
\delta\left(q^{\mu}q_{\mu}-m^{2}\right)\delta(\left(q^{\prime\mu}q_{\mu}^{\prime}-m^{2}\right) & = & \delta\left(k^{\mu}k_{\mu}+\frac{1}{4}u^{\mu}u_{\mu}-m^{2}+k^{\mu}u_{\mu}\right)\delta\left(k^{\mu}k_{\mu}+\frac{1}{4}u^{\mu}u_{\mu}-m^{2}-k^{\mu}u_{\mu}\right)\nonumber \\
 & = & \frac{1}{2}\delta\left(k^{\mu}k_{\mu}+\frac{1}{4}u^{\mu}u_{\mu}-m^{2}\right)\delta\left(k^{\mu}u_{\mu}\right),
\end{eqnarray}
On the other hand, we have to deal with the step functions in the
Wigner function (\ref{eq:free fermion Wigner function}). The product
of two step functions can be rewritten as
\begin{equation}
\theta(x)\theta(y)=\theta(x+y)\theta(x+y-|x-y|),
\end{equation}
So we obtain
\begin{eqnarray}
\theta(q^{0})\theta(q^{\prime0}) & = & \theta(k^{0})\theta\left(k^{0}-\left|\frac{1}{2}u^{0}\right|\right),\nonumber \\
\theta(-q^{0})\theta(-q^{\prime0}) & = & \theta(-k^{0})\theta\left(-k^{0}-\left|\frac{1}{2}u^{0}\right|\right),\nonumber \\
\theta(q^{0})\theta(-q^{\prime0}) & = & \theta(u^{0})\theta\left(\frac{1}{2}u^{0}-\left|k^{0}\right|\right),\nonumber \\
\theta(-q^{0})\theta(q^{\prime0}) & = & \theta(-u^{0})\theta\left(-\frac{1}{2}u^{0}-\left|k^{0}\right|\right).\label{eq:product of two theta functions}
\end{eqnarray}
Since $q^{\mu}$ and $q^{\prime\mu}$ are fixed on the mass-shell,
their zeroth component is $q^{0}=\pm E_{\mathbf{q}}$ and $q^{\prime0}=\pm E_{\mathbf{q}^{\prime}}$.
Thus we can check the following relations between the absolute values
of $\left|k^{0}\right|$ and $\left|u^{0}\right|$ %
\begin{equation}
\begin{cases}
\frac{1}{2}\left|u^{0}\right|<\left|k^{0}\right|, & \mathrm{sgn}(q^{0})\mathrm{sgn}(q^{\prime0})=1,\\
\frac{1}{2}\left|u^{0}\right|>\left|k^{0}\right|, & \mathrm{sgn}(q^{0})\mathrm{sgn}(q^{\prime0})=-1,
\end{cases}
\end{equation}
Using these relations we find that the products of two step functions
in Eq. (\ref{eq:product of two theta functions}) can be simplified
as
\begin{eqnarray}
\theta(q^{0})\theta(q^{\prime0}) & = & \theta(k^{0}),\nonumber \\
\theta(-q^{0})\theta(-q^{\prime0}) & = & \theta(-k^{0}),\nonumber \\
\theta(q^{0})\theta(-q^{\prime0}) & = & \theta(u^{0}),\nonumber \\
\theta(-q^{0})\theta(q^{\prime0}) & = & \theta(-u^{0}).
\end{eqnarray}
Using the new variables $k^{\mu}$, $u^{\mu}$, the Wigner function
can be put into the form %
\begin{eqnarray}
W(x,p) & = & \int\frac{d^{4}u}{(2\pi)^{6}}\sum_{ss^{\prime}}\exp\left(iu^{\mu}x_{\mu}\right)\delta\left(p^{\mu}p_{\mu}+\frac{1}{4}u^{\mu}u_{\mu}-m^{2}\right)\delta\left(p^{\mu}u_{\mu}\right)\sqrt{\left|p^{0}+\frac{1}{2}u^{0}\right|\left|p^{0}-\frac{1}{2}u^{0}\right|}\nonumber \\
 &  & \times\left[\theta(p^{0})\bar{\psi}_{s}^{(+)}\left(\mathbf{p}+\frac{1}{2}\mathbf{u}\right)\otimes\psi_{s^{\prime}}^{(+)}\left(\mathbf{p}-\frac{1}{2}\mathbf{u}\right)\left\langle \Omega\left|\hat{a}_{\mathbf{p}+\frac{1}{2}\mathbf{u},s}^{\dagger}\hat{a}_{\mathbf{p}-\frac{1}{2}\mathbf{u},s^{\prime}}\right|\Omega\right\rangle \right.\nonumber \\
 &  & \qquad+\theta(-p^{0})\bar{\psi}_{s}^{(-)}\left(-\mathbf{p}-\frac{1}{2}\mathbf{u}\right)\otimes\psi_{s^{\prime}}^{(-)}\left(-\mathbf{p}+\frac{1}{2}\mathbf{u}\right)\left\langle \Omega\left|\hat{b}_{-\mathbf{p}-\frac{1}{2}\mathbf{u},s}\hat{b}_{-\mathbf{p}+\frac{1}{2}\mathbf{u},s^{\prime}}^{\dagger}\right|\Omega\right\rangle \nonumber \\
 &  & \qquad+\theta(u^{0})\bar{\psi}_{s}^{(+)}\left(\mathbf{p}+\frac{1}{2}\mathbf{u}\right)\otimes\psi_{s^{\prime}}^{(-)}\left(-\mathbf{p}+\frac{1}{2}\mathbf{u}\right)\left\langle \Omega\left|\hat{a}_{\mathbf{p}+\frac{1}{2}\mathbf{u},s}^{\dagger}\hat{b}_{-\mathbf{p}+\frac{1}{2}\mathbf{u},s^{\prime}}^{\dagger}\right|\Omega\right\rangle \nonumber \\
 &  & \qquad\left.+\theta(-u^{0})\bar{\psi}_{s}^{(-)}\left(-\mathbf{p}-\frac{1}{2}\mathbf{u}\right)\otimes\psi_{s^{\prime}}^{(+)}\left(\mathbf{p}-\frac{1}{2}\mathbf{u}\right)\left\langle \Omega\left|\hat{b}_{-\mathbf{p}-\frac{1}{2}\mathbf{u},s}\hat{a}_{\mathbf{p}-\frac{1}{2}\mathbf{u},s^{\prime}}\right|\Omega\right\rangle \right].\nonumber \\
\label{eq:free Wigner function momentum form}
\end{eqnarray}
Note that the last two lines contribute if and only if there is mixture
between the fermion state and the anti-fermion state. Since we choose
to neglect collisions between particles, processes such as pair-production
or pair-annihilation are not included yet. Then the last two lines
in Eq. (\ref{eq:free Wigner function momentum form}) will be dropped
in future discussions.

We consider a fermionic system where eigenstates of fermions with
different momenta are not mixed together. Then the expectation values
in Eq. (\ref{eq:free Wigner function momentum form}) contain a delta
function of $\mathbf{u}$,
\begin{eqnarray}
\left\langle \Omega\left|\hat{a}_{\mathbf{p}+\frac{1}{2}\mathbf{u},s}^{\dagger}\hat{a}_{\mathbf{p}-\frac{1}{2}\mathbf{u},s^{\prime}}\right|\Omega\right\rangle  & = & (2\pi)^{3}\delta^{(3)}(\mathbf{u})f_{ss^{\prime}}^{(+)}(\mathbf{p}),\nonumber \\
\left\langle \Omega\left|\hat{b}_{-\mathbf{p}-\frac{1}{2}\mathbf{u},s}\hat{b}_{-\mathbf{p}+\frac{1}{2}\mathbf{u},s^{\prime}}^{\dagger}\right|\Omega\right\rangle  & = & (2\pi)^{3}\delta^{(3)}(\mathbf{u})\left[1-f_{s^{\prime}s}^{(-)}(-\mathbf{p})\right],
\end{eqnarray}
where $f_{ss^{\prime}}^{(+)}(\mathbf{p})$ and $f_{ss^{\prime}}^{(-)}(-\mathbf{p})$
are distribution functions of fermions and anti-fermions, respectively.
Since we do not consider any spin interaction, the energy states are
degenerate with respect to the spin direction. If the polarization
of the system is not parallel to the spin quantization direction,
these distribution functions are then not diagonal with respect to
$ss^{\prime}$. The delta function $\delta^{(3)}(\mathbf{u})$, together
with $\delta(p^{\mu}u_{\mu})$ in Eq. (\ref{eq:free Wigner function momentum form}),
gives a four-dimensional delta function,
\begin{equation}
\delta^{(3)}(\mathbf{u})\delta(p^{\mu}u_{\mu})=\delta^{(3)}(\mathbf{u})\delta(p^{0}u^{0})=\frac{1}{\left|p^{0}\right|}\delta^{(4)}(u).
\end{equation}
Thus we can carry out the integration over $d^{4}u$ and obtain the
following Wigner function
\begin{eqnarray}
 &  & W(x,p)=\frac{1}{(2\pi)^{3}}\sum_{ss^{\prime}}\delta(p^{\mu}p_{\mu}-m^{2})\nonumber \\
 &  & \qquad\times\left\{ \theta(p^{0})\bar{\psi}_{s}^{(+)}\left(\mathbf{p}\right)\otimes\psi_{s^{\prime}}^{(+)}\left(\mathbf{p}\right)f_{ss^{\prime}}^{(+)}(\mathbf{p})+\theta(-p^{0})\bar{\psi}_{s}^{(-)}\left(-\mathbf{p}\right)\otimes\psi_{s^{\prime}}^{(-)}\left(-\mathbf{p}\right)\left[1-f_{s^{\prime}s}^{(-)}(-\mathbf{p})\right]\right\} .\nonumber \\
\end{eqnarray}
In this formula, the Wigner function is independent of the space-time
coordinates $x^{\mu}$. This is because from the beginning we have
assumed that the fermions are described by plane waves, which are
homogeneous with respect to $x^{\mu}$.

As we discuss in Appendix \ref{sec:Wave-packet-description}, the
plane wave cannot describe a quantum particle which is located at
a given spatial point. According to the uncertainty principle, the
momentum uncertainty for the plane wave is zero, $\sigma_{p}=0$,
thus its conjugate variable, the uncertainty of spatial position is
infinity, $\sigma_{x}=\infty$. In order to introduce the $x$-dependence
into distribution functions, we adopt the wave-packet description,
as shown in Eq. (\ref{eq:wave packet state}). This wave packet describes
quantum particles at given center positions and average momenta. The
expectation value of $a_{\mathbf{p}+\frac{1}{2}\mathbf{u},s}^{\dagger}a_{\mathbf{p}-\frac{1}{2}\mathbf{u},s}$
in a wave-packet state is
\begin{eqnarray}
 &  & \left\langle \mathbf{p}^{\prime},s,+\left|\hat{a}_{\mathbf{p}+\frac{1}{2}\mathbf{u},s}^{\dagger}\hat{a}_{\mathbf{p}-\frac{1}{2}\mathbf{u},s^{\prime}}\right|\mathbf{p}^{\prime},s,+\right\rangle \nonumber \\
 & = & \frac{1}{N^{2}}\int\frac{d^{3}\mathbf{p}_{1}d^{3}\mathbf{p}_{2}}{(2\pi)^{6}}\exp\left[-\frac{(\mathbf{p}^{\prime}-\mathbf{p}_{1})^{2}+(\mathbf{p}^{\prime}-\mathbf{p}_{2})^{2}}{4\sigma_{p}^{2}}\right]\left\langle 0\left|\hat{a}_{\mathbf{p}_{2},s}\hat{a}_{\mathbf{p}+\frac{1}{2}\mathbf{u},s}^{\dagger}\hat{a}_{\mathbf{p}-\frac{1}{2}\mathbf{u},s}\hat{a}_{\mathbf{p}_{1},s}^{\dagger}\right|0\right\rangle \nonumber \\
 & = & \frac{1}{N^{2}}\int d^{3}\mathbf{p}_{1}d^{3}\mathbf{p}_{2}\exp\left[-\frac{(\mathbf{p}^{\prime}-\mathbf{p}_{1})^{2}+(\mathbf{p}^{\prime}-\mathbf{p}_{2})^{2}}{4\sigma_{p}^{2}}\right]\delta^{(3)}(\mathbf{p}+\frac{1}{2}\mathbf{u}-\mathbf{p}_{2})\delta^{(3)}(\mathbf{p}-\frac{1}{2}\mathbf{u}-\mathbf{p}_{1})\nonumber \\
 & = & \frac{1}{N^{2}}\exp\left[-\frac{(\mathbf{p}^{\prime}-\mathbf{p})^{2}+\frac{1}{4}\mathbf{u}^{2}}{2\sigma_{p}^{2}}\right].
\end{eqnarray}
In general, since the whole system is made of many wave packets, it
is reasonable to expect that the expectation values in Eq. (\ref{eq:free Wigner function momentum form})
are given by a distribution functions which depends on the parameters
$\mathbf{p}$, $\mathbf{u}$, $s$, and $s^{\prime}$,
\begin{eqnarray}
\left\langle \Omega\left|\hat{a}_{\mathbf{p}+\frac{1}{2}\mathbf{u},s}^{\dagger}\hat{a}_{\mathbf{p}-\frac{1}{2}\mathbf{u},s^{\prime}}\right|\Omega\right\rangle  & = & f_{ss^{\prime}}^{(+)}(\mathbf{p},\mathbf{u}),\nonumber \\
\left\langle \Omega\left|\hat{b}_{-\mathbf{p}-\frac{1}{2}\mathbf{u},s}\hat{b}_{-\mathbf{p}+\frac{1}{2}\mathbf{u},s^{\prime}}^{\dagger}\right|\Omega\right\rangle  & = & (2\pi)^{3}\delta^{(3)}(\mathbf{u})-f_{s^{\prime}s}^{(-)}(-\mathbf{p},\mathbf{u}),\label{eq:distribution w.r.t p and u}
\end{eqnarray}
where the first term in the second line, e.g. $(2\pi)^{3}\delta^{(3)}(\mathbf{u})$,
comes from the anti-commutator of $\hat{b}_{-\mathbf{p}-\frac{1}{2}\mathbf{u},s}$
and $\hat{b}_{-\mathbf{p}+\frac{1}{2}\mathbf{u},s^{\prime}}^{\dagger}$.
Here $f_{ss^{\prime}}^{(+)}(\mathbf{p},\mathbf{u})$ and $f_{s^{\prime}s}^{(-)}(-\mathbf{p},\mathbf{u})$
are functions determined by the state of the system $\left|\Omega\right\rangle $.
Inserting these expectation values back into the Wigner function (\ref{eq:free Wigner function momentum form})
we obtain
\begin{eqnarray}
W(x,p) & = & \int\frac{d^{4}u}{(2\pi)^{6}}\sum_{ss^{\prime}}\exp\left[iu^{\mu}x_{\mu}\right]\delta\left(p^{\mu}p_{\mu}+\frac{1}{4}u^{\mu}u_{\mu}-m^{2}\right)\delta\left(p^{\mu}u_{\mu}\right)\sqrt{(p^{0})^{2}-\frac{1}{4}(u^{0})^{2}}\nonumber \\
 &  & \times\left[\theta(p^{0})\bar{\psi}_{s}^{(+)}\left(\mathbf{p}+\frac{1}{2}\mathbf{u}\right)\otimes\psi_{s^{\prime}}^{(+)}\left(\mathbf{p}-\frac{1}{2}\mathbf{u}\right)f_{ss^{\prime}}^{(+)}(\mathbf{p},\mathbf{u}\right.)\nonumber \\
 &  & \qquad\left.-\theta(-p^{0})\bar{\psi}_{s}^{(-)}\left(-\mathbf{p}-\frac{1}{2}\mathbf{u}\right)\otimes\psi_{s^{\prime}}^{(-)}\left(-\mathbf{p}+\frac{1}{2}\mathbf{u}\right)f_{s^{\prime}s}^{(-)}(-\mathbf{p},\mathbf{u})\right]\nonumber \\
 &  & +\frac{1}{(2\pi)^{3}}\sum_{ss^{\prime}}\delta\left(p^{\mu}p_{\mu}-m^{2}\right)\theta(-p^{0})\bar{\psi}_{s}^{(-)}(-\mathbf{p})\otimes\psi_{s^{\prime}}^{(-)}(-\mathbf{p}).
\end{eqnarray}
Note that in general the uncertainty in momentum is small, which means
the spread of the wave packet in momentum space is not large. So we
can expect that the functions $f_{ss^{\prime}}^{(+)}(\mathbf{p},\mathbf{u})$
and $f_{s^{\prime}s}^{(-)}(-\mathbf{p},\mathbf{u})$ are narrow with
respect to $\mathbf{u}$. The Wigner function can then be expanded
in terms of the small variable $\mathbf{u}$ and higher-order terms
can be dropped. The wavefunction part is expanded as follows
\begin{eqnarray}
 &  & \bar{\psi}_{s}^{(+)}\left(\mathbf{p}+\frac{1}{2}\mathbf{u}\right)\otimes\psi_{s^{\prime}}^{(+)}\left(\mathbf{p}-\frac{1}{2}\mathbf{u}\right)\simeq\bar{\psi}_{s}^{(+)}(\mathbf{p})\otimes\psi_{s^{\prime}}^{(+)}(\mathbf{p})\nonumber \\
 &  & \qquad\qquad+\frac{1}{2}\mathbf{u}\cdot\left\{ \left[\boldsymbol{\nabla}_{\mathbf{p}}\bar{\psi}_{s}^{(+)}(\mathbf{p})\right]\otimes\psi_{s^{\prime}}^{(+)}(\mathbf{p})-\bar{\psi}_{s}^{(+)}(\mathbf{p})\otimes\boldsymbol{\nabla}_{\mathbf{p}}\psi_{s^{\prime}}^{(+)}(\mathbf{p})\right\} +\mathcal{O}(\mathbf{u}^{2}).\label{eq:u-expansion of wave function}
\end{eqnarray}
Inserting (\ref{eq:u-expansion of wave function}) into the Wigner
function, the leading-order term is
\begin{eqnarray}
W^{(0)}(x,p) & = & \frac{1}{(2\pi)^{3}}\delta(p^{\mu}p_{\mu}-m^{2})\sum_{ss^{\prime}}\left\{ \theta(p^{0})\bar{\psi}_{s}^{(+)}(\mathbf{p})\otimes\psi_{s^{\prime}}^{(+)}(\mathbf{p})f_{ss^{\prime}}^{(+)}(x,\mathbf{p})\right.\nonumber \\
 &  & \qquad\left.+\theta(-p^{0})\bar{\psi}_{s}^{(-)}(-\mathbf{p})\otimes\psi_{s^{\prime}}^{(-)}(-\mathbf{p})\left[1-f_{s^{\prime}s}^{(-)}(x,-\mathbf{p})\right]\right\} .\label{eq:Wigner function for free case}
\end{eqnarray}
On the other hand, the first-order correction in the expansion (\ref{eq:u-expansion of wave function})
contributes to the Wigner function as
\begin{eqnarray}
W^{(1)}(x,p) & = & \frac{1}{2}\delta(p^{\mu}p_{\mu}-m^{2})\theta(p^{0})\sum_{ss^{\prime}}\nonumber \\
 &  & \qquad\times\left\{ \left[\boldsymbol{\nabla}_{\mathbf{p}}\bar{\psi}_{s}^{(+)}(\mathbf{p})\right]\otimes\psi_{s^{\prime}}^{(+)}(\mathbf{p})-\bar{\psi}_{s}^{(+)}(\mathbf{p})\otimes\boldsymbol{\nabla}_{\mathbf{p}}\psi_{s^{\prime}}^{(+)}(\mathbf{p})\right\} \cdot i\boldsymbol{\nabla}_{\mathbf{x}}f_{ss^{\prime}}^{(+)}(x,\mathbf{p})\nonumber \\
 &  & +\frac{1}{2}\delta(p^{\mu}p_{\mu}-m^{2})\theta(-p^{0})\sum_{ss^{\prime}}\nonumber \\
 &  & \qquad\times\left\{ \left[\boldsymbol{\nabla}_{\mathbf{p}}\bar{\psi}_{s}^{(-)}(-\mathbf{p})\right]\otimes\psi_{s^{\prime}}^{(-)}(-\mathbf{p})-\bar{\psi}_{s}^{(-)}(-\mathbf{p})\otimes\boldsymbol{\nabla}_{\mathbf{p}}\psi_{s^{\prime}}^{(-)}(-\mathbf{p})\right\} \cdot i\boldsymbol{\nabla}_{\mathbf{x}}f_{s^{\prime}s}^{(-)}(x,-\mathbf{p}).\nonumber \\
\label{eq:free Wigner function first order}
\end{eqnarray}
Here we have defined the semi-distribution functions,
\begin{equation}
f_{ss^{\prime}}^{(\pm)}(x,\mathbf{p})=\int\frac{d^{4}u}{(2\pi)^{3}}\delta\left(u^{0}-\frac{\mathbf{p}\cdot\mathbf{u}}{p^{0}}\right)f_{ss^{\prime}}^{(\pm)}(\mathbf{p},\mathbf{u})\exp\left(iu^{\mu}x_{\mu}\right).\label{eq:local particle distribution}
\end{equation}
In the first-order part, we have replaced $\mathbf{u}f_{ss^{\prime}}^{(\pm)}(x,\mathbf{p})$
by the spatial derivative $i\boldsymbol{\nabla}_{\mathbf{x}}f_{s^{\prime}s}^{(\pm)}(x,\mathbf{p})$,
thus $W^{(1)}(x,p)$ is of first order in the spatial gradients of
the distributions $f_{ss^{\prime}}^{(\pm)}(x,\mathbf{p})$. If we
consider classical particles, the semi-distributions $f_{ss^{\prime}}^{(\pm)}(x,\mathbf{p})$
can be interpreted as the classical distributions of fermions or anti-fermions
at the phase space point $\left\{ t,\mathbf{x},\mathbf{p}\right\} $.
Making a comparison with the results from the semi-classical expansion,
which will be done in Sec. \ref{sec:Semi-classical-expansion}, we
identify $W^{(0)}(x,p)$ as of zeroth order in $\hbar$ and $W^{(1)}(x,p)$
as of first order in $\hbar$. The first order contribution $W^{(1)}(x,p)$
can be calculated using Eq. (\ref{eq:free Wigner function first order}),
since the wavefunctions in this equation have already been derived.
But actually the calculation is too complicated, so that, in the following
part of this subsection, we will only compute the leading order contribution
$W^{(0)}(x,p)$.

\subsubsection{Components of the Wigner function}

The Wigner function can be calculated via inserting Eq. (\ref{sol:free wave functions})
into Eq. (\ref{eq:Wigner function for free case}). In the following
we will decompose the Wigner function as shown in Eq. (\ref{def:Wigner function decomposition})
and compute the different components.

Before we do so, we first discuss the transformation matrix $S_{\mathbf{p}}$
in Eq. (\ref{def:transformation matrix Sp}), which is useful because
the matrices $\sqrt{p_{\mu}\bar{\sigma}^{\mu}}$ and $\sqrt{p_{\mu}\sigma^{\mu}}$
in the wavefunction are defined with $S_{\mathbf{p}}$ as shown in
Eq. (\ref{def:short notations}). Under the transformation $S_{\mathbf{p}}$,
the Pauli matrices transform as %
\begin{eqnarray}
S_{\mathbf{p}}^{\dagger}\sigma^{x}S_{\mathbf{p}} & = & \frac{1}{\left|\mathbf{p}\right|}\left[\frac{p^{x}p^{z}}{\sqrt{(p^{x})^{2}+(p^{y})^{2}}}\sigma^{x}-\frac{p^{y}\left|\mathbf{p}\right|}{\sqrt{(p^{x})^{2}+(p^{y})^{2}}}\sigma^{y}+p^{x}\sigma^{z}\right],\nonumber \\
S_{\mathbf{p}}^{\dagger}\sigma^{y}S_{\mathbf{p}} & = & \frac{1}{\left|\mathbf{p}\right|}\left[\frac{p^{y}p^{z}}{\sqrt{(p^{x})^{2}+(p^{y})^{2}}}\sigma^{x}+\frac{p^{x}\left|\mathbf{p}\right|}{\sqrt{(p^{x})^{2}+(p^{y})^{2}}}\sigma^{y}+p^{y}\sigma^{z}\right],\nonumber \\
S_{\mathbf{p}}^{\dagger}\sigma^{z}S_{\mathbf{p}} & = & \frac{1}{\left|\mathbf{p}\right|}\left[-\sqrt{(p^{x})^{2}+(p^{y})^{2}}\sigma^{x}+p^{z}\sigma^{z}\right].\label{eq:S^dagger sigma S}
\end{eqnarray}
Multiplying each equation by $S_{\mathbf{p}}$ on the left and $S_{\mathbf{p}}^{\dagger}$
on the right, and using $S_{\mathbf{p}}^{\dagger}S_{\mathbf{p}}=S_{\mathbf{p}}S_{\mathbf{p}}^{\dagger}=\mathbb{I}_{2}$,
one obtains
\begin{eqnarray}
\sigma^{x} & = & \frac{1}{\left|\mathbf{p}\right|}\left[\frac{p^{x}p^{z}}{\sqrt{(p^{x})^{2}+(p^{y})^{2}}}S_{\mathbf{p}}\sigma^{x}S_{\mathbf{p}}^{\dagger}-\frac{p^{y}\left|\mathbf{p}\right|}{\sqrt{(p^{x})^{2}+(p^{y})^{2}}}S_{\mathbf{p}}\sigma^{y}S_{\mathbf{p}}^{\dagger}+p^{x}S_{\mathbf{p}}\sigma^{z}S_{\mathbf{p}}^{\dagger}\right],\nonumber \\
\sigma^{y} & = & \frac{1}{\left|\mathbf{p}\right|}\left[\frac{p^{y}p^{z}}{\sqrt{(p^{x})^{2}+(p^{y})^{2}}}S_{\mathbf{p}}\sigma^{x}S_{\mathbf{p}}^{\dagger}+\frac{p^{x}\left|\mathbf{p}\right|}{\sqrt{(p^{x})^{2}+(p^{y})^{2}}}S_{\mathbf{p}}\sigma^{y}S_{\mathbf{p}}^{\dagger}+p^{y}S_{\mathbf{p}}\sigma^{z}S_{\mathbf{p}}^{\dagger}\right],\nonumber \\
\sigma^{z} & = & \frac{1}{\left|\mathbf{p}\right|}\left[-\sqrt{(p^{x})^{2}+(p^{y})^{2}}S_{\mathbf{p}}\sigma^{x}S_{\mathbf{p}}^{\dagger}+p^{z}S_{\mathbf{p}}\sigma^{z}S_{\mathbf{p}}^{\dagger}\right].
\end{eqnarray}
From this we can obtain that
\begin{eqnarray}
S_{\mathbf{p}}\sigma^{x}S_{\mathbf{p}}^{\dagger} & = & p^{z}\frac{\mathbf{p}\cdot\boldsymbol{\sigma}}{\left|\mathbf{p}\right|\sqrt{(p^{x})^{2}+(p^{y})^{2}}}-\frac{\left|\mathbf{p}\right|}{\sqrt{(p^{x})^{2}+(p^{y})^{2}}}\sigma^{z},\nonumber \\
S_{\mathbf{p}}\sigma^{y}S_{\mathbf{p}}^{\dagger} & = & -\frac{\sigma^{x}p^{y}-\sigma^{y}p^{x}}{\sqrt{(p^{x})^{2}+(p^{y})^{2}}},\nonumber \\
S_{\mathbf{p}}\sigma^{z}S_{\mathbf{p}}^{\dagger} & = & \frac{\mathbf{p}\cdot\boldsymbol{\sigma}}{\left|\mathbf{p}\right|}.\label{eq:S sigma S^dagger}
\end{eqnarray}
These properties for the Pauli matrices will help us when we compute
the axial-vector and tensor components of the Wigner function.

The Wigner function in Eq. (\ref{eq:Wigner function for free case})
will now be decomposed in terms of the generators of the Clifford
algebra $\Gamma_{i}=\{1,-i\gamma^{5},\gamma^{\mu},\gamma^{\mu}\gamma^{5},\sigma^{\mu\nu}\}$
as in Eq. (\ref{def:Wigner function decomposition}). The expansion
coefficients are calculated via Eq. (\ref{eq:reproduce components of Wigner funtion}),
and the traces in Eq. (\ref{eq:reproduce components of Wigner funtion})
are given by
\begin{eqnarray}
\text{Tr}\left[\Gamma_{i}W^{(0)}(x,p)\right] & = & \frac{1}{(2\pi)^{3}}\delta(p^{\mu}p_{\mu}-m^{2})\sum_{ss^{\prime}}\left\{ \theta(p^{0})\bar{\psi}_{s}^{(+)}(\mathbf{p})\Gamma_{i}\psi_{s^{\prime}}^{(+)}(\mathbf{p})f_{ss^{\prime}}^{(+)}(x,\mathbf{p})\right.\nonumber \\
 &  & \qquad\left.+\theta(-p^{0})\bar{\psi}_{s}^{(-)}(-\mathbf{p})\Gamma_{i}\psi_{s^{\prime}}^{(-)}(-\mathbf{p})\left[1-f_{s^{\prime}s}^{(-)}(x,-\mathbf{p})\right]\right\} .\label{eq:Wigner function for free case-1}
\end{eqnarray}
We observe that Eq. (\ref{eq:Wigner function for free case-1}) consists
of a fermion part and an anti-fermion part. We first focus on the
fermion part and then the anti-fermion part can be derived in the
same way. The key point is to calculate
\begin{equation}
\bar{\psi}_{s}^{(+)}(\mathbf{p})\Gamma_{i}\psi_{s^{\prime}}^{(+)}(\mathbf{p}),\label{eq:psibar Gamma psi}
\end{equation}
where $\psi_{s}^{(+)}(\mathbf{p})$ is the single particle wavefunction
in momentum space, which is given in Eq. (\ref{sol:free wave functions}).
The scalar and pseudoscalar parts can be derived directly by inserting
Eq. (\ref{sol:free wave functions}) into Eq. (\ref{eq:psibar Gamma psi})
and using the relations (\ref{eq:product of Sqrt p=00005Csigma}),%
\begin{eqnarray}
\bar{\psi}_{s}^{(+)}(\mathbf{p})\psi_{s^{\prime}}^{(+)}(\mathbf{p}) & = & 2m\delta_{ss^{\prime}},\nonumber \\
-i\bar{\psi}_{s}^{(+)}(\mathbf{p})\gamma^{5}\psi_{s^{\prime}}^{(+)}(\mathbf{p}) & = & 0.
\end{eqnarray}
Now we focus on the vector part, the calculation of the zeroth component
is straightforward %
\begin{equation}
\bar{\psi}_{s}^{(+)}(\mathbf{p})\gamma^{0}\psi_{s^{\prime}}^{(+)}(\mathbf{p})=2E_{\mathbf{p}}\delta_{ss^{\prime}},
\end{equation}
while the spatial components read
\begin{eqnarray}
\bar{\psi}_{s}^{(+)}(\mathbf{p})\boldsymbol{\gamma}\psi_{s^{\prime}}^{(+)}(\mathbf{p}) & = & \xi_{s}^{\dagger}\left(\sqrt{p_{\mu}\bar{\sigma}^{\mu}}\boldsymbol{\sigma}\sqrt{p_{\mu}\bar{\sigma}^{\mu}}-\sqrt{p_{\mu}\sigma^{\mu}}\boldsymbol{\sigma}\sqrt{p_{\mu}\sigma^{\mu}}\right)\xi_{s}\nonumber \\
 & = & 2\xi_{s}^{\dagger}S_{\mathbf{p}}\left(\begin{array}{cc}
\left|\mathbf{p}\right|(S_{\mathbf{p}}^{\dagger}\boldsymbol{\sigma}S_{\mathbf{p}})_{11} & 0\\
0 & -\left|\mathbf{p}\right|(S_{\mathbf{p}}^{\dagger}\boldsymbol{\sigma}S_{\mathbf{p}})_{22}
\end{array}\right)S_{\mathbf{p}}^{\dagger}\xi_{s}.\label{eq:vector component of free fermion}
\end{eqnarray}
In the last step of Eq. (\ref{eq:vector component of free fermion})
we have used the definitions in (\ref{def:short notations}) for $\sqrt{p_{\mu}\bar{\sigma}^{\mu}}$
and $\sqrt{p_{\mu}\sigma^{\mu}}$. Here $S_{\mathbf{p}}^{\dagger}\boldsymbol{\sigma}S_{\mathbf{p}}$
is a $2\times2$ matrix and the subscript labels different elements
of this matrix. From the transformation properties of the Pauli matrices
in Eq. (\ref{eq:S^dagger sigma S}) we obtain
\begin{equation}
(S_{\mathbf{p}}^{\dagger}\boldsymbol{\sigma}S_{\mathbf{p}})_{11}=-(S_{\mathbf{p}}^{\dagger}\boldsymbol{\sigma}S_{\mathbf{p}})_{22}=\frac{\mathbf{p}}{\left|\mathbf{p}\right|}.
\end{equation}
Inserting this into Eq. (\ref{eq:vector component of free fermion}),
we have
\begin{equation}
\bar{\psi}_{s}^{(+)}(\mathbf{p})\boldsymbol{\gamma}\psi_{s^{\prime}}^{(+)}(\mathbf{p})=2\mathbf{p}\delta_{ss^{\prime}}.
\end{equation}
On the other hand, the axial-vector components of Eq. (\ref{eq:psibar Gamma psi})
are given by %
\begin{eqnarray}
\bar{\psi}_{s}^{(+)}(\mathbf{p})\gamma^{0}\gamma^{5}\psi_{s^{\prime}}^{(+)}(\mathbf{p}) & = & 2\xi_{s}^{\dagger}\mathbf{p}\cdot\boldsymbol{\sigma}\xi_{s^{\prime}},\nonumber \\
\bar{\psi}_{s}^{(+)}(\mathbf{p})\boldsymbol{\gamma}\gamma^{5}\psi_{s^{\prime}}^{(+)}(\mathbf{p}) & = & 2\xi_{s}^{\dagger}S_{\mathbf{p}}\left(\begin{array}{cc}
E_{\mathbf{p}}(S_{\mathbf{p}}^{\dagger}\boldsymbol{\sigma}S_{\mathbf{p}})_{11} & m(S_{\mathbf{p}}^{\dagger}\boldsymbol{\sigma}S_{\mathbf{p}})_{12}\\
m(S_{\mathbf{p}}^{\dagger}\boldsymbol{\sigma}S_{\mathbf{p}})_{21} & E_{\mathbf{p}}(S_{\mathbf{p}}^{\dagger}\boldsymbol{\sigma}S_{\mathbf{p}})_{22}
\end{array}\right)S_{\mathbf{p}}^{\dagger}\xi_{s^{\prime}}.
\end{eqnarray}
Since we do not have a universal formula for $S_{\mathbf{p}}^{\dagger}\boldsymbol{\sigma}S_{\mathbf{p}}$,
we have to calculate different components one by one. The computations
are straightforward using the explicit expressions of $S_{\mathbf{p}}^{\dagger}\boldsymbol{\sigma}S_{\mathbf{p}}$
and $S_{\mathbf{p}}\boldsymbol{\sigma}S_{\mathbf{p}}^{\dagger}$ in
Eqs. (\ref{eq:S^dagger sigma S}) and (\ref{eq:S sigma S^dagger}).
The final result reads,%
\begin{equation}
\bar{\psi}_{s}^{(+)}(\mathbf{p})\boldsymbol{\gamma}\gamma^{5}\psi_{s^{\prime}}^{(+)}(\mathbf{p})=2\xi_{s}^{\dagger}\left(m\boldsymbol{\sigma}+\frac{\mathbf{p}\cdot\boldsymbol{\sigma}}{E_{\mathbf{p}}+m}\mathbf{p}\right)\xi_{s^{\prime}}.
\end{equation}
The tensor component of Eq. (\ref{eq:psibar Gamma psi}) is given
by %
\begin{eqnarray}
\bar{\psi}_{s}^{(+)}(\mathbf{p})\sigma^{0i}\psi_{s^{\prime}}^{(+)}(\mathbf{p}) & = & 2i\xi_{s}^{\dagger}S_{\mathbf{p}}\left(\begin{array}{cc}
0 & -\left|\mathbf{p}\right|(S_{\mathbf{p}}^{\dagger}\sigma^{i}S_{\mathbf{p}})_{12}\\
\left|\mathbf{p}\right|(S_{\mathbf{p}}^{\dagger}\sigma^{i}S_{\mathbf{p}})_{21} & 0
\end{array}\right)S_{\mathbf{p}}^{\dagger}\xi_{s^{\prime}},\nonumber \\
\bar{\psi}_{s}^{(+)}(\mathbf{p})\sigma^{ij}\psi_{s^{\prime}}^{(+)}(\mathbf{p}) & = & 2\epsilon^{ijk}\xi_{s}^{\dagger}S_{\mathbf{p}}\left(\begin{array}{cc}
m(S_{\mathbf{p}}^{\dagger}\sigma^{k}S_{\mathbf{p}})_{11} & E_{\mathbf{p}}(S_{\mathbf{p}}^{\dagger}\sigma^{k}S_{\mathbf{p}})_{12}\\
E_{\mathbf{p}}(S_{\mathbf{p}}^{\dagger}\sigma^{k}S_{\mathbf{p}})_{21} & m(S_{\mathbf{p}}^{\dagger}\sigma^{k}S_{\mathbf{p}})_{22}
\end{array}\right)S_{\mathbf{p}}^{\dagger}\xi_{s^{\prime}},
\end{eqnarray}
where $i,j,k=1,2,3$. Again these terms are calculated using the properties
in Eqs. (\ref{eq:S^dagger sigma S}) and (\ref{eq:S sigma S^dagger})
and the results are %
\begin{equation}
\bar{\psi}_{s}^{(+)}(\mathbf{p})\sigma^{0i}\psi_{s^{\prime}}^{(+)}(\mathbf{p})=-2\epsilon^{ijk}p^{j}\xi_{s}^{\dagger}\sigma^{k}\xi_{s^{\prime}},
\end{equation}
and
\begin{equation}
\bar{\psi}_{s}^{(+)}(\mathbf{p})\sigma^{ij}\psi_{s^{\prime}}^{(+)}(\mathbf{p})=2\epsilon^{ijk}\left(E_{\mathbf{p}}\xi_{s}^{\dagger}\sigma^{k}\xi_{s}-\frac{p^{k}}{E_{\mathbf{p}}+m}\xi_{s}^{\dagger}\mathbf{p}\cdot\boldsymbol{\sigma}\xi_{s}\right).
\end{equation}
As a conclusion, we now collect all results from the above calculations,
where we have written the vector, axial-vector, and tensor components
in a covariant form,
\begin{eqnarray}
\bar{\psi}_{s}^{(+)}(\mathbf{p})\psi_{s^{\prime}}^{(+)}(\mathbf{p}) & = & 2m\delta_{ss^{\prime}},\nonumber \\
-i\bar{\psi}_{s}^{(+)}(\mathbf{p})\gamma^{5}\psi_{s^{\prime}}^{(+)}(\mathbf{p}) & = & 0,\nonumber \\
\bar{\psi}_{s}^{(-)}(\mathbf{p})\gamma^{\mu}\psi_{s^{\prime}}^{(-)}(\mathbf{p}) & = & 2p^{\mu}\delta_{ss^{\prime}},\nonumber \\
\bar{\psi}_{s}^{(+)}(\mathbf{p})\gamma^{\mu}\gamma^{5}\psi_{s^{\prime}}^{(+)}(\mathbf{p}) & = & 2\xi_{s}^{\dagger}n^{\mu}(\mathbf{p})\xi_{s^{\prime}},\nonumber \\
\bar{\psi}_{s}^{(+)}(\mathbf{p})\sigma^{\mu\nu}\psi_{s^{\prime}}^{(+)}(\mathbf{p}) & = & -\frac{2}{m}\epsilon^{\mu\nu\alpha\beta}p_{\alpha}\xi_{s}^{\dagger}n_{\beta}(\mathbf{p})\xi_{s^{\prime}},\label{eq:fermion barpsi Gamma psi}
\end{eqnarray}
where $p^{0}=E_{\mathbf{p}}$ is the on-shell energy and we have defined
a vector for the spin polarization
\begin{equation}
n^{\mu}(\mathbf{p})\equiv\left(\mathbf{p}\cdot\boldsymbol{\sigma},\ m\boldsymbol{\sigma}+\frac{\mathbf{p}\cdot\boldsymbol{\sigma}}{E_{\mathbf{p}}+m}\mathbf{p}\right)^{T}.\label{eq:polarization operator}
\end{equation}
In spin space, the scalar, pseudoscalar and vector parts are diagonalized,
while the axial-vector and tensor parts depend on $\xi_{s}^{\dagger}n^{\mu}(\mathbf{p})\xi_{s^{\prime}}$,
which is in general not diagonal. This is because generally the spin
quantization direction is different from the spin polarization direction.
In the last part of this section we will discuss the effect of different
choices for the spin quantization direction.

The antiparticle contributions can be computed repeating above calculations.
An easier way is to use the relation between the particle and antiparticle
wavefunctions
\begin{equation}
\psi_{s}^{(-)}(\mathbf{p})=-\gamma^{5}\psi_{s}^{(+)}(\mathbf{p}),
\end{equation}
so we have
\begin{equation}
\bar{\psi}_{s}^{(-)}(-\mathbf{p})\Gamma_{i}\psi_{s^{\prime}}^{(-)}(-\mathbf{p})=-\bar{\psi}_{s}^{(+)}(-\mathbf{p})\gamma^{5}\Gamma_{i}\gamma^{5}\psi_{s^{\prime}}^{(+)}(-\mathbf{p}),
\end{equation}
Substitute $\Gamma_{i}$ with different matrices $\{1,-i\gamma^{5},\gamma^{\mu},\gamma^{\mu}\gamma^{5},\sigma^{\mu\nu}\}$,
we obtain
\begin{eqnarray}
\bar{\psi}_{s}^{(-)}(-\mathbf{p})\psi_{s^{\prime}}^{(-)}(-\mathbf{p}) & = & -2m\delta_{ss^{\prime}},\nonumber \\
-i\bar{\psi}_{s}^{(-)}(\mathbf{p})\gamma^{5}\psi_{s^{\prime}}^{(-)}(-\mathbf{p}) & = & 0,\nonumber \\
\bar{\psi}_{s}^{(-)}(-\mathbf{p})\gamma^{\mu}\psi_{s^{\prime}}^{(-)}(-\mathbf{p}) & = & -2p^{\mu}\delta_{ss^{\prime}},\nonumber \\
\bar{\psi}_{s}^{(-)}(-\mathbf{p})\gamma^{\mu}\gamma^{5}\psi_{s^{\prime}}^{(-)}(-\mathbf{p}) & = & 2\xi_{s}^{\dagger}\hat{n}^{\mu}(-\mathbf{p})\xi_{s^{\prime}},\nonumber \\
\bar{\psi}_{s}^{(-)}(-\mathbf{p})\sigma^{\mu\nu}\psi_{s^{\prime}}^{(-)}(-\mathbf{p}) & = & -\frac{2}{m}\epsilon^{\mu\nu\alpha\beta}p_{\alpha}\xi_{s}^{\dagger}\hat{n}_{\beta}(-\mathbf{p})\xi_{s^{\prime}},\label{eq:antifermion barpsi Gamma psi}
\end{eqnarray}
where $p^{0}=-E_{\mathbf{p}}$ is the on-shell energy for the anti-fermions.

Inserting the fermion contributions in (\ref{eq:fermion barpsi Gamma psi})
and the anti-fermion contributions in (\ref{eq:antifermion barpsi Gamma psi})
into Eq. (\ref{eq:Wigner function for free case-1}), we derive different
components of the Wigner function,
\begin{eqnarray}
\mathcal{F}^{(0)} & = & m\frac{2\delta(p^{2}-m^{2})}{(2\pi)^{3}}\sum_{s}\left\{ \theta(p^{0})f_{ss}^{(+)}(x,\mathbf{p})-\theta(-p^{0})\left[1-f_{ss}^{(-)}(x,-\mathbf{p})\right]\right\} ,\nonumber \\
\mathcal{P}^{(0)} & = & 0,\nonumber \\
\mathcal{V}_{\mu}^{(0)} & = & p_{\mu}\frac{2\delta(p^{2}-m^{2})}{(2\pi)^{3}}\sum_{s}\left\{ \theta(p^{0})f_{ss}^{(+)}(x,\mathbf{p})-\theta(-p^{0})\left[1-f_{ss}^{(-)}(x,-\mathbf{p})\right]\right\} ,\nonumber \\
\mathcal{A}_{\mu}^{(0)} & = & \frac{2\delta(p^{2}-m^{2})}{(2\pi)^{3}}\sum_{ss^{\prime}}\left\{ \theta(p^{0})\xi_{s}^{\dagger}n^{\mu}(\mathbf{p})\xi_{s^{\prime}}f_{ss^{\prime}}^{(+)}(x,\mathbf{p})+\theta(-p^{0})\xi_{s}^{\dagger}n^{\mu}(-\mathbf{p})\xi_{s^{\prime}}\left[1-f_{s^{\prime}s}^{(-)}(x,-\mathbf{p})\right]\right\} ,\nonumber \\
\mathcal{S}_{\mu\nu}^{(0)} & = & -\frac{2\delta(p^{2}-m^{2})}{(2\pi)^{3}}\frac{1}{m}\epsilon_{\mu\nu\alpha\beta}p^{\alpha}\nonumber \\
 &  & \times\sum_{ss^{\prime}}\left\{ \theta(p^{0})\xi_{s}^{\dagger}n^{\beta}(\mathbf{p})\xi_{s^{\prime}}f_{ss^{\prime}}^{(+)}(x,\mathbf{p})+\theta(-p^{0})\xi_{s}^{\dagger}n^{\beta}(-\mathbf{p})\xi_{s^{\prime}}\left[1-f_{s^{\prime}s}^{(-)}(x,-\mathbf{p})\right]\right\} .
\end{eqnarray}
The Wigner function is then recovered by Eq. (\ref{def:Wigner function decomposition}).
Note that the above results are of zeroth order in spatial gradients
of the distribution function. Higher-order terms in spatial gradients
are calculated via Eq. (\ref{eq:free Wigner function first order})
but will not be done here, because it is too complicated. Now we define
functions which can be interpreted as the net fermion density and
polarization, respectively,
\begin{eqnarray}
V^{(0)}(x,p) & \equiv & \frac{2}{(2\pi)^{3}}\sum_{s}\left\{ \theta(p^{0})f_{ss}^{(+)}(x,\mathbf{p})-\theta(-p^{0})\left[1-f_{ss}^{(-)}(x,-\mathbf{p})\right]\right\} ,\nonumber \\
n^{(0)\mu}(x,p) & \equiv & \frac{2}{(2\pi)^{3}}\sum_{ss^{\prime}}\left\{ \theta(p^{0})\xi_{s}^{\dagger}n^{\mu}(\mathbf{p})\xi_{s^{\prime}}f_{ss^{\prime}}^{(+)}(x,\mathbf{p})+\theta(-p^{0})\xi_{s}^{\dagger}n^{\mu}(-\mathbf{p})\xi_{s^{\prime}}\left[1-f_{s^{\prime}s}^{(-)}(x,-\mathbf{p})\right]\right\} .\nonumber \\
\label{def:definition of V and n}
\end{eqnarray}
Then the components of the Wigner function at the zeroth order in
spatial gradients read
\begin{eqnarray}
\mathcal{F}^{(0)} & = & \delta(p^{2}-m^{2})mV^{(0)}(x,p),\nonumber \\
\mathcal{P}^{(0)} & = & 0,\nonumber \\
\mathcal{V}_{\mu}^{(0)} & = & \delta(p^{2}-m^{2})p_{\mu}V^{(0)}(x,p),\nonumber \\
\mathcal{A}_{\mu}^{(0)} & = & \delta(p^{2}-m^{2})n_{\mu}^{(0)}(x,p),\nonumber \\
\mathcal{S}_{\mu\nu}^{(0)} & = & -\delta(p^{2}-m^{2})\frac{1}{m}\epsilon_{\mu\nu\alpha\beta}p^{\alpha}n^{(0)\beta}(x,p).\label{sol:free paritlce Wigner function}
\end{eqnarray}
These results agree with the ones from the semi-classical expansion
\cite{Vasak:1987um,Fang:2016uds,Weickgenannt:2019dks}. Note that
the results are independent of the choice of the spin quantization
direction. Different spin quantizations are related by rotations in
spin space. Both $f_{ss^{\prime}}^{(\pm)}(x,\mathbf{p})$ and $\xi_{s}^{\dagger}n^{\mu}(\mathbf{p})\xi_{s^{\prime}}$
depend on the quantization direction but $V^{(0)}$ and $n^{(0)\mu}$
only depend on the trace in spin space, which are invariant under
spin-rotations \cite{DeGroot:1980dk}. The solutions in Eq. (\ref{sol:free paritlce Wigner function})
are all on the normal mass shell $p^{2}=m^{2}$ because we have not
considered any electromagnetic field. In the semi-classical expansion
discussed in Sec. \ref{sec:Semi-classical-expansion} we will clearly
show that the normal mass shell is shifted by the spin-electromagnetic
coupling.

\subsubsection{Diagonalization of distributions}

According to Eq. (\ref{eq:distribution w.r.t p and u}), we can find
the relation between the function $f_{ss^{\prime}}^{(+)}(\mathbf{p},\mathbf{u})$
and its complex conjugate,
\begin{equation}
\left[f_{ss^{\prime}}^{(\pm)}(\mathbf{p},\mathbf{u})\right]^{\ast}=f_{s^{\prime}s}^{(\pm)}(\mathbf{p},-\mathbf{u}).
\end{equation}
Then a relation between the distribution functions $f_{ss^{\prime}}^{(\pm)}(x,\mathbf{p})$
in Eq. (\ref{eq:local particle distribution}) and their complex conjugates
can be derived,
\begin{equation}
\left[f_{ss^{\prime}}^{(\pm)}(x,\mathbf{p})\right]^{\ast}=\int\frac{d^{4}u}{(2\pi)^{3}}\delta\left(u^{0}-\frac{\mathbf{p}\cdot\mathbf{u}}{p^{0}}\right)f_{s^{\prime}s}^{(\pm)}(\mathbf{p},-\mathbf{u})\exp\left(-iu^{\mu}x_{\mu}\right)=f_{s^{\prime}s}^{(\pm)}(x,\mathbf{p}),\label{eq:Hermitian of distribution function}
\end{equation}
Here in the second step we have made a replacement $u^{\mu}\rightarrow-u^{\mu}$.
The distribution $f_{ss^{\prime}}^{(\pm)}(x,\mathbf{p})$ is actually
the $ss^{\prime}$ element of a $2\times2$ matrix distribution $f^{(\pm)}(x,\mathbf{p})$
in spin space. So the relation (\ref{eq:Hermitian of distribution function})
indicates that $f^{(\pm)}(x,\mathbf{p})$ is a Hermitian matrix, which
can be diagonalized by a unitary transformation. The unitary transformation
can be interpreted as a rotation of the spin quantization direction
\cite{DeGroot:1980dk}.

We take the fermion part $f^{(+)}(x,\mathbf{p})$ as an example to
show the procedure of diagonalizing the distribution functions $f^{(\pm)}(x,\mathbf{p})$.
Note that any 2-dimensional Hermitian matrix can be parameterized
using the Pauli matrices $\boldsymbol{\sigma}$ together with the
unit matrix,
\begin{equation}
f^{(+)}(x,\mathbf{p})=a\,\mathbb{I}_{2}+\mathbf{b}\cdot\boldsymbol{\sigma}.
\end{equation}
Here $a$ and $\mathbf{b}$ are real functions of $x^{\mu}$ and $\mathbf{p}$.
The matrix $\mathbf{b}\cdot\boldsymbol{\sigma}$ has eigenvalues $\pm\left|\mathbf{b}\right|$
with $\left|\mathbf{b}\right|\equiv\sqrt{\mathbf{b}^{2}}$ is the
length of the 3-vector $\mathbf{b}$. Assuming that the corresponding
eigenvectors are $\overrightarrow{d}_{\pm}$, which are 2-dimensional
column vectors and satisfy
\begin{equation}
(\mathbf{b}\cdot\boldsymbol{\sigma})\overrightarrow{d}_{\pm}=\pm\left|\mathbf{b}\right|\overrightarrow{d}_{\pm}.
\end{equation}
Then the distribution function can be diagonalized as
\begin{equation}
\tilde{f}_{s}^{(+)}(x,\mathbf{p})\delta_{rs}=\sum_{r^{\prime}s^{\prime}}(D^{\dagger})_{rr^{\prime}}f_{r^{\prime}s^{\prime}}^{(+)}(x,\mathbf{p})D_{s^{\prime}s},\label{eq:diagonalize of f_ss}
\end{equation}
where the transformation matrix $D$ is a $2\times2$ matrix in spin
space and constructed from the eigenvectors $\overrightarrow{d}_{\pm}$,
\begin{equation}
D\equiv\left(\begin{array}{cc}
\overrightarrow{d}_{+} & \overrightarrow{d}_{-}\end{array}\right).
\end{equation}
The new distribution functions are then given by
\begin{equation}
\tilde{f}_{\pm}^{(+)}(x,\mathbf{p})=a\pm\left|\mathbf{b}\right|.
\end{equation}
In general, due to the fact that $a$ and $\mathbf{b}$ are defined
locally, the transformation matrix $D$ should be a function of $\{x^{\mu},\mathbf{p}\}$.
We also rotate the wavefunctions and define the following local ones,
\begin{equation}
\tilde{\psi}_{s}^{(+)}(x,\mathbf{p})\equiv\sum_{s^{\prime}}(D^{\dagger})_{ss^{\prime}}\psi_{s^{\prime}}^{(+)}(\mathbf{p}).
\end{equation}
Note that these new basis functions are still normalized because the
transformation is unitary. The plane-wavefunctions $\psi_{s^{\prime}}^{(+)}(\mathbf{p})$
are given in Eq. (\ref{sol:free wave functions}), from which the
new wavefunctions are obtained,
\begin{equation}
\tilde{\psi}_{s}^{(+)}(x,\mathbf{p})=\left(\begin{array}{c}
\sqrt{p_{\mu}\sigma^{\mu}}\tilde{\xi}_{s}\\
\sqrt{p_{\mu}\bar{\sigma}^{\mu}}\tilde{\xi}_{s}
\end{array}\right),\label{eq:new wave function psi_tilde}
\end{equation}
where
\begin{equation}
\tilde{\xi}_{s}\equiv\sum_{s^{\prime}}(D^{\dagger})_{ss^{\prime}}\xi_{s^{\prime}}.\label{eq:rotation of spinor}
\end{equation}
Since $s$ and $s^{\prime}$ label the spin state parallel or anti-parallel
to a given quantization direction, the transformation matrix $D$
is then interpreted as the $SU(2)$ representation of a rotation of
the quantization direction. Analogously, we can take similar procedure
for anti-particles, and finally the Wigner function (\ref{eq:Wigner function for free case})
can be put into the form
\begin{eqnarray}
W^{(0)}(x,p) & = & \frac{1}{(2\pi)^{3}}\delta(p^{\mu}p_{\mu}-m^{2})\sum_{s}\left\{ \theta(p^{0})\bar{\tilde{\psi}}_{s}^{(+)}(x,\mathbf{p})\otimes\tilde{\psi}_{s}^{(+)}(x,\mathbf{p})\tilde{f}_{s}^{(+)}(x,\mathbf{p})\right.\nonumber \\
 &  & \qquad\left.+\theta(-p^{0})\bar{\tilde{\psi}}_{s}^{(-)}(x,-\mathbf{p})\otimes\tilde{\psi}_{s}^{(-)}(x,-\mathbf{p})\left[1-\tilde{f}_{s}^{(-)}(x,-\mathbf{p})\right]\right\} ,
\end{eqnarray}
where the anti-particle parts are diagonalized as
\begin{eqnarray}
\tilde{f}_{s}^{(-)}(x,-\mathbf{p})\delta_{rs} & = & \sum_{r^{\prime}s^{\prime}}(\bar{D}^{\dagger})_{rr^{\prime}}f_{r^{\prime}s^{\prime}}^{(-)}(x,-\mathbf{p})\bar{D}_{s^{\prime}s},\nonumber \\
\tilde{\psi}_{s}^{(-)}(x,-\mathbf{p}) & = & \sum_{s^{\prime}}(\bar{D}^{\dagger})_{ss^{\prime}}\psi_{s^{\prime}}^{(-)}(-\mathbf{p}),
\end{eqnarray}
with the transformation matrix $\bar{D}$ is a function of $\left\{ x^{\mu},\mathbf{p}\right\} $.

The redefinition of $\tilde{\xi}_{s}$ in Eq. (\ref{eq:rotation of spinor})
corresponds to a new spin quantization direction. If we assume before
the transformation $\xi_{+}=(1,0)^{T}$, $\xi_{-}=(0,1)^{T}$, the
new spinors then read
\begin{equation}
\tilde{\xi}_{+}=\overrightarrow{d}_{+},\ \ \tilde{\xi}_{-}=\overrightarrow{d}_{-},
\end{equation}
which are eigenvectors of $\mathbf{b}\cdot\boldsymbol{\sigma}$ with
eigenvalues $\pm\left|\mathbf{b}\right|$. This indicates that the
new quantization direction is the direction of $\mathbf{b}$. The
components of the Wigner function are computed from Eq. (\ref{eq:free fermion Wigner function}),
where $V^{(0)}(x,p)$ and $n^{(0)\mu}(x,p)$ are given by Eq. (\ref{def:definition of V and n}).
The rotation of the spin quantization direction does not change the
trace of the matrix distribution $f^{(+)}(x,\mathbf{p})$, i.e., the
following relation holds in any case,
\begin{equation}
\sum_{s}f_{ss}^{(+)}(x,\mathbf{p})=\sum_{s}\tilde{f}_{s}^{(+)}(x,\mathbf{p}).
\end{equation}
Thus the function $V^{(0)}(x,p)$ can be expressed in terms of the
new distribution functions
\begin{equation}
V^{(0)}(x,p)=\frac{2}{(2\pi)^{3}}\sum_{s}\left\{ \theta(p^{0})\tilde{f}_{s}^{(+)}(x,\mathbf{p})-\theta(-p^{0})\left[1-\tilde{f}_{s}^{(-)}(x,-\mathbf{p})\right]\right\} .
\end{equation}
Meanwhile, the polarization part reads%
\begin{equation}
\sum_{ss^{\prime}}\xi_{s}^{\dagger}n^{\mu}(\mathbf{p})\xi_{s^{\prime}}f_{ss^{\prime}}^{(+)}(x,\mathbf{p})=\sum_{s}\tilde{\xi}_{s}^{\dagger}n^{\mu}(\mathbf{p})\tilde{\xi}_{s}\ \tilde{f}_{s}^{(+)}(x,\mathbf{p}),
\end{equation}
where $n^{\mu}(\mathbf{p})$ is given by Eq. (\ref{eq:polarization operator}).
Note that $\tilde{\xi}_{s}^{\dagger}\boldsymbol{\sigma}\tilde{\xi}_{s}=s\frac{\mathbf{b}}{\left|\mathbf{b}\right|}$
because the new spinors $\tilde{\xi}_{\pm}$ are now eigenvectors
of $\mathbf{b}\cdot\boldsymbol{\sigma}$. Thus the right-hand-side
of the above equation can be computed and we obtain
\begin{equation}
\sum_{ss^{\prime}}\xi_{s}^{\dagger}n^{\mu}(\mathbf{p})\xi_{s^{\prime}}f_{ss^{\prime}}^{(+)}(x,\mathbf{p})=\frac{1}{\left|\mathbf{b}\right|}\left(\mathbf{p}\cdot\mathbf{b},\ m\mathbf{b}+\frac{\mathbf{p}\cdot\mathbf{b}}{E_{\mathbf{p}}+m}\mathbf{p}\right)^{T}\sum_{s}s\tilde{f}_{s}^{(+)}(x,\mathbf{p}).
\end{equation}
Similar results can be done for anti-fermions. Finally the function
$n^{(0)\mu}(x,p)$, defined in Eq. (\ref{def:definition of V and n}),
becomes
\begin{eqnarray}
n^{(0)\mu}(x,p) & = & \left[\theta(p^{0})n_{0}^{\mu}(\mathbf{p},\mathbf{b}^{(+)})-\theta(p^{0})n_{0}^{\mu}(-\mathbf{p},\mathbf{b}^{(-)})\right]\nonumber \\
 &  & \times\frac{2}{(2\pi)^{3}}\sum_{s}s\left\{ \theta(p^{0})\tilde{f}_{s}^{(+)}(x,\mathbf{p})-\theta(-p^{0})\left[1-\tilde{f}_{s}^{(-)}(x,-\mathbf{p})\right]\right\} ,
\end{eqnarray}
where $\mathbf{b}^{(+)}$ represents the diagonalization parameter
for fermions while $\mathbf{b}^{(-)}$ is that for anti-fermions.
Here we defined
\begin{equation}
n_{0}^{\mu}(\mathbf{p},\mathbf{b})\equiv\frac{1}{\left|\mathbf{b}\right|}\left(\mathbf{p}\cdot\mathbf{b},\ m\mathbf{b}+\frac{\mathbf{p}\cdot\mathbf{b}}{E_{\mathbf{p}}+m}\mathbf{p}\right)^{T}.
\end{equation}
If we boost to the rest frame of the particles with momentum $\mathbf{p}$,
the above polarization direction is $n_{0}^{\mu}\propto\left(0,\mathbf{b}\right)^{T}$.
Thus $\mathbf{b}^{(\pm)}$ can be identified as the spin polarization
direction in the rest frame of fermions and anti-fermions respectively.
In general the polarization of fermions can be different from that
of anti-fermions, i.e., $\mathbf{b}^{(+)}$ can be different from
$\mathbf{b}^{(-)}$.

\subsection{Free fermions with chiral imbalance \label{subsec:Free-with-chiral}}

In this subsection we will study a system of free fermions with a
non-vanishing chiral chemical potential. Since for massive fermions
the helicity is not a conserved quantity, the chiral chemical potential
$\mu_{5}$ is no longer a well-defined conjugate variable of the axial
charge. However, on a time scale which is much smaller than the one
for varing axial charges, one can still use $\mu_{5}$ to describe
a thermal equilibrium system. We work with an effective theory where
the chemical potentials $\mu$ and $\mu_{5}$ are introduced in the
Dirac equation as self-energy corrections. The effective Lagrangian
reads,
\begin{equation}
\mathcal{L}=\bar{\psi}\left(i\gamma_{\mu}\partial_{x}^{\mu}-m\mathbb{I}_{4}\right)\psi+\mu\psi^{\dagger}\psi+\mu_{5}\psi^{\dagger}\gamma^{5}\psi.
\end{equation}
We can find the similar treatment in the Nambu-Jona-Lasinio model
\cite{Klevansky:1992qe} or other QCD effective theories with topological
charge \cite{Kharzeev:2007tn,Fukushima:2008xe}. The Dirac equation
is then given by
\begin{equation}
\left(i\gamma_{\mu}\partial_{x}^{\mu}-m\mathbb{I}_{4}+\mu\gamma^{0}+\mu_{5}\gamma^{0}\gamma^{5}\right)\psi=0.
\end{equation}
In general the mass $m$, the chemical potential $\mu$, and the chiral
chemical potential $\mu_{5}$ are dynamical quantities which depend
on the space-time coordinates, but here we assume all these variables
are constants. Under this assumption, $\partial_{x}^{\mu}$ commute
with the Hamiltonian, so we can define a conserved 4-momentum $p^{\mu}$.

\subsubsection{Plane-wave solutions with chiral imbalance}

Analogous to the case in the previous subsection, we first take a
Fourier transformation of the Dirac equation. Then in momentum space,
the Dirac field satisfies the following equation
\begin{equation}
(p^{0}+\mu)\psi(p)=[\gamma^{0}\boldsymbol{\gamma}\cdot\mathbf{p}+m\gamma^{0}-\mu_{5}\gamma^{5}]\psi(p).\label{eq:Dirac p equation with mu}
\end{equation}
Here the chemical potential $\mu$ shifts the energy levels. We now
define
\begin{equation}
\tilde{\psi}(\mathbf{p})=\left(\mathbb{I}_{2}\otimes S_{\mathbf{p}}^{\dagger}\right)\psi(\mathbf{p}),\label{eq:tilde psi and psi}
\end{equation}
where the matrix $S_{\mathbf{p}}$ is the transformation matrix in
Eq. (\ref{def:transformation matrix Sp}) which diagonalizes $\mathbf{p}\cdot\boldsymbol{\sigma}$
as shown in Eq. (\ref{eq:diagonalization of sigma cdot p}). With
this definition, Eq. (\ref{eq:Dirac p equation with mu}) is put into
the following form, %
\begin{equation}
(p^{0}+\mu)\tilde{\psi}(\mathbf{p})=\left(\begin{array}{cccc}
-\left|\mathbf{p}\right|+\mu_{5} & 0 & m & 0\\
0 & \left|\mathbf{p}\right|+\mu_{5} & 0 & m\\
m & 0 & \left|\mathbf{p}\right|-\mu_{5} & 0\\
0 & m & 0 & -\left|\mathbf{p}\right|-\mu_{5}
\end{array}\right)\tilde{\psi}(p).
\end{equation}
This can be treated as an eigenvalue problem, with $p^{0}+\mu$ being
the eigenvalue of the coefficient matrix on the right-hand-side while
$\tilde{\psi}(p)$ is the corresponding eigenstate. Direct forward
calculations give the eigenvalues
\begin{equation}
p^{0}=-\mu\pm E_{\mathbf{p},s},
\end{equation}
with $E_{\mathbf{p},s}=\sqrt{m^{2}+(\left|\mathbf{p}\right|-s\mu_{5})^{2}}$
and $s=\pm$. The eigenstates corresponding to positive energies $p^{0}+\mu=E_{\mathbf{p},s}$
are given by
\begin{equation}
\tilde{\psi}_{+}^{(+)}(\mathbf{p})=\left(\begin{array}{c}
\sqrt{E_{\mathbf{p},+}-\left(\left|\mathbf{p}\right|-\mu_{5}\right)}\\
0\\
\sqrt{E_{\mathbf{p},+}+\left(\left|\mathbf{p}\right|-\mu_{5}\right)}\\
0
\end{array}\right),\ \ \tilde{\psi}_{-}^{(+)}(\mathbf{p})=\left(\begin{array}{c}
0\\
\sqrt{E_{\mathbf{p},-}+\left(\left|\mathbf{p}\right|+\mu_{5}\right)}\\
0\\
\sqrt{E_{\mathbf{p},-}-\left(\left|\mathbf{p}\right|+\mu_{5}\right)}
\end{array}\right),\label{eq:fermion wave function with chiral}
\end{equation}
while the eigenstates for negative energies $p^{0}+\mu=-E_{\mathbf{p},s}$
read
\begin{equation}
\tilde{\psi}_{+}^{(-)}(\mathbf{p})=\left(\begin{array}{c}
-\sqrt{E_{\mathbf{p},+}+\left|\mathbf{p}\right|-\mu_{5}}\\
0\\
\sqrt{E_{\mathbf{p},+}-\left(\left|\mathbf{p}\right|-\mu_{5}\right)}\\
0
\end{array}\right),\ \ \tilde{\psi}_{-}^{(-)}(\mathbf{p})=\left(\begin{array}{c}
0\\
-\sqrt{E_{\mathbf{p},-}-\left(\left|\mathbf{p}\right|+\mu_{5}\right)}\\
0\\
\sqrt{E_{\mathbf{p},-}+\left(\left|\mathbf{p}\right|+\mu_{5}\right)}
\end{array}\right).\label{eq:antifermion wave function with chiral}
\end{equation}
The wavefunctions in Eqs. (\ref{eq:fermion wave function with chiral})
and (\ref{eq:antifermion wave function with chiral}) are normalized
\begin{equation}
\tilde{\psi}_{s_{2}}^{(s_{1})\dagger}(\mathbf{p})\tilde{\psi}_{s_{2}^{\prime}}^{(s_{1}^{\prime})}(\mathbf{p})=2E_{\mathbf{p},s_{2}}\delta_{s_{1}s_{1}^{\prime}}\delta_{s_{2}s_{2}^{\prime}}.\label{eq:orthonarmality of psi with chiral}
\end{equation}
Here they are normalized to the corresponding eigenenergies in order
to smoothly reproduce the normalization relations without chiral chemical
potential in Eq. (\ref{eq:normalization of psi pm}). The wavefunctions
in coordinate space are then computed by adding a Fourier factor
\begin{equation}
\psi_{s_{2}}^{(s_{1})}(x,\mathbf{p})=e^{i\mu t}\exp\left(-i\,s_{1}E_{\mathbf{p},s_{2}}t+i\mathbf{p}\cdot\mathbf{x}\right)\left(\mathbb{I}_{2}\otimes S_{\mathbf{p}}\right)\tilde{\psi}_{s_{2}}^{(s_{1})}(\mathbf{p}),\label{eq:single particle wave function}
\end{equation}
In the solution (\ref{eq:single particle wave function}), $s_{1}=\pm$
labels fermions ($+$) or anti-fermions ($-$). The states with $s_{1}=+$
are interpreted as fermions with the kinetic momentum $\mathbf{p}$
while $s_{1}=-$ as anti-fermions with the kinetic momentum $-\mathbf{p}$.
Meanwhile, $s_{2}=\pm$ does not have an explicit meaning in the massive
case. But in the massless limit it parameterizes the chirality.

The wavefunctions in Eqs. (\ref{eq:fermion wave function with chiral}),
(\ref{eq:antifermion wave function with chiral}) are superpositions
of the LH states and the RH ones. In the massless limit, the eigenenergies
are given by $E_{\mathbf{p},s}=(\left|\mathbf{p}\right|-s\mu_{5})\mathrm{sgn}(\left|\mathbf{p}\right|-s\mu_{5})$,
where $\mathrm{sgn}$ is the sign function. Meanwhile, the wavefunction
$\tilde{\psi}_{+}^{(+)}(\mathbf{p})$ in Eq. (\ref{eq:fermion wave function with chiral})
reduces to the following expression,
\begin{equation}
\tilde{\psi}_{+}^{(+)}(\mathbf{p})=\left|\left|\mathbf{p}\right|-\mu_{5}\right|\left(\begin{array}{c}
\sqrt{1-\mathrm{sgn}(\left|\mathbf{p}\right|-\mu_{5})}\\
0\\
\sqrt{1+\mathrm{sgn}(\left|\mathbf{p}\right|-\mu_{5})}\\
0
\end{array}\right),
\end{equation}
which represents a RH wavefunction if $\left|\mathbf{p}\right|-\mu_{5}$
is positive, and a LH wavefunction if $\left|\mathbf{p}\right|-\mu_{5}$
is negative. Similar discussion can be done for other functions in
Eqs. (\ref{eq:fermion wave function with chiral}), (\ref{eq:antifermion wave function with chiral}).
We conclude that if $\left|\mathbf{p}\right|-\mu_{5}>0$, $\tilde{\psi}_{+}^{(+)}(\mathbf{p})$
and $\tilde{\psi}_{-}^{(-)}(\mathbf{p})$ are RH while $\tilde{\psi}_{-}^{(+)}(\mathbf{p})$,
and $\tilde{\psi}_{+}^{(-)}(\mathbf{p})$ are LH. On the other hand,
if $\left|\mathbf{p}\right|-\mu_{5}<0$, $\tilde{\psi}_{+}^{(+)}(\mathbf{p})$
and $\tilde{\psi}_{-}^{(-)}(\mathbf{p})$ are LH while $\tilde{\psi}_{-}^{(+)}(\mathbf{p})$
and $\tilde{\psi}_{+}^{(-)}(\mathbf{p})$ are RH.

On the other hand, we can consider the limit $\mu_{5}\rightarrow0$,
which corresponds to a state where the chiral symmetry is restored.
We find that the eigenenergies are now independent of $s$, which
read $E_{\mathbf{p},s}=E_{\mathbf{p}}=\sqrt{m^{2}+\left|\mathbf{p}\right|^{2}}$.
The states $\tilde{\psi}_{\pm}^{(+)}(\mathbf{p})$ then have the same
eigenenergy $p^{0}=-\mu+E_{\mathbf{p}}$ while $\tilde{\psi}_{\pm}^{(-)}(\mathbf{p})$
have eigenenergy $p^{0}=-\mu-E_{\mathbf{p}}$. The wavefunctions in
Eqs. (\ref{eq:fermion wave function with chiral}) and (\ref{eq:antifermion wave function with chiral})
reduce to the following forms in this limit,
\begin{eqnarray}
 &  & \tilde{\psi}_{+}^{(+)}(\mathbf{p})=\left(\begin{array}{c}
\sqrt{E_{\mathbf{p}}-\left|\mathbf{p}\right|}\\
0\\
\sqrt{E_{\mathbf{p}}+\left|\mathbf{p}\right|}\\
0
\end{array}\right),\ \ \tilde{\psi}_{-}^{(+)}(\mathbf{p})=\left(\begin{array}{c}
0\\
\sqrt{E_{\mathbf{p}}+\left|\mathbf{p}\right|}\\
0\\
\sqrt{E_{\mathbf{p}}-\left|\mathbf{p}\right|}
\end{array}\right),\nonumber \\
 &  & \tilde{\psi}_{-}^{(-)}(\mathbf{p})=\left(\begin{array}{c}
0\\
-\sqrt{E_{\mathbf{p}}-\left|\mathbf{p}\right|}\\
0\\
\sqrt{E_{\mathbf{p}}+\left|\mathbf{p}\right|}
\end{array}\right),\ \ \tilde{\psi}_{+}^{(-)}(\mathbf{p})=\left(\begin{array}{c}
-\sqrt{E_{\mathbf{p}}+\left|\mathbf{p}\right|}\\
0\\
\sqrt{E_{\mathbf{p}}-\left|\mathbf{p}\right|}\\
0
\end{array}\right),
\end{eqnarray}
where all these functions are normalized to $2E_{\mathbf{p}}$ respectively.
For a nonzero mass, all these states are superposition of RH and LH
spinors. But in the massless case, the energy $E_{\mathbf{p}}=\left|\mathbf{p}\right|$
and the subscript labels the helicity. In order to show the coincidence
with the plane-wave solution in Eq. (\ref{sol:free wave functions}),
we form a linear combination between the states with the same eigenenergies,
\begin{eqnarray*}
\tilde{\xi}_{1}\tilde{\psi}_{+}^{(+)}(\mathbf{p})\pm\tilde{\xi}_{2}\tilde{\psi}_{-}^{(+)}(\mathbf{p}) & \rightarrow & \tilde{\psi}_{\pm}^{(+)}(\mathbf{p}),\\
-\left[\tilde{\xi}_{1}\tilde{\psi}_{+}^{(-)}(\mathbf{p})\pm\tilde{\xi}_{2}\tilde{\psi}_{-}^{(-)}(\mathbf{p})\right] & \rightarrow & \tilde{\psi}_{\pm}^{(-)}(-\mathbf{p}).
\end{eqnarray*}
These wavefunctions read,
\begin{equation}
\tilde{\psi}_{\pm}^{(+)}(\mathbf{p})=\left(\begin{array}{c}
\sqrt{E_{\mathbf{p}}-\left|\mathbf{p}\right|}\tilde{\xi}_{1}\\
\pm\sqrt{E_{\mathbf{p}}+\left|\mathbf{p}\right|}\tilde{\xi}_{2}\\
\sqrt{E_{\mathbf{p}}+\left|\mathbf{p}\right|}\tilde{\xi}_{1}\\
\pm\sqrt{E_{\mathbf{p}}-\left|\mathbf{p}\right|}\tilde{\xi}_{2}
\end{array}\right),\ \ \tilde{\psi}_{\pm}^{(-)}(-\mathbf{p})=\left(\begin{array}{c}
\sqrt{E_{\mathbf{p}}+\left|\mathbf{p}\right|}\tilde{\xi}_{1}\\
\pm\sqrt{E_{\mathbf{p}}-\left|\mathbf{p}\right|}\tilde{\xi}_{2}\\
-\sqrt{E_{\mathbf{p}}-\left|\mathbf{p}\right|}\tilde{\xi}_{1}\\
\mp\sqrt{E_{\mathbf{p}}+\left|\mathbf{p}\right|}\tilde{\xi}_{2}
\end{array}\right).
\end{equation}
We further demand that
\begin{equation}
\xi_{s}\equiv S_{\mathbf{p}}\left(\begin{array}{c}
\tilde{\xi}_{1}\\
\pm\tilde{\xi}_{2}
\end{array}\right),
\end{equation}
and using Eq. (\ref{def:short notations}) and Eq. (\ref{eq:tilde psi and psi})
we finally obtain
\begin{equation}
\psi_{s}^{(+)}(\mathbf{p})=\left(\begin{array}{c}
\sqrt{p_{\mu}\sigma^{\mu}}\xi_{s}\\
\sqrt{p_{\mu}\bar{\sigma}^{\mu}}\xi_{s}
\end{array}\right),\ \ \psi_{s}^{(-)}(-\mathbf{p})=\left(\begin{array}{c}
\sqrt{p_{\mu}\bar{\sigma}^{\mu}}\xi_{s}\\
-\sqrt{p_{\mu}\sigma^{\mu}}\xi_{s}
\end{array}\right),
\end{equation}
which agree with the previous results (\ref{sol:free wave functions}).
Thus we conclude that in the presence of a constant chiral chemical
potential $\mu_{5}$, the single-particle wavefunctions in (\ref{eq:single particle wave function})
when setting $\mu_{5}=0$, up to a linear combination, coincide with
the solutions without $\mu_{5}$ in the previous subsection.

\subsubsection{Chiral quantization}

Analogous to the case without chiral imbalance, the fermionic field
can be quantized using the single-particle wavefunctions with finite
$\mu_{5}$, which are given in Eq. (\ref{eq:single particle wave function}),
\begin{equation}
\hat{\psi}(x)=e^{i\mu t}\sum_{s}\int\frac{d^{3}\mathbf{p}}{(2\pi)^{3}\sqrt{2E_{\mathbf{p},s}}}e^{i\mathbf{p}\cdot\mathbf{x}}\left(\mathbb{I}_{2}\otimes S_{\mathbf{p}}\right)\left[e^{-iE_{\mathbf{p},s}t}\tilde{\psi}_{s}^{(+)}(\mathbf{p})\hat{a}_{\mathbf{p},s}+e^{iE_{\mathbf{p},s}t}\tilde{\psi}_{s}^{(-)}(\mathbf{p})\hat{b}_{-\mathbf{p},s}^{\dagger}\right],\label{eq:field operator with =00005Cmu_5}
\end{equation}
where the creation and annihilation operators satisfy the following
canonical anti-commutation relations
\begin{equation}
\left\{ \hat{a}_{\mathbf{p},s},\,\hat{a}_{\mathbf{p}^{\prime},s^{\prime}}^{\dagger}\right\} =\left\{ \hat{b}_{\mathbf{p},s},\,\hat{b}_{\mathbf{p}^{\prime},s^{\prime}}^{\dagger}\right\} =(2\pi)^{3}\delta^{(3)}(\mathbf{p}-\mathbf{p}^{\prime})\delta_{ss^{\prime}},\label{eq:anticommutation chiral case}
\end{equation}
while all other anti-commutators vanish. In order to check whether
the fermionic field is correctly quantized, we calculate the anti-commutator
for the field operator $\hat{\psi}$ and its Hermitian conjugate $\hat{\psi}^{\dagger}$,
\begin{equation}
\left\{ \hat{\psi}_{\alpha}(t,\mathbf{x}),\hat{\psi}_{\beta}^{\dagger}(t,\mathbf{x}^{\prime})\right\} =\delta_{\alpha\beta}\delta^{(3)}(\mathbf{x}-\mathbf{x}^{\prime}),
\end{equation}
where $\alpha,\beta$ label components of the Dirac field. Furthermore,
other anti-commutators, such as $\left\{ \psi_{\alpha}(t,\mathbf{x}),\psi_{\beta}(t,\mathbf{x}^{\prime})\right\} $
and $\left\{ \psi_{\alpha}^{\dagger}(t,\mathbf{x}),\psi_{\beta}^{\dagger}(t,\mathbf{x}^{\prime})\right\} $,
vanish because they do not contain any nonzero anti-commutator (\ref{eq:anticommutation chiral case}).
Note that the field operator in Eq. (\ref{eq:field operator with =00005Cmu_5})
recovers the one in Eq. (\ref{def:quantized free field}) if we take
$\mu_{5}=0$.

The Hamilton operator $\hat{H}$ is now given by
\begin{equation}
\hat{H}=\int d^{3}\mathbf{x}\,\hat{\psi}^{\dagger}(-i\gamma^{0}\boldsymbol{\gamma}\cdot\boldsymbol{\nabla}_{\mathbf{x}}-m-\mu-\mu_{5}\gamma^{5})\hat{\psi}=\hat{H}_{0}-\mu\hat{N}-\mu_{5}\hat{N}_{5},
\end{equation}
where $\hat{H}_{0}=\psi^{\dagger}(-i\gamma^{0}\boldsymbol{\gamma}\cdot\boldsymbol{\nabla}_{\mathbf{x}}-m)\psi$
is the free fermion Hamiltonian, $\hat{N}=\psi^{\dagger}\psi$ is
the net particle number operator and $\hat{N}_{5}=\psi^{\dagger}\gamma^{5}\psi$
is the axial-charge operator. Inserting the quantized field operator
(\ref{eq:field operator with =00005Cmu_5}) into the Hamiltonian,
we obtain
\begin{equation}
\hat{H}=\sum_{s}\int\frac{d^{3}\mathbf{p}}{(2\pi)^{3}}\left[(E_{\mathbf{p},s}-\mu)\hat{a}_{\mathbf{p},s}^{\dagger}\hat{a}_{\mathbf{p},s}+(E_{\mathbf{p},s}+\mu)\left(\hat{b}_{-\mathbf{p},s}^{\dagger}\hat{b}_{-\mathbf{p},s}-1\right)\right].\label{eq:quantized Hamiltonian}
\end{equation}
From the above expression we observe that the lowest energy state
is no longer empty. In the lowest energy state, all the states with
$E_{\mathbf{p},s}<\mu$ are occupied, which agrees with our expectation.
The chemical potential means the system has non-vanishing net fermion
number. And the lowest energy state is reached when the thermal temperature
is zero and the fermions occupy all states below the Fermi surface.
On the other hand, the momentum operator is %
\begin{equation}
\hat{\mathbf{P}}=\int d^{3}\mathbf{x}\,\hat{\psi}^{\dagger}\left(-i\boldsymbol{\nabla}_{\mathbf{x}}\right)\hat{\psi}=\sum_{s}\int\frac{d^{3}\mathbf{p}}{(2\pi)^{3}}\mathbf{p}\left(\hat{a}_{\mathbf{p},s}^{\dagger}\hat{a}_{\mathbf{p},s}+\hat{b}_{\mathbf{p},s}^{\dagger}\hat{b}_{\mathbf{p},s}\right).
\end{equation}
The Hamiltonian and momentum operators indicate that $\hat{a}_{\mathbf{p},s}^{\dagger}$
plays as the creation operator of a fermion with the momentum $\mathbf{p}$
and the energy $E_{\mathbf{p},s}-\mu$ while $\hat{b}_{\mathbf{p},s}^{\dagger}$
creates an anti-fermion with the same momentum $\mathbf{p}$ and the
energy $E_{\mathbf{p},s}+\mu$.

\subsubsection{Wigner function}

Inserting the field operator (\ref{eq:field operator with =00005Cmu_5})
into the definition (\ref{def:Wigner function}) of the Wigner function,
one obtains %
\begin{eqnarray}
W(x,p) & = & \sum_{ss^{\prime}}\int\frac{dt^{\prime}d^{3}\mathbf{x}^{\prime}}{(2\pi)^{4}}\int\frac{d^{3}\mathbf{q}d^{3}\mathbf{q}^{\prime}}{(2\pi)^{6}\sqrt{2E_{\mathbf{q},s}}\sqrt{2E_{\mathbf{q}^{\prime},s^{\prime}}}}\exp\left[i\mathbf{x}^{\prime}\cdot\left(\mathbf{p}-\frac{\mathbf{q}+\mathbf{q}^{\prime}}{2}\right)+i\mathbf{x}\cdot\left(\mathbf{q}-\mathbf{q}^{\prime}\right)\right]\nonumber \\
 &  & \times\left\{ \exp\left[-it^{\prime}\left(p^{0}+\mu-\frac{E_{\mathbf{q}^{\prime},s^{\prime}}+E_{\mathbf{q},s}}{2}\right)+it\left(E_{\mathbf{q}^{\prime},s^{\prime}}-E_{\mathbf{q},s}\right)\right]\right.\nonumber \\
 &  & \ \ \ \ \times\tilde{\psi}_{s^{\prime}}^{(+)\dagger}(\mathbf{q}^{\prime})\left(\mathbb{I}_{2}\otimes S_{\mathbf{q}^{\prime}}^{\dagger}\right)\gamma^{0}\otimes\left[\left(\mathbb{I}_{2}\otimes S_{\mathbf{q}}\right)\tilde{\psi}_{s}^{(+)}(\mathbf{q})\right]\left\langle \Omega\left|\hat{a}_{\mathbf{q}^{\prime},s^{\prime}}^{\dagger}\hat{a}_{\mathbf{q},s}\right|\Omega\right\rangle \nonumber \\
 &  & \ \ +\exp\left[-it^{\prime}\left(p^{0}+\mu+\frac{E_{\mathbf{q}^{\prime},s^{\prime}}+E_{\mathbf{q},s}}{2}\right)-it\left(E_{\mathbf{q}^{\prime},s^{\prime}}-E_{\mathbf{q},s}\right)\right]\nonumber \\
 &  & \ \ \ \ \left.\times\tilde{\psi}_{s^{\prime}}^{(-)\dagger}(\mathbf{q}^{\prime})\left(\mathbb{I}_{2}\otimes S_{\mathbf{q}^{\prime}}^{\dagger}\right)\gamma^{0}\otimes\left[\left(\mathbb{I}_{2}\otimes S_{\mathbf{q}}\right)\tilde{\psi}_{s}^{(-)}(\mathbf{q})\right]\left\langle \Omega\left|\hat{b}_{-\mathbf{q}^{\prime},s^{\prime}}\hat{b}_{-\mathbf{q},s}^{\dagger}\right|\Omega\right\rangle \right\} ,
\end{eqnarray}
where we have dropped terms of $a_{\mathbf{q}^{\prime},s^{\prime}}^{\dagger}b_{-\mathbf{q},s}^{\dagger}$
or $b_{-\mathbf{q}^{\prime},s^{\prime}}a_{\mathbf{q},s}$. These terms
represent the mixtures between fermion states and anti-fermion states,
which are not considered in this thesis. Analogous to the discussion
in subsection \ref{subsec:Free-fermions}, we introduce the average
and relative momenta,
\begin{equation}
\mathbf{k}=\frac{\mathbf{q}+\mathbf{q}^{\prime}}{2},\ \ \mathbf{u}=\mathbf{q}-\mathbf{q}^{\prime}.
\end{equation}
The integration measure is then invariant under the momentum redefinition,
\begin{equation}
d^{3}\mathbf{q}d^{3}\mathbf{q}^{\prime}=d^{3}\mathbf{k}d^{3}\mathbf{u}.
\end{equation}
In the Wigner function, the integration over $d^{3}\mathbf{x}^{\prime}$
gives a 3-dimensional delta-function for the momentum, while the integration
over $dt^{\prime}$ gives a delta-function for the energy. After a
straightforward calculation, we obtain
\begin{eqnarray}
W(x,p) & = & \sum_{ss^{\prime}}\int\frac{d^{3}\mathbf{u}}{(2\pi)^{6}\sqrt{2E_{\mathbf{p}+\mathbf{u}/2,s}}\sqrt{2E_{\mathbf{p}-\mathbf{u}/2,s^{\prime}}}}e^{i\mathbf{u}\cdot\mathbf{x}}\nonumber \\
 &  & \times\left\{ \delta\left(p^{0}+\mu-\frac{E_{\mathbf{p}-\mathbf{u}/2,s^{\prime}}+E_{\mathbf{p}+\mathbf{u}/2,s}}{2}\right)\exp\left[it\left(E_{\mathbf{p}-\mathbf{u}/2,s^{\prime}}-E_{\mathbf{p}+\mathbf{u}/2,s}\right)\right]\right.\nonumber \\
 &  & \qquad\qquad\times\tilde{\psi}_{s^{\prime}}^{(+)\dagger}\left(\mathbf{p}-\frac{\mathbf{u}}{2}\right)\left(\mathbb{I}_{2}\otimes S_{\mathbf{p}-\frac{\mathbf{u}}{2}}^{\dagger}\right)\gamma^{0}\otimes\left[\left(\mathbb{I}_{2}\otimes S_{\mathbf{p}+\frac{\mathbf{u}}{2}}\right)\tilde{\psi}_{s}^{(+)}\left(\mathbf{p}+\frac{\mathbf{u}}{2}\right)\right]\nonumber \\
 &  & \qquad\qquad\times\left\langle \Omega\left|a_{\mathbf{p}-\frac{\mathbf{u}}{2},s^{\prime}}^{\dagger}a_{\mathbf{p}+\frac{\mathbf{u}}{2},s}\right|\Omega\right\rangle \nonumber \\
 &  & \qquad+\delta\left(p^{0}+\mu+\frac{E_{\mathbf{p}-\mathbf{u}/2,s^{\prime}}+E_{\mathbf{p}+\mathbf{u}/2,s}}{2}\right)\exp\left[-it\left(E_{\mathbf{p}-\mathbf{u}/2,s^{\prime}}-E_{\mathbf{p}+\mathbf{u}/2,s}\right)\right]\nonumber \\
 &  & \qquad\qquad\times\tilde{\psi}_{s^{\prime}}^{(-)\dagger}\left(\mathbf{p}-\frac{\mathbf{u}}{2}\right)\left(\mathbb{I}_{2}\otimes S_{\mathbf{p}-\frac{\mathbf{u}}{2}}^{\dagger}\right)\gamma^{0}\otimes\left[\left(\mathbb{I}_{2}\otimes S_{\mathbf{p}+\frac{\mathbf{u}}{2}}\right)\tilde{\psi}_{s}^{(-)}\left(\mathbf{p}+\frac{\mathbf{u}}{2}\right)\right]\nonumber \\
 &  & \qquad\qquad\times\left.\left\langle \Omega\left|b_{-\mathbf{p}+\frac{\mathbf{u}}{2},s^{\prime}}b_{-\mathbf{p}-\frac{\mathbf{u}}{2},s}^{\dagger}\right|\Omega\right\rangle \right\} .
\end{eqnarray}
Again we adopt the wave-packet prescription and assume that the expectation
value is given by some distribution function,
\begin{eqnarray}
\left\langle \Omega\left|a_{\mathbf{p}-\frac{\mathbf{u}}{2},s^{\prime}}^{\dagger}a_{\mathbf{p}+\frac{\mathbf{u}}{2},s}\right|\Omega\right\rangle  & = & f_{s}^{(+)}(\mathbf{p},\mathbf{u})\delta_{ss^{\prime}},\nonumber \\
\left\langle \Omega\left|b_{-\mathbf{p}+\frac{\mathbf{u}}{2},s^{\prime}}b_{-\mathbf{p}-\frac{\mathbf{u}}{2},s}^{\dagger}\right|\Omega\right\rangle  & = & (2\pi)^{3}\delta^{(3)}(\mathbf{u})\delta_{ss^{\prime}}-f_{s}^{(-)}(-\mathbf{p},\mathbf{u})\delta_{ss^{\prime}}.\label{eq:definition of distributions with chiral}
\end{eqnarray}
Here the presence of $\delta_{ss^{\prime}}$ takes into account that
states with different $s$ have different energy shells. We further
define the distribution function in phase space,
\begin{equation}
f_{s}^{(\pm)}(x,\mathbf{p})\equiv\int\frac{d^{3}\mathbf{u}}{(2\pi)^{3}}e^{i\mathbf{u}\cdot\mathbf{x}}\exp\left[\pm it\left(E_{\mathbf{p}-\mathbf{u}/2,s}-E_{\mathbf{p}+\mathbf{u}/2,s}\right)\right]f_{s}^{(\pm)}(\mathbf{p},\mathbf{u}).
\end{equation}
Analogous to what we did for the free fermion case in subsection \ref{subsec:Free-fermions},
we expand the Wigner function in terms of $\mathbf{u}$. Since we
adopt the wave-packet prescription, the relative momentum $\mathbf{u}$
contributes if it is much smaller than the width of the wave packet.
Thus $\mathbf{u}$ is treated as a small variable and the Wigner function
at leading order in spatial gradient reads,
\begin{eqnarray}
W^{(0)}(x,p) & = & \sum_{s}\frac{1}{(2\pi)^{3}}\left\{ \delta\left(p^{0}+\mu-E_{\mathbf{p},s}\right)W_{s}^{(+)}(\mathbf{p})f_{s}^{(+)}(x,\mathbf{p})\right.\nonumber \\
 &  & \qquad\left.+\delta\left(p^{0}+\mu+E_{\mathbf{p},s}\right)W_{s}^{(-)}(\mathbf{p})\left[1-f_{s}^{(-)}(x,-\mathbf{p})\right]\right\} ,\label{eq:Wigner function with chiral imbalance}
\end{eqnarray}
where we have defined the following terms for contributions from fermions
or anti-fermions,
\begin{equation}
W_{s}^{(\pm)}(\mathbf{p})\equiv\frac{1}{2E_{\mathbf{p},s}}\tilde{\psi}_{s}^{(\pm)\dagger}(\mathbf{p})\left(\mathbb{I}_{2}\otimes S_{\mathbf{p}}^{\dagger}\right)\gamma^{0}\otimes\left[\left(\mathbb{I}_{2}\otimes S_{\mathbf{p}}\right)\tilde{\psi}_{s}^{(\pm)}(\mathbf{p})\right].
\end{equation}
Here the single-particle wavefunctions $\tilde{\psi}_{s}^{(r)}(\mathbf{p})$
are listed in Eqs. (\ref{eq:fermion wave function with chiral}) and
(\ref{eq:antifermion wave function with chiral}), meanwhile the transformation
matrices $S_{\mathbf{p}}$ and $S_{\mathbf{p}}^{\dagger}$ are given
in Eq. (\ref{def:transformation matrix Sp}). The delta-function for
the energy can be written in a covariant form
\begin{equation}
\frac{1}{2E_{\mathbf{p},s}}\delta\left(p^{0}+\mu-rE_{\mathbf{p},s}\right)=\theta[r(p^{0}+\mu)]\delta\left[(p^{0}+\mu)^{2}-(\left|\mathbf{p}\right|-s\mu_{5})^{2}-m^{2}\right],
\end{equation}
which recovers with the normal on-shell condition when setting $\mu=\mu_{5}=0$.

Then the next step is to insert the single-particle wavefunctions
(\ref{eq:fermion wave function with chiral}) and (\ref{eq:antifermion wave function with chiral})
into Eq. (\ref{eq:Wigner function with chiral imbalance}) and calculate
16 components of the Wigner function. First we focus on the fermion
part, $\tilde{\psi}_{s}^{(+)}(\mathbf{p})$, which can be written
in terms of the Kronecker product of two column vectors,
\begin{equation}
\tilde{\psi}_{s}^{(+)}(\mathbf{p})=\left(\begin{array}{c}
\sqrt{E_{\mathbf{p},+}-\left(\left|\mathbf{p}\right|-\mu_{5}\right)}\\
\sqrt{E_{\mathbf{p},+}+\left(\left|\mathbf{p}\right|-\mu_{5}\right)}
\end{array}\right)\otimes\left(\begin{array}{c}
1\\
0
\end{array}\right),\ \ \tilde{\psi}_{-}^{(+)}(\mathbf{p})=\left(\begin{array}{c}
\sqrt{E_{\mathbf{p},-}+\left(\left|\mathbf{p}\right|+\mu_{5}\right)}\\
\sqrt{E_{\mathbf{p},-}-\left(\left|\mathbf{p}\right|+\mu_{5}\right)}
\end{array}\right)\otimes\left(\begin{array}{c}
0\\
1
\end{array}\right).
\end{equation}
Then using the property of the Kronecker product in Eq. (\ref{eq:mixed-product}),
we obtain %
\begin{eqnarray}
W_{s}^{(+)}(\mathbf{p}) & = & \frac{1}{2E_{\mathbf{p},s}}\left[\left(\begin{array}{cc}
m & E_{\mathbf{p},s}-s\left(\left|\mathbf{p}\right|-s\mu_{5}\right)\\
E_{\mathbf{p},s}+s\left(\left|\mathbf{p}\right|-s\mu_{5}\right) & m
\end{array}\right)\right]\nonumber \\
 &  & \qquad\otimes\left[\left(\begin{array}{cc}
\delta_{s+} & \delta_{s-}\end{array}\right)S_{\mathbf{p}}^{\dagger}\otimes S_{\mathbf{p}}\left(\begin{array}{c}
\delta_{s+}\\
\delta_{s-}
\end{array}\right)\right].
\end{eqnarray}
The explicit forms of $S_{\mathbf{p}}$ and $S_{\mathbf{p}}^{\dagger}$
are given in Eq. (\ref{def:transformation matrix Sp}). Then after
some complicated but straightforward calculation we obtain %
\begin{equation}
\left(\begin{array}{cc}
\delta_{s+} & \delta_{s-}\end{array}\right)S_{\mathbf{p}}^{\dagger}\otimes S_{\mathbf{p}}\left(\begin{array}{c}
\delta_{s+}\\
\delta_{s-}
\end{array}\right)=\frac{1}{2}\mathbb{I}_{2}+s\frac{\mathbf{p}\cdot\boldsymbol{\sigma}}{2\left|\mathbf{p}\right|}.\label{eq:temp relation}
\end{equation}
With the help of the Kronecker-product of the gamma matrices in Eq.
(\ref{eq:relation between gamma matrices and Pauli}), we obtain %
\begin{eqnarray}
W_{s}^{(+)}(\mathbf{p}) & = & \frac{1}{4E_{\mathbf{p},s}}\left[m\left(\mathbb{I}_{4}+s\frac{1}{2\left|\mathbf{p}\right|}\epsilon^{ijk}\sigma^{ij}\mathbf{p}^{k}\right)+E_{\mathbf{p},s}\left(\gamma^{0}-s\frac{1}{\left|\mathbf{p}\right|}\mathbf{p}\cdot\gamma^{5}\boldsymbol{\gamma}\right)\right.\nonumber \\
 &  & \qquad\left.+\left(\left|\mathbf{p}\right|-s\mu_{5}\right)\left(s\gamma^{5}\gamma^{0}-\frac{1}{\left|\mathbf{p}\right|}\mathbf{p}\cdot\boldsymbol{\gamma}\right)\right].
\end{eqnarray}
An analogous calculation can be done for anti-fermions, which gives%
\begin{eqnarray}
W_{s}^{(-)}(\mathbf{p}) & = & \frac{1}{2E_{\mathbf{p},s}}\left[\left(\begin{array}{cc}
-m & E_{\mathbf{p},s}+s\left(\left|\mathbf{p}\right|-s\mu_{5}\right)\\
E_{\mathbf{p},s}-s\left(\left|\mathbf{p}\right|-s\mu_{5}\right) & -m
\end{array}\right)\right]\nonumber \\
 &  & \qquad\otimes\left[\left(\begin{array}{cc}
\delta_{s+} & \delta_{s-}\end{array}\right)S_{\mathbf{p}}^{\dagger}\otimes S_{\mathbf{p}}\left(\begin{array}{c}
\delta_{s+}\\
\delta_{s-}
\end{array}\right)\right].
\end{eqnarray}
Then using Eqs. (\ref{eq:temp relation}) and (\ref{eq:relation between gamma matrices and Pauli})
we can express $W_{s}^{(-)}(\mathbf{p})$ in gamma matrices,
\begin{eqnarray}
W_{s}^{(-)}(\mathbf{p}) & = & \frac{1}{4E_{\mathbf{p},s}}\left[-m\left(\mathbb{I}_{4}+s\frac{1}{2\left|\mathbf{p}\right|}\epsilon^{ijk}\sigma^{ij}\mathbf{p}^{k}\right)+E_{\mathbf{p},s}\left(\gamma^{0}-s\frac{1}{\left|\mathbf{p}\right|}\mathbf{p}\cdot\gamma^{5}\boldsymbol{\gamma}\right)\right.\nonumber \\
 &  & \left.-\left(\left|\mathbf{p}\right|-s\mu_{5}\right)\left(s\gamma^{5}\gamma^{0}-\frac{1}{\left|\mathbf{p}\right|}\mathbf{p}\cdot\boldsymbol{\gamma}\right)\right].
\end{eqnarray}
Inserting these back into the Wigner function in Eq. (\ref{eq:Wigner function with chiral imbalance})
and taking the trace after multiplying with $\Gamma_{i}=\{\mathbb{I}_{4},\ i\gamma^{5},\ \gamma^{\mu},\ \gamma^{5}\gamma^{\mu},\ \frac{1}{2}\sigma^{\mu\nu}\}$,
we can extract different components of the Wigner function,
\begin{eqnarray}
\mathcal{F} & = & mV(x,p),\nonumber \\
\mathcal{P} & = & 0,\nonumber \\
\mathcal{V}^{0} & = & (p^{0}+\mu)V(x,p),\nonumber \\
\boldsymbol{\mathcal{V}} & = & \mathbf{p}\left[V(x,\mathbf{p})-\frac{\mu_{5}}{\left|\mathbf{p}\right|}A(x,p)\right],\nonumber \\
\mathcal{A}^{0} & = & \left|\mathbf{p}\right|\left[A(x,p)-\frac{\mu_{5}}{\left|\mathbf{p}\right|}V(x,p)\right],\nonumber \\
\boldsymbol{\mathcal{A}} & = & \frac{\mathbf{p}}{\left|\mathbf{p}\right|}(p^{0}+\mu)A(x,p),\nonumber \\
\mathcal{S}^{0i} & = & 0,\nonumber \\
\mathcal{S}^{ij} & = & \frac{m}{\left|\mathbf{p}\right|}\epsilon^{ijk}\mathbf{p}^{k}A(x,p),\label{eq:Wigner function with chiral}
\end{eqnarray}
where we have defined
\begin{eqnarray}
V(x,p) & \equiv & \frac{2}{(2\pi)^{3}}\sum_{s}\delta\left[(p^{0}+\mu)^{2}-(\left|\mathbf{p}\right|-s\mu_{5})^{2}-m^{2}\right]\nonumber \\
 &  & \qquad\times\left\{ \theta(p^{0}+\mu)f_{s}^{(+)}(x,\mathbf{p})-\theta[-(p^{0}+\mu)]\left[1-f_{s}^{(-)}(x,-\mathbf{p})\right]\right\} ,\nonumber \\
A(x,p) & \equiv & \frac{2}{(2\pi)^{3}}\sum_{s}\delta\left[(p^{0}+\mu)^{2}-(\left|\mathbf{p}\right|-s\mu_{5})^{2}-m^{2}\right]\nonumber \\
 &  & \qquad\times s\left\{ \theta(p^{0}+\mu)f_{s}^{(+)}(x,\mathbf{p})-\theta[-(p^{0}+\mu)]\left[1-f_{s}^{(-)}(x,-\mathbf{p})\right]\right\} .
\end{eqnarray}
Note that the presence of $\mu$ and $\mu_{5}$ breaks the Lorentz
covariance of the Wigner function. That is why in (\ref{eq:Wigner function with chiral}),
we separately listed $\mathcal{V}^{0}$ and $\boldsymbol{\mathcal{V}}$
instead of a four-vector $\mathcal{V}^{\mu}$. For the same reason,
the axial-vector $\mathcal{A}^{\mu}$ is separated into $\mathcal{A}^{0}$
and $\boldsymbol{\mathcal{A}}$, while the tensor $\mathcal{S}^{\mu\nu}$
is separated into the electric-like components $\mathcal{S}^{0i}$
and the magnetic-like components $\mathcal{S}^{ij}$.

In the case without chiral imbalance, we have four undetermined functions,
$V^{(0)}(x,p)$ and $n^{(0)\mu}(x,p)$, which are defined in (\ref{def:definition of V and n}).
(Here $p_{\mu}n^{(0)\mu}=0$ thus $n^{(0)\mu}$ has only three independent
components.) However, in (\ref{eq:Wigner function with chiral}),
we only have two functions, $V(x,p)$ and $A(x,p)$. This loss of
degrees of freedom is attributed to the spin degeneracy. In the case
without $\mu_{5}$, the energy states are degenerate with respect
to spin. In the particle's rest frame, its spin can take arbitrary
spatial direction. However, a finite chiral chemical potential $\mu_{5}$
breaks the spin degeneracy, so the eigenstates, given by Eqs. (\ref{eq:fermion wave function with chiral})
and (\ref{eq:antifermion wave function with chiral}), have fixed
spin directions, i.e. along the direction of $\mathbf{p}$. Hence
in (\ref{eq:Wigner function with chiral}) the polarization density
$\boldsymbol{\mathcal{A}}$ is parallel to $\mathbf{p}$. This is
different to the case $\mu_{5}=0$, where the polarization density
$\boldsymbol{\mathcal{A}}$ can point in any direction. The reason
for this difference is because we forbid the mixture between different
energy levels. In Eq. (\ref{eq:definition of distributions with chiral})
we assume that the expectation values of $a_{\mathbf{p}-\frac{\mathbf{u}}{2},s^{\prime}}^{\dagger}a_{\mathbf{p}+\frac{\mathbf{u}}{2},s}$
and $b_{-\mathbf{p}+\frac{\mathbf{u}}{2},s^{\prime}}b_{-\mathbf{p}-\frac{\mathbf{u}}{2},s}^{\dagger}$
are proportional to $\delta_{ss^{\prime}}$, because states with different
$s$ are supported on different energy levels. If we allow the mixture
between different spin states, the final result would have more degrees
of freedom and thus is expected to coincide with the result in subsection
\ref{subsec:Free-fermions} in the limit $\mu_{5}\rightarrow0$.

\subsection{Fermions in a constant magnetic field \label{subsec:Fermions-in-const-B}}

\subsubsection{Dirac equation}

In a constant magnetic field, the transverse momentum of a particle
is discrete while the momentum along the direction of the magnetic
field stays continuous. The eigenenergies are given by the well-known
Landau energy levels
\begin{equation}
E_{p^{z}}^{(n)}=\sqrt{m^{2}+(p^{z})^{2}+2nB_{0}},\label{eq:well known Landau levels}
\end{equation}
where $B_{0}$ is the strength of the magnetic field and $n=0,1,2,\cdots$
label the Landau levels. Note that the electric charge has been absorbed
into the field. We can rewrite the quantum number $n$ as $n=n^{\prime}+\frac{1}{2}+\frac{1}{2}s$,
with $n^{\prime}=0,1,2,\cdots$ denotes the orbital quantum number
and $s=\pm$ represents the spin direction. Then the lowest Landau
level $n=0$ corresponds to $n^{\prime}=0$ and $s=-$, which means
the particles in the lowest Landau level $n=0$ have a definite spin
direction. According to the principle of minimum energy, the spins
of fermions with positive charges are parallel, while those of negatively
charged anti-fermions are anti-parallel, to the direction of the magnetic
field. Meanwhile, higher Landau levels $n>0$ are degenerate for $n^{\prime}=n$,
$s=-$ and $n^{\prime}=n-1$, $s=+$, which means these levels are
2-fold degenerate.

In this section we will consider fermions in a constant magnetic field.
Since the magnetic field is not Lorentz-covariant itself, we should
choose a specific frame. The choice of the frame will break the covariance.
We also consider finite $\mu$ and $\mu_{5}$. The Dirac equation
reads in this case%
\begin{equation}
i\frac{\partial}{\partial t}\psi=[\gamma^{0}\boldsymbol{\gamma}\cdot(-i\boldsymbol{\nabla}-\boldsymbol{\mathbb{A}})+m\gamma^{0}-\mu-\mu_{5}\gamma^{5}]\psi.\label{eq:Dirac-equation}
\end{equation}
Here $\mu$ and $\mu_{5}$ are assumed to be constant. We can read
off the Hamilton operator from the above equation
\begin{equation}
\hat{H}=\gamma^{0}\boldsymbol{\gamma}\cdot(-i\boldsymbol{\nabla}-\boldsymbol{\mathbb{A}})+m\gamma^{0}-\mu-\mu_{5}\gamma^{5}.\label{eq:Hamiltonian operator in magnetic field}
\end{equation}
Without loss of generality, the magnetic field is taken in the z-direction.
The magnetic field and gauge potential are
\begin{eqnarray}
\mathbf{B} & = & B_{0}\mathbf{e}_{z},\nonumber \\
\boldsymbol{\mathbb{A}} & = & -B_{0}y\mathbf{e}_{x},\label{eq:def-vector-potential}
\end{eqnarray}
where the field strength being a positive constant $B_{0}>0$. This
choice of the gauge potential is known as the Landau gauge. Another
widely used gauge is the symmetric gauge with $\boldsymbol{\mathbb{A}}=\frac{1}{2}\mathbf{B}\times\mathbf{r}$
but here we adopt the Landau gauge because under this gauge the wavefunctions
take the simplest form. The Wigner function will only depend on the
magnetic field and then be independent fromthe choice of gauge.

Since the gauge potential $\boldsymbol{\mathbb{A}}$ only depends
on the $y$-coordinate in the Landau gauge, while the mass $m$, the
chemical potential $\mu$, and the chiral chemical potential $\mu_{5}$
are assumed to be constant, one can check that the spatial derivatives
$\frac{\partial}{\partial x}$ and $\frac{\partial}{\partial z}$
commute with the Hamiltonian $\hat{H}$ in Eq. (\ref{eq:Hamiltonian operator in magnetic field}).
This indicates that we can introduce the momenta $p^{x}$ and $p^{z}$
as conserved variables. The solution of the Dirac equation (\ref{eq:Dirac-equation})
can be cast into the Fourier mode
\begin{equation}
\psi(t,\mathbf{x})=\int\frac{dp^{x}dp^{z}}{(2\pi)^{2}}e^{-iEt+ip^{x}x+ip^{z}z}\psi(p^{x},p^{z},y).\label{eq:solution}
\end{equation}
Here we adopt the Weyl representation, i.e., the Dirac spinor can
be decomposed into the LH and RH Pauli spinors,
\begin{equation}
\psi(p^{x},p^{z},y)=\left(\begin{array}{c}
\chi_{L}(p^{x},p^{z},y)\\
\chi_{R}(p^{x},p^{z},y)
\end{array}\right).\label{eq:wave function in magnetic field}
\end{equation}
Inserting Eq. (\ref{eq:solution}) into the Dirac equation (\ref{eq:Dirac-equation}),
one obtains the equations for the LH and RH spinors,%
\begin{eqnarray}
\left[E+\sigma^{1}(p^{x}+B_{0}y)+\sigma^{2}(-i\frac{\partial}{\partial y})+\sigma^{3}p^{z}+\mu-\mu_{5}\right]\chi_{L}(p^{x},p^{z},y) & = & m\chi_{R}(p^{x},p^{z},y),\nonumber \\
\left[E-\sigma^{1}(p^{x}+B_{0}y)-\sigma^{2}(-i\frac{\partial}{\partial y})-\sigma^{3}p^{z}+\mu+\mu_{5}\right]\chi_{R}(p^{x},p^{z},y) & = & m\chi_{L}(p^{x},p^{z},y).
\end{eqnarray}
For massless fermions $m=0$, the LH spinor is decoupled from the
RH one, which indicates that particles are either LH or RH. But in
the massive case there is a mixture between $\chi_{L}$ and $\chi_{R}$,
so the chirality is no longer a good quantum number. Inserting the
explicit formula for the Pauli matrices into the equations of $\chi_{R,L}$,
we obtain the matrix form, %
\begin{eqnarray}
\left(\begin{array}{cc}
E+\mu-\mu_{5}+p^{z} & -\frac{\partial}{\partial y}+(p^{x}+B_{0}y)\\
\frac{\partial}{\partial y}+(p^{x}+B_{0}y) & E+\mu-\mu_{5}-p^{z}
\end{array}\right)\chi_{L}(p^{x},p^{z},y) & = & m\chi_{R}(p^{x},p^{z},y),\nonumber \\
\left(\begin{array}{cc}
E+\mu+\mu_{5}-p^{z} & \frac{\partial}{\partial y}-(p^{x}+B_{0}y)\\
-\frac{\partial}{\partial y}-(p^{x}+B_{0}y) & E+\mu+\mu_{5}+p^{z}
\end{array}\right)\chi_{R}(p^{x},p^{z},y) & = & m\chi_{L}(p^{x},p^{z},y).
\end{eqnarray}
In order to make the formula simpler, we introduce the creation and
annihilation operators
\begin{eqnarray}
\hat{a} & = & \frac{1}{\sqrt{2B_{0}}}\left[\frac{\partial}{\partial y}+(p^{x}+B_{0}y)\right],\nonumber \\
\hat{a}^{\dagger} & = & \frac{1}{\sqrt{2B_{0}}}\left[-\frac{\partial}{\partial y}+(p^{x}+B_{0}y)\right].
\end{eqnarray}
They are, respectively, the creation and annihilation operators of
a harmonic oscillator around an equilibrium point $p^{x}/B_{0}$ with
the frequency $\omega=\sqrt{B_{0}}$. We can check that they satisfy
the commutation relation $\left[\hat{a},\hat{a}^{\dagger}\right]=1$.
But we should note that here $\hat{a}^{\dagger}$ is not the Hermitian
transpose of $\hat{a}$. Using these operators, the equations for
the Pauli spinors read
\begin{eqnarray}
\left(\begin{array}{cc}
E+\mu-\mu_{5}+p^{z} & \sqrt{2B_{0}}\hat{a}^{\dagger}\\
\sqrt{2B_{0}}\hat{a} & E+\mu-\mu_{5}-p^{z}
\end{array}\right)\chi_{L}(p^{x},p^{z},y) & = & m\chi_{R}(p^{x},p^{z},y),\nonumber \\
\left(\begin{array}{cc}
E+\mu+\mu_{5}-p^{z} & -\sqrt{2B_{0}}\hat{a}^{\dagger}\\
-\sqrt{2B_{0}}\hat{a} & E+\mu+\mu_{5}+p^{z}
\end{array}\right)\chi_{R}(p^{x},p^{z},y) & = & m\chi_{L}(p^{x},p^{z},y).\label{eq:equation of Chi_RL}
\end{eqnarray}
Inserting the second line into the first line or vise versa, one derives
the equation for RH or LH spinors by eliminating $\chi_{L}$ or $\chi_{R}$%
\begin{equation}
\left(\begin{array}{cc}
(E+\mu)^{2}-\Lambda^{-} & 2\mu_{5}\sqrt{2B_{0}}\hat{a}^{\dagger}\\
2\mu_{5}\sqrt{2B_{0}}\hat{a} & (E+\mu)^{2}-\Lambda^{+}
\end{array}\right)\chi_{R,L}(p^{x},p^{z},y)=0,\label{eq:eq-chi-RL-decoupled}
\end{equation}
where
\begin{equation}
\Lambda^{\pm}\equiv m^{2}+(p^{z}\pm\mu_{5})^{2}+2B_{0}\left(\hat{a}^{\dagger}\hat{a}+\frac{1}{2}\right)\pm B_{0}
\end{equation}
is the energy squared of particles with spin parallel or anti-parallel
to the magnetic field. We see that the RH and LH spinors satisfy the
same differential equation, thus we can solve one of them and derive
the other through the relation (\ref{eq:equation of Chi_RL}). Note
that if the chiral chemical potential vanishes, $\mu_{5}=0$, the
off-diagonal terms in Eq. (\ref{eq:eq-chi-RL-decoupled}) also vanish
and a straightforward calculation gives the eigenenergy
\begin{equation}
E_{s_{1}s_{2}}=-\mu+s_{1}\sqrt{m^{2}+(p^{z})^{2}+2B_{0}\left(\hat{a}^{\dagger}\hat{a}+\frac{1}{2}\right)+s_{2}B_{0}}\,.
\end{equation}
The term $2B_{0}\left(\hat{a}^{\dagger}\hat{a}+\frac{1}{2}\right)$
is the transverse energy squared, which comes from the coupling between
the magnetic field and orbital angular momentum. The last term in
the square root, $\pm B_{0}$, is the spin-magnetic coupling. This
energy level agrees with the well-known Landau levels in Eq. (\ref{eq:well known Landau levels}).

In the case of finite $\mu_{5}$, the off-diagonal terms in Eq. (\ref{eq:eq-chi-RL-decoupled})
take non-vanishing values. In order to solve Eq. (\ref{eq:eq-chi-RL-decoupled}),
we choose the basis of the harmonic oscillator, i.e., eigenstates
of $\hat{a}^{\dagger}\hat{a}$, %
\begin{equation}
\phi_{n}(p^{x},y)=\left(\frac{B_{0}}{\pi}\right)^{1/4}\frac{1}{\sqrt{2^{n}n!}}\exp\left[-\frac{B_{0}}{2}\left(y+\frac{p^{x}}{B_{0}}\right)^{2}\right]H_{n}\left[\sqrt{B_{0}}\left(y+\frac{p^{x}}{B_{0}}\right)\right].\label{eq:def-phi}
\end{equation}
Here $H_{n}$ are the Hermite polynomials. One can check the completeness
and orthonormality of $\phi_{n}$ as
\begin{eqnarray}
\sum_{n=0}^{\infty}\phi_{n}(p^{x},y)\phi_{n}(p^{x\prime},y^{\prime}) & = & \delta\left(y+\frac{p^{x}}{B_{0}}-y^{\prime}-\frac{p^{x\prime}}{B_{0}}\right),\nonumber \\
\int dy\phi_{n}(p^{x},y)\phi_{n'}(p^{x},y) & = & \delta_{nn'}.\label{eq:orthonormality}
\end{eqnarray}
When acting on the basis functions $\phi_{n}(p^{x},y)$, the operators
$\hat{a}^{\dagger}$ and $\hat{a}$ raise or decrease the quantum
number $n$,
\begin{eqnarray}
\hat{a}\phi_{n}(p^{x},y) & = & \sqrt{n}\phi_{n-1}(p^{x},y),\nonumber \\
\hat{a}^{\dagger}\phi_{n}(p^{x},y) & = & \sqrt{n+1}\phi_{n+1}(p^{x},y).\label{eq:raise or decrease quantum number}
\end{eqnarray}
Due to the completeness of $\phi_{n}$, the spinors can be expanded
as
\begin{equation}
\chi_{R/L}(p^{x},p^{z},y)=\sum_{n=0}^{\infty}\left(\begin{array}{c}
c_{n}(p^{x},p^{z})\\
d_{n}(p^{x},p^{z})
\end{array}\right)\phi_{n}(p^{x},y),\label{eq:decompose of Chi_RL}
\end{equation}
where all the $y$ dependence was put into  $\phi_{n}$. Then, from
Eq. (\ref{eq:eq-chi-RL-decoupled}) we derive
\begin{equation}
\sum_{n=0}^{\infty}\left(\begin{array}{cc}
\left[(E+\mu)^{2}-\lambda_{n}^{-}\right]\phi_{n}(p_{x},y) & 2\mu_{5}\sqrt{2(n+1)B_{0}}\phi_{n+1}(p_{x},y)\\
2\mu_{5}\sqrt{2nB_{0}}\phi_{n-1}(p_{x},y) & \left[(E+\mu)^{2}-\lambda_{n+1}^{+}\right]\phi_{n}(p_{x},y)
\end{array}\right)\left(\begin{array}{c}
c_{n}\\
d_{n}
\end{array}\right)=0,
\end{equation}
where
\begin{equation}
\lambda_{n}^{\pm}\equiv m^{2}+(p^{z}\pm\mu_{5})^{2}+2nB_{0}.
\end{equation}
Using the orthonormality conditions in Eq. (\ref{eq:orthonormality})
we can derive the equations for the coefficients $c_{n}$ and $d_{n}$%
\begin{eqnarray}
\left[(E+\mu)^{2}-\lambda_{0}^{-}\right]c_{0} & = & 0,\nonumber \\
2\mu_{5}\sqrt{2nB_{0}}c_{n}+\left[(E+\mu)^{2}-\lambda_{n}^{+}\right]d_{n-1} & = & 0,\ \ n>0,\nonumber \\
\left[(E+\mu)^{2}-\lambda_{n}^{-}\right]c_{n}+2\mu_{5}\sqrt{2nB_{0}}d_{n-1} & = & 0,\ \ n>0.\label{eq:eq-coefficients}
\end{eqnarray}
Here the coefficient $c_{0}$ decouples from all others, while $c_{n}$
always couples with $d_{n-1}$ for any $n\geq1$.

\subsubsection{Lowest Landau level}

\begin{figure}
\includegraphics[width=8cm]{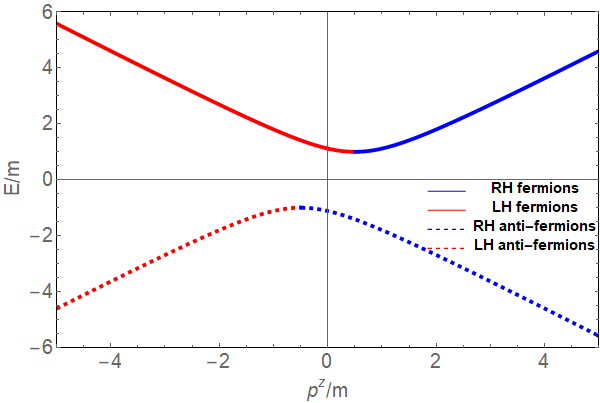}

\caption{\label{fig:Lowest-Landau-energy-level}Energy spectrum for fermions
(solid line with $E>0$) and anti-fermions (dashed line with $E<0$)
in the lowest Landau level. We take the mass as the energy unit. The
chiral chemical potential is taken to be $\mu_{5}/m=0.5$. RH particles
are shown in red, while LH in blue.}
\end{figure}

The lowest Landau level is given by demanding a non-vanishing $c_{0}$.
From the first line in Eq. (\ref{eq:eq-coefficients}), we obtain
the eigenenergies $E=-\mu\pm E_{p^{z}}^{(0)}$, there the upper/lower
sign labels fermions/anti-fermions. %
The energy of the lowest Landau level is
\begin{equation}
E_{p^{z}}^{(0)}=\sqrt{m^{2}+(p^{z}-\mu_{5})^{2}}\,.\label{eq:sol-level-E0}
\end{equation}
Inserting the eigenenergy $E=-\mu\pm E_{p^{z}}^{(0)}$ into the second
and third lines of Eq. (\ref{eq:eq-coefficients}), we obtain $c_{n}=d_{n-1}=0$
for any $n>0$. %
Then from the decomposition (\ref{eq:decompose of Chi_RL}), one can
construct the unnormalized eigenspinor for the lowest Landau level
\begin{equation}
\chi^{(0)}(p^{x},p^{z},y)=\left(\begin{array}{c}
c_{0}(p^{x},p^{z})\\
0
\end{array}\right)\phi_{0}(p^{x},y).
\end{equation}
Since the functions $\phi_{0}(p^{x},y)$ satisfies the orthonormality
relation in Eq. (\ref{eq:orthonormality}), we demand that the spinor
satisfies the normalization condition $\int dy\chi^{(0)\dagger}\chi^{(0)}=1$.
With the help of Eq. (\ref{eq:orthonormality}), we find that $c_{0}(p^{x},p^{z})=1$,
so the normalized eigenspinor is independent of the longitudinal momentum
$p^{z}$, which agrees with the case without chemical potentials.
The normalized eigenspinor for the lowest Landau level is
\begin{equation}
\chi^{(0)}(p^{x},y)=\left(\begin{array}{c}
1\\
0
\end{array}\right)\phi_{0}(p^{x},y).\label{eq:sol-chi-0}
\end{equation}
The lower component is zero, so this state is occupied by a particle
with the spin along the positive z-direction or an anti-particle with
the spin along the negative z-direction. Recalling that the z-direction
is the direction of the magnetic field, we see that the spin configuration
in the lowest Landau level ensures the lowest spin-magnetic coupling.
Since the LH and RH spinors satisfy the same equation (\ref{eq:eq-chi-RL-decoupled}),
we take $\chi_{L}^{(0)}(p^{x},y)=\chi^{(0)}(p^{x},y)$ without loss
of generality. The RH spinor is then derived from Eq. (\ref{eq:equation of Chi_RL}),
\begin{equation}
\chi_{R}^{(0)}(p^{x},p^{z},y)=\frac{E+\mu-\mu_{5}+p^{z}}{m}\chi^{(0)}(p^{x},y).
\end{equation}
Here the energy takes the eigenvalue for the lowest Landau level $E=-\mu\pm E_{p^{z}}^{(0)}$.
Then the wavefunction in Eq. (\ref{eq:wave function in magnetic field})
becomes
\begin{equation}
\psi_{r}^{(0)}(p^{x},p^{z},y)=\frac{1}{N_{r}}\left(\begin{array}{c}
1\\
\frac{E_{r}+\mu-\mu_{5}+p^{z}}{m}
\end{array}\right)\otimes\chi^{(0)}(p^{x},y),
\end{equation}
where $E_{r}=rE_{p^{z}}^{(0)}-\mu$ with $r=\pm$, and the symbol
$\otimes$ represents the tensor product of two matrices. The normalization
factor $N_{r}$ is determined by the normalization condition for the
wavefunction,
\begin{equation}
\int dy\ \psi_{r}^{(0)\dagger}(p^{x},p^{z},y)\psi_{r^{\prime}}^{(0)}(p^{x},p^{z},y)=\delta_{rr^{\prime}},\label{eq:normalization condition for LLL}
\end{equation}
which gives%
\begin{equation}
N_{r}=\frac{1}{\sqrt{E_{p^{z}}^{(0)}-r(p^{z}-\mu_{5})}}.
\end{equation}
Inserting $N_{r}$ into the solution we obtain the normalized wavefunction
\begin{equation}
\psi_{r}^{(0)}(p^{x},p^{z},y)=\frac{1}{\sqrt{2E_{p^{z}}^{(0)}}}\left(\begin{array}{c}
\sqrt{E_{p^{z}}^{(0)}-r(p^{z}-\mu_{5})}\\
r\sqrt{E_{p^{z}}^{(0)}+r(p^{z}-\mu_{5})}
\end{array}\right)\otimes\chi^{(0)}(p^{x},y).\label{eq:wave function LLL}
\end{equation}
The terms in the square root in the solution are always positive because
\begin{equation}
E_{p^{z}}^{(0)}=\sqrt{m^{2}+(p^{z}-\mu_{5})^{2}}\geq\left|p^{z}-\mu_{5}\right|.
\end{equation}
In the massless limit we obtain $E_{p^{z}}^{(0)}=\left|p^{z}-\mu_{5}\right|$,
and the wavefunction reduces to the chiral one,
\begin{equation}
\psi_{r}(p^{x},p^{z},y)=\begin{cases}
\left(\begin{array}{c}
0\\
\chi^{(0)}(p^{x},y)
\end{array}\right), & r\,\mathrm{sgn}(p^{z}-\mu_{5})>0,\\
\left(\begin{array}{c}
\chi^{(0)}(p^{x},y)\\
0
\end{array}\right), & r\,\mathrm{sgn}(p^{z}-\mu_{5})<0.
\end{cases}
\end{equation}
Here $r=\pm$ represent fermions and anti-fermions respectively. The
fermion states with $p^{z}>\mu_{5}$ are RH, while states with $p^{z}<\mu_{5}$
are LH. On the other hand, for anti-fermion states, $p^{z}>\mu_{5}$
corresponds to LH, while $p^{z}<\mu_{5}$ corresponds to RH. We plot
the energy spectrum $\pm E_{p^{z}}^{(0)}$ as a function of $p^{z}$
in Fig. \ref{fig:Lowest-Landau-energy-level}. In the Figure, the
x-axis represents the dimensionless longitudinal momentum $p^{z}/m$
and the y-axis represents the dimensionless energy $E/m$. The branch
with the positive energy is for fermions while the one with the negative
energy is for anti-fermions. We use the blue color for fermions/anti-fermions
with RH chirality while the red color for LH chirality. We observe
an energy gap $2m$ between fermions and anti-fermions induced by
the mass. There is also a gap $2\mu_{5}$ in the $x$-direction is
induced by the chiral chemical potential. %

The lowest Landau level is related to the CME. The fermions fill in
the positive-energy states from the lowest one. Hence fermions are
more likely to have positive $p^{z}$ because the energy spectrum
is not symmetric. On the hand, anti-fermions are more likely to have
negative longitudinal momentum $p^{z}$, which can be observed from
the dashed line in Fig. \ref{fig:Lowest-Landau-energy-level}. Therefore
there will be a net fermion current along the positive $z$-direction,
i.e., the direction of the magnetic field. Later on we will show that
the higher Landau levels do not contribute to the CME because they
are symmetric in $p^{z}$.

\subsubsection{Higher Landau levels}

Similar to the lowest Landau level, the higher Landau levels are obtained
by demanding a non-vanishing $c_{n}$ with $n>0$. According to Eq.
(\ref{eq:eq-coefficients}), in the presence of a non-vanishing $\mu_{5}$,
the coefficient $c_{n}$ is always coupled to the coefficient $d_{n-1}$.
Eliminating $d_{n-1}$ we obtain an equation for $c_{n}$,%
\begin{equation}
\left\{ \left[(E+\mu)^{2}-\lambda_{n}^{+}\right]\left[(E+\mu)^{2}-\lambda_{n}^{-}\right]-8nB_{0}\mu_{5}^{2}\right\} c_{n}=0.
\end{equation}
In order to have a non-trivial $c_{n}$, the coefficient must vanish,
which gives the eigenenergies are $E=-\mu+s_{1}E_{p^{z}s_{2}}^{(n)}$,
where
\begin{equation}
E_{p^{z}s}^{(n)}=\sqrt{m^{2}+\left[\sqrt{(p^{z})^{2}+2nB_{0}}-s\mu_{5}\right]^{2}}\label{eq:sol-level-En}
\end{equation}
is the energy of the $n$-th Landau level, with $n>0$ and $s=\pm$.
The coefficient $c_{m}$ with $m\neq n$ must vanish because $c_{m}$
and $c_{n}$ correspond to different eigenenergies.

In the massless limit and assuming $\sqrt{(p^{z})^{2}+2nB_{0}}\gg\left|\mu_{5}\right|$
(this is possible because $\mu_{5}$ labels the chiral imbalance which
in general should be a small variable compared to the momentum), we
have the eigenenergies
\begin{equation}
E=r\sqrt{(p^{z})^{2}+2nB_{0}}-(\mu+rs\mu_{5}).
\end{equation}
Note that $r=\pm$ label states with the positive ($+$) and negative
($-$) energy. In the massless case, $\mu+\mu_{5}$ is the chemical
potential for RH particles while $\mu-\mu_{5}$ is the one for LH
ones, so the product $rs$ denotes the chirality. The parameter $s$
labels the helicity because fermions ($r=+$) with the RH helicity
($s=+$) and anti-fermions ($r=-$) with the LH helicity ($s=-$)
have the RH chirality ($rs=+$), or vice versa. In the case $\mu_{5}=0$,
the eigenenergies in Eqs. (\ref{eq:sol-level-E0}) and (\ref{eq:sol-level-En})
reproduce the well-known Landau energy levels in (\ref{eq:well known Landau levels}).
The higher Landau levels $E_{p^{z}s}^{(n)}$ in Eq. (\ref{eq:sol-level-En})
are degenerate for $s=\pm$ and any $n>0$.

\begin{figure}
\includegraphics[width=8cm]{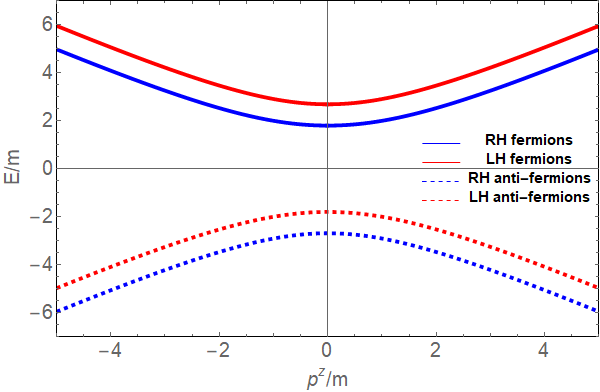}

\caption{\label{fig:Energy-spectrum-for-higher-Landau }Energy spectrum for
fermions (solid lines with $E>0$) and anti-fermions (dashed lines
with $E<0$) in the Landau level $n=1$. The mass $m$ is taken to
be the unit of the energy and momentum. The magnetic field strength
is chosen to be $B_{0}/m^{2}=2$ and the chiral chemical potential
$\mu_{5}/m=0.5$. We use the blue color for particles with the RH
chirality and the red color for those with the LH chirality. The curves
are even functions of $p^{z}$.}
\end{figure}

In Fig. \ref{fig:Energy-spectrum-for-higher-Landau } we plot the
energy spectrum for the first Landau level $n=1$. In this figure
we use the blue color to label the branches which in the massless
limit reproduce the states of the RH chirality and use the red color
for the LH chirality. We observe a gap between LH and RH branches,
which is attributed to a finite $\mu_{5}$. Note that the energy spectrum
is symmetric for flipping the sign of the longitudinal momentum $p^{z}\leftrightarrow-p^{z}$.
Thus, if the distribution only depends on the energy spectrum, the
number of particles moving in the positive $z$-direction equals to
that moving in the negative $z$-direction. The corresponding currents
cancel with each other and there is no macroscopic current for the
Landau levels with $n>0$.

Now we derive the wavefunction of the $n$-th Landau level. Inserting
the eigenenergy into Eq. (\ref{eq:eq-coefficients}) we obtain a relation
between $c_{n}$ and $d_{n-1}$. All other coefficients $c_{m},\ d_{m-1}$
with $m\neq n$ have to vanish. According to the expansion in Eq.
(\ref{eq:decompose of Chi_RL}), we obtain the unnormalized Pauli
spinors %
\begin{equation}
\chi_{s}^{(n)}(p^{x},p^{z},y)=c_{n}(p^{x},p^{z})\left(\begin{array}{c}
\sqrt{2nB_{0}}\phi_{n}(p^{x},y)\\
\left(s\sqrt{(p^{z})^{2}+2nB_{0}}-p^{z}\right)\phi_{n-1}(p^{x},y)
\end{array}\right).
\end{equation}
Again we demand the orthonormality condition $\int dy\chi_{s}^{(n)\dagger}\chi_{s^{\prime}}^{(n)}=\delta_{ss^{\prime}}$
to determine the coefficient $c_{n}$. The normalized eigenspinors
read
\begin{equation}
\chi_{s}^{(n)}(p^{x},p^{z},y)=\frac{1}{\sqrt{2\sqrt{(p^{z})^{2}+2nB_{0}}}}\left(\begin{array}{c}
\sqrt{\sqrt{(p^{z})^{2}+2nB_{0}}+sp^{z}}\phi_{n}(p^{x},y)\\
s\sqrt{\sqrt{(p^{z})^{2}+2nB_{0}}-sp^{z}}\phi_{n-1}(p^{x},y)
\end{array}\right).\label{eq:sol-chi-n}
\end{equation}
Note that the spinors $\chi_{s}^{(n)}(p^{x},p^{z},y)$ are real functions
because a) $\phi_{n}(p^{x},y)$ are real; b) the magnetic field strength
$B_{0}$ is positive; c) $\sqrt{(p^{z})^{2}+2nB_{0}}>\left|p^{z}\right|$.
If the chiral chemical potential vanish, $\mu_{5}=0$, according to
Eq. (\ref{eq:eq-coefficients}), the state with $c_{n}\neq0$ and
the one with $d_{n-1}\neq0$ have the same energy $E=-\mu\pm\sqrt{\lambda^{\pm}}$
with $\lambda^{\pm}=m^{2}+(p^{z})^{2}+2nB_{0}$. Since $c_{n}\neq0$
corresponds to a spin-up state and $d_{n-1}\neq0$ corresponds to
a spin-down state, we conclude that the higher Landau levels are two-fold
degenerate with respect to spin if $\mu_{5}=0$. But for a finite
$\mu_{5}$, the eigenenergy $E=-\mu+s_{1}E_{p^{z}s_{2}}^{(n)}$ depends
on $s_{1}$ and $s_{2}$, while the eigenspinors in Eq. (\ref{eq:sol-chi-n})
are mixture of spin-up and spin-down states. Using the orthonormality
condition for $\phi_{n}$ in Eq. (\ref{eq:orthonormality}) we can
check that the spinors for higher Landau levels satisfy the orthonormality
condition, and they are also orthogonal to the one for the lowest
Landau level,
\begin{eqnarray}
\int dy\chi_{s}^{(n)\dagger}(p^{x},p^{z},y)\chi^{(0)}(p^{x},y) & = & 0,\nonumber \\
\int dy\chi_{s^{\prime}}^{(n^{\prime})\dagger}(p^{x},p^{z},y)\chi_{s}^{(n)}(p^{x},p^{z},y) & = & \delta_{ss^{\prime}}\delta_{nn^{\prime}}.\label{eq:orthonomality-chi}
\end{eqnarray}

The wavefunction in momentum space can be obtained following the procedure
for the lowest Landau level. First the LH spinor is assumed to take
the form given in Eq. (\ref{eq:sol-chi-n}), $\chi_{L,s}^{(n)}(p^{x},p^{z},y)=\chi_{s}^{(n)}(p^{x},p^{z},y)$.
Then the RH spinor can be derived from Eq. (\ref{eq:equation of Chi_RL}),
\begin{equation}
\chi_{R,s}^{(n)}(p^{x},p^{z},y)=\frac{1}{m}\left(\begin{array}{cc}
E+\mu-\mu_{5}+p^{z} & \sqrt{2B_{0}}\hat{a}^{\dagger}\\
\sqrt{2B_{0}}\hat{a} & E+\mu-\mu_{5}-p^{z}
\end{array}\right)\chi_{s}^{(n)}(p^{x},p^{z},y),
\end{equation}
where $E=rE_{p^{z}s}^{(n)}-\mu$ which depends on $s$ and $n$. Here
$r=\pm$ label the fermion ($+$) and anti-fermion ($-$). Inserting
the solution (\ref{eq:sol-chi-n}) into the above equation, and using
Eq. (\ref{eq:raise or decrease quantum number}) to deal with $\hat{a}^{\dagger}$
and $\hat{a}$, we obtain the RH spinor and then the unnormalized
wavefunction %
\begin{equation}
\psi_{rs}^{(n)}(p^{x},p^{z},y)=\frac{1}{mN_{rs}^{(n)}}\left(\begin{array}{c}
1\\
rE_{p^{z}s}^{(n)}+s\sqrt{(p^{z})^{2}+2nB_{0}}-\mu_{5}
\end{array}\right)\otimes\chi_{s}^{(n)}(p^{x},p^{z},y).
\end{equation}
After proper normalization, we obtain %
\begin{equation}
\psi_{rs}^{(n)}(p^{x},p^{z},y)=\frac{1}{\sqrt{2E_{p_{z}s}^{(n)}}}\left(\begin{array}{c}
r\sqrt{E_{p^{z}s}^{(n)}+r\mu_{5}-rs\sqrt{p_{z}^{2}+2nB_{0}}}\\
\sqrt{E_{p^{z}s}^{(n)}-r\mu_{5}+rs\sqrt{p_{z}^{2}+2nB_{0}}}
\end{array}\right)\otimes\chi_{s}^{(n)}(p^{x},p^{z},y).\label{eq:wave function HLL}
\end{equation}
Again, due to the non-zero mass, the terms in square roots are always
positive, and the wavefunction is real. With the help of Eq. (\ref{eq:orthonormality}),
we can check that the wavefunctions satisfy the orthonormality conditions
\begin{eqnarray}
\int dy\psi_{r^{\prime}s}^{(n)\dagger}(p^{x},p^{z},y)\psi_{r}^{(0)}(p^{x},y) & = & 0,\nonumber \\
\int dy\psi_{r^{\prime}s^{\prime}}^{(n^{\prime})\dagger}(p^{x},p^{z},y)\psi_{rs}^{(n)}(p^{x},p^{z},y) & = & \delta_{rr^{\prime}}\delta_{ss^{\prime}}\delta_{nn^{\prime}}.\label{eq:orthonomality-psi}
\end{eqnarray}
Note that $p^{x}$ in the wavefunctions (\ref{eq:wave function LLL})
and (\ref{eq:wave function HLL}) is the momentum in the $x$ direction,
but it also determines the center position in the $y$ direction.
The obtained wavefunctions are plane waves in the $x$ and $z$ directions
but have a finite extent in the $y$ direction because of the harmonic
oscillator eigenfunctions $\phi_{n}(p^{x},y)$. One can have a more
realistic description of a quantum particle with given position and
momentum by constructing wave packets by superposition of the single
particle wavefunctions with different $p^{x}$ and $p^{z}$. In this
thesis, the wave-packet description will be used in computing the
Wigner function.

Now we briefly discuss the density of state. We consider a finite
volume $l_{x}\times l_{y}\times l_{z}$ with periodic boundary conditions.
Then $p^{x}$ takes discrete values,
\begin{equation}
p^{x}=\frac{2\pi n_{x}}{l_{x}},\ \ n_{x}=\cdots,-1,\,0,\,1,\,2,\cdots.
\end{equation}
All the wavefunctions are constructed from the harmonic oscillator
wavefunction $\phi_{n}(p^{x},y)$, and the center position of the
harmonic oscillator is $y=-p^{x}/B_{0}$. In order to make sure this
center position is located inside the area considered, we demand
\begin{equation}
0\leq-\frac{p^{x}}{B_{0}}\leq l_{y},
\end{equation}
from which we obtain
\begin{equation}
-\frac{B_{0}l_{x}l_{y}}{2\pi}\leq n^{x}\leq0.
\end{equation}
So the density of state is $B_{0}l_{x}l_{y}/(2\pi)$ for given $p^{z}$
and $n,\,r,\,s$. This result is consistent with our knowledge about
Landau levels.

\subsubsection{Landau quantization}

We have given in (\ref{eq:wave function LLL}) and (\ref{eq:wave function HLL})
the wavefunctions for the lowest Landau level and higher Landau levels
respectively. Using these wavefunctions, the Dirac operator can be
quantized as
\begin{eqnarray}
\hat{\psi}(t,\mathbf{x}) & = & e^{i\mu t}\sum_{n,s}\int\frac{dp^{x}dp^{z}}{(2\pi)^{2}}e^{ip^{x}x+ip^{z}z}\left[\exp\left(-iE_{p^{z}s}^{(n)}t\right)\psi_{+,s}^{(n)}(p^{x},p^{z},y)\hat{a}_{s}^{(n)}(p^{x},p^{z})\right.\nonumber \\
 &  & \qquad\left.+\exp\left(iE_{p^{z}s}^{(n)}t\right)\psi_{-,s}^{(n)}(p^{x},p^{z},y)\hat{b}_{s}^{(n)\dagger}(-p^{x},-p^{z})\right],\label{eq:def-field-operator}
\end{eqnarray}
where we have defined
\begin{equation}
\sum_{n,s}f_{s}^{(n)}\equiv f^{(0)}+\sum_{s=\pm}\sum_{n>0}f_{s}^{(n)},
\end{equation}
for any function $f_{s}^{(n)}$ which depends on the helicity index
$s$ and Landau level $n$. The particles in the lowest Landau level
always have the fixed spin. Here $\hat{a}_{s}^{(n)}(p^{x},p^{z})$
is the annihilation operator for a fermion in the $n$-th Landau level
with $p^{x}$, $p^{z}$, and $s$. Similarly, $\hat{b}_{s}^{(n)\dagger}(p^{x},p^{z})$
is the creation operator for an anti-fermion in the $n$-th Landau
level with the same $p^{x}$, $p^{z}$, and $s$. We observe that
the contribution from the chemical potential $\mu$ to the field operator
is an overall factor $e^{i\mu t}$, while the chiral chemical potential
$\mu_{5}$ enters the energy levels $E_{p^{z}s}^{(n)}$. We further
assume that the creation and annihilation operators satisfy the following
anti-commutation relations
\begin{eqnarray}
\left\{ \hat{a}_{p^{x}p^{z}s}^{(n)},\hat{a}_{q^{x}q^{z}s^{\prime}}^{(n^{\prime})\dagger}\right\}  & = & (2\pi)^{2}\delta(p^{x}-q^{x})\delta(p^{z}-q^{z})\delta_{nn^{\prime}}\delta_{ss^{\prime}},\nonumber \\
\left\{ \hat{b}_{p^{x}p^{z}s}^{(n)},\hat{b}_{q^{x}q^{z}s^{\prime}}^{(n^{\prime})\dagger}\right\}  & = & (2\pi)^{2}\delta(p^{x}-q^{x})\delta(p^{z}-q^{z})\delta_{nn^{\prime}}\delta_{ss^{\prime}},
\end{eqnarray}
with all other anti-commutators vanishing. These relations are straightforward
extensions of the free case, but it is reasonable because we can derive
the following equal-time anti-commutation relations for the field
operators %
\begin{eqnarray}
 &  & \left\{ \hat{\psi}_{\alpha}(t,\mathbf{x}),\hat{\psi}_{\beta}^{\dagger}(t,\mathbf{x}^{\prime})\right\} =\delta_{\alpha\beta}\delta^{(3)}(\mathbf{x}-\mathbf{x}^{\prime}),\nonumber \\
 &  & \left\{ \hat{\psi}_{\alpha}(t,\mathbf{x}),\hat{\psi}_{\beta}(t,\mathbf{x}^{\prime})\right\} =\left\{ \hat{\psi}_{\alpha}^{\dagger}(t,\mathbf{x}),\hat{\psi}_{\beta}^{\dagger}(t,\mathbf{x}^{\prime})\right\} =0,
\end{eqnarray}
where $\alpha$ and $\beta$ are indices of the Dirac spinors.

Since we have computed the eigenstates of the Hamilton operator and
used them to quantize the Dirac field in Eq. (\ref{eq:def-field-operator}),
it is straightforward to rewrite the Hamiltonian using the creation
and annihilation operators,
\begin{eqnarray}
\hat{H} & = & \sum_{n,s}\int\frac{dp^{x}dp^{z}}{(2\pi)^{2}}\left[\left(E_{p^{z}s}^{(n)}-\mu\right)a_{s}^{(n)\dagger}(p^{x},p^{z})a_{s}^{(n)}(p^{x},p^{z})\right.\nonumber \\
 &  & \qquad\left.-\left(E_{p^{z}s}^{(n)}+\mu\right)b_{s}^{(n)}(-p^{x},-p^{z})b_{s}^{(n)\dagger}(-p^{x},-p^{z})\right].\label{eq:quantized Hamiltonian in magnetic}
\end{eqnarray}
The momentum operators are given by %
\begin{eqnarray}
\hat{P}_{x} & = & \sum_{n,s}\int\frac{dp^{x}dp^{z}}{(2\pi)^{2}}p^{x}\left[a_{s}^{(n)\dagger}(p^{x},p^{z})a_{s}^{(n)}(p^{x},p^{z})-b_{s}^{(n)}(p^{x},p^{z})b_{s}^{(n)\dagger}(p^{x},p^{z})\right],\nonumber \\
\hat{P}_{z} & = & \sum_{n,s}\int\frac{dp^{x}dp^{z}}{(2\pi)^{2}}p^{z}\left[a_{s}^{(n)\dagger}(p^{x},p^{z})a_{s}^{(n)}(p^{x},p^{z})-b_{s}^{(n)}(p^{x},p^{z})b_{s}^{(n)\dagger}(p^{x},p^{z})\right].
\end{eqnarray}
In these calculations we have used the orthonormality conditions in
Eqs. (\ref{eq:normalization condition for LLL}) and (\ref{eq:orthonomality-psi}).

\subsubsection{Wigner function }

We have derived the wavefunctions in Eqs. (\ref{eq:wave function LLL})
and (\ref{eq:wave function HLL}) and the field operator is quantized
in Eq. (\ref{eq:def-field-operator}). Inserting the field operator
into the definition of the Wigner function (\ref{def:Wigner function})
we obtain %
\begin{eqnarray}
W(x,p) & = & \sum_{n,s}\sum_{n^{\prime}s^{\prime}}\int\frac{dy^{\prime}}{(2\pi)}\int\frac{dq_{1}^{x}dq_{1}^{z}dq_{2}^{x}dq_{2}^{z}}{(2\pi)^{4}}\delta\left(p^{x}-B_{0}y-\frac{q_{1}^{x}+q_{2}^{x}}{2}\right)\delta\left(p^{z}-\frac{q_{1}^{z}+q_{2}^{z}}{2}\right)\nonumber \\
 &  & \times\exp\left[i(q_{1}^{x}-q_{2}^{x})x+i(q_{1}^{z}-q_{2}^{z})z+ip^{y}y^{\prime}\right]\nonumber \\
 &  & \times\Biggl\{\left\langle a_{s^{\prime}}^{(n^{\prime})\dagger}(q_{2}^{x},q_{2}^{z})a_{s}^{(n)}(q_{1}^{x}q_{1}^{z})\right\rangle \bar{\psi}_{+,s^{\prime}}^{(n^{\prime})}\left(q_{2}^{x},q_{2}^{z},y+\frac{y^{\prime}}{2}\right)\otimes\psi_{+,s}^{(n)}\left(q_{1}^{x},q_{1}^{z},y-\frac{y^{\prime}}{2}\right)\nonumber \\
 &  & \qquad\times\exp\left[i\left(E_{q_{2}^{z}s^{\prime}}^{(n^{\prime})}-E_{q_{1}^{z}s}^{(n)}\right)t\right]\delta\left(p^{0}+\mu-\frac{E_{q_{2}^{z}s^{\prime}}^{(n^{\prime})}+E_{q_{1}^{z}s}^{(n)}}{2}\right)\nonumber \\
 &  & \quad+\left\langle b_{s^{\prime}}^{(n^{\prime})}(-q_{2}^{x},-q_{2}^{z})b_{s}^{(n)\dagger}(-q_{1}^{x},-q_{1}^{z})\right\rangle \bar{\psi}_{-,s^{\prime}}^{(n^{\prime})}\left(q_{2}^{x},q_{2}^{z},y+\frac{y^{\prime}}{2}\right)\otimes\psi_{-,s}^{(n)}\left(q_{1}^{x},q_{1}^{z},y-\frac{y^{\prime}}{2}\right)\nonumber \\
 &  & \qquad\times\exp\left[-i\left(E_{q_{2}^{z}s^{\prime}}^{(n^{\prime})}-E_{q_{1}^{z}s}^{(n)}\right)t\right]\delta\left(p^{0}+\mu+\frac{E_{q_{2}^{z}s^{\prime}}^{(n^{\prime})}+E_{q_{1}^{z}s}^{(n)}}{2}\right)\Biggr\},
\end{eqnarray}
where $y$ is the y-component of the four-vector $x^{\mu}$, and $y^{\prime}$
is an integration variable. Here we have dropped mixing terms of fermions
and anti-fermions, i.e., $a_{q_{2}^{x}q_{2}^{z}s^{\prime}}^{(n^{\prime})\dagger}b_{-q_{1}^{x},-q_{1}^{z},s}^{(n)\dagger}$
and $b_{-q_{2}^{x},-q_{2}^{z},s^{\prime}}^{(n^{\prime})}a_{q_{1}^{x}q_{1}^{z}s}^{(n)}$.
These terms only contribute if there is a mixture between the fermion
and anti-fermion state. Analogous to what we did in the case of free
fermions in subsection \ref{subsec:Free-fermions}, we define the
average and relative momenta,
\begin{equation}
k^{x}=\frac{q_{1}^{x}+q_{2}^{x}}{2},\ \ k^{z}=\frac{q_{1}^{z}+q_{2}^{z}}{2},\ \ u^{x}=q_{1}^{x}-q_{2}^{x},\ \ u^{z}=q_{1}^{z}-q_{2}^{z}.
\end{equation}
Using these new variables, the integration measure stays unchanged
\begin{equation}
dq_{1}^{x}dq_{1}^{z}dq_{2}^{x}dq_{2}^{z}=dk^{x}dk^{z}du^{x}du^{z}.
\end{equation}
Then the Wigner function reads
\begin{eqnarray}
W(x,p) & = & \sum_{n,s}\sum_{n^{\prime}s^{\prime}}\int\frac{dy^{\prime}}{(2\pi)}\int\frac{dk^{x}dk^{z}du^{x}du^{z}}{(2\pi)^{4}}\delta\left(p^{x}-B_{0}y-k^{x}\right)\delta\left(p^{z}-k^{z}\right)\exp\left(iu^{x}x+iu^{z}z+ip^{y}y^{\prime}\right)\nonumber \\
 &  & \times\Biggl\{\left\langle a_{s^{\prime}}^{(n^{\prime})\dagger}\left(k^{x}-\frac{u^{x}}{2},\,k^{z}-\frac{u^{z}}{2}\right)a_{s}^{(n)}\left(k^{x}+\frac{u^{x}}{2},\,k^{z}+\frac{u^{z}}{2}\right)\right\rangle \nonumber \\
 &  & \qquad\times\bar{\psi}_{+,s^{\prime}}^{(n^{\prime})}\left(k^{x}-\frac{u^{x}}{2},\,k^{z}-\frac{u^{z}}{2},\,y+\frac{y^{\prime}}{2}\right)\otimes\psi_{+,s}^{(n)}\left(k^{x}+\frac{u^{x}}{2},\,k^{z}+\frac{u^{z}}{2},\,y-\frac{y^{\prime}}{2}\right)\nonumber \\
 &  & \qquad\times\exp\left[i\left(E_{k^{z}-\frac{1}{2}u^{z},s^{\prime}}^{(n^{\prime})}-E_{k^{z}+\frac{1}{2}u^{z},s}^{(n)}\right)t\right]\delta\left(p^{0}+\mu-\frac{E_{k^{z}-\frac{1}{2}u^{z},s^{\prime}}^{(n^{\prime})}+E_{k^{z}+\frac{1}{2}u^{z},s}^{(n)}}{2}\right)\nonumber \\
 &  & \quad+\left\langle b_{s^{\prime}}^{(n^{\prime})}\left(-k^{x}+\frac{u^{x}}{2},\,-k^{z}+\frac{u^{z}}{2}\right)b_{s}^{(n)\dagger}\left(-k^{x}-\frac{u^{x}}{2},\,-k^{z}-\frac{u^{z}}{2}\right)\right\rangle \nonumber \\
 &  & \qquad\times\bar{\psi}_{-,s^{\prime}}^{(n^{\prime})}\left(k^{x}-\frac{u^{x}}{2},\,k^{z}-\frac{u^{z}}{2},\,y+\frac{y^{\prime}}{2}\right)\otimes\psi_{-,s}^{(n)}\left(k^{x}+\frac{u^{x}}{2},\,k^{z}+\frac{u^{z}}{2},\,y-\frac{y^{\prime}}{2}\right)\nonumber \\
 &  & \qquad\times\exp\left[-i\left(E_{k^{z}-\frac{1}{2}u^{z},s^{\prime}}^{(n^{\prime})}-E_{k^{z}+\frac{1}{2}u^{z},s}^{(n)}\right)t\right]\delta\left(p^{0}+\mu+\frac{E_{k^{z}-\frac{1}{2}u^{z},s^{\prime}}^{(n^{\prime})}+E_{k^{z}+\frac{1}{2}u^{z},s}^{(n)}}{2}\right)\Biggr\}.\nonumber \\
\end{eqnarray}
The Wigner function now is a two-point correlation function in momentum
space. We adopt the wave-packet description and assume the expectation
values are given by some distribution functions that will be determined
later,
\begin{eqnarray}
\left\langle a_{s^{\prime}}^{(n^{\prime})\dagger}\left(k^{x}-\frac{u^{x}}{2},\,k^{z}-\frac{u^{z}}{2}\right)a_{s}^{(n)}\left(k^{x}+\frac{u^{x}}{2},\,k^{z}+\frac{u^{z}}{2}\right)\right\rangle  & = & \delta_{ss^{\prime}}\delta_{nn^{\prime}}f_{s}^{(+)(n)}(k^{x},k^{z},u^{x},u^{z}),\nonumber \\
\left\langle b_{s^{\prime}}^{(n^{\prime})}\left(-k^{x}+\frac{u^{x}}{2},\,-k^{z}+\frac{u^{z}}{2}\right)b_{s}^{(n)\dagger}\left(-k^{x}-\frac{u^{x}}{2},\,-k^{z}-\frac{u^{z}}{2}\right)\right\rangle  & = & \delta_{ss^{\prime}}\delta_{nn^{\prime}}\left[(2\pi)^{2}\delta(u^{x})\delta(u^{z})\right.\nonumber \\
 &  & \quad\left.-f_{s}^{(-)(n)}(-k^{x},-k^{z},u^{x},u^{z})\right].\nonumber \\
\end{eqnarray}
The expectation values are proportional to the Kronecker-deltas, $\delta_{ss^{\prime}}\delta_{nn^{\prime}}$
because we assume the wave packets are constructed by states at the
same Landau level $n$ with the same helicity $s$. In the free fermion
case, energies are two-fold degenerated for the spin, but for non-zero
$\mu_{5}$, this spin degeneracy disappears and all the quantum states
are not degenerate any more. Inserting these expectation values back
to the Wigner function we obtain
\begin{eqnarray}
W(x,p) & = & \sum_{n,s}\int\frac{dy^{\prime}}{(2\pi)}\int\frac{du^{x}du^{z}}{(2\pi)^{4}}\exp\left(iu^{x}x+iu^{z}z+ip^{y}y^{\prime}\right)\nonumber \\
 &  & \times\Biggl\{ f_{s}^{(+)(n)}(p^{x}-B_{0}y,p^{z},u^{x},u^{z})\nonumber \\
 &  & \qquad\times\exp\left[i\left(E_{p^{z}-\frac{1}{2}u^{z},s}^{(n)}-E_{p^{z}+\frac{1}{2}u^{z},s}^{(n)}\right)t\right]\delta\left(p^{0}+\mu-\frac{E_{p^{z}-\frac{1}{2}u^{z},s}^{(n)}+E_{p^{z}+\frac{1}{2}u^{z},s}^{(n)}}{2}\right)\nonumber \\
 &  & \qquad\times\bar{\psi}_{+,s}^{(n)}\left(p^{x}-B_{0}y-\frac{u^{x}}{2},\,p^{z}-\frac{u^{z}}{2},\,y+\frac{y^{\prime}}{2}\right)\nonumber \\
 &  & \qquad\otimes\psi_{+,s}^{(n)}\left(p^{x}-B_{0}y+\frac{u^{x}}{2},\,p^{z}+\frac{u^{z}}{2},\,y-\frac{y^{\prime}}{2}\right)\nonumber \\
 &  & \quad+\left[(2\pi)^{2}\delta(u^{x})\delta(u^{z})-f_{s}^{(-)(n)}(-p^{x}+B_{0}y,-p^{z},u^{x},u^{z})\right]\nonumber \\
 &  & \qquad\times\exp\left[-i\left(E_{p^{z}-\frac{1}{2}u^{z},s}^{(n)}-E_{p^{z}+\frac{1}{2}u^{z},s}^{(n)}\right)t\right]\delta\left(p^{0}+\mu+\frac{E_{p^{z}-\frac{1}{2}u^{z},s}^{(n)}+E_{p^{z}+\frac{1}{2}u^{z},s}^{(n)}}{2}\right)\nonumber \\
 &  & \qquad\times\bar{\psi}_{-,s}^{(n)}\left(p^{x}-B_{0}y-\frac{u^{x}}{2},\,p^{z}-\frac{u^{z}}{2},\,y+\frac{y^{\prime}}{2}\right)\nonumber \\
 &  & \qquad\otimes\psi_{-,s}^{(n)}\left(p^{x}-B_{0}y+\frac{u^{x}}{2},\,p^{z}+\frac{u^{z}}{2},\,y-\frac{y^{\prime}}{2}\right)\Biggr\}.
\end{eqnarray}
Assuming that the relative momenta $u^{x}$ and $u^{z}$ are small
we can expand $E_{p^{z}\pm\frac{1}{2}u^{z},s}^{(n)}$ as well as the
wavefunctions in $u^{x}$ and $u^{z}$
\begin{equation}
E_{p^{z}+\frac{1}{2}u^{z},s}^{(n)}\equiv E_{p^{z},s}^{(n)}+\mathcal{O}(u).
\end{equation}
The local distributions are defined as
\begin{eqnarray}
f_{s}^{(+)(n)}(p^{x},p^{z},\mathbf{x}) & \equiv & \int\frac{du^{x}du^{z}}{(2\pi)^{2}}\exp\left(iu^{x}x+iu^{z}z\right)f_{s}^{(+)(n)}(p^{x}-B_{0}y,p^{z},u^{x},u^{z}),\nonumber \\
f_{s}^{(-)(n)}(-p^{x},-p^{z},\mathbf{x}) & \equiv & \int\frac{du^{x}du^{z}}{(2\pi)^{2}}\exp\left(iu^{x}x+iu^{z}z\right)f_{s}^{(-)(n)}(-p^{x}+B_{0}y,p^{z},u^{x},u^{z}).\label{eq:def-distribution for nth Landau level}
\end{eqnarray}
Then the Wigner function at the leading order in the spatial gradient
reads %
{} %
\begin{eqnarray}
W(x,p) & = & \frac{1}{(2\pi)^{3}}\sum_{n,s}f_{s}^{(+)(n)}(p^{x},p^{z},\mathbf{x})W_{+,s}^{(n)}(\mathbf{p})\delta\left(p^{0}+\mu-E_{p^{z},s}^{(n)}\right)\nonumber \\
 &  & +\frac{1}{(2\pi)^{3}}\sum_{n,s}\left[1-f_{s}^{(-)(n)}(-p^{x},-p^{z},\mathbf{x})\right]W_{-,s}^{(n)}(\mathbf{p})\delta\left(p^{0}+\mu+E_{p^{z},s}^{(n)}\right),\label{eq:Wigner function in magnetic field}
\end{eqnarray}
where the contribution from the $n$-th Landau level is
\begin{equation}
W_{rs}^{(n)}(\mathbf{p})=\frac{1}{(2\pi)^{3}}\int dy^{\prime}\exp\left(ip_{y}y^{\prime}\right)\bar{\psi}_{rs}^{(n)}\left(p_{x},\,p_{z},\,\frac{y^{\prime}}{2}\right)\otimes\psi_{rs}^{(n)}\left(p_{x},\,p_{z},\,-\frac{y^{\prime}}{2}\right).\label{eq:Wigner function nrs}
\end{equation}
Here we have used the property $\phi_{n}\left(p_{x}-eBy,\,y-\frac{y^{\prime}}{2}\right)=\phi_{n}\left(p_{x},-\frac{y^{\prime}}{2}\right)$
and the fact that the dependence of $\psi_{rs}^{(n)}$ on $p_{x}$
and $y$ only appears in the eigenfunctions $\phi_{n}$. The contribution
from fermions is separated from that of anti-fermions. The distributions
$f_{s}^{(\pm)(n)}$ turn out to be locally defined, which also depend
on $n$, $s$, $p^{x}$, and $p^{z}$. The Dirac-delta functions in
Eq. (\ref{eq:Wigner function in magnetic field}) can be written by
an on-shell condition multiplied with a step function,
\begin{equation}
\delta\left(p_{0}+\mu-rE_{p^{z},s}^{(n)}\right)=2E_{p^{z},s}^{(n)}\delta\left\{ (p_{0}+\mu)^{2}-[E_{p^{z},s}^{(n)}]^{2}\right\} \theta[r(p_{0}+\mu)].\label{eq:rel-delta-square}
\end{equation}

The lowest Landau level $n=0$ does not depend on $s$. The wavefunctions
are given in Eq. (\ref{eq:wave function LLL}). Inserting the wavefunctions
into Eq. (\ref{eq:Wigner function nrs}), we obtain %
\begin{equation}
W_{r}^{(0)}(\mathbf{p})=\frac{1}{(2\pi)^{3}}\frac{1}{4E_{p_{z}}^{(0)}}\left[rm\mathbb{I}_{2}+E_{p_{z}}^{(0)}\sigma^{1}-r(p_{z}-\mu_{5})i\sigma^{2}\right]\otimes\left(\mathbb{I}_{2}+\sigma^{3}\right)I_{00}(p^{x},p^{y}),
\end{equation}
where $I_{ij}(p^{x},p^{y})$ is defined in Eq. (\ref{eq:definition of I_ij}).
The tensor product of two Pauli matrices can be written in terms of
gamma matrices, as shown in Eq. (\ref{eq:relation between gamma matrices and Pauli}).
Thus we obtain
\begin{equation}
W_{r}^{(0)}(\mathbf{p})=\frac{r}{(2\pi)^{3}4E_{p_{z}}^{(0)}}I_{00}(p^{x},p^{y})\left[m(\mathbb{I}_{4}+\sigma^{12})+rE_{p_{z}}^{(0)}(\gamma^{0}-\gamma^{5}\gamma^{3})-(p^{z}-\mu_{5})(\gamma^{3}-\gamma^{5}\gamma^{0})\right].
\end{equation}
Here $I_{00}(p^{x},p^{y})$ is calculated in Eq. (\ref{eq:def-Lambda-0}).
Similarly, using the wavefunctions for higher Landau levels in Eq.
(\ref{eq:wave function HLL}), we can compute the contributions of
the higher Landau levels %
\begin{eqnarray}
W_{rs}^{(n)}(\mathbf{p}) & = & \frac{1}{(2\pi)^{3}}\frac{1}{2E_{p^{z}s}^{(n)}}\left[rm\mathbb{I}_{2}+E_{p^{z}s}^{(n)}\sigma^{1}+(i\sigma^{2})r(\mu_{5}-s\sqrt{(p^{z})^{2}+2nB_{0}})\right]\nonumber \\
 &  & \otimes\frac{1}{2\sqrt{(p^{z})^{2}+2nB_{0}}}\left(\begin{array}{cc}
(\sqrt{(p^{z})^{2}+2nB_{0}}+sp^{z})I_{nn} & s\sqrt{2nB_{0}}I_{n-1,n}\\
s\sqrt{2nB_{0}}I_{n,n-1} & (\sqrt{(p^{z})^{2}+2nB_{0}}-sp^{z})I_{n-1,n-1}
\end{array}\right).\nonumber \\
\label{eq:Wnrs as tensor product}
\end{eqnarray}
The matrix in the second line of Eq. (\ref{eq:Wnrs as tensor product})
can be decomposed in terms of the unit matrix and Pauli matrices,
\begin{eqnarray}
 &  & \left(\begin{array}{cc}
(\sqrt{(p^{z})^{2}+2nB_{0}}+sp^{z})I_{nn} & s\sqrt{2nB_{0}}I_{n-1,n}\\
s\sqrt{2nB_{0}}I_{n,n-1} & (\sqrt{(p^{z})^{2}+2nB_{0}}-sp^{z})I_{n-1,n-1}
\end{array}\right)\nonumber \\
 & = & \left(\sqrt{(p^{z})^{2}+2nB_{0}}\frac{I_{nn}+I_{n-1,n-1}}{2}+sp^{z}\frac{I_{nn}-I_{n-1,n-1}}{2}\right)\mathbb{I}_{2}\\
 &  & +s\sqrt{2nB_{0}}\frac{I_{n-1,n}+I_{n,n-1}}{2}\sigma^{1}+s\sqrt{2nB_{0}}\frac{I_{n-1,n}-I_{n,n-1}}{2}i\sigma^{2}\nonumber \\
 &  & +\left(\sqrt{(p^{z})^{2}+2nB_{0}}\frac{I_{nn}-I_{n-1,n-1}}{2}+sp^{z}\frac{I_{nn}+I_{n-1,n-1}}{2}\right)\sigma^{3}.
\end{eqnarray}
Note that these functions are independent of the choice of gauge.
We start from the Landau gauge where $p^{x}$ is a well-defined momentum
while $p^{y}$ is not. But Eq. (\ref{eq:def-Lambda-n}) only depends
on $p_{T}$, where $p^{x}$ and $p^{y}$ have the same importance.
The functions $I_{ij}(p^{x},p^{y})$ are computed in Eq. (\ref{eq:result of I_mn}).
Using the results (\ref{eq:relation between I_mn and Lambda_pm}),
the Wigner function for the higher Landau levels $n>0$ can be written
as
\begin{eqnarray}
W_{rs}^{(n)}(\mathbf{p}) & = & \frac{r}{(2\pi)^{3}4E_{p^{z}s}^{(n)}}\left\{ \left[m\mathbb{I}_{4}+rE_{p^{z}s}^{(n)}\gamma^{0}+(s\sqrt{(p^{z})^{2}+2nB_{0}}-\mu_{5})\gamma^{5}\gamma^{0}\right]\right.\nonumber \\
 &  & \qquad\times\left(\Lambda_{+}^{(n)}(p_{T})+s\frac{p^{z}}{\sqrt{(p^{z})^{2}+2nB_{0}}}\Lambda_{-}^{(n)}(p_{T})\right)\nonumber \\
 &  & \quad+\left[m\sigma^{12}-rE_{p^{z}s}^{(n)}\gamma^{5}\gamma^{3}+(\mu_{5}-s\sqrt{(p^{z})^{2}+2nB_{0}})\gamma^{3}\right]\nonumber \\
 &  & \qquad\times\left(\Lambda_{-}^{(n)}(p_{T})+s\frac{p^{z}}{\sqrt{(p^{z})^{2}+2nB_{0}}}\Lambda_{+}^{(n)}(p_{T})\right)\nonumber \\
 &  & \quad+\left[m\left(\sigma^{23}p^{x}+\sigma^{31}p^{y}\right)-rE_{p^{z}s}^{(n)}\left(\gamma^{5}\gamma^{1}p^{x}+\gamma^{5}\gamma^{2}p^{y}\right)\right.\nonumber \\
 &  & \qquad\left.\left.+(\mu_{5}-s\sqrt{(p^{z})^{2}+2nB_{0}})\left(\gamma^{1}p^{x}+\gamma^{2}p^{y}\right)\right]\times s\frac{2nB_{0}}{p_{T}^{2}\sqrt{(p^{z})^{2}+2nB_{0}}}\Lambda_{+}^{(n)}(p_{T})\right\} ,\nonumber \\
\label{eq:sol-W-nrs}
\end{eqnarray}
where we have defined a new function for $n>0$. Different components
of the Wigner function can be extracted using the trace properties
in Eq. (\ref{eq:reproduce components of Wigner funtion}),
\begin{eqnarray}
\left(\begin{array}{c}
\mathcal{G}_{1}\\
\mathcal{G}_{2}
\end{array}\right) & = & \left[\sum_{n=0}V_{n}e_{1}^{(n)}+\sum_{n>0}A_{n}\frac{1}{\sqrt{(p^{z})^{2}+2nB_{0}}}\left(p^{z}e_{2}^{(n)}+\sqrt{2nB_{0}}e_{3}^{(n)}\right)\right]\left(\begin{array}{c}
m\\
p_{0}+\mu
\end{array}\right),\nonumber \\
\mathcal{G}_{3} & = & (p^{z}-\mu_{5})V_{0}e_{1}^{(0)}+\sum_{n>0}\left[\sqrt{(p^{z})^{2}+2nB_{0}}A_{n}-\mu_{5}V_{n}\right]e_{1}^{(n)}\nonumber \\
 &  & \qquad+\sum_{n>0}\left[V_{n}-\frac{\mu_{5}}{\sqrt{(p^{z})^{2}+2nB_{0}}}A_{n}\right]\left(p^{z}e_{2}^{(n)}+\sqrt{2nB_{0}}e_{3}^{(n)}\right),\nonumber \\
\mathcal{G}_{4} & = & 0,\label{eq:sol-Wigner-function}
\end{eqnarray}
where the basis vectors $e_{1}^{(n)},\ e_{2}^{(n)}$, and $e_{3}^{(n)}$
are defined in Eq. (\ref{eq:def-basis}) and we have defined two functions
for $n>0$ %
{}
\begin{eqnarray}
V_{n} & \equiv & \frac{2}{(2\pi)^{3}}\sum_{s}\delta\left\{ (p_{0}+\mu)^{2}-[E_{p^{z}s}^{(n)}]^{2}\right\} \nonumber \\
 &  & \qquad\times\left\{ f_{s}^{(+)(n)}(p^{x},p^{z},\mathbf{x})\theta(p^{0}+\mu)+\left[f_{s}^{(-)(n)}(-p^{x},-p^{z},\mathbf{x})-1\right]\theta(-p_{0}-\mu)\right\} ,\nonumber \\
A_{n} & \equiv & \frac{2}{(2\pi)^{3}}\sum_{s}s\delta\left\{ (p_{0}+\mu)^{2}-[E_{p^{z}s}^{(n)}]^{2}\right\} \nonumber \\
 &  & \qquad\times\left\{ f_{s}^{(+)(n)}(p^{x},p^{z},\mathbf{x})\theta(p^{0}+\mu)+\left[f_{s}^{(-)(n)}(-p^{x},-p^{z},\mathbf{x})-1\right]\theta(-p_{0}-\mu)\right\} ,\label{eq:def-VnAn}
\end{eqnarray}
and
\begin{eqnarray}
V_{0} & = & \frac{2}{(2\pi)^{3}}\delta\left\{ (p_{0}+\mu)^{2}-[E_{p_{z}}^{(0)}]^{2}\right\} \nonumber \\
 &  & \qquad\times\left\{ f^{(+)(0)}(p^{x},p^{z},\mathbf{x})\theta(p_{0}+\mu)+\left[f^{(-)(0)}(-p^{x},-p^{z},\mathbf{x})-1\right]\theta(-p_{0}-\mu)\right\} .\label{eq:def-V0}
\end{eqnarray}
In Eq. (\ref{eq:sol-Wigner-function}) we have separated the 16 components
of the Wigner function into four groups $\mathcal{G}_{i}$ with $i=1,2,3,4$
as shown in Eq. (\ref{def:definition of G_i}). From the solutions
(\ref{eq:sol-Wigner-function}) we observe that the pseudoscalar codensate
$\mathcal{P}$ and the electric dipole-moment $\boldsymbol{\mathcal{T}}$
vanish and the remaining components can be decomposed into different
Landau levels while there is no mixture among different levels.

The Wigner function computed in this subsection is useful for studying
the physics in strong magnetic fields. In the Landau-level description,
we find the eigenstates of the Dirac equation, thus thermal equilibrium
can be well defined. This allows us to calculate various physical
quantities under the assumption of thermal equilibrium. We will show
in Sec. \ref{sec:Physical quantities} that the CME and CSE can be
correctly obtained from the Wigner function (\ref{eq:sol-Wigner-function}).
In the limit of sufficiently strong magnetic fields, all particles
will stay in the lowest Landau level. The system then reaches a fully
polarized state, which means that the average polarization of spin-$1/2$
particles can reach its maximum value $1/2$. In Sec. \ref{sec:Physical quantities}
we will also calculate the average polarization and the results agree
with those expected.

\subsection{Fermions in an electric field\label{subsec:Fermions-in-electric}}

In this subsection we will focus on a system in a pure electric field.
If the electric field is large enough, fermion-antifermion pairs can
be excited from the vacuum, which is known as Schwinger pair production
\cite{Schwinger:1951nm}. Since this process is a time-evolution problem,
in this subsection we will use the equal-time Wigner function for
convenience. The Schwinger process can be analytically solved in the
case of a constant electric field $\mathbf{E}(t,\mathbf{x})=E_{0}\mathbf{e}^{z}$
and in the case of a Sauter-type field $\mathbf{E}(t,\mathbf{x})=E_{0}\cosh^{-2}(t/\tau)\mathbf{e}^{z}$,
where $\mathbf{e}^{z}$ is the unit vector in the $z$ direction.
Both cases are homogeneous in space, while the Sauter-type field depends
on the time $t$. Analytically solving the Sauter-type field requires
knowledge about special functions, thus in this subsection we will
solve this case numerically instead of analytically. The numerical
calculation for a Sauter-type field also provides an universal approach
for solving the Wigner function in a time-dependent but spatially
homogeneous electric field. In this subsection we analytically solve
the Wigner function in a constant electric field.

\subsubsection{Asymptotic condition}

In order to solve a time-evolution problem, we need the equations
of motions and an asymptotic condition. The equations of motions for
the equal-time Wigner function have been derived in subsection \ref{subsec:Equal-time-formula}
while the asymptotic condition can be chosen as the Wigner function
for vanishing electric field.

Since the Schwinger pair-production process does not depend on the
spin, we choose to neglect the spin of the particles. According to
the calculations for the free-particle case in subsection \ref{subsec:Free-fermions},
where the Wigner function is given in Eq. (\ref{eq:free fermion Wigner function}),
we obtain
\begin{equation}
\left(\begin{array}{c}
\mathcal{F}(x,p)\\
\mathcal{V}_{\mu}(x,p)
\end{array}\right)=\left(\begin{array}{c}
m\\
p_{\mu}
\end{array}\right)\delta(p^{2}-m^{2})V(x,p),\label{eq:Wigner function in free case}
\end{equation}
with
\begin{equation}
V(x,p)\equiv\frac{2d_{s}}{(2\pi)^{3}}\left\{ \theta(p^{0})f^{(+)}(x,\mathbf{p})-\theta(-p^{0})\left[1-f^{(-)}(x,-\mathbf{p})\right]\right\} .
\end{equation}
Here $d_{s}$ is the spin degeneracy, for spin-1/2 fermions we have
$d_{s}=2$. Other components $\mathcal{P}$, $\mathcal{A}_{\mu}$,
and $\mathcal{S}_{\mu\nu}$ vanish as desired  because we have neglected
spin effects. In general the distribution functions $f^{(\pm)}$ should
be space-time dependent. But if a spatial inhomogenity is taken into
account, the Wigner function cannot be analytically computed. In this
subsection we assume that the electric field as well as the distributions
$f^{(\pm)}$ are independent of spatial coordinates, which indicates
that the whole system has translationally invariance in space.

First we derive the equal-time Wigner function from the covariant
one by integrating over energy $p^{0}$. The mass-shell delta-function
in Eq. (\ref{eq:Wigner function in free case}) can be integrated
out and we obtain
\begin{equation}
\left(\begin{array}{c}
\mathcal{F}(\mathbf{p})\\
\boldsymbol{\mathcal{V}}(\mathbf{p})
\end{array}\right)=\frac{1}{E_{\mathbf{p}}}\left(\begin{array}{c}
m\\
\mathbf{p}
\end{array}\right)C_{1}(\mathbf{p}),\label{eq:equilibrium solution for FV}
\end{equation}
and
\begin{equation}
\mathcal{V}^{0}(\mathbf{p})\equiv C_{2}(\mathbf{p}),\label{eq:equilibrium solution for V0}
\end{equation}
where we have defined $C_{1}(\mathbf{p})\equiv\int dp^{0}\ E_{\mathbf{p}}V(p)\delta(p^{2}-m^{2})$
and $C_{2}(\mathbf{p})\equiv\int dp^{0}\ p^{0}V(p)\delta(p^{2}-m^{2})$,
respectively. More explicitly, the integration can be performed and
yields
\begin{eqnarray}
C_{1}(\mathbf{p}) & = & \frac{2}{(2\pi)^{3}}\left[f^{(+)}(\mathbf{p})+f^{(-)}(-\mathbf{p})-1\right],\nonumber \\
C_{2}(\mathbf{p}) & = & \frac{2}{(2\pi)^{3}}\left[f^{(+)}(\mathbf{p})-f^{(-)}(-\mathbf{p})+1\right].\label{eq:definition of C_1 C_2}
\end{eqnarray}
The last term $\pm1$ in Eq. (\ref{eq:definition of C_1 C_2}) represents
the contribution from the vacuum, which aries because the operators
in the Wigner function are not normal-ordered. If the Wigner function
is defined with a normal ordering, the vacuum contribution vanishes
and the function $C_{1}(\mathbf{p})$ is the sum of fermion and anti-fermion
distributions. Meanwhile, $C_{2}(\mathbf{p})$ is the net fermion
number density.

\subsubsection{Equations of motion}

In a time-dependent but spatially homogeneous electric field, the
equations of motions (\ref{eq:equation of motion equal-time component})
take a simple form. The operators used in these equations, which are
defined in Eq. (\ref{eq:generalized operators for equal-t formula}),
have the following expressions, where the electric field is assumed
to be in the $z$-direction.
\begin{eqnarray}
D_{t} & = & \partial_{t}+E(t)\partial_{p^{z}},\nonumber \\
\mathbf{D}_{\mathbf{x}} & = & 0,\nonumber \\
\boldsymbol{\Pi} & = & \mathbf{p}.
\end{eqnarray}
Here we have dropped spatial derivatives because the whole system
is translationally invariant. Due to the translation invariance, we
find that the 16 components of the Wigner function can be divided
into several subgroups. The members in each group are coupled together
according to Eq. (\ref{eq:equation of motion equal-time component}).
It is a good feature that the component $\mathcal{V}^{0}(t,\mathbf{p})$
decouples from all other components, which satisfy the following equation
\begin{equation}
D_{t}\mathcal{V}^{0}(t,\mathbf{p})=0.
\end{equation}
Since the net charge density in coordinate space can be derived from
$\mathcal{V}^{0}(t,\mathbf{p})$ by integrating over $d^{3}\mathbf{p}$,
the above equation is nothing but the conservation law of the net
charge density. Taking the solution in Eq. (\ref{eq:equilibrium solution for V0})
at time $t_{0}$, the solution for $\mathcal{V}^{0}(t,\mathbf{p})$
reads
\begin{equation}
\mathcal{V}^{0}(t,\mathbf{p})=C_{2}\left(\mathbf{p}-\int_{t_{0}}^{t}dt^{\prime}E(t^{\prime})\mathbf{e}^{z}\right),
\end{equation}
where $\mathbf{e}^{z}$ is the unit vector along the electric field
direction. This solution reflects the overall acceleration of fermions
in an electric field, with $-\int_{t_{0}}^{t}dt^{\prime}E(t^{\prime})\mathbf{e}^{z}$
is the momentum shift due to the electric field.

Meanwhile, the equations for $\mathcal{P}$, $\mathcal{A}^{0}$, and
$\boldsymbol{\mathcal{S}}$ decouple from others and thus form a closed
subsystem. Since these components are all zero when the electric field
vanishes, they will remain zeros even after the electric field is
turned on. The rest ten components, $\mathcal{F}$, $\boldsymbol{\mathcal{V}}$,
$\boldsymbol{\mathcal{A}}$, $\boldsymbol{\mathcal{T}}$, form another
subsystem, which satisfy the equations of motion in a matrix form
\begin{equation}
D_{t}\boldsymbol{w}(t,\mathbf{p})=M(\mathbf{p})\boldsymbol{w}(t,\mathbf{p}),\label{eq:EOM in constant electric field}
\end{equation}
where the column vector $\boldsymbol{w}(t,\mathbf{p})\equiv(\mathcal{F},\boldsymbol{\mathcal{V}},\boldsymbol{\mathcal{A}},\boldsymbol{\mathcal{T}})^{T}$
has ten elements and $M(\mathbf{p})$ is a $10\times10$ coefficient
matrix
\begin{equation}
M(\mathbf{p})=2\left(\begin{array}{cccc}
0 & 0 & 0 & \mathbf{p}^{T}\\
0 & 0 & \mathbf{p}^{\times} & -m\mathbb{I}_{3}\\
0 & \mathbf{p}^{\times} & 0 & 0\\
-\mathbf{p} & m\mathbb{I}_{3} & 0 & 0
\end{array}\right).
\end{equation}
The initial condition at a given time $t_{0}$ for the equations of
motion in (\ref{eq:EOM in constant electric field}) is taken to be
the solution (\ref{eq:equilibrium solution for FV}) without the electric
field. Based on the fact that all fermions will be accelerated in
the electric field, we make the following ansatz for $\boldsymbol{w}(t,\mathbf{p})$,
\begin{equation}
\boldsymbol{w}(t,\mathbf{p})=C_{1}\left(\mathbf{p}-\int_{t_{0}}^{t}dt^{\prime}E(t^{\prime})\mathbf{e}^{z}\right)\sum_{i=1}^{10}\chi_{i}(t,\mathbf{p})\boldsymbol{e}_{i}(\mathbf{p}).\label{eq:solution of Wigner function in Electric}
\end{equation}
Here the overall factor $C_{1}\left(\mathbf{p}-\int_{t_{0}}^{t}dt^{\prime}E(t^{\prime})\mathbf{e}^{z}\right)$
is constructed from the distribution function for fermions with the
momentum $\mathbf{p}-\int_{t_{0}}^{t}dt^{\prime}E(t^{\prime})\mathbf{e}^{z}$
and that for anti-fermions with the opposite momentum $-\mathbf{p}+\int_{t_{0}}^{t}dt^{\prime}E(t^{\prime})\mathbf{e}^{z}$.
Thus we observe that particles are accelerated along the direction
of electric field, while antiparticles are accelerated along the opposite
direction. Meanwhile, we take ten basis vectors $\boldsymbol{e}_{i}(\mathbf{p})$
because $\boldsymbol{w}(t,\mathbf{p})$ has ten components. The basis
vectors are assumed to be time-independent, while the expanding coefficients
$\chi_{i}(t,\mathbf{p})$ are time-dependent. The first three basis
vectors read
\begin{equation}
\boldsymbol{e}_{1}=\left(\begin{array}{c}
0\\
\mathbf{e}_{z}\\
\boldsymbol{0}\\
\boldsymbol{0}
\end{array}\right),\ \ \boldsymbol{e}_{2}(\mathbf{p}_{T})=\frac{1}{m_{T}}\left(\begin{array}{c}
m\\
\mathbf{p}_{T}\\
\boldsymbol{0}\\
\boldsymbol{0}
\end{array}\right),\ \ \boldsymbol{e}_{3}(\mathbf{p}_{T})=\frac{1}{m_{T}}\left(\begin{array}{c}
0\\
\boldsymbol{0}\\
\mathbf{e}^{z}\times\mathbf{p}_{T}\\
-m\mathbf{e}^{z}
\end{array}\right),\label{eq:basis functions 123}
\end{equation}
where $m_{T}\equiv\sqrt{m^{2}+\mathbf{p}_{T}^{2}}$ is the transverse
mass, which ensures that these vectors are properly normalized and
orthogonal to each other $\boldsymbol{e}_{i}\cdot\boldsymbol{e}_{j}=\delta_{ij}$.
Since they are independent of $t$ and $p^{z}$, we have $D_{t}\boldsymbol{e}_{i}=0$
for $i=1,2,3$. We can check that these basis vectors form a closed
sub-Hilbert space under the operator $M(\mathbf{p})$,
\begin{equation}
M(\mathbf{p})\left(\begin{array}{c}
\boldsymbol{e}_{1}\\
\boldsymbol{e}_{2}\\
\boldsymbol{e}_{3}
\end{array}\right)=2\left(\begin{array}{ccc}
0 & 0 & -m_{T}\\
0 & 0 & p^{z}\\
m_{T} & -p^{z} & 0
\end{array}\right)\left(\begin{array}{c}
\boldsymbol{e}_{1}\\
\boldsymbol{e}_{2}\\
\boldsymbol{e}_{3}
\end{array}\right).\label{eq:Matrix on e123}
\end{equation}
Note that the initial condition, i.e., the Wigner function when the
electric field vanishes, stays in such a subspace, the system will
be in this subspace at later time of the evolution. The other basis
vectors $\boldsymbol{e}_{i}(\mathbf{p})$, $i=4,5,\cdots,10$, in
Eq. (\ref{eq:solution of Wigner function in Electric}), are not necessary
because the first three are sufficient to describe the time evolution.
The evolution of the coefficients $\chi_{i}(t,\mathbf{p})$, $i=1,2,3$
are then derived from Eqs. (\ref{eq:EOM in constant electric field}),
(\ref{eq:solution of Wigner function in Electric}), and (\ref{eq:Matrix on e123}),
\begin{equation}
D_{t}\left(\begin{array}{c}
\chi_{1}\\
\chi_{2}\\
\chi_{3}
\end{array}\right)(t,\mathbf{p})=2\left(\begin{array}{ccc}
0 & 0 & m_{T}\\
0 & 0 & -p^{z}\\
-m_{T} & p^{z} & 0
\end{array}\right)\left(\begin{array}{c}
\chi_{1}\\
\chi_{2}\\
\chi_{3}
\end{array}\right)(t,\mathbf{p}).\label{eq:EOM of chi_i}
\end{equation}
This system of partial differential equations is equivalent to the
well-known Vlasov equation for pair production in quantum kinetic
theory \cite{Hebenstreit:2010vz}. Once the functions $\chi_{i}$
are solved from Eq. (\ref{eq:EOM of chi_i}), the Wigner function
can be reproduced by inserting $\chi_{i}$ and Eq. (\ref{eq:basis functions 123})
into Eq. (\ref{eq:solution of Wigner function in Electric}).

\subsubsection{Solutions for a Sauter-type field \label{subsec:Solutions-for-Sauter-type}}

In order to solve Eq. (\ref{eq:EOM of chi_i}), we first need an initial
condition. One naive choice is, assuming that the electric field does
not exist before time $t_{0}$,
\begin{equation}
\chi_{1}(t_{0},\mathbf{p})=\frac{p^{z}}{E_{\mathbf{p}}},\ \ \chi_{2}(t_{0},\mathbf{p})=\frac{m_{T}}{E_{\mathbf{p}}},\ \ \chi_{3}(t_{0},\mathbf{p})=0.\label{eq:initial condition for E}
\end{equation}
This initial condition corresponds to a field which is suddenly switched
on at $t_{0}$, i.e., a time-dependent electric field as
\begin{equation}
E(t)=\theta(t-t_{0})E(t).
\end{equation}
Such an initial condition is useful when dealing with a field which
vanishes when $t\rightarrow t_{0}$. For example, for a Sauter-type
field $E(t)=E_{0}\cosh^{-2}(t/\tau)$, we can specify the solution
(\ref{eq:initial condition for E}) for $t_{0}\rightarrow-\infty$
and the system evolves with time according to Eq. (\ref{eq:EOM of chi_i}).

\begin{figure}
\includegraphics[width=8cm]{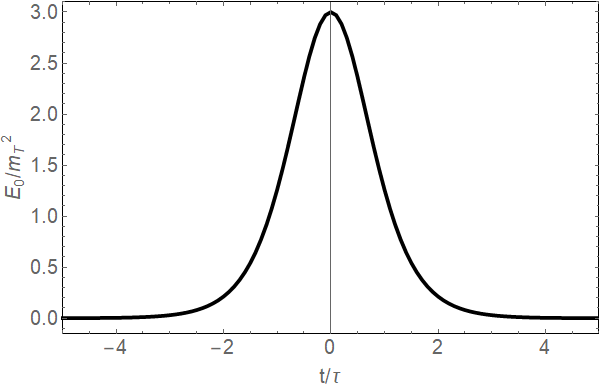}

\caption{\label{fig:Sauter-type field}The time dependence of a Sauter-type
electric field $E(t)=E_{0}\cosh^{-2}(t/\tau)$. Here we take the transverse
mass $m_{T}$ as the energy unit and the peak value of the electric
field is taken to be $3m_{T}^{2}$. }
\end{figure}

We now take the Sauter-type electric field $E(t)=E_{0}\cosh^{-2}(t/\tau)$
as an example. In Fig. \ref{fig:Sauter-type field} we plot the time
dependence of the field strength. The Sauter-type field can be used
to describe a pulse, which converges to zero in the limit $t\rightarrow\pm\infty$.
We define the canonical momentum $q^{z}$ as
\begin{equation}
q^{z}=p^{z}-E_{0}\tau\left[\tanh(t/\tau)+1\right],
\end{equation}
which ensures that $q^{z}=p^{z}$ in the limit $t\rightarrow-\infty$.
Then we substitute the kinetic momentum $p^{z}$ in the operator $D_{t}$
by the canonical one $q^{z}$ and obtain
\begin{equation}
\left[\partial_{t}+E_{0}\cosh^{-2}(t/\tau)\partial_{p^{z}}\right]\chi_{i}(t,\mathbf{p})=\frac{d}{dt}\chi_{i}(t,\mathbf{p}_{T},q^{z}).
\end{equation}
The equations of motions (\ref{eq:EOM of chi_i}) now transform into
ordinary differential equations,
\begin{equation}
\frac{d}{dt}\left(\begin{array}{c}
\chi_{1}\\
\chi_{2}\\
\chi_{3}
\end{array}\right)=2\left(\begin{array}{ccc}
0 & 0 & m_{T}\\
0 & 0 & -q^{z}-E_{0}\tau\left[\tanh(t/\tau)+1\right]\\
-m_{T} & q^{z}+E_{0}\tau\left[\tanh(t/\tau)+1\right] & 0
\end{array}\right)\left(\begin{array}{c}
\chi_{1}\\
\chi_{2}\\
\chi_{3}
\end{array}\right),\label{eq:EOM of chi_i-1}
\end{equation}
with initial conditions
\begin{equation}
\lim_{t_{0}\rightarrow-\infty}\chi_{1}(t_{0},\mathbf{p}_{T},q^{z})=\frac{q^{z}}{E_{\mathbf{p}}},\ \ \lim_{t_{0}\rightarrow-\infty}\chi_{2}(t_{0},\mathbf{p}_{T},q^{z})=\frac{m_{T}}{E_{\mathbf{p}}},\ \ \lim_{t_{0}\rightarrow-\infty}\chi_{3}(t_{0},\mathbf{p}_{T},q^{z})=0.\label{eq:initial condition for E-1}
\end{equation}
This system of ordinary differential equations can be easily solved
using the finite-difference method. One can also see Ref. \cite{Hebenstreit:2010vz}
for the analytical solution from quantum kinetic theory.

\begin{figure}
\includegraphics[width=8cm]{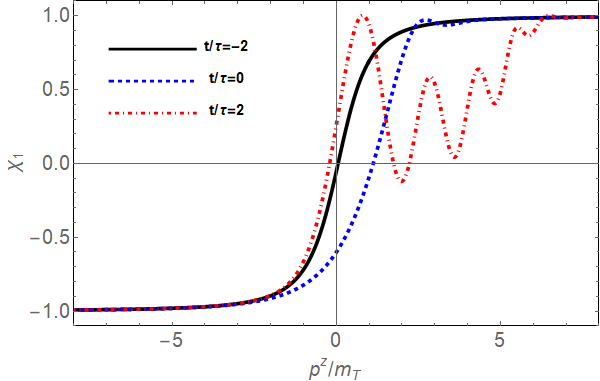}\caption{\label{fig:pz-dependence-of-chi1}The $p^{z}$-dependence of $\chi_{1}$
at times $t=-2\tau$ (solid line), $t=0$ (dashed line), and $t=2\tau$
(dot-dashed line). }
\end{figure}

\begin{figure}
\includegraphics[width=8cm]{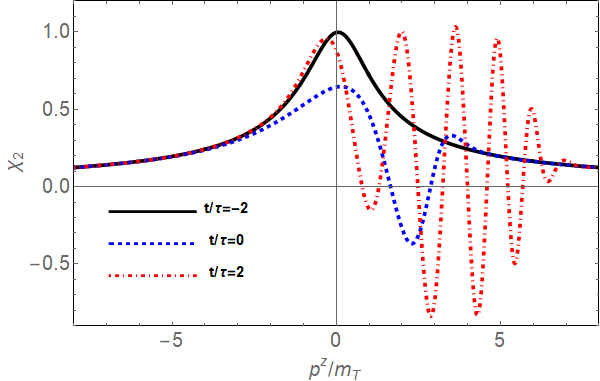}\caption{\label{fig:pz-dependence-of-chi2}The $p^{z}$-dependence of $\chi_{2}$
at timets $t=-2\tau$ (solid line), $t=0$ (dashed line), and $t=2\tau$
(dot-dashed line).}
\end{figure}

\begin{figure}
\includegraphics[width=8cm]{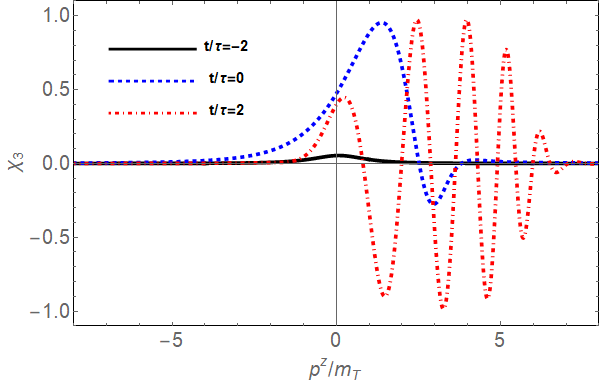}\caption{\label{fig:pz-dependence-of-chi3}The $p^{z}$-dependence of $\chi_{3}$
at times $t=-2\tau$ (solid line), $t=0$ (dashed line), and $t=2\tau$
(dot-dashed line). }
\end{figure}

As an example, we take the transverse mass $m_{T}=\sqrt{m^{2}+\mathbf{p}_{T}^{2}}$
as the energy unit and $\tau=1/m_{T}$ as the time unit. The peak
value of the Sauter-type electric field is chosen to be $E_{0}/m_{T}^{2}=3$.
In Figs. \ref{fig:pz-dependence-of-chi1}, \ref{fig:pz-dependence-of-chi2},
and \ref{fig:pz-dependence-of-chi3} we plot the $p^{z}$-dependence
of $\chi_{1}$, $\chi_{2}$, and $\chi_{3}$, respectively, at several
times, $t=-2\tau$, $0$, and $2\tau$. We emphasize that in these
figures the $x$-axis is the kinetic momentum $p^{z}$. According
to our calculation, even though the electric field strength turns
to zero in the limit $t\rightarrow+\infty$, the functions $\chi_{1}$,
$\chi_{2}$, and $\chi_{3}$ cannot reach stationary states. Instead,
these functions will oscillate and the oscillations become more and
more pronounced at later times. In Sec. \ref{subsec:Pair-production}
we will clearly see that the oscillation does not contribute to the
pair-production rate, and the pair spectrum finally reaches a stationary
state.

\subsubsection{Solutions in a constant electric field. }

However, the initial conditions in Eq. (\ref{eq:initial condition for E})
do not work for a constant electric field $E(t)=E_{0}$. Since a constant
field is not integrable, the momentum shift $\int_{t_{0}}^{t}dt^{\prime}E(t^{\prime})\mathbf{e}_{z}$
will be infinitely large if we take the limit $t_{0}\rightarrow-\infty$.
From the physical point of view, the fermions can collide with each
other and the kinetic energy will be converted to the thermal energy
through collisions. The collision processes will retard the movement
of particles and the system would finally reach a new balance state
in the electric field. If the system has a boundary, the particles
would accumulate near the boundary and the chemical potential $\mu$
then becomes spatially-dependent. Finally the force from the Pauli
exclusion principle, i.e., the effect from the gradient of $\mu$,
will cancel with that from the electric field. If the system is infinitely
large, the system would reach a state with a collective charge current,
and more fermions (assumed to have positive charge) moving in the
direction of the electric field. In this case, the current can be
independent of the spatial coordinate and so does the distribution.

Here we assume that the system is described by spatial independent
distribution functions at time $t_{0}$ and we focus on a short period
after this moment. The system w deviates from the initial state during
this period because of pair production. Then our goal is to find a
solution which coincides with Eq. (\ref{eq:equilibrium solution for FV})
when the electric field vanishes,
\begin{equation}
\left.\left(\begin{array}{c}
\chi_{1}\\
\chi_{2}\\
\chi_{3}
\end{array}\right)(t,\mathbf{p})\right|_{E_{0}\rightarrow0}=\frac{1}{E_{\mathbf{p}}}\left(\begin{array}{c}
p^{z}\\
m_{T}\\
0
\end{array}\right).\label{eq:limit E0->0}
\end{equation}
The Wigner function in a constant electric field is then given by
Eqs. (\ref{eq:solution of Wigner function in Electric}) and (\ref{eq:basis functions 123}),
where the coefficients $\chi_{i}$ with $i=1,2,3$ are solved from
Eq. (\ref{eq:EOM of chi_i}) with the condition (\ref{eq:limit E0->0}),
while other $\chi_{i}$ with $i=4,5,\cdots10$ are zeros. From quantum
kinetic theory one can obtain the following solution \cite{Hebenstreit:2010vz},
\begin{equation}
\left(\begin{array}{c}
\chi_{1}\\
\chi_{2}\\
\chi_{3}
\end{array}\right)(\mathbf{p})=\left(\begin{array}{c}
d_{1}\left(\eta,\ \sqrt{\frac{2}{E_{0}}}p^{z}\right)\\
\frac{m_{T}}{\sqrt{2E_{0}}}d_{2}\left(\eta,\ \sqrt{\frac{2}{E_{0}}}p^{z}\right)\\
\frac{m_{T}}{\sqrt{2E_{0}}}d_{3}\left(\eta,\ \sqrt{\frac{2}{E_{0}}}p^{z}\right)
\end{array}\right),
\end{equation}
where $\eta\equiv m_{T}^{2}/E_{0}$ is the dimensionless transverse
mass square. One can check that this solution satisfies Eq. (\ref{eq:EOM of chi_i})
and the constraint (\ref{eq:limit E0->0}). Here the auxiliary functions
$d_{1}$, $d_{2}$, and $d_{3}$ are defined in Eq. (\ref{eq:auxiliary functions-1}).
Then the Wigner function can be reproduced using Eq. (\ref{eq:solution of Wigner function in Electric}),
\begin{eqnarray}
\mathcal{F} & = & \frac{m}{\sqrt{2E_{0}}}d_{2}\left(\eta,\ \sqrt{\frac{2}{E_{0}}}p^{z}\right)C_{1}\left(\mathbf{p}-E_{0}t\mathbf{e}_{z}\right),\nonumber \\
\mathcal{P} & = & 0,\nonumber \\
\mathcal{V}^{0} & = & C_{2}\left(\mathbf{p}-E_{0}t\mathbf{e}_{z}\right),\nonumber \\
\boldsymbol{\mathcal{V}}_{T} & = & \frac{\mathbf{p}_{T}}{\sqrt{2E_{0}}}d_{2}\left(\eta,\ \sqrt{\frac{2}{E_{0}}}p^{z}\right)C_{1}\left(\mathbf{p}-E_{0}t\mathbf{e}_{z}\right),\nonumber \\
\mathcal{V}^{z} & = & d_{1}\left(\eta,\ \sqrt{\frac{2}{E_{0}}}p^{z}\right)C_{1}\left(\mathbf{p}-E_{0}t\mathbf{e}_{z}\right),\nonumber \\
\mathcal{A}^{0} & = & 0,\nonumber \\
\boldsymbol{\mathcal{A}} & = & \frac{\mathbf{e}_{z}\times\mathbf{p}_{T}}{\sqrt{2E_{0}}}d_{3}\left(\eta,\ \sqrt{\frac{2}{E_{0}}}p^{z}\right)C_{1}\left(\mathbf{p}-E_{0}t\mathbf{e}_{z}\right),\nonumber \\
\boldsymbol{\mathcal{T}} & = & -\frac{m\mathbf{e}_{z}}{\sqrt{2E_{0}}}d_{3}\left(\eta,\ \sqrt{\frac{2}{E_{0}}}p^{z}\right)C_{1}\left(\mathbf{p}-E_{0}t\mathbf{e}_{z}\right),\nonumber \\
\boldsymbol{\mathcal{S}} & = & 0.\label{eq:Wigner function in pure electric}
\end{eqnarray}
When taking the limit $E_{0}\rightarrow0$, the Wigner function recovers
the results in Eqs. (\ref{eq:equilibrium solution for FV}) and (\ref{eq:equilibrium solution for V0}).
At the moment $t=0$, the existing particles are assumed to produce
distributions which are determined by $C_{1}(\mathbf{p})$ and $C_{2}(\mathbf{p})$.
Note that due to the lack of collisions, the solutions in Eq. (\ref{eq:Wigner function in pure electric})
can only be used to describe a short period after $t=0$, i.e., for
times smaller than the mean free time.

\subsection{Fermions in constant parallel electromagnetic fields \label{subsec:Fermions-in-parallel-EB}}

\subsubsection{Asymptotic condition}

In subsection \ref{subsec:Fermions-in-const-B}, we have derived the
Wigner function in a constant magnetic field. In this subsection we
will add an electric field, which is assumed to be parallel to the
magnetic field. Both the electric field and the magnetic field are
chosen to be constant so that the problem can be analytically solved.
Similar to the case of a pure electric field, the case in this subsection
is a time-evolution problem. Particles in constant magnetic field
are described by the Landau levels, which is used as the initial condition
for the time evolution. Taking the chiral chemical potential $\mu_{5}=0$,
and integrating over energy $p^{0}$, we obtain the following equal-time
Wigner function from Eq. (\ref{eq:sol-Wigner-function}),
\begin{eqnarray}
\mathcal{G}_{1} & = & \sum_{n=0}\frac{m}{E_{p^{z}}^{(n)}}C_{1}^{(n)}(p^{z})e_{1}^{(n)}(p_{T}),\nonumber \\
\mathcal{G}_{2} & = & \sum_{n=0}C_{2}^{(n)}(p^{z})e_{1}^{(n)}(p_{T}),\nonumber \\
\mathcal{G}_{3} & = & \frac{p^{z}}{E_{p^{z}}^{(0)}}C_{1}^{(0)}(p^{z})e_{1}^{(0)}(p_{T})+\sum_{n>0}\frac{1}{E_{p^{z}}^{(n)}}C_{1}^{(n)}(p^{z})\left[p^{z}e_{2}^{(n)}(p_{T})+\sqrt{2nB_{0}}e_{3}^{(n)}(p_{T})\right],\nonumber \\
\mathcal{G}_{4} & = & 0.\label{eq:equal-t Wigner function in magnetic}
\end{eqnarray}
Here $\mathcal{G}_{i}$ are constructed from the Wigner function as
shown in Eq. (\ref{def:definition of G_i}), $C_{1}^{(n)}\equiv\int dp^{0}E_{p^{z}}^{(n)}V^{(n)}$
and $C_{2}^{(n)}\equiv\int dp^{0}(p^{0}+\mu)V^{(n)}$. The basis vectors
$e_{1,2,3}^{(0)}(\mathbf{p}_{T})$ are defined in Appendix \ref{sec:Auxiliary-functions}.
The function $V^{(n)}$ is defined in (\ref{eq:def-VnAn}) and (\ref{eq:def-V0}),
from which we obtain an explicit relation between $C_{1}^{(n)},\ C_{2}^{(n)}$
and $f^{(\pm)(n)}$,
\begin{eqnarray}
C_{1}^{(n)}(p^{z}) & = & \frac{2-\delta_{n0}}{(2\pi)^{3}}\left[f^{(+)(n)}(p^{z})+f^{(-)(n)}(p^{z})-1\right],\nonumber \\
C_{2}^{(n)}(p^{z}) & = & \frac{2-\delta_{n0}}{(2\pi)^{3}}\left[f^{(+)(n)}(p^{z})-f^{(-)(n)}(p^{z})+1\right].\label{eq:equilibrium C1C2}
\end{eqnarray}
Up to a vacuum contribution, the auxiliary function $C_{1}^{(n)}(p^{z})$
is the sum of the fermion and anti-fermion distribution in the $n$-th
Landau level, while $C_{2}^{(n)}(p^{z})$ is the difference. In general
the distributions depend on $p^{x}$, $p^{z}$, and $\mathbf{x}$,
where $p^{x}-B_{0}y$ plays a role as the center position in the $y$-direction.
Here we choose to neglect the spatial dependence of the distributions.
Thus, there is no dependence on $\mathbf{x}$ and $p^{x}$ in Eq.
(\ref{eq:equilibrium C1C2}). The pre-factor $2-\delta_{n0}$ is the
spin degeneracy for the $n$-th Landau level. We find that the equal-time
Wigner function, whose components are given in Eq. (\ref{eq:equal-t Wigner function in magnetic}),
is a sum over different Landau levels. Later on we will show that
in the presence of a constant electric field, different Landau levels
evolve independently.

\subsubsection{Equations of motion}

In the presence of constant electromagnetic fields, we assume that
the whole system is spatially homogeneous so that the spatial derivative
$\boldsymbol{\nabla}_{\mathbf{x}}$ can be dropped. The operators
in Eq. (\ref{eq:generalized operators for equal-t formula}) are now
given by
\begin{eqnarray}
D_{t} & = & \partial_{t}+E_{0}\partial_{p^{z}},\nonumber \\
\mathbf{D}_{\mathbf{x}} & = & B_{0}\mathbf{e}_{z}\times\boldsymbol{\nabla}_{\mathbf{p}},\nonumber \\
\boldsymbol{\Pi} & = & \mathbf{p},
\end{eqnarray}
where $E_{0}$ and $B_{0}$ are strengths of electric and magnetic
field, respectively. Then the matrix operators $M_{1}$ and $M_{2}$,
defined in Eq. (\ref{def:matrices M1M2}), take the following forms,
\begin{equation}
M_{1}=\left(\begin{array}{cccc}
0 & 2p^{x} & 2p^{y} & 2p^{z}\\
2p^{x} & 0 & 0 & \hbar B_{0}\partial_{p^{x}}\\
2p^{y} & 0 & 0 & \hbar B_{0}\partial_{p^{y}}\\
2p^{z} & -\hbar B_{0}\partial_{p^{x}} & -\hbar B_{0}\partial_{p^{y}} & 0
\end{array}\right),\ \ M_{2}=\left(\begin{array}{cccc}
0 & -\hbar B_{0}\partial_{p^{y}} & \hbar B_{0}\partial_{p^{x}} & 0\\
-\hbar B_{0}\partial_{p^{y}} & 0 & -2p^{z} & 2p^{y}\\
\hbar B_{0}\partial_{p^{x}} & 2p^{z} & 0 & -2p^{x}\\
0 & -2p^{y} & 2p^{x} & 0
\end{array}\right).
\end{equation}
For the lowest Landau level, only the basis vector $e_{1}^{(0)}(p_{T})$
defined in (\ref{eq:def-basis}), contributes to the solution in (\ref{eq:equal-t Wigner function in magnetic}).
Furthermore, we can check that $e_{1}^{(0)}(p_{T})$ is an eigenvector
of the operators $D_{t}$, $M_{1}$, and $M_{2}$,
\begin{equation}
D_{t}e_{1}^{(0)}(p_{T})=0,\ \ M_{1}e_{1}^{(0)}(p_{T})=2p^{z}e_{1}^{(0)}(p_{T}),\ \ M_{2}e_{1}^{(0)}(p_{T})=0.
\end{equation}
Thus we only need $e_{1}^{(0)}(p_{T})$ to describe the dynamics of
the lowest Landau level.

For higher Landau levels, the initial Wigner function in (\ref{eq:equal-t Wigner function in magnetic})
contains all three basis vectors $e_{i}^{(n)}(\mathbf{p}_{T})$, $i=1,2,3$
and $n>0$, which are defined in Eq. (\ref{eq:def-basis}). But we
can check that these basis vectors are not closed under the operator
$M_{2}$. In order to construct a closed Hilbert space under the operators
$D_{t}$, $M_{1}$, and $M_{2}$, we need another basis vector, i.e.,
$e_{4}^{(n)}(\mathbf{p}_{T})$ defined in Eq. (\ref{eq:def-basis}).
Acting the matrix operators $M_{1}$ and $M_{2}$ on these basis vectors,
we obtain
\begin{eqnarray}
M_{1}e_{i}^{(n)}(\mathbf{p}_{T}) & = & \sum_{j=1}^{4}(c_{1}^{(n)})_{ij}^{T}\,e_{j}^{(n)}(\mathbf{p}_{T}),\nonumber \\
M_{2}e_{i}^{(n)}(\mathbf{p}_{T}) & = & \sum_{j=1}^{4}(c_{2}^{(n)})_{ij}^{T}\,e_{j}^{(n)}(\mathbf{p}_{T}),\label{eq:matrix operators act on basis}
\end{eqnarray}
where the coefficients are
\begin{equation}
c_{1}^{(n)}=2\left(\begin{array}{cccc}
0 & p^{z} & \sqrt{2nB_{0}} & 0\\
p^{z} & 0 & 0 & 0\\
\sqrt{2nB_{0}} & 0 & 0 & 0\\
0 & 0 & 0 & 0
\end{array}\right),\ \ c_{2}^{(n)}=-2\left(\begin{array}{cccc}
0 & 0 & 0 & 0\\
0 & 0 & 0 & \sqrt{2nB_{0}}\\
0 & 0 & 0 & -p^{z}\\
0 & -\sqrt{2nB_{0}} & p^{z} & 0
\end{array}\right).
\end{equation}
Note that in Eq. (\ref{eq:matrix operators act on basis}), we have
used the transposes of $c_{1}^{(n)}$ and $c_{2}^{(n)}$ for convenience
of further calculations. Due to the fact that the basis vectors $e_{i}^{(n)}(\mathbf{p}_{T})$,
$i=1,2,3,4$, are independent of $t$ and $p^{z}$, we find
\begin{equation}
D_{t}e_{i}^{(n)}(\mathbf{p}_{T})=0.
\end{equation}
When the electric field vanishes, the equal-time Wigner function in
Eq. (\ref{eq:equal-t Wigner function in magnetic}) can be expressed
in terms of the basis vectors $e_{1}^{(0)}(p_{T})$ and $e_{i}^{(n)}(\mathbf{p}_{T})$.
Analogous to the case in the previous subsection, we take the Wigner
function in a constant magnetic field as an asymptotic condition when
$E_{0}\rightarrow0$. Then it is straightforward to conclude that
the Wigner function will stay in the Hilbert space formed by $e_{1}^{(0)}(p_{T})$
and $e_{i}^{(n)}(\mathbf{p}_{T})$ because this space is closed for
all operators $D_{t}$, $M_{1}$ and $M_{2}$ in the equations of
motion (\ref{eq:EOM of 4 groups}). We thus decompose the Wigner function
as
\begin{equation}
\mathcal{G}_{i}(t,\mathbf{p})=f_{i}^{(0)}(t,p^{z})e_{1}^{(0)}(p_{T})+\sum_{n>0}\sum_{j=1}^{4}f_{ij}^{(n)}(t,p^{z})e_{j}^{(n)}(\mathbf{p}_{T}),\label{eq:decomposition of G_i}
\end{equation}
where $i=1,2,3,4$. Note that since we focus on a constant magnetic
field, the basis vectors are independent to time and all the time-dependence
is put into the coefficients $f_{i}^{(0)}$ and $f_{ij}^{(n)}$. We
also find that the transverse momentum is separated in Eq. (\ref{eq:decomposition of G_i}).
Inserting the decomposition (\ref{eq:decomposition of G_i}) into
the equations of motion (\ref{eq:EOM of 4 groups}), and then using
the orthonormality conditions in Eqs. (\ref{eq:normality e_1^0}),
(\ref{eq:orthogonal e_1^0 e_i^n}), and (\ref{eq:orthonormality e_i^m e_j^n})
to separate the coefficients of different basis vectors, we derive
the equations of motions for $f_{i}^{(0)}(t,p^{z})$ and $f_{ij}^{(n)}(t,p^{z})$.
For the lowest Landau level, the equations of motions read
\begin{equation}
D_{t}\left(\begin{array}{c}
f_{1}^{(0)}\\
f_{2}^{(0)}\\
f_{3}^{(0)}\\
f_{4}^{(0)}
\end{array}\right)(t,p^{z})=2\left(\begin{array}{cccc}
0 & 0 & 0 & p^{z}\\
0 & 0 & 0 & 0\\
0 & 0 & 0 & -m\\
-p^{z} & 0 & m & 0
\end{array}\right)\left(\begin{array}{c}
f_{1}^{(0)}\\
f_{2}^{(0)}\\
f_{3}^{(0)}\\
f_{4}^{(0)}
\end{array}\right)(t,p^{z}),\label{eq:EOM of lowest Landau level}
\end{equation}
while for the higher Landau levels we obtain
\begin{equation}
D_{t}\left(\begin{array}{c}
\boldsymbol{f}_{1}^{(n)}\\
\boldsymbol{f}_{2}^{(n)}\\
\boldsymbol{f}_{3}^{(n)}\\
\boldsymbol{f}_{4}^{(n)}
\end{array}\right)(t,p^{z})=\left(\begin{array}{cccc}
0 & 0 & 0 & c_{1}^{(n)}\\
0 & 0 & -c_{2}^{(n)} & 0\\
0 & -c_{2}^{(n)} & 0 & -2m\mathbb{I}_{4}\\
-c_{1}^{(n)} & 0 & 2m\mathbb{I}_{4} & 0
\end{array}\right)\left(\begin{array}{c}
\boldsymbol{f}_{1}^{(n)}\\
\boldsymbol{f}_{2}^{(n)}\\
\boldsymbol{f}_{3}^{(n)}\\
\boldsymbol{f}_{4}^{(n)}
\end{array}\right)(t,p^{z}),\label{eq:EOM of higher Landau levels}
\end{equation}
where $\boldsymbol{f}_{i}^{(n)}\equiv(f_{i1}^{(n)},f_{i2}^{(n)},f_{i3}^{(n)},f_{i4}^{(n)})^{T}$
is a four-dimensional column vector. From these equations we observe
that different Landau levels decouple from each other and thus evolve
separately.

\subsubsection{Lowest Landau level}

In the lowest Landau level, the spin of a positive charged particle
is parallel to the magnetic field. Meanwhile, the higher Landau levels
are 2-fold degenerate with respect to spin. Thus the lowest Landau
level is special and needs a careful treatment. The equations of motion
for the lowest Landau level (\ref{eq:EOM of lowest Landau level})
are obviously distinct from those for the higher Landau levels (\ref{eq:EOM of higher Landau levels}).
We note that in Eq. (\ref{eq:EOM of lowest Landau level}), the equation
for $f_{2}^{(0)}$ decouples from the others, which gives
\begin{equation}
D_{t}f_{2}^{(0)}(t,p^{z})=0.\label{eq:EOM for f^0_2}
\end{equation}
Since the net fermion number at the lowest Landau level is given by
\begin{equation}
n^{(0)}=\int d^{3}\mathbf{p}\,f_{2}^{(0)}(t,p^{z})\Lambda_{+}^{(0)}(p_{T}),
\end{equation}
we find that the equation for $f_{2}^{(0)}$ correspond to the conservation
of $n^{(0)}$, i.e.,
\begin{equation}
\partial_{t}n^{(0)}=\int d^{3}\mathbf{p}\,\left[D_{t}f_{2}^{(0)}(t,p^{z})\right]\Lambda_{+}^{(0)}(p_{T})=0,
\end{equation}
Here we have integrated the $p^{z}$-derivative by parts and dropped
the boundary term. Equation (\ref{eq:EOM for f^0_2}), together with
the asymptotic condition $\left.f_{2}^{(0)}(t,p^{z})\right|_{E_{0}\rightarrow0}=C_{2}^{(0)}(p^{z})$,
give the following specific solution
\begin{equation}
f_{2}^{(0)}(t,p^{z})=C_{2}^{(0)}(p^{z}-E_{0}t).\label{eq:specific solution for f_2^0}
\end{equation}
It describes the overall acceleration of particles along the direction
of the electric field. Note that due to the absence of collisions,
the particle distribution will be far from the initial one after a
long time period. But in reality, collisions prevent the acceleration
and the system will stay near the thermal equilibrium and the specific
solution is only suitable to describe the physics for $t<t_{\text{relax}}$,
where $t_{\text{relax}}$ is the relaxation time of the system.

The other three functions, $f_{1}^{(0)}$, $f_{3}^{(0)}$ and $f_{4}^{(4)}$
can be parametrized as
\begin{equation}
\left\{ f_{1}^{(0)},f_{3}^{(0)},f_{4}^{(4)}\right\} =\left\{ \chi_{1}^{(0)},\chi_{2}^{(0)},\chi_{3}^{(0)}\right\} C_{1}^{(0)}(p^{z}-E_{0}t),\label{eq:relation f134 with chi123}
\end{equation}
where $C_{1}^{(0)}$ is defined in Eq. (\ref{eq:equilibrium C1C2}).
Here the canonical momentum $p^{z}-E_{0}t$ again reflects the acceleration
of particles. Comparing with Eq. (\ref{eq:equal-t Wigner function in magnetic}),
we obtain the asymptotic condition when the electric field vanishes,
\begin{equation}
\left.\left(\begin{array}{c}
\chi_{1}^{(0)}\\
\chi_{2}^{(0)}\\
\chi_{3}^{(0)}
\end{array}\right)(t,p^{z})\right|_{E_{0}\rightarrow0}=\frac{1}{E_{p^{z}}^{(0)}}\left(\begin{array}{c}
m\\
p^{z}\\
0
\end{array}\right).\label{eq:asymptotic condition of chi_123^0}
\end{equation}
The equations of motion for $\chi_{1}^{(0)},\chi_{2}^{(0)},\chi_{3}^{(0)}$
are derived from Eq. (\ref{eq:EOM of lowest Landau level}) by using
the fact that $D_{t}C_{1}^{(0)}(p^{z}-E_{0}t)=0$,
\begin{equation}
D_{t}\left(\begin{array}{c}
\chi_{1}^{(0)}\\
\chi_{2}^{(0)}\\
\chi_{3}^{(0)}
\end{array}\right)(t,p^{z})=2\left(\begin{array}{ccc}
0 & 0 & p^{z}\\
0 & 0 & -m\\
-p^{z} & m & 0
\end{array}\right)\left(\begin{array}{c}
\chi_{1}^{(0)}\\
\chi_{2}^{(0)}\\
\chi_{3}^{(0)}
\end{array}\right)(t,p^{z}).\label{eq:EOM of chi_123^0}
\end{equation}
Comparing Eq. (\ref{eq:EOM of chi_123^0}) and the asymptotic condition
in Eq. (\ref{eq:asymptotic condition of chi_123^0}) with Eqs. (\ref{eq:EOM of chi_i})
and (\ref{eq:limit E0->0}), we find that they are exactly the same
if we substitute $\chi_{1}\rightarrow\chi_{2}^{(0)}$,$\chi_{2}\rightarrow\chi_{1}^{(0)}$,$\chi_{3}\rightarrow-\chi_{3}^{(0)}$,
and $m_{T}\rightarrow m$ in Eqs. (\ref{eq:EOM of chi_i}) and (\ref{eq:limit E0->0}).
This indicates that the pair production in the lowest Landau level
and in a pure electric field are controlled by the same system of
partial differential equations. The solution for Eq. (\ref{eq:EOM of chi_123^0})
in a constant electric field is straightforward,
\begin{equation}
\left(\begin{array}{c}
\chi_{1}^{(0)}\\
\chi_{2}^{(0)}\\
\chi_{3}^{(0)}
\end{array}\right)(p^{z})=\left(\begin{array}{c}
\frac{m}{\sqrt{2E_{0}}}d_{2}(\eta^{(0)},\sqrt{\frac{2}{E_{0}}}p^{z})\\
d_{1}(\eta^{(0)},\sqrt{\frac{2}{E_{0}}}p^{z})\\
-\frac{m}{\sqrt{2E_{0}}}d_{3}(\eta^{(0)},\sqrt{\frac{2}{E_{0}}}p^{z})
\end{array}\right),
\end{equation}
where $\eta^{(0)}\equiv m^{2}/E_{0}$ and $d_{i}$ are defined in
Eq. (\ref{eq:auxiliary functions-1}). The functions $f_{1}^{(0)}$,
$f_{3}^{(0)}$ and $f_{4}^{(4)}$ can be reproduced using Eq. (\ref{eq:relation f134 with chi123}).
To summarize, we liste all the functions for the Lowest Landau level,
\begin{equation}
\left(\begin{array}{c}
f_{1}^{(0)}\\
f_{2}^{(0)}\\
f_{3}^{(0)}\\
f_{4}^{(0)}
\end{array}\right)(t,p^{z})=\left(\begin{array}{c}
\frac{m}{\sqrt{2E_{0}}}d_{2}(\eta^{(0)},\sqrt{\frac{2}{E_{0}}}p^{z})C_{1}^{(0)}(p^{z}-E_{0}t)\\
C_{2}^{(0)}(p^{z}-E_{0}t)\\
d_{1}(\eta^{(0)},\sqrt{\frac{2}{E_{0}}}p^{z})C_{1}^{(0)}(p^{z}-E_{0}t)\\
-\frac{m}{\sqrt{2E_{0}}}d_{3}(\eta^{(0)},\sqrt{\frac{2}{E_{0}}}p^{z})C_{1}^{(0)}(p^{z}-E_{0}t)
\end{array}\right).\label{eq:solutions for LLL}
\end{equation}
By inserting them into Eq. (\ref{eq:decomposition of G_i}) one can
obtain the contribution of the lowest Landau level to the Wigner function,
which will be given later.

\subsubsection{Higher Landau levels}

For all the higher Landau levels, the equations of motion in Eq. (\ref{eq:EOM of higher Landau levels})
take the same form for different $n$. Note that these equations contain
16 functions, $f_{ij}^{(n)}$ with $i,j=1,2,3,4$, for each $n$.
Solving such a system seems to be very difficult, thus we first analyze
the relation between $f_{ij}^{(n)}$, which is listed in Table \ref{tab:Coupling-relations}.
Using this table we can divide the 16 functions into several subgroups.
For example, we start from $f_{11}^{(n)}$, which directly couples
with $f_{42}^{(n)}$ and $f_{43}^{(n)}$. Then $f_{42}^{(n)}$ couples
with $f_{32}^{(n)}$, while $f_{43}^{(n)}$ couples with $f_{33}^{(n)}$.
Furthermore, $f_{32}^{(n)}$ and $f_{33}^{(n)}$ couple with $f_{24}^{(n)}$.
Thus, these six functions, $\left\{ f_{11}^{(n)},f_{24}^{(n)},f_{32}^{(n)},f_{33}^{(n)},f_{42}^{(n)},f_{43}^{(n)}\right\} $,
form one subgroup because every member only couples with other members
in this group. Analogously, we can find other subgroups from Table
\ref{tab:Coupling-relations}: $\left\{ f_{12}^{(n)},f_{13}^{(n)},f_{31}^{(n)},f_{41}^{(n)}\right\} $,
$\left\{ f_{14}^{(n)}\right\} $, $\left\{ f_{21}^{(n)}\right\} $,
$\left\{ f_{22}^{(n)},f_{23}^{(n)},f_{34}^{(n)},f_{44}^{(n)}\right\} $.
Note that not all of these subgroups contribute to the Wigner function,
which can be understood as follows: according to the solution in Eq.
(\ref{eq:equal-t Wigner function in magnetic}), when the electric
field vanishes $E_{0}\rightarrow0$, the non-vanishing functions are
$f_{11}^{(n)}$, $f_{21}^{(n)}$, $f_{32}^{(n)}$, and $f_{33}^{(n)}$.
During the time evolution, only the terms coupled with them, i.e.,
$\left\{ f_{21}^{(n)}\right\} $ and $\left\{ f_{11}^{(n)},f_{24}^{(n)},f_{32}^{(n)},f_{33}^{(n)},f_{42}^{(n)},f_{43}^{(n)}\right\} $,
can have non-trivial solutions, while other terms will stay zero.

\begin{table}
\begin{tabular}{|c|c|c|c|c|c|c|c|c|c|c|c|c|c|c|c|c|}
\hline
 & $f_{11}^{(n)}$ & $f_{12}^{(n)}$ & $f_{13}^{(n)}$ & $f_{14}^{(n)}$ & $f_{21}^{(n)}$ & $f_{22}^{(n)}$ & $f_{23}^{(n)}$ & $f_{24}^{(n)}$ & $f_{31}^{(n)}$ & $f_{32}^{(n)}$ & $f_{33}^{(n)}$ & $f_{34}^{(n)}$ & $f_{41}^{(n)}$ & $f_{42}^{(n)}$ & $f_{43}^{(n)}$ & $f_{44}^{(n)}$\tabularnewline
\hline
\hline
$f_{11}^{(n)}$ &  &  &  &  &  &  &  &  &  &  &  &  &  & $\checkmark$  & $\checkmark$  & \tabularnewline
\hline
$f_{12}^{(n)}$ &  &  &  &  &  &  &  &  &  &  &  &  & $\checkmark$  &  &  & \tabularnewline
\hline
$f_{13}^{(n)}$ &  &  &  &  &  &  &  &  &  &  &  &  & $\checkmark$  &  &  & \tabularnewline
\hline
$f_{14}^{(n)}$ &  &  &  &  &  &  &  &  &  &  &  &  &  &  &  & \tabularnewline
\hline
$f_{21}^{(n)}$ &  &  &  &  &  &  &  &  &  &  &  &  &  &  &  & \tabularnewline
\hline
$f_{22}^{(n)}$ &  &  &  &  &  &  &  &  &  &  &  & $\checkmark$  &  &  &  & \tabularnewline
\hline
$f_{23}^{(n)}$ &  &  &  &  &  &  &  &  &  &  &  & $\checkmark$  &  &  &  & \tabularnewline
\hline
$f_{24}^{(n)}$ &  &  &  &  &  &  &  &  &  & $\checkmark$  & $\checkmark$  &  &  &  &  & \tabularnewline
\hline
$f_{31}^{(n)}$ &  &  &  &  &  &  &  &  &  &  &  &  & $\checkmark$  &  &  & \tabularnewline
\hline
$f_{32}^{(n)}$ &  &  &  &  &  &  &  & $\checkmark$  &  &  &  &  &  & $\checkmark$  &  & \tabularnewline
\hline
$f_{33}^{(n)}$ &  &  &  &  &  &  &  & $\checkmark$  &  &  &  &  &  &  & $\checkmark$  & \tabularnewline
\hline
$f_{34}^{(n)}$ &  &  &  &  &  & $\checkmark$  & $\checkmark$  &  &  &  &  &  &  &  &  & $\checkmark$ \tabularnewline
\hline
$f_{41}^{(n)}$ &  & $\checkmark$  & $\checkmark$  &  &  &  &  &  & $\checkmark$  &  &  &  &  &  &  & \tabularnewline
\hline
$f_{42}^{(n)}$ & $\checkmark$  &  &  &  &  &  &  &  &  & $\checkmark$  &  &  &  &  &  & \tabularnewline
\hline
$f_{43}^{(n)}$ & $\checkmark$  &  &  &  &  &  &  &  &  &  & $\checkmark$  &  &  &  &  & \tabularnewline
\hline
$f_{44}^{(n)}$ &  &  &  &  &  &  &  &  &  &  &  & $\checkmark$  &  &  &  & \tabularnewline
\hline
\end{tabular}

\caption{\label{tab:Coupling-relations}Coupling relations among $f_{ij}^{(n)}$,
$i,j=1,2,3,4$. The table describes the coupling for each pair of
functions. Here $\checkmark$ means that we can find one equation
in Eq. (\ref{eq:EOM of higher Landau levels}) which contains both
of these two functions. On the other hand, blank  means we cannot
find such an equation in Eq. (\ref{eq:EOM of higher Landau levels}). }

\end{table}

First we focus on the function $f_{21}^{(n)}$. It decouples from
all the other functions and the corresponding equation reads $D_{t}f_{21}^{(n)}(t,p^{z})=0$.
Analogous to the case in the lowest Landau level, this equation gives
nothing but the conservation of the net fermion number in each Landau
levels. Its specific solution is
\begin{equation}
f_{21}^{(n)}(t,p^{z})=C_{2}^{(n)}(p^{z}-E_{0}t),\label{eq:solution of f^n_21}
\end{equation}
where $C_{2}^{(n)}$ is defined in Eq. (\ref{eq:equilibrium C1C2}).
Again this solution describes the overall acceleration of charged
particles, and at $t=0$ the system is described by $C_{1}^{(n)}(p^{z})$
and $C_{2}^{(n)}(p^{z})$.

As mentioned above, $\left\{ f_{11}^{(n)},f_{24}^{(n)},f_{32}^{(n)},f_{33}^{(n)},f_{42}^{(n)},f_{43}^{(n)}\right\} $
form one subgroup for the equations of motion. They can be further
decoupled by introducing a linear recombination,
\begin{equation}
\left(\begin{array}{cc}
g_{1}^{(n)} & g_{3}^{(n)}\\
g_{4}^{(n)} & g_{2}^{(n)}
\end{array}\right)=\frac{1}{m^{(n)}}\left(\begin{array}{cc}
m & \sqrt{2nB_{0}}\\
\sqrt{2nB_{0}} & -m
\end{array}\right)\left(\begin{array}{cc}
f_{11}^{(n)} & f_{24}^{(n)}\\
f_{33}^{(n)} & f_{42}^{(n)}
\end{array}\right),
\end{equation}
where we define $m^{(n)}\equiv\sqrt{m^{2}+2nB_{0}}$ as the effective
mass in the $n$-th Landau level. The transformation matrix is unitary,
so the inverse transformation reads
\begin{equation}
\left(\begin{array}{cc}
f_{11}^{(n)} & f_{24}^{(n)}\\
f_{33}^{(n)} & f_{42}^{(n)}
\end{array}\right)=\frac{1}{m^{(n)}}\left(\begin{array}{cc}
m & \sqrt{2nB_{0}}\\
\sqrt{2nB_{0}} & -m
\end{array}\right)\left(\begin{array}{cc}
g_{1}^{(n)} & g_{3}^{(n)}\\
g_{4}^{(n)} & g_{2}^{(n)}
\end{array}\right).\label{eq:relation between f_ij and g_i}
\end{equation}
Then from the equations of motion (\ref{eq:EOM of higher Landau levels})
we obtain the following two groups of equations,
\begin{equation}
D_{t}\left(\begin{array}{c}
g_{1}^{(n)}\\
g_{2}^{(n)}\\
f_{32}^{(n)}
\end{array}\right)(t,p^{z})=2\left(\begin{array}{ccc}
0 & -p^{z} & 0\\
p^{z} & 0 & -m^{(n)}\\
0 & m^{(n)} & 0
\end{array}\right)\left(\begin{array}{c}
g_{1}^{(n)}\\
g_{2}^{(n)}\\
f_{32}^{(n)}
\end{array}\right)(t,p^{z}),\label{eq:equation for g12 f32}
\end{equation}
and
\begin{equation}
D_{t}\left(\begin{array}{c}
g_{3}^{(n)}\\
g_{4}^{(n)}\\
f_{43}^{(n)}
\end{array}\right)(t,p^{z})=2\left(\begin{array}{ccc}
0 & -p^{z} & 0\\
p^{z} & 0 & m^{(n)}\\
0 & -m^{(n)} & 0
\end{array}\right)\left(\begin{array}{c}
g_{3}^{(n)}\\
g_{4}^{(n)}\\
f_{43}^{(n)}
\end{array}\right)(t,p^{z}).
\end{equation}
When the electric field vanishes, we can calculate $g_{i}^{(n)}$
and the result reads,
\begin{equation}
\left.\left(\begin{array}{c}
g_{1}^{(n)}\\
g_{2}^{(n)}\\
f_{32}^{(n)}
\end{array}\right)\right|_{E_{0}\rightarrow0}=\frac{1}{E_{p^{z}}^{(n)}}\left(\begin{array}{c}
m^{(n)}\\
0\\
p^{z}
\end{array}\right)C_{1}^{(n)}(p^{z}),\ \ \left.\left(\begin{array}{c}
g_{3}^{(n)}\\
g_{4}^{(n)}\\
f_{43}^{(n)}
\end{array}\right)\right|_{E_{0}\rightarrow0}=\left(\begin{array}{c}
0\\
0\\
0
\end{array}\right).
\end{equation}
The equations for $g_{3}^{(n)}$, $g_{4}^{(n)}$, and $f_{43}^{(n)}$
will have trivial solutions, i.e., all of three stay zero even after
the electric field is turned on. Therefore, for the higher Landau
levels we only need to focus on $g_{1}^{(n)}$, $g_{2}^{(n)}$, and
$f_{32}^{(n)}$. We take the overall acceleration ansatz and parameterize
them as
\begin{equation}
\left\{ g_{1}^{(n)},g_{2}^{(n)},f_{32}^{(n)}\right\} =\left\{ \chi_{1}^{(n)},\chi_{2}^{(n)},\chi_{3}^{(n)}\right\} C_{1}^{(n)}(p^{z}-E_{0}t).\label{eq:relation between g_12 and chi_123}
\end{equation}
Since $D_{t}C_{1}^{(n)}(p^{z}-E_{0}t)=0$, the equations of motion
for $\chi_{1}^{(n)}$, $\chi_{2}^{(n)}$, and $\chi_{3}^{(n)}$ can
be derived from Eq. (\ref{eq:equation for g12 f32})
\begin{equation}
D_{t}\left(\begin{array}{c}
\chi_{1}^{(n)}\\
\chi_{2}^{(n)}\\
\chi_{3}^{(n)}
\end{array}\right)(t,p^{z})=2\left(\begin{array}{ccc}
0 & -p^{z} & 0\\
p^{z} & 0 & -m^{(n)}\\
0 & m^{(n)} & 0
\end{array}\right)\left(\begin{array}{c}
\chi_{1}^{(n)}\\
\chi_{2}^{(n)}\\
\chi_{3}^{(n)}
\end{array}\right)(t,p^{z}),
\end{equation}
with the asymptotic condition
\begin{equation}
\left.\left(\begin{array}{c}
\chi_{1}^{(n)}\\
\chi_{2}^{(n)}\\
\chi_{3}^{(n)}
\end{array}\right)(t,p^{z})\right|_{E_{0}\rightarrow0}=\frac{1}{E_{p^{z}}^{(n)}}\left(\begin{array}{c}
m^{(n)}\\
0\\
p^{z}
\end{array}\right).
\end{equation}
The equations and asymptotic conditions coincide with Eqs. (\ref{eq:EOM of chi_i})
and (\ref{eq:limit E0->0}) if we make the replacements $\chi_{1}\rightarrow\chi_{3}^{(n)}$,
$\chi_{2}\rightarrow\chi_{1}^{(n)}$, $\chi_{3}\rightarrow\chi_{2}^{(n)}$,
and $m_{T}\rightarrow m^{(n)}$. The solution for the case of a constant
electric field is then straightforward to obtain,
\begin{equation}
\left(\begin{array}{c}
\chi_{1}^{(n)}\\
\chi_{2}^{(n)}\\
\chi_{3}^{(n)}
\end{array}\right)(t,p^{z})=\left(\begin{array}{c}
\frac{m^{(n)}}{\sqrt{2E_{0}}}d_{2}\left(\eta^{(n)},\sqrt{\frac{2}{E_{0}}}p^{z}\right)\\
\frac{m^{(n)}}{\sqrt{2E_{0}}}d_{3}\left(\eta^{(n)},\sqrt{\frac{2}{E_{0}}}p^{z}\right)\\
d_{1}\left(\eta^{(n)},\sqrt{\frac{2}{E_{0}}}p^{z}\right)
\end{array}\right),
\end{equation}
where $\eta^{(n)}\equiv(m^{2}+2nB_{0})/E_{0}$ is the dimensionless
effective mass squared. Inserting the solutions into Eq. (\ref{eq:relation between g_12 and chi_123})
and then using the inverse transformation in Eq. (\ref{eq:relation between f_ij and g_i}),
one obtains the non-vanishing functions,
\begin{eqnarray}
\left(\begin{array}{c}
f_{11}^{(n)}\\
f_{33}^{(n)}
\end{array}\right) & = & \left(\begin{array}{c}
m\\
\sqrt{2nB_{0}}
\end{array}\right)\frac{1}{\sqrt{2E_{0}}}d_{2}\left(\eta^{(n)},\sqrt{\frac{2}{E_{0}}}p^{z}\right)C_{1}^{(n)}(p^{z}-E_{0}t),\nonumber \\
\left(\begin{array}{c}
f_{24}^{(n)}\\
f_{42}^{(n)}
\end{array}\right) & = & \left(\begin{array}{c}
\sqrt{2nB_{0}}\\
-m
\end{array}\right)\frac{1}{\sqrt{2E_{0}}}d_{3}\left(\eta^{(n)},\sqrt{\frac{2}{E_{0}}}p^{z}\right)C_{1}^{(n)}(p^{z}-E_{0}t),\nonumber \\
f_{32}^{(n)} & = & d_{1}\left(\eta^{(n)},\sqrt{\frac{2}{E_{0}}}p^{z}\right)C_{1}^{(n)}(p^{z}-E_{0}t),\label{eq:solution of f_ij HLL}
\end{eqnarray}
together with $f_{21}^{(n)}$ listed in Eq. (\ref{eq:solution of f^n_21}).
The remaining ten of $f_{ij}^{(n)}$ are zero.

\subsubsection{Wigner function}

In the above parts of this subsection, we have solved the Wigner function
in parallel electromagnetic fields by properly choosing basis functions.
Finally we obtained a system of partial differential equations, which
is the same as the one in a pure constant electric field. The only
difference is, in a electric field, the equations depends on the magnitude
of the transverse momentum $p_{T}$, while in parallel electromagnetic
fields we have to replace $p_{T}$ by the quantized momentum $p_{T}\rightarrow\sqrt{2nB_{0}}$.
For convenience of future works, we list all the components of Wigner
function in the following. These components are obtained by inserting
the solutions (\ref{eq:solutions for LLL}), (\ref{eq:solution of f^n_21}),
and (\ref{eq:solution of f_ij HLL}) into Eq. (\ref{eq:decomposition of G_i}).
The four groups $\mathcal{G}_{i}$, $i=1,2,3,4$, defined in Eq. (\ref{def:definition of G_i}),
are given by
\begin{eqnarray}
\left(\begin{array}{c}
\mathcal{F}\\
\boldsymbol{\mathcal{S}}
\end{array}\right) & = & \frac{m}{\sqrt{2E_{0}}}\sum_{n=0}d_{2}\left(\eta^{(n)},\sqrt{\frac{2}{E_{0}}}p^{z}\right)C_{1}^{(n)}(p^{z}-E_{0}t)\left(\begin{array}{c}
\Lambda_{+}^{(n)}(p_{T})\\
0\\
0\\
\Lambda_{-}^{(n)}(p_{T})
\end{array}\right),\nonumber \\
\left(\begin{array}{c}
\mathcal{P}\\
\boldsymbol{\mathcal{T}}
\end{array}\right) & = & -\frac{m}{\sqrt{2E_{0}}}\sum_{n=0}d_{3}\left(\eta^{(n)},\sqrt{\frac{2}{E_{0}}}p^{z}\right)C_{1}^{(n)}(p^{z}-E_{0}t)\left(\begin{array}{c}
\Lambda_{-}^{(n)}(p_{T})\\
0\\
0\\
\Lambda_{+}^{(n)}(p_{T})
\end{array}\right),\nonumber \\
\left(\begin{array}{c}
\mathcal{V}^{0}\\
\boldsymbol{\mathcal{A}}
\end{array}\right) & = & \sum_{n=0}C_{2}^{(n)}(p^{z}-E_{0}t)\left(\begin{array}{c}
\Lambda_{+}^{(n)}(p_{T})\\
0\\
0\\
\Lambda_{-}^{(n)}(p_{T})
\end{array}\right)\nonumber \\
 &  & \ \ \ \ +\frac{1}{\sqrt{2E_{0}}}\sum_{n>0}d_{3}\left(\eta^{(n)},\sqrt{\frac{2}{E_{0}}}p^{z}\right)C_{1}^{(n)}(p^{z}-E_{0}t)\frac{2nB_{0}}{p_{T}^{2}}\Lambda_{+}^{(n)}(p_{T})\left(\begin{array}{c}
0\\
-p^{y}\\
p^{x}\\
0
\end{array}\right),\nonumber \\
\left(\begin{array}{c}
\mathcal{A}^{0}\\
\boldsymbol{\mathcal{V}}
\end{array}\right) & = & \sum_{n=0}d_{1}\left(\eta^{(n)},\sqrt{\frac{2}{E_{0}}}p^{z}\right)C_{1}^{(n)}(p^{z}-E_{0}t)\left(\begin{array}{c}
\Lambda_{-}^{(n)}(p_{T})\\
0\\
0\\
\Lambda_{+}^{(n)}(p_{T})
\end{array}\right)\nonumber \\
 &  & \ \ \ \ +\frac{1}{\sqrt{2E_{0}}}\sum_{n>0}d_{2}\left(\eta^{(n)},\sqrt{\frac{2}{E_{0}}}p^{z}\right)C_{1}^{(n)}(p^{z}-E_{0}t)\frac{2nB_{0}}{p_{T}^{2}}\Lambda_{+}^{(n)}(p_{T})\left(\begin{array}{c}
0\\
p^{x}\\
p^{y}\\
0
\end{array}\right),\nonumber \\
\label{eq:Wigner function in parallel EM fields}
\end{eqnarray}
where $\eta^{(n)}\equiv(m^{2}+2nB_{0})/E_{0}$ is the dimensionless
effective mass squared. The functions $C_{1}^{(n)}$, $C_{2}^{(n)}$,
$d_{i}$ with $i=1,2,3$, and $\Lambda_{\pm}^{(n)}(p_{T})$ are defined
in (\ref{eq:equilibrium C1C2}), (\ref{eq:auxiliary functions-1}),
and (\ref{eq:def-Lambda-n}), respectively. If the magnetic field
is sufficiently small, the sum over all Landau levels can be done
using Eqs. (\ref{eq:weak field limit-1}) and (\ref{eq:weak field limit-2}).
The discrete Landau levels are then replaced by the continuous transverse
momentum squared $p_{T}^{2}$ and (\ref{eq:Wigner function in parallel EM fields})
reproduce the results (\ref{eq:Wigner function in pure electric})
in a contant electric field.

\newpage{}$\ $

\newpage{}

\section{Semi-classical expansion\label{sec:Semi-classical-expansion}}

\subsection{Introduction to the $\hbar$ expansion}

In Sec. \ref{sec:Analytically-solvable-cases} we have shown several
cases in which the Wigner function has an analytically solution. The
semi-classical expansion is a more general approach which can be used
for a general space-time dependent field. In the semi-classical expansion,
we make a Taylor expansions for the Wigner function, all the operators,
as well as all the equations in powers of the reduced Planck's constant
$\hbar$, and then solve the equations order by order. The Wigner
function, taking its scalar component as an example, is expanded as
follows,
\begin{equation}
\mathcal{F}=\sum_{n=0}^{\infty}\hbar^{n}\mathcal{F}^{(n)}.
\end{equation}
Here we use the superscript $(n)$ to label different orders in $\hbar$.
The operators in Eq. (\ref{def:operators Kmu with hbar}) are expanded
as
\begin{eqnarray}
 &  & \Pi^{\mu}=\sum_{n=0}^{\infty}\hbar^{2n}\Pi^{(2n)\mu}=p^{\mu}-\frac{\hbar}{2}\sum_{n=0}^{\infty}\frac{(-1)^{n}}{(2n+3)(2n+1)!}\Delta^{2n+1}F^{\mu\nu}(x)\partial_{p\nu},\nonumber \\
 &  & \nabla^{\mu}=\sum_{n=0}^{\infty}\hbar^{2n}\nabla^{(2n)\mu}=\partial_{x}^{\mu}-\sum_{n=0}^{\infty}\frac{(-1)^{n}}{(2n+1)!}\Delta^{2n}F^{\mu\nu}(x)\partial_{p\nu},\label{eq:operators after hbar expansion}
\end{eqnarray}
where the spatial-derivative $\partial_{x\alpha}$ in the product
$\Delta\equiv\frac{\hbar}{2}\partial_{p}^{\alpha}\partial_{x\alpha}$
acts only on the electromagnetic field tensor $F^{\mu\nu}(x)$. The
other operators, $\Re K^{2}$, $\Im K^{2}$, $\Re K_{\mu\nu}$, and
$\Im K_{\mu\nu}$ can be written in terms of $\Pi^{\mu}$ and $\nabla^{\mu}$
as shown in (\ref{eq:real and imaginary parts of second order operators}).
The leading-order contributions to these operators are
\begin{eqnarray}
\Pi^{\mu} & = & p^{\mu}+\mathcal{O}(\hbar^{2}),\nonumber \\
\nabla^{\mu} & = & \partial_{x}^{\mu}-F^{\mu\nu}\partial_{p\nu}+\mathcal{O}(\hbar^{2}),\nonumber \\
\Re K^{2} & = & p^{2}+\mathcal{O}(\hbar^{2}),\nonumber \\
\Im K^{2} & = & \hbar p_{\mu}(\partial_{x}^{\mu}-F^{\mu\nu}\partial_{p,\nu})+\mathcal{O}(\hbar^{2}),\nonumber \\
\Re K_{\mu\nu} & = & -\frac{\hbar^{2}}{2}(\partial_{x}^{\alpha}F_{\mu\nu})\partial_{p\alpha}+\mathcal{O}(\hbar^{3}),\nonumber \\
\Im K_{\mu\nu} & = & -\hbar F_{\mu\nu}+\mathcal{O}(\hbar^{2}),\label{eq:semi-classical operators}
\end{eqnarray}

The reduced Planck's constant $\hbar$ labels the strength of the
spin-electromagnetic coupling. For example, the quantum of the spin-angular
momentum in a given direction is $\pm\hbar/2$ for a spin-1/2 particle.
Thus the method of the semi-classical expansion, in some sense, is
the Taylor expansion of the spin effect. In the zeroth order of $\hbar$,
a particle can be treated as a spinless classical one. The first order
in $\hbar$ gives the leading-order correction from the spin. In this
section we truncate at the linear order in $\hbar$ because the equations
will be more and more complicated and hard to solve in higher orders.
In this section we will preform a semi-classical expansion and then
in Sec. \ref{sec:Physical quantities} we will compare the physical
quantities calculated using the semi-classical expansion with those
from analytical calculations.

The semi-classical expansion works well if and only if high order
contributions in $\hbar$ are much smaller than the lower order ones.
This requirement is ensured by the following inequality, which is
derived from Eq. (\ref{eq:operators after hbar expansion})
\begin{equation}
\left(\frac{\hbar}{2}\partial_{p}^{\alpha}\partial_{x\alpha}\right)^{2n+2}F^{\mu\nu}(x)W(x,p)\ll\left(\frac{\hbar}{2}\partial_{p}^{\alpha}\partial_{x\alpha}\right)^{2n}F^{\mu\nu}(x)W(x,p).
\end{equation}
Assuming that the fluctuations of the electromagnetic field are significant
over a typical spatial scale $\Delta R$, while the Wigner function
fluctuates over a typical momentum scale $\Delta P$, we demand that
\begin{equation}
\Delta R\Delta P\gg\hbar.\label{eq:DeltaR DeltaP}
\end{equation}
In the unit of $\mathrm{MeV\cdot fm}$, the value of the reduced Planck's
constant is $\hbar=197\ \mathrm{MeV\cdot fm}$. If we consider cosmic
systems such as a neutron star, the typical spatial scale is large
enough to ensure that the semi-classical expansion is valid. If we
consider heavy-ion collisions such as Au+Au collisions at 200 GeV/A
at RHIC, the typical momentum is several GeV while the typical spatial
scale is several fm, which means Eq. (\ref{eq:DeltaR DeltaP}) can
also be satisfied. Therefore the method discussed in this section
could be useful for both cosmic and microscopic systems. On the other
hand, the zeroth-order part of the Dirac-form equation (\ref{eq:Dirac equation for Wigner})
for the Wigner function is $(\gamma^{\mu}p_{\mu}-m)W$, while the
first-order part is $\frac{i\hbar}{2}\gamma^{\mu}\partial_{x\mu}W$.
Thus we demand
\begin{equation}
\left|\hbar\gamma^{\mu}\partial_{x\mu}W\right|\ll m\left|W\right|.
\end{equation}
to ensure that the $\hbar$-order correction is much smaller than
the zeroth-order contribution. Note that $\hbar/m$ is the Compton
wave length, thus the above condition means the wave length of macroscopic
fluctuations should be much larger than the Compton wave length.

The semi-classical expansion is widely used in recent years \cite{Vasak:1987um,Hidaka:2016yjf,Gao:2017gfq,Huang:2018wdl,Gao:2019znl,Weickgenannt:2019dks,Hattori:2019ahi,Wang:2019moi}.
In this section we will solve the Wigner function up to order $\hbar$
and derive the corresponding kinetic equation at the same order. Higher
order contributions can be solved employing a similar procedure but
the results would be too complicated for further analysis and thus
are not listed in this thesis.

\subsection{Massless case}

As we discussed in Sec. \ref{sec:Overview-of-Wigner}, in the massless
case the system of partial differential equations is much simpler
because the vector and axial-vector components $\mathcal{V}^{\mu}$
and $\mathcal{A}^{\mu}$ are decoupled from the others. Meanwhile,
$\mathcal{V}^{\mu}$ is equivalent to $\mathcal{A}^{\mu}$ because
of the chiral symmetry of massless fermions. In this section we will
solve $\mathcal{V}^{\mu}$ and $\mathcal{A}^{\mu}$ up to order $\hbar$.
First we form a linear combination and define the LH and RH currents
$\mathcal{J}_{\pm}^{\mu}$ as in Eq. (\ref{def:left- and right-handed currents}).
Inserting the operators in Eq. (\ref{eq:semi-classical operators})
into the on-shell equation (\ref{eq:on-shell equation for f_pm})
and the constraint equations (\ref{eq:equations of currents massless}),
at the zeroth order in $\hbar$ we have
\begin{eqnarray}
p_{\mu}\mathcal{J}_{\chi}^{(0)\mu} & = & 0,\nonumber \\
p_{\mu}\mathcal{J}_{\chi,\nu}^{(0)}-p_{\nu}\mathcal{J}_{\chi,\mu}^{(0)} & = & 0,\nonumber \\
p^{2}\mathcal{J}_{\chi}^{(0)\mu} & = & 0.\label{eq:zeroth order equations of J=00005Cchi}
\end{eqnarray}
Here $\mathcal{J}_{\chi}^{(0)\mu}$ represents the zeroth-order part
of $\mathcal{J}_{\chi}^{\mu}$, where $\chi=\pm$ labeling the chirality.
The last line ensures that $\mathcal{J}_{\chi}^{(0)\mu}$ should be
on the mass-shell $p^{2}=0$ otherwise we would obtain the trivial
solution $\mathcal{J}_{\chi}^{(0)\mu}=0$. The general nontrivial
zeroth-order solution reads
\begin{equation}
\mathcal{J}_{\chi}^{(0)\mu}=p^{\mu}f_{\chi}^{(0)}\delta(p^{2}).\label{eq:massless zeroth order}
\end{equation}
Here the distribution $f_{\chi}^{(0)}$ is still undetermined, but
it should not have a singularity on the mass-shell $p^{2}=0$. The
zeroth-order currents are parallel to $p^{\mu}$, which agrees with
our expectation: the spins of chiral fermions are always parallel
or anti-parallel to their momentum, so the fermion number current
and axial-charge current of massless particles are both proportional
to $p^{\mu}$, and so is their linear combination.

At the first order in $\hbar$, Eqs. (\ref{eq:equations of currents massless})
and (\ref{eq:on-shell equation for f_pm}) can be separated into two
groups, one of which only depends on the zeroth-order function $\mathcal{J}_{\chi}^{(0)\mu}$,
\begin{equation}
\nabla_{\mu}^{(0)}\mathcal{J}_{\chi}^{(0)\mu}=0,\label{eq:massless zeroth order kin}
\end{equation}
and the other group depends on the first-order function $\mathcal{J}_{\chi}^{(1)\mu}$,
\begin{eqnarray}
p_{\mu}\mathcal{J}_{\chi}^{(1)\mu} & = & 0,\nonumber \\
p_{\mu}\mathcal{J}_{\chi,\nu}^{(1)}-p_{\nu}\mathcal{J}_{\chi,\mu}^{(1)}+\frac{\chi}{2}\epsilon_{\mu\nu\alpha\beta}\nabla^{(0)\alpha}\mathcal{J}_{\chi}^{(0)\beta} & = & 0,\nonumber \\
p^{2}\mathcal{J}_{\chi}^{(1)\mu}+\frac{\chi}{2}\epsilon^{\mu\nu\alpha\beta}F_{\alpha\beta}\mathcal{J}_{\chi\nu}^{(0)} & = & 0.\label{eq:massless first order equations}
\end{eqnarray}
Inserting the zeroth-order solution (\ref{eq:massless zeroth order})
into Eq. (\ref{eq:massless zeroth order kin}) we obtain the kinetic
equations for $f_{\chi}^{(0)}$ ,
\begin{equation}
\delta(p^{2})p^{\mu}\nabla_{\mu}^{(0)}f_{\chi}^{(0)}=0,\label{eq:massless zeroth order kin-1}
\end{equation}
which agrees with the collisionless Boltzmann-Vlasov equation \cite{DeGroot:1980dk}.
The first-order current $\mathcal{J}_{\chi}^{(1)\mu}$ can be assumed
to have a solution of the form
\begin{equation}
\mathcal{J}_{\chi}^{(1)\mu}=j_{\chi}^{\mu}\delta(p^{2})+\chi\tilde{F}^{\mu\nu}p_{\nu}f_{\chi}^{(0)}\delta^{\prime}(p^{2}),\label{eq:massless first order}
\end{equation}
where the derivative of the Dirac-delta function is $\delta^{\prime}(x)=-\delta(x)/x$,
which can be proved using the method of integrating by parts. Using
$\tilde{F}^{\mu\nu}=\frac{1}{2}\epsilon^{\mu\nu\alpha\beta}F_{\alpha\beta}$
we can check that the solution (\ref{eq:massless first order}) automatically
satisfies the third line of Eq. (\ref{eq:massless first order equations})
for an arbitrary $j_{\chi}^{\mu}$ . Inserting the solution (\ref{eq:massless first order})
into the first and second lines of Eq. (\ref{eq:massless first order equations})
we obtain the following relations
\begin{eqnarray}
0 & = & \delta(p^{2})p_{\mu}j_{\chi}^{\mu},\nonumber \\
0 & = & \delta(p^{2})\left[p_{\mu}j_{\chi,\nu}-p_{\nu}j_{\chi,\mu}+\frac{\chi}{2}\epsilon_{\mu\nu\alpha\beta}\nabla^{(0)\alpha}\left(p^{\beta}f_{\chi}^{(0)}\right)\right]\nonumber \\
 &  & +\chi\delta^{\prime}(p^{2})\left(p_{\mu}\tilde{F}_{\nu\alpha}p^{\alpha}-p_{\nu}\tilde{F}_{\mu\alpha}p^{\alpha}-\epsilon_{\mu\nu\alpha\beta}p^{\beta}F^{\alpha\gamma}p_{\gamma}\right)f_{\chi}^{(0)}.\label{eq:massless first order constraint}
\end{eqnarray}
Here the last line can be simplified using the Schouten identity,
\begin{equation}
p_{\mu}\epsilon_{\nu\alpha\beta\gamma}+p_{\nu}\epsilon_{\alpha\beta\gamma\mu}+p_{\alpha}\epsilon_{\beta\gamma\mu\nu}+p_{\beta}\epsilon_{\gamma\mu\nu\alpha}+p_{\gamma}\epsilon_{\mu\nu\alpha\beta}=0.\label{eq:Schouten identity-1}
\end{equation}
This identity holds because in 4-dimensional Minkowski space, the
indices can take the values $1-4$, thus at least two of the indices
$\mu\nu\alpha\beta\gamma$ are identical. Considering without loss
of generality the case $\mu=\nu$, the Levi-Civita symbols $\epsilon_{\beta\gamma\mu\nu}$,
$\epsilon_{\gamma\mu\nu\alpha}$, and $\epsilon_{\mu\nu\alpha\beta}$
vanish, and the remaining two terms cancel with each other due to
the anti-symmetric property of the Levi-Civita symbol. In the case
that three or more indices are equal to each other, all Levi-Civita
symbols are vanishing. With the help of the Schouten identity, the
term which multiplies $\delta^{\prime}(p^{2})$ can be simplified
as
\begin{eqnarray}
 &  & p_{\mu}\tilde{F}_{\nu\alpha}p^{\alpha}-p_{\nu}\tilde{F}_{\mu\alpha}p^{\alpha}-\epsilon_{\mu\nu\alpha\beta}p^{\beta}F^{\alpha\gamma}p_{\gamma}\nonumber \\
 & = & \frac{1}{2}(p_{\mu}\epsilon_{\nu\alpha\beta\gamma}+p_{\nu}\epsilon_{\alpha\beta\gamma\mu})p^{\alpha}F^{\beta\gamma}-\epsilon_{\mu\nu\alpha\beta}p^{\beta}F^{\alpha\gamma}p_{\gamma}\nonumber \\
 & = & -\frac{1}{2}(p_{\alpha}\epsilon_{\beta\gamma\mu\nu}+p_{\beta}\epsilon_{\gamma\mu\nu\alpha}+p_{\gamma}\epsilon_{\mu\nu\alpha\beta})p^{\alpha}F^{\beta\gamma}-\epsilon_{\mu\nu\alpha\beta}p^{\beta}F^{\alpha\gamma}p_{\gamma}\nonumber \\
 & = & -p^{2}\tilde{F}_{\mu\nu},
\end{eqnarray}
Inserting back into Eq. (\ref{eq:massless first order constraint})
one obtains two constraints
\begin{eqnarray}
0 & = & \delta(p^{2})p_{\mu}j_{\chi}^{\mu},\nonumber \\
0 & = & \delta(p^{2})\left[p_{\mu}j_{\chi,\nu}-p_{\nu}j_{\chi,\mu}+\frac{\chi}{2}\epsilon_{\mu\nu\alpha\beta}p^{\beta}\nabla^{(0)\alpha}f_{\chi}^{(0)}\right].
\end{eqnarray}
On account of Eq. (\ref{eq:Schouten identity-1}) and using Eq. (\ref{eq:massless zeroth order kin-1}),
the general solution reads
\begin{equation}
j_{\chi,\mu}=p_{\mu}f_{\chi}^{(1)}+\frac{\chi}{2(p\cdot u)}\epsilon_{\mu\nu\alpha\beta}p^{\nu}u^{\alpha}\nabla^{(0)\beta}f_{\chi}^{(0)},
\end{equation}
where $u^{\mu}$ is an arbitrary reference vector with $p\cdot u\neq0$.
Substituting $j_{\chi}^{\mu}$ into Eq. (\ref{eq:massless first order}),
we obtain
\begin{equation}
\mathcal{J}_{\chi}^{(1)\mu}=\left[p^{\mu}f_{\chi}^{(1)}+\frac{\chi}{2(p\cdot u)}\epsilon^{\mu\nu\alpha\beta}p_{\nu}u_{\alpha}\nabla_{\beta}^{(0)}f_{\chi}^{(0)}\right]\delta(p^{2})+\chi\tilde{F}^{\mu\nu}p_{\nu}f_{\chi}^{(0)}\delta^{\prime}(p^{2}).\label{eq:massless first order-1}
\end{equation}
Here we again demand that the distribution $f_{\chi}^{(1)}$ is non-singular
at $p^{2}=0$. The first-order solution derived here agrees with previous
results \cite{Hidaka:2016yjf,Gao:2017gfq,Huang:2018wdl}.

At the order $\hbar^{2}$, Eqs. (\ref{eq:on-shell equation for f_pm})
and (\ref{eq:equations of currents massless}) contain the second-order
current $\mathcal{J}_{\chi}^{(2)\mu}$. Since we only focus on the
leading two orders of the solution, the equations for $\mathcal{J}_{\chi}^{(2)\mu}$
will be neglected. Only one equation is independent of $\mathcal{J}_{\chi}^{(2)\mu}$
at the order $\hbar^{2}$,
\begin{equation}
\nabla_{\mu}^{(0)}\mathcal{J}_{\chi}^{(1)\mu}=0.\label{eq:massless first order kin}
\end{equation}
The kinetic equation for $f_{\chi}^{(1)}$ is then derived by substituting
$\mathcal{J}_{\chi}^{(1)\mu}$ in Eq. (\ref{eq:massless first order-1})
into the above equation,%
\begin{eqnarray}
0 & = & \delta(p^{2})\left\{ p^{\mu}\nabla_{\mu}^{(0)}f_{\chi}^{(1)}+\frac{\chi}{2}\left(\nabla_{\mu}^{(0)}\frac{u_{\alpha}}{p\cdot u}\right)\epsilon^{\mu\nu\alpha\beta}p_{\nu}\nabla_{\beta}^{(0)}f_{\chi}^{(0)}\right.\nonumber \\
 &  & \qquad\left.-\frac{\chi}{2}\epsilon^{\mu\nu\alpha\beta}F_{\mu\nu}\nabla_{\beta}^{(0)}f_{\chi}^{(0)}+\frac{\chi}{4(p\cdot u)}\epsilon^{\mu\nu\alpha\beta}p_{\nu}u_{\alpha}\left[\nabla_{\mu}^{(0)},\nabla_{\beta}^{(0)}\right]f_{\chi}^{(0)}\right\} \nonumber \\
 &  & +\chi\delta^{\prime}(p^{2})\left[\nabla_{\mu}^{(0)}\left(\tilde{F}^{\mu\nu}p_{\nu}f_{\chi}^{(0)}\right)-\frac{1}{p\cdot u}\epsilon^{\mu\nu\alpha\beta}F_{\mu\gamma}p^{\gamma}p_{\nu}u_{\alpha}\nabla_{\beta}^{(0)}f_{\chi}^{(0)}\right]\nonumber \\
 &  & -2\chi\tilde{F}^{\mu\nu}p_{\nu}F_{\mu\alpha}p^{\alpha}f_{\chi}^{(0)}\delta^{\prime\prime}(p^{2}).
\end{eqnarray}
The Schouten identity (\ref{eq:Schouten identity-1}) is then used
for further simplification. We also use the following relation
\begin{equation}
\tilde{F}^{\mu\nu}p_{\nu}F_{\mu\alpha}p^{\alpha}=\frac{1}{4}p^{2}F_{\beta\gamma}\tilde{F}^{\beta\gamma},
\end{equation}
and the properties of the Dirac-delta function
\begin{eqnarray}
p^{2}\delta^{\prime}(p^{2}) & = & -\delta(p^{2}),\nonumber \\
p^{2}\delta^{\prime\prime}(p^{2}) & = & -2\delta^{\prime}(p^{2}).
\end{eqnarray}
After simplification we finally obtain
\begin{eqnarray}
0 & = & \delta(p^{2})\left\{ p^{\mu}\nabla_{\mu}^{(0)}f_{\chi}^{(1)}+\frac{\chi}{2}\left(\nabla_{\mu}^{(0)}\frac{u_{\alpha}}{p\cdot u}\right)\epsilon^{\mu\nu\alpha\beta}p_{\nu}\nabla_{\beta}^{(0)}f_{\chi}^{(0)}+\frac{\chi}{2(p\cdot u)}p_{\nu}u_{\alpha}(\partial_{x\gamma}\tilde{F}^{\nu\alpha})\partial_{p}^{\gamma}f_{\chi}^{(0)}\right\} \nonumber \\
 &  & \qquad+\chi\delta^{\prime}(p^{2})\left[(\partial_{x\mu}\tilde{F}^{\mu\nu})p_{\nu}f_{\chi}^{(0)}-\frac{1}{p\cdot u}\tilde{F}^{\mu\nu}p_{\mu}u_{\nu}p^{\alpha}\nabla_{\alpha}^{(0)}f_{\chi}^{(0)}\right],\label{eq:massless first order kin-1}
\end{eqnarray}
which is the kinetic equation for the first-order distribution function
$f_{\chi}^{(1)}$.

Collecting the zeroth- and first-order solutions, we find that the
LH and RH currents in the massless case are given by
\begin{equation}
\mathcal{J}_{\chi}^{\mu}=\left[p^{\mu}f_{\chi}+\frac{\hbar\chi}{2(p\cdot u)}\epsilon^{\mu\nu\alpha\beta}p_{\nu}u_{\alpha}\nabla_{\beta}^{(0)}f_{\chi}\right]\delta(p^{2})+\chi\hbar\tilde{F}^{\mu\nu}p_{\nu}f_{\chi}\delta^{\prime}(p^{2})+\mathcal{O}(\hbar^{2}),
\end{equation}
where
\begin{equation}
f_{\chi}\equiv f_{\chi}^{(0)}+\hbar f_{\chi}^{(1)}+\mathcal{O}(\hbar^{2}),
\end{equation}
is the full distribution function for RH ($\chi=+$) or LH ($\chi=-$)
particles, which depends on the phase-space position $\{x^{\mu},p^{\mu}\}$.
Note that the full distribution $f_{\chi}$ contains contributions
of all orders in $\hbar$, but higher order terms should be much smaller
than the leading two orders, which ensures the validity of the semi-classical
expansion. The kinetic equation for $f_{\chi}$ can be derived from
the zeroth order one in (\ref{eq:massless zeroth order kin-1}) and
the first order one in (\ref{eq:massless first order kin-1}),
\begin{eqnarray}
0 & = & \delta(p^{2})\left\{ p^{\mu}\nabla_{\mu}^{(0)}f_{\chi}+\frac{\chi\hbar}{2}\left(\nabla_{\mu}^{(0)}\frac{u_{\alpha}}{p\cdot u}\right)\epsilon^{\mu\nu\alpha\beta}p_{\nu}\nabla_{\beta}^{(0)}f_{\chi}+\frac{\chi\hbar}{2(p\cdot u)}p_{\nu}u_{\alpha}(\partial_{x\gamma}\tilde{F}^{\nu\alpha})\partial_{p}^{\gamma}f_{\chi}\right\} \nonumber \\
 &  & \qquad+\chi\hbar\delta^{\prime}(p^{2})\left[(\partial_{x\mu}\tilde{F}^{\mu\nu})p_{\nu}f_{\chi}-\frac{1}{p\cdot u}\tilde{F}^{\mu\nu}p_{\mu}u_{\nu}p^{\alpha}\nabla_{\alpha}^{(0)}f_{\chi}\right]+\mathcal{O}(\hbar^{2}).\label{eq:massless CKT}
\end{eqnarray}
The kinetic equation agrees with the result of Refs. \cite{Hidaka:2016yjf,Gao:2017gfq,Huang:2018wdl}.
In the classical limit $\hbar\rightarrow0$, only the first term survives
and the equation is reduced to the well-known Boltzmann-Vlasov equation
\cite{DeGroot:1980dk}. The vector and axial-vector currents can be
recovered from the LH and RH currents as
\begin{equation}
\mathcal{V}^{\mu}=\frac{1}{2}\sum_{\chi=\pm}\mathcal{J}_{\chi}^{\mu},\ \ \mathcal{A}^{\mu}=\frac{1}{2}\sum_{\chi=\pm}\chi\mathcal{J}_{\chi}^{\mu}.
\end{equation}
We define the scalar distribution $V$ and the axial distribution
$A$ as
\begin{equation}
V\equiv\frac{1}{2}\sum_{\chi=\pm}f_{\chi},\ \ A\equiv\frac{1}{2}\sum_{\chi=\pm}\chi f_{\chi}.
\end{equation}
Then we obtain
\begin{eqnarray}
\mathcal{V}^{\mu} & = & \left[p^{\mu}V+\frac{\hbar}{2(p\cdot u)}\epsilon^{\mu\nu\alpha\beta}p_{\nu}u_{\alpha}\nabla_{\beta}^{(0)}A\right]\delta(p^{2})+\hbar\tilde{F}^{\mu\nu}p_{\nu}A\delta^{\prime}(p^{2})+\mathcal{O}(\hbar^{2}),\nonumber \\
\mathcal{A}^{\mu} & = & \left[p^{\mu}A+\frac{\hbar}{2(p\cdot u)}\epsilon^{\mu\nu\alpha\beta}p_{\nu}u_{\alpha}\nabla_{\beta}^{(0)}V\right]\delta(p^{2})+\hbar\tilde{F}^{\mu\nu}p_{\nu}V\delta^{\prime}(p^{2})+\mathcal{O}(\hbar^{2}),\label{sol:massless V and A}
\end{eqnarray}
where the distributions $V$ and $A$ in general depend on the coordinates
$\{x^{\mu},p^{\mu}\}$ in phase space. Here the vector $u^{\mu}$
plays the role of the reference frame, and $V$ , $A$ are identified
as the net fermion and axial-charge distributions in the frame $u^{\mu}$,
respectively \cite{Hidaka:2016yjf,Huang:2018wdl,Gao:2018jsi}. Note
that $V$ and $A$ should depend on the choice of $u^{\mu}$ so that
the whole currents $\mathcal{V}^{\mu}$, $\mathcal{A}^{\mu}$ are
independent of $u^{\mu}$. In the massless case, the dependence on
the reference frame is the result of the side-jump effect \cite{Chen:2014cla,Hidaka:2016yjf}.
In the next subsections we will prove that in the massive case the
reference frame can be chosen as the rest frame of the particle and
then the solution will not depend on the auxiliary vector $u^{\mu}$.

\subsection{Massive case 1: taking vector and axial-vector components as basis\label{subsec:Massive-case-1}}

In this subsection we will focus on the massive case $m\neq0$. As
discussed in Sec. \ref{sec:Overview-of-Wigner}, we can take either
the vector and axial-vector components $\mathcal{V}^{\mu}$ and $\mathcal{A}^{\mu}$
or the rest ones $\mathcal{F}$, $\mathcal{P}$, and $\mathcal{S}^{\mu\nu}$
as basis functions. In this subsection, $\mathcal{V}^{\mu}$ and $\mathcal{A}^{\mu}$
are taken as basis, while $\mathcal{F}$, $\mathcal{P}$, and $\mathcal{S}^{\mu\nu}$
are derived from $\mathcal{V}^{\mu}$ and $\mathcal{A}^{\mu}$ as
shown in Eq. (\ref{eq:FPS from VA}).

In the massive case, $\mathcal{J}_{\chi}^{\mu}$ defined in (\ref{def:left- and right-handed currents})
no longer have definite chirality but we still use $\mathcal{J}_{\chi}^{\mu}$
to represent the linear combination of $\mathcal{V}^{\mu}$ and $\mathcal{A}^{\mu}$.
We first insert the expanded operators in Eq. (\ref{eq:semi-classical operators})
into the on-shell condition (\ref{eq:on-shell equation for f_pm}).
The zeroth- and the first-order parts read
\begin{eqnarray}
\left(p^{2}-m^{2}\right)\mathcal{J}_{\chi}^{(0)\mu} & = & 0,\nonumber \\
\left(p^{2}-m^{2}\right)\mathcal{J}_{\chi}^{(1)}+\chi\tilde{F}_{\mu\nu}\mathcal{J}_{\chi}^{(0)\nu} & = & 0.
\end{eqnarray}
The general nontrivial solution of these on-shell conditions is
\begin{eqnarray}
\mathcal{J}_{\chi}^{(0)\mu} & = & f_{\chi}^{(0)\mu}\delta(p^{2}-m^{2}),\nonumber \\
\mathcal{J}_{\chi}^{(1)\mu} & = & f_{\chi}^{(1)\mu}\delta(p^{2}-m^{2})+\chi\tilde{F}^{\mu\nu}f_{\chi\nu}^{(0)}\delta^{\prime}(p^{2}-m^{2}),
\end{eqnarray}
where $f_{\pm}^{(0)\mu}$ and $f_{\pm}^{(1)\mu}$ are arbitrary functions
which are non-singular at $p^{2}-m^{2}=0$. We find that the zeroth-order
solutions are on the mass-shell $p^{2}=m^{2}$, while the first-order
ones have off-shell contributions. From the definition of $\mathcal{J}_{\chi}^{\mu}$
in (\ref{def:left- and right-handed currents}) we can recover the
vector and axial-vector currents. Up to the order $\hbar$, we obtain
\begin{eqnarray}
\mathcal{V}^{\mu} & = & \delta(p^{2}-m^{2})\sum_{\chi=\pm}\left(f_{\chi}^{(0)\mu}+\hbar f_{\chi}^{(1)\mu}\right)+\hbar\tilde{F}^{\mu\nu}\delta^{\prime}(p^{2}-m^{2})\sum_{\chi=\pm}\chi f_{\chi\nu}^{(0)}+\mathcal{O}(\hbar^{2}),\nonumber \\
\mathcal{A}^{\mu} & = & \delta(p^{2}-m^{2})\sum_{\chi=\pm}\chi\left(f_{\chi}^{(0)\mu}+\hbar f_{\chi}^{(1)\mu}\right)+\hbar\tilde{F}^{\mu\nu}\delta^{\prime}(p^{2}-m^{2})\sum_{\chi=\pm}f_{\chi\nu}^{(0)}+\mathcal{O}(\hbar^{2}).\label{eq:solution of VA}
\end{eqnarray}
The solutions should satisfy Eq. (\ref{eq:constraints for VA}), which
gives several constraints on the functions $f_{\chi}^{(0)\mu}$ and
$f_{\chi}^{(1)\mu}$. Up to the first order in $\hbar$, the last
line of Eq. (\ref{eq:constraints for VA}) reads
\begin{equation}
p_{[\mu}\left(\mathcal{V}_{\nu]}^{(0)}+\hbar\mathcal{V}_{\nu]}^{(1)}\right)+\frac{\hbar}{2}\epsilon_{\mu\nu\alpha\beta}\nabla^{(0)\alpha}\mathcal{A}^{(0)\beta}=\mathcal{O}(\hbar^{2}).\label{eq:massive constraint}
\end{equation}
Substituting the vector and axial-vector currents (\ref{eq:solution of VA})
into Eq. (\ref{eq:massive constraint}), we obtain
\begin{eqnarray}
 &  & \delta(p^{2}-m^{2})p_{[\mu}\sum_{\chi=\pm}\left(f_{\chi\nu]}^{(0)}+\hbar f_{\chi\nu]}^{(1)}\right)+\hbar p_{[\mu}\tilde{F}_{\nu]\alpha}\delta^{\prime}(p^{2}-m^{2})\sum_{\chi=\pm}\chi f_{\chi}^{(0)\alpha}\nonumber \\
 &  & \qquad\qquad\qquad\qquad\qquad\qquad+\frac{\hbar}{2}\epsilon_{\mu\nu\alpha\beta}\nabla^{(0)\alpha}\left[\delta(p^{2}-m^{2})\sum_{\chi=\pm}\chi f_{\chi}^{(0)\beta}\right]=0.
\end{eqnarray}
Contracting this equation with $p^{\mu}$ and taking out different
orders in $\hbar$ gives
\begin{eqnarray}
0 & = & \delta(p^{2}-m^{2})\left[m^{2}\sum_{\chi=\pm}f_{\chi\nu}^{(0)}-p_{\nu}p^{\mu}\sum_{\chi=\pm}f_{\chi\mu}^{(0)}\right],\nonumber \\
0 & = & \delta(p^{2}-m^{2})\left[m^{2}\sum_{\chi=\pm}f_{\chi\nu}^{(1)}-p_{\nu}p^{\mu}\sum_{\chi=\pm}f_{\chi\mu}^{(1)}+\frac{1}{2}\epsilon_{\mu\nu\alpha\beta}p^{\mu}\nabla^{(0)\alpha}\sum_{\chi=\pm}\chi f_{\chi}^{(0)\beta}\right]\nonumber \\
 &  & +\delta^{\prime}(p^{2}-m^{2})\left[p^{2}\tilde{F}_{\nu\alpha}\sum_{\chi=\pm}\chi f_{\chi}^{(0)\alpha}-p_{\nu}p^{\mu}\tilde{F}_{\mu\alpha}\sum_{\chi=\pm}\chi f_{\chi}^{(0)\alpha}-p^{\mu}\epsilon_{\mu\nu\alpha\beta}F^{\alpha\gamma}p_{\gamma}\sum_{\chi=\pm}\chi f_{\chi}^{(0)\beta}\right].\nonumber \\
\label{eq:massive equation for f_chi}
\end{eqnarray}
Now we define the distribution functions and polarization vectors
using $f_{\chi}^{(0)\mu}$ and $f_{\chi}^{(1)\mu}$,
\begin{eqnarray}
V^{(0)}\equiv\frac{1}{m^{2}}p_{\mu}\sum_{\chi=\pm}f_{\chi}^{(0)\mu}, &  & V^{(1)}\equiv\frac{1}{m^{2}}p_{\mu}\sum_{\chi=\pm}f_{\chi}^{(1)\mu},\nonumber \\
n^{(0)\mu}\equiv\frac{1}{m}\sum_{\chi=\pm}\chi f_{\chi}^{(0)\mu}, &  & n^{(1)\mu}\equiv\frac{1}{m}\sum_{\chi=\pm}\chi f_{\chi}^{(1)\mu}.
\end{eqnarray}
Using these new functions, the solutions of Eq. (\ref{eq:massive equation for f_chi})
can be expressed as %
\begin{eqnarray}
\sum_{\chi=\pm}f_{\chi}^{(0)\mu} & = & p^{\mu}V^{(0)}+(p^{2}-m^{2})g^{(0)\mu},\nonumber \\
\sum_{\chi=\pm}f_{\chi}^{(1)\mu} & = & p^{\mu}V^{(1)}+\frac{1}{2m}\epsilon^{\mu\nu\alpha\beta}p_{\nu}\nabla_{\alpha}^{(0)}n_{\beta}^{(0)}+\frac{p^{2}}{m(p^{2}-m^{2})}\tilde{F}^{\mu\nu}n_{\nu}^{(0)}\nonumber \\
 &  & -\frac{1}{m(p^{2}-m^{2})}p^{\mu}\tilde{F}_{\alpha\beta}p^{\alpha}n^{(0)\beta}+\frac{1}{m(p^{2}-m^{2})}\epsilon^{\mu\nu\alpha\beta}p_{\nu}F_{\alpha\gamma}p^{\gamma}n_{\beta}^{(0)}+(p^{2}-m^{2})g^{(1)\mu},\nonumber \\
\end{eqnarray}
where we have used $\delta^{\prime}(x)=-\delta(x)/x$. Here $g_{\mu}^{(0)}$
and $g_{\mu}^{(1)}$ are functions which are non-singular on the mass-shell
$p^{2}-m^{2}$. Since they are multiplied with $p^{2}-m^{2}$, they
will not contribute to the vector component. In terms of $V^{(0)}$,
$V^{(1)}$, $n^{(0)\mu}$, and $n^{(1)\mu}$, the solutions for $\mathcal{V}^{\mu}$
and $\mathcal{A}^{\mu}$ in Eq. (\ref{eq:solution of VA}) are given
by %
\begin{eqnarray}
\mathcal{V}^{\mu} & = & \delta(p^{2}-m^{2})\left[p^{\mu}\left(V^{(0)}+\hbar V^{(1)}\right)+\frac{\hbar}{2m}\epsilon^{\mu\nu\alpha\beta}\nabla_{\alpha}^{(0)}(p_{\nu}n_{\beta}^{(0)})\right]\nonumber \\
 &  & +\hbar\delta^{\prime}(p^{2}-m^{2})\left(\frac{1}{m}\epsilon^{\mu\nu\alpha\beta}p_{\alpha}n_{\beta}^{(0)}F_{\nu\gamma}p^{\gamma}+\frac{1}{2m}p^{\mu}\epsilon^{\nu\alpha\beta\gamma}p_{\nu}n_{\alpha}^{(0)}F_{\beta\gamma}\right),\nonumber \\
\mathcal{A}^{\mu} & = & \delta(p^{2}-m^{2})\left(mn^{(0)\mu}+\hbar mn^{(1)\mu}-\hbar\tilde{F}^{\mu\nu}g_{\nu}^{(0)}\right)+\hbar\tilde{F}^{\mu\nu}p_{\nu}V^{(0)}\delta^{\prime}(p^{2}-m^{2}).
\end{eqnarray}
Now we define the resummed distribution $V$ and the resummed polarization
$n^{\mu}$,
\begin{eqnarray}
V & \equiv & V^{(0)}+\hbar V^{(1)}+\mathcal{O}(\hbar^{2}),\nonumber \\
n^{\mu} & \equiv & n^{(0)\mu}+\hbar n^{(1)\mu}-\frac{\hbar}{m}\tilde{F}^{\mu\nu}g_{\nu}^{(0)}+\mathcal{O}(\hbar^{2}),
\end{eqnarray}
and for simplicity we also define the following dipole-moment tensor,
\begin{equation}
\Sigma^{\mu\nu}\equiv-\frac{1}{m}\epsilon^{\mu\nu\alpha\beta}p_{\alpha}n_{\beta}.\label{eq:relation between Sigma and n}
\end{equation}
Then the solutions for $\mathcal{V}^{\mu}$ and $\mathcal{A}^{\mu}$
can be rewritten as
\begin{eqnarray}
\mathcal{V}^{\mu} & = & \delta(p^{2}-m^{2})\left[p^{\mu}V+\frac{\hbar}{2}\nabla_{\nu}^{(0)}\Sigma^{\mu\nu}\right]-\hbar\delta^{\prime}(p^{2}-m^{2})\left(\Sigma^{\mu\alpha}F_{\alpha\beta}p^{\beta}+\frac{1}{2}p^{\mu}\Sigma^{\alpha\beta}F_{\alpha\beta}\right)+\mathcal{O}(\hbar^{2}),\nonumber \\
\mathcal{A}^{\mu} & = & \delta(p^{2}-m^{2})mn^{\mu}+\hbar\tilde{F}^{\mu\nu}p_{\nu}V\delta^{\prime}(p^{2}-m^{2})+\mathcal{O}(\hbar^{2}).\label{eq:massive solutions VA}
\end{eqnarray}
Inserting them into Eq. (\ref{eq:FPS from VA}) we obtain the scalar,
pseudo-scalar and tensor components $\mathcal{F}$, $\mathcal{P}$,
and $\mathcal{S}^{\mu\nu}$ %

\begin{eqnarray}
\mathcal{F} & = & m\left[\delta(p^{2}-m^{2})V-\frac{\hbar}{2}\Sigma^{\mu\nu}F_{\mu\nu}\delta^{\prime}(p^{2}-m^{2})\right]+\mathcal{O}(\hbar^{2}),\nonumber \\
\mathcal{P} & = & -\delta(p^{2}-m^{2})\frac{\hbar}{2}\nabla^{\mu}n_{\mu}+\hbar F^{\mu\nu}p_{\nu}n_{\mu}\delta^{\prime}(p^{2}-m^{2})+\mathcal{O}(\hbar^{2}),\nonumber \\
\mathcal{S}_{\mu\nu} & = & \delta(p^{2}-m^{2})\left(m\Sigma_{\mu\nu}-\frac{\hbar}{2m}p_{[\mu}\nabla_{\nu]}V\right)-m\hbar F_{\mu\nu}V\delta^{\prime}(p^{2}-m^{2})+\mathcal{O}(\hbar^{2}).\label{eq:massive solutions FPS}
\end{eqnarray}
Here the undetermined functions are $V$ and $n^{\mu}$. In the classical
limit $\hbar\rightarrow0$, we have
\begin{eqnarray}
\mathcal{V}^{\mu} & \rightarrow & p^{\mu}V\delta(p^{2}-m^{2}),\nonumber \\
\mathcal{A}^{\mu} & \rightarrow & mn^{\mu}\delta(p^{2}-m^{2}).
\end{eqnarray}
Thus $V$ can be interpreted as the fermion number distribution and
$n^{\mu}$ as the polarization density. Substituting the solutions
(\ref{eq:massive solutions VA}) into the second line of Eq. (\ref{eq:constraints for VA}),
one obtains a constraint for $n^{\mu}$,
\begin{equation}
\delta(p^{2}-m^{2})p_{\mu}n^{\mu}=\mathcal{O}(\hbar^{2}),
\end{equation}
which can be identified as a requirement for the spin: for massive
particles, their spins must be perpendicular to their momenta. On
the other hand, the kinetic equation for $V$ can be derived from
the first line of Eq. (\ref{eq:constraints for VA}). Up to the order
$\hbar^{2}$, we obtain
\begin{equation}
\hbar\nabla^{(0)\mu}\left(\mathcal{V}_{\mu}^{(0)}+\hbar\mathcal{V}_{\mu}^{(1)}\right)=\mathcal{O}(\hbar^{2}).
\end{equation}
Replacing the vector components by the solution (\ref{eq:massive solutions VA}),
we obtain the following kinetic equation %
\begin{equation}
\delta(p^{2}-m^{2})\left[p^{\mu}\nabla_{\mu}^{(0)}V+\frac{\hbar}{4}(\partial_{x\alpha}F_{\mu\nu})\partial_{p}^{\alpha}\Sigma^{\mu\nu}\right]-\frac{\hbar}{2}\delta^{\prime}(p^{2}-m^{2})F_{\alpha\beta}p^{\mu}\nabla_{\mu}^{(0)}\Sigma^{\alpha\beta}=0.\label{eq:kinetic equation for V}
\end{equation}
In the classical limit $\hbar\rightarrow0$, this kinetic equation
is reduced to the Boltzmann-Vlasov equation \cite{DeGroot:1980dk}.
On the other hand, an addition kinetic equation is necessary to determine
$n^{\mu}$, which can be derived from the last line of Eq. (\ref{eq:constraints for VA})
by contracting with $\epsilon^{\rho\sigma\mu\nu}p_{\sigma}$. We also
have another approach to derive the kinetic equation for $n^{\mu}$.
In Sec. \ref{sec:Overview-of-Wigner} we have listed all Vlasov equations
in Eq. (\ref{eq:decomposed Vlasov}), where the one for the axial-vector
component reads,
\begin{equation}
p_{\nu}\nabla^{(0)\nu}\mathcal{A}_{\mu}-F_{\mu\nu}\mathcal{A}^{\nu}-\frac{\hbar}{2}(\partial_{x}^{\alpha}\tilde{F}_{\mu\nu})\partial_{p\alpha}\mathcal{V}^{\nu}=\mathcal{O}(\hbar^{2}).\label{eq:massive kinetic VA}
\end{equation}
Substituting $\mathcal{V}^{\mu}$ and $\mathcal{A}^{\mu}$ by those
in Eq. (\ref{eq:massive solutions VA}), Eq. (\ref{eq:massive kinetic VA})
gives the following kinetic equation for $n^{\mu}$, %
\begin{equation}
\delta(p^{2}-m^{2})\left[p_{\nu}\nabla^{(0)\nu}n_{\mu}-F_{\mu\nu}n^{\nu}-\frac{\hbar}{2m}(\partial_{x}^{\alpha}\tilde{F}_{\mu\nu})p^{\nu}\partial_{p\alpha}V\right]+\frac{\hbar}{m}\delta^{\prime}(p^{2}-m^{2})\tilde{F}_{\mu\alpha}p^{\alpha}p_{\nu}\nabla^{(0)\nu}V=0.\label{eq:kinetic equation for n_mu}
\end{equation}
In the classical limit $\hbar\rightarrow0$, it reproduces the Bargmann-Michel-Telegdi
equation \cite{Bargmann:1959gz} for the classical spin precession
in an electromagnetic field.

\subsection{Massive case 2: taking scalar, pseudo-scalar, and tensor components
as basis\label{subsec:Massive-case-2}}

In this subsection we will take the scalar, pseudo-scalar, and tensor
components $\mathcal{F}$, $\mathcal{P}$, and $\mathcal{S}^{\mu\nu}$
as basis functions and solve the Wigner function in the massive case.
The remaining components, the vector and axial-vector ones $\mathcal{V}^{\mu}$,
$\mathcal{A}^{\mu}$ are then given by Eq. (\ref{eq:VA from FPS}).
We start from the zeroth order in $\hbar$. At this order, the on-shell
conditions (\ref{eq:decomposed on-shell}) read
\begin{eqnarray}
\left(p^{2}-m^{2}\right)\mathcal{F}^{(0)} & = & 0,\nonumber \\
\left(p^{2}-m^{2}\right)\mathcal{P}^{(0)} & = & 0,\nonumber \\
\left(p^{2}-m^{2}\right)\mathcal{S}_{\mu\nu}^{(0)} & = & 0,
\end{eqnarray}
which means that the zeroth-order functions are on the normal mass
shell $p^{2}-m^{2}=0$. The corresponding constraint conditions (\ref{eq:constraints of FPS})
are
\begin{eqnarray}
p^{\nu}\mathcal{S}_{\nu\mu}^{(0)} & = & 0,\nonumber \\
p_{\mu}\mathcal{P}^{(0)} & = & 0.\label{eq:zeroth order constraint}
\end{eqnarray}
Then we obtain the following general solutions
\begin{eqnarray}
\mathcal{F}^{(0)} & = & mV^{(0)}\delta(p^{2}-m^{2}),\nonumber \\
\mathcal{P}^{(0)} & = & 0,\nonumber \\
\mathcal{S}_{\mu\nu}^{(0)} & = & m\Sigma_{\mu\nu}^{(0)}\delta(p^{2}-m^{2}),\label{eq:massive zeroth order FPS}
\end{eqnarray}
where $V^{(0)}$ and $\Sigma_{\mu\nu}^{(0)}$ are now arbitrary functions
of the phase-space position $\{x^{\mu},p^{\mu}\}$. In order to satisfy
the on-shell condition, $V^{(0)}$ and $\Sigma_{\mu\nu}^{(0)}$ should
not have any singularities for an on-shell momentum $p^{2}=m^{2}$.
We also demand that
\begin{equation}
\delta(p^{2}-m^{2})\Sigma^{(0)\mu\nu}p_{\nu}=0,\label{eq:constraint for Sigma 0}
\end{equation}
in order to satisfy the constraint condition for $\mathcal{S}_{\mu\nu}^{(0)}$
in Eq. (\ref{eq:zeroth order constraint}). Since the Wigner function
has the dimension of the energy, we find that both $V^{(0)}$ and
$\Sigma_{\mu\nu}^{(0)}$ are dimensionless. Recalling that $\mathcal{F}$
is interpreted as the mass density, $V^{(0)}$ is then identified
as the zeroth-order fermion number distribution. And $\Sigma_{\mu\nu}^{(0)}$
is the dimensionless zeroth-order dipole-moment tensor. Due to the
second line of the constraint equation (\ref{eq:zeroth order constraint}),
the pseudoscalar component vanishes at the zeroth order in $\hbar$.

The first-order on-shell conditions (\ref{eq:decomposed on-shell})
read
\begin{eqnarray}
\left(p^{2}-m^{2}\right)\mathcal{F}^{(1)}-\frac{1}{2}F_{\mu\nu}\mathcal{S}^{(0)\mu\nu} & = & 0,\nonumber \\
\left(p^{2}-m^{2}\right)\mathcal{P}^{(1)}-\frac{1}{4}\epsilon_{\mu\nu\alpha\beta}F^{\mu\nu}\mathcal{S}^{(0)\alpha\beta} & = & 0,\nonumber \\
\left(p^{2}-m^{2}\right)\mathcal{S}_{\mu\nu}^{(1)}-F_{\mu\nu}\mathcal{F}^{(0)}+\frac{1}{2}\epsilon_{\mu\nu\alpha\beta}F^{\alpha\beta}\mathcal{P}^{(0)} & = & 0.
\end{eqnarray}
Inserting the zeroth-order solutions (\ref{eq:massive zeroth order FPS}),
we obtain the following general solutions at the first order in $\hbar$,
\begin{eqnarray}
\mathcal{F}^{(1)} & = & m\left[V^{(1)}\delta(p^{2}-m^{2})-\frac{1}{2}F_{\mu\nu}\Sigma^{(0)\mu\nu}\delta^{\prime}(p^{2}-m^{2})\right],\nonumber \\
\mathcal{P}^{(1)} & = & m\left[G^{(1)}\delta(p^{2}-m^{2})-\frac{1}{2}\tilde{F}_{\mu\nu}\Sigma^{(0)\mu\nu}\delta^{\prime}(p^{2}-m^{2})\right],\nonumber \\
\mathcal{S}_{\mu\nu}^{(1)} & = & m\left[\Sigma_{\mu\nu}^{(1)}\delta(p^{2}-m^{2})-F_{\mu\nu}V^{(0)}\delta^{\prime}(p^{2}-m^{2})\right],\label{eq:massive first order FPS}
\end{eqnarray}
where $\tilde{F}_{\mu\nu}\equiv\frac{1}{2}\epsilon_{\mu\nu\alpha\beta}F^{\alpha\beta}$
is the dual field tensor. Here $V^{(1)}$, $G^{(1)}$, and $\Sigma_{\mu\nu}^{(1)}$
are to be determined, which are dimensionless and non-singular on
the mass-shell $p^{2}=m^{2}$. The first-order part of constraint
equations (\ref{eq:constraints of FPS}) read
\begin{eqnarray}
\frac{1}{2}\nabla_{\mu}^{(0)}\mathcal{F}^{(0)}+p^{\nu}\mathcal{S}_{\nu\mu}^{(1)} & = & 0,\nonumber \\
p_{\mu}\mathcal{P}^{(1)}+\frac{1}{4}\epsilon_{\mu\nu\alpha\beta}\nabla^{(0)\nu}\mathcal{S}^{(0)\alpha\beta} & = & 0.
\end{eqnarray}
Substituting the zeroth-order and first-order functions by the general
solutions in (\ref{eq:massive zeroth order FPS}) and (\ref{eq:massive first order FPS}),
we obtain %
\begin{equation}
\delta(p^{2}-m^{2})\left[p^{\nu}\Sigma_{\mu\nu}^{(1)}-\frac{1}{2}\nabla_{\mu}^{(0)}V^{(0)}\right]=0,\label{eq:constraint for Sigma 1}
\end{equation}
and
\begin{eqnarray}
0 & = & \delta(p^{2}-m^{2})\left[p_{\mu}G^{(1)}+\frac{1}{4}\epsilon_{\mu\nu\alpha\beta}\nabla^{(0)\nu}\Sigma^{(0)\alpha\beta}\right]\nonumber \\
 &  & -\frac{1}{4}\left[p_{\mu}\epsilon_{\alpha\beta\nu\gamma}+p_{\gamma}\epsilon_{\mu\nu\alpha\beta}-p_{\nu}\epsilon_{\mu\gamma\alpha\beta}\right]F^{\nu\gamma}\Sigma^{(0)\alpha\beta}\delta^{\prime}(p^{2}-m^{2}).\label{eq:constaint G1}
\end{eqnarray}
Then using the Schouten identity in (\ref{eq:Schouten identity-1})
to simplify the second line of Eq. (\ref{eq:constaint G1}) and we
obtain a constraint equation for $G^{(1)}$
\begin{equation}
\delta(p^{2}-m^{2})\left[p_{\mu}G^{(1)}+\frac{1}{4}\epsilon_{\mu\nu\alpha\beta}\nabla^{(0)\nu}\Sigma^{(0)\alpha\beta}\right]+\tilde{F}_{\mu\nu}\Sigma^{(0)\nu\alpha}p_{\alpha}\delta^{\prime}(p^{2}-m^{2})=0,\label{eq:constraint equation for G1}
\end{equation}
which leads to a solution
\begin{equation}
G^{(1)}=-\frac{1}{4m^{2}}\epsilon_{\mu\nu\alpha\beta}p^{\mu}\nabla^{(0)\nu}\Sigma^{(0)\alpha\beta}+\frac{1}{m^{2}(p^{2}-m^{2})}p^{\mu}\tilde{F}_{\mu\nu}\Sigma^{(0)\nu\alpha}p_{\alpha}.
\end{equation}
Since the zeroth-order dipole-moment tensor $\Sigma^{(0)\nu\alpha}$
satisfies Eq. (\ref{eq:constraint for Sigma 0}), we find that $G^{(1)}$
is not singular on the mass-shell $p^{2}-m^{2}=0$, which agrees with
our requirement. Then we can prove that the pseudo-scalar component
$\mathcal{P}^{(1)}$ can be written as %
\begin{equation}
\mathcal{P}^{(1)}=\frac{1}{4m}\nabla^{(0)\mu}\left[\epsilon_{\mu\nu\alpha\beta}p^{\nu}\Sigma^{(0)\alpha\beta}\delta(p^{2}-m^{2})\right].
\end{equation}
Thus up to the order $\hbar$, the undetermined functions are $V^{(0)}$,
$V^{(1)}$, $\Sigma_{\mu\nu}^{(0)}$, and $\Sigma_{\mu\nu}^{(1)}$.

Now we define the resummed functions
\begin{eqnarray}
V & \equiv & V^{(0)}+\hbar V^{(1)}+\mathcal{O}(\hbar^{2}),\nonumber \\
\Sigma_{\mu\nu} & \equiv & \Sigma_{\mu\nu}^{(0)}+\hbar\Sigma_{\mu\nu}^{(1)}+\frac{\hbar}{2m^{2}}p_{[\mu}\nabla_{\nu]}^{(0)}V+\mathcal{O}(\hbar^{2}).\label{eq:resummed functions}
\end{eqnarray}
Here in the definition of the resummed dipole-moment tensor we add
an additional term $\frac{\hbar}{2m^{2}}p_{[\mu}\nabla_{\nu]}^{(0)}V$
so that the final results are comparable with the ones in the previous
subsection. In terms of these resummed functions, the up-to-$\hbar$-order
solutions for the scalar, pseudo-scalar, and tensor components are
written as
\begin{eqnarray}
\mathcal{F} & = & m\left[V\delta(p^{2}-m^{2})-\frac{\hbar}{2}F_{\mu\nu}\Sigma^{\mu\nu}\delta^{\prime}(p^{2}-m^{2})\right]+\mathcal{O}(\hbar^{2}),\nonumber \\
\mathcal{P} & = & \frac{\hbar}{4m}\nabla^{(0)\mu}\left[\epsilon_{\mu\nu\alpha\beta}p^{\nu}\Sigma^{\alpha\beta}\delta(p^{2}-m^{2})\right]+\mathcal{O}(\hbar^{2}),\nonumber \\
\mathcal{S}_{\mu\nu} & = & \delta(p^{2}-m^{2})\left(m\Sigma_{\mu\nu}-\frac{\hbar}{2m}p_{[\mu}\nabla_{\nu]}^{(0)}V\right)-m\hbar F_{\mu\nu}V\delta^{\prime}(p^{2}-m^{2})+\mathcal{O}(\hbar^{2}).\label{eq:order hbar solutions of FPS}
\end{eqnarray}
The resummed dipole-moment tensor $\Sigma_{\mu\nu}$ satisfies the
following constraint equation, which is derived from Eqs. (\ref{eq:constraint for Sigma 0})
and (\ref{eq:constraint for Sigma 1})
\begin{equation}
\delta(p^{2}-m^{2})\left(\Sigma_{\mu\nu}p^{\nu}-\frac{\hbar}{2m^{2}}p_{\mu}p^{\nu}\nabla_{\nu}^{(0)}V\right)=\mathcal{O}(\hbar^{2}).\label{eq:constraint resummed Sigma}
\end{equation}
On the other hand, from Eq. (\ref{eq:decomposed Vlasov}) we obtain
the Vlasov equations for $\mathcal{F}$ and $\mathcal{S}_{\mu\nu}$,
\begin{eqnarray}
\hbar p_{\mu}\nabla^{(0)\mu}\mathcal{F}+\frac{\hbar^{2}}{4}(\partial_{x}^{\alpha}F_{\mu\nu})\partial_{p\alpha}\mathcal{S}^{\mu\nu} & = & \mathcal{O}(\hbar^{3}).\nonumber \\
\hbar p_{\alpha}\nabla^{(0)\alpha}\mathcal{S}_{\mu\nu}-\hbar F_{\ [\mu}^{\alpha}\mathcal{S}_{\nu]\alpha}+\frac{\hbar^{2}}{2}(\partial_{x}^{\alpha}F_{\mu\nu})\partial_{p\alpha}\mathcal{F}-\frac{\hbar^{2}}{4}\epsilon_{\mu\nu\alpha\beta}(\partial_{x}^{\gamma}F^{\alpha\beta})\partial_{p\gamma}\mathcal{P} & = & \mathcal{O}(\hbar^{3}).
\end{eqnarray}
Here the Vlasov equation for the pseudo-scalar component $\mathcal{P}$
is not listed because it does not contain any new undetermined function.
Inserting the solutions (\ref{eq:order hbar solutions of FPS}) into
the above Vlasov equations, we obtain %
{}
\begin{eqnarray}
\delta(p^{2}-m^{2})\left[p_{\mu}\nabla^{(0)\mu}V+\frac{\hbar}{4}(\partial_{x}^{\alpha}F_{\mu\nu})\partial_{p\alpha}\Sigma^{\mu\nu}\right]-\frac{\hbar}{2}\delta^{\prime}(p^{2}-m^{2})F_{\mu\nu}p_{\alpha}\nabla^{(0)\alpha}\Sigma^{\mu\nu} & = & \mathcal{O}(\hbar^{2}).\nonumber \\
\delta(p^{2}-m^{2})\left[p_{\alpha}\nabla^{(0)\alpha}\bar{\Sigma}_{\mu\nu}-F_{\ [\mu}^{\alpha}\bar{\Sigma}_{\nu]\alpha}+\frac{\hbar}{2}(\partial_{x}^{\alpha}F_{\mu\nu})\partial_{p\alpha}V\right]-\delta^{\prime}(p^{2}-m^{2})\hbar F_{\mu\nu}p_{\alpha}\nabla^{(0)\alpha}V & = & \mathcal{O}(\hbar^{2}).\nonumber \\
\label{eq:massive kinetic equations}
\end{eqnarray}
where in the second line $\bar{\Sigma}_{\mu\nu}\equiv\Sigma_{\mu\nu}-\frac{\hbar}{2m^{2}}p_{[\mu}\nabla_{\nu]}^{(0)}V$.
This redefinition does not introduce new functions but it makes the
Vlasov equations more concise. With the help of the Vlasov equation
for $V$, we find that the constraint equation (\ref{eq:constraint resummed Sigma})
for $\Sigma_{\mu\nu}$ can be further simplified,
\begin{equation}
\delta(p^{2}-m^{2})\Sigma_{\mu\nu}p^{\nu}=\mathcal{O}(\hbar^{2}).\label{eq:constraint equation for Sigma}
\end{equation}
This means that the dipole-moment tensor is perpendicular to the momentum.
The vector and axial-vector components are calculated using Eq. (\ref{eq:VA from FPS}).
Up to the order $\hbar$, we obtain
\begin{eqnarray}
\mathcal{V}_{\mu} & = & \delta(p^{2}-m^{2})\left[p_{\mu}V+\frac{\hbar}{2}\nabla^{(0)\nu}\Sigma_{\mu\nu}\right]-\hbar\delta^{\prime}(p^{2}-m^{2})\left[\frac{1}{2}F_{\alpha\beta}\Sigma^{\alpha\beta}p_{\mu}+\Sigma_{\mu\nu}F^{\nu\alpha}p_{\alpha}\right]+\mathcal{O}(\hbar^{2}),\nonumber \\
\mathcal{A}_{\mu} & = & -\frac{1}{2}\epsilon_{\mu\nu\alpha\beta}p^{\nu}\Sigma^{\alpha\beta}\delta(p^{2}-m^{2})+\hbar\tilde{F}_{\mu\nu}p^{\nu}V\delta^{\prime}(p^{2}-m^{2})+\mathcal{O}(\hbar^{2}).\label{eq:VA from FPS-1}
\end{eqnarray}
The solutions in (\ref{eq:order hbar solutions of FPS}) and (\ref{eq:VA from FPS-1})
provide all components of the Wigner function. Defining
\begin{equation}
n_{\mu}\equiv-\frac{1}{2m}\epsilon_{\mu\nu\alpha\beta}p^{\nu}\Sigma^{\alpha\beta},
\end{equation}
the axial-vector component can be written as
\begin{equation}
\mathcal{A}_{\mu}=mn_{\mu}\delta(p^{2}-m^{2})+\hbar\tilde{F}_{\mu\nu}p^{\nu}V\delta^{\prime}(p^{2}-m^{2})+\mathcal{O}(\hbar^{2}).
\end{equation}
The kinetic equation for $n_{\mu}$ can be derived by acting the operator
$p_{\alpha}\nabla^{(0)\alpha}$ onto the definition of $n_{\mu}$,
and then using the second line of Eq. (\ref{eq:massive kinetic equations}).
A carefully calculation reproduces Eq. (\ref{eq:kinetic equation for n_mu}).
Hence the results in this subsection, i.e., Eqs. (\ref{eq:order hbar solutions of FPS}),
(\ref{eq:VA from FPS-1}), and (\ref{eq:massive kinetic equations}),
coincide with the results in the previous subsection \ref{subsec:Massive-case-1},
i.e., Eqs. (\ref{eq:massive solutions VA}), (\ref{eq:massive solutions FPS}),
(\ref{eq:kinetic equation for V}), and (\ref{eq:kinetic equation for n_mu}).

In the classical limit $\hbar\rightarrow0$, the solutions (\ref{eq:order hbar solutions of FPS})
and (\ref{eq:VA from FPS-1}) coincide with the results from the first-principle
calculations in Sec. \ref{sec:Analytically-solvable-cases}. So the
analytical solutions give a constructive suggestion for the undetermined
functions. In practice, one assumes that the undetermined functions
takes their equilibrium form, i.e., they are solutions of the collisionless
Boltzmann-Vlasov equation. However, the equilibrium form of $V$ at
order $\hbar$ is still under debate. In Ref. \cite{Weickgenannt:2019dks}
we proposed a possible equilibrium distribution but it can only be
used in very limited cases. This is because we have neglected the
momentum dependence of the dipole-moment tensor. But in real cases,
the dipole-moment tensor should be computed from its kinetic equation
and thus in general depends on $x^{\mu}$ and $p^{\mu}$. A self-consistent
treatment for the kinetic equations in Eq. (\ref{eq:massive kinetic equations})
has not been done yet.

\subsection{Ambiguity of functions\label{subsec:Ambiguity-of-functions}}

Comparing the results in subsection \ref{subsec:Massive-case-1} with
that in subsection \ref{subsec:Massive-case-2}, we find that even
though in these subsections we start from different points, the final
solutions as well as the corresponding kinetic equations and constraint
equations are exactly the same. This agrees with our expectation that
the Wigner function should have only one solution. Although the Wigner
function has 16 components, the solutions up to the first order in
$\hbar$ only depends on four functions $V$ and $n_{\mu}$ (note
here that $n_{\mu}$ is a 4-vector which is perpendicular to $p_{\mu}$,
so it only has three components). In this section we will first analyze
the Wigner function as an eigenvalue problem, which will clearly show
why there are only four independent degrees of freedom. Then we will
discuss the ambiguity of the undetermined functions, where we find
some transformations which change the basis functions without changing
the Wigner function. We will also show in this subsection how to smoothly
reproduce the massless results from the massive ones.

\subsubsection{eigenvalue problem}

We first focus on the Dirac-form equation for the Wigner function
in Eq. (\ref{eq:Dirac equation for Wigner}). The leading two orders
in $\hbar$ read
\begin{equation}
(\gamma^{\mu}p_{\mu}-m)\left(W^{(0)}+\hbar W^{(1)}\right)=-\frac{i\hbar}{2}\gamma^{\mu}\nabla_{\mu}^{(0)}W^{(0)}+\mathcal{O}(\hbar^{2}).\label{eq:eigen-value equation}
\end{equation}
Here we have moved the spatial gradient term to the right-hand-side
of the equation. Since the Wigner function has 16 components as shown
in Eq. (\ref{def:Wigner function decomposition}), we now put these
components in a column vector as follows
\begin{equation}
w(x,p)\equiv\left(\mathcal{F},\,\mathcal{P},\,\mathcal{V}^{0},\:\mathcal{V}^{1},\:\mathcal{V}^{2},\:\mathcal{V}^{3},\,\mathcal{A}^{0},\,\mathcal{A}^{1},\,\mathcal{A}^{2},\,\mathcal{A}^{3},\,\mathcal{S}^{01},\,\mathcal{S}^{02},\,\mathcal{S}^{03},\,\mathcal{S}^{23},\,\mathcal{S}^{31},\,\mathcal{S}^{12}\right)^{T}.
\end{equation}
These component can be derived from the Wigner function by multiplying
with $\Gamma_{i}=\{\mathbb{I}_{4},\ i\gamma^{5},\ \gamma^{\mu},\ \gamma^{5}\gamma^{\mu},\ \frac{1}{2}\sigma^{\mu\nu}\}$
and then taking the trace. Then Eq. (\ref{eq:eigen-value equation})
can be written as
\begin{equation}
(M-m\mathbb{I}_{16})\,w(x,p)=\hbar\delta w,\label{eq:eigen-value equation-1}
\end{equation}
where $\mathbb{I}_{16}$ is a $16$-dimensional unit matrix, $\delta w$
represents the order-$\hbar$ correction that can be calculated from
the left-hand-side of Eq. (\ref{eq:eigen-value equation}). In this
section we only focus on the properties of the solution, thus we will
not list the exact formula for $\delta w$. The vector $w(x,p)$ contains
both the zeroth order and first order contributions. The coefficient
matrix $M$ is a $16\times16$ complex matrix given by %
{\footnotesize{}
\begin{equation}
M=\left(\begin{array}{cccccccccccccccc}
0 & 0 & p^{0} & -p_{x} & -p_{y} & -p_{z} & 0 & 0 & 0 & 0 & 0 & 0 & 0 & 0 & 0 & 0\\
0 & 0 & 0 & 0 & 0 & 0 & ip^{0} & -ip_{x} & -ip_{y} & -ip_{z} & 0 & 0 & 0 & 0 & 0 & 0\\
p^{0} & 0 & 0 & 0 & 0 & 0 & 0 & 0 & 0 & 0 & -ip_{x} & -ip_{y} & -ip_{z} & 0 & 0 & 0\\
p_{x} & 0 & 0 & 0 & 0 & 0 & 0 & 0 & 0 & 0 & -ip^{0} & 0 & 0 & 0 & -ip_{z} & ip_{y}\\
p_{y} & 0 & 0 & 0 & 0 & 0 & 0 & 0 & 0 & 0 & 0 & -ip^{0} & 0 & ip_{z} & 0 & -ip_{x}\\
p_{z} & 0 & 0 & 0 & 0 & 0 & 0 & 0 & 0 & 0 & 0 & 0 & -ip^{0} & -ip_{y} & ip_{x} & 0\\
0 & -ip^{0} & 0 & 0 & 0 & 0 & 0 & 0 & 0 & 0 & 0 & 0 & 0 & p_{x} & p_{y} & p_{z}\\
0 & -ip_{x} & 0 & 0 & 0 & 0 & 0 & 0 & 0 & 0 & 0 & -p_{z} & p_{y} & p^{0} & 0 & 0\\
0 & -ip_{y} & 0 & 0 & 0 & 0 & 0 & 0 & 0 & 0 & p_{x} & 0 & -p_{x} & 0 & p^{0} & 0\\
0 & -ip_{z} & 0 & 0 & 0 & 0 & 0 & 0 & 0 & 0 & -p_{y} & p_{x} & 0 & 0 & 0 & p^{0}\\
0 & 0 & -ip_{x} & ip^{0} & 0 & 0 & 0 & 0 & -p_{z} & p_{y} & 0 & 0 & 0 & 0 & 0 & 0\\
0 & 0 & -ip_{y} & 0 & ip^{0} & 0 & 0 & p_{z} & 0 & -p_{x} & 0 & 0 & 0 & 0 & 0 & 0\\
0 & 0 & -ip_{z} & 0 & 0 & ip^{0} & 0 & -p_{y} & p_{x} & 0 & 0 & 0 & 0 & 0 & 0 & 0\\
0 & 0 & 0 & 0 & ip_{z} & -ip_{y} & -p_{x} & p^{0} & 0 & 0 & 0 & 0 & 0 & 0 & 0 & 0\\
0 & 0 & 0 & -ip_{z} & 0 & ip_{x} & -p_{y} & 0 & p^{0} & 0 & 0 & 0 & 0 & 0 & 0 & 0\\
0 & 0 & 0 & ip_{y} & -ip_{x} & 0 & -p_{z} & 0 & 0 & p^{0} & 0 & 0 & 0 & 0 & 0 & 0
\end{array}\right).
\end{equation}
}The solution of Eq. (\ref{eq:eigen-value equation-1}) can be decomposed
into one special solution and several general solutions, where the
general solutions are solved by taking $\delta w\rightarrow0$. In
the limit $\delta w\rightarrow0$, Eq. (\ref{eq:eigen-value equation-1})
is the characteristic equation for the matrix $M$, with $m$ the
eigenvalue and $w$ the corresponding eigenvector. This characteristic
equation has a nontrivial solution if and only if the determinant
of its coefficient matrix vanishes
\begin{equation}
\det(M-m\mathbb{I}_{16})=0,
\end{equation}
which gives
\begin{equation}
(p^{2}-m^{2})^{8}=0.
\end{equation}
So the matrix $M$ has eight positive eigenvalues $m=\sqrt{p^{2}}$
and eight negative ones $m=-\sqrt{p^{2}}$. In real cases the particle's
mass is positive, thus the negative eigenvalues are non-physical.
For the positive eigenvalues, we can find the following eigenvectors
{\small{}
\begin{eqnarray}
v_{1} & = & \left(\begin{array}{cccccccccccccccc}
m, & 0, & p^{0}, & p_{x}, & p_{y}, & p_{z}, & 0, & 0, & 0, & 0, & 0, & 0, & 0, & 0, & 0, & 0\end{array}\right),\nonumber \\
v_{2} & = & \left(\begin{array}{cccccccccccccccc}
0, & im, & 0, & 0, & 0, & 0, & p^{0}, & p_{x}, & p_{y}, & p_{z}, & 0, & 0, & 0, & 0, & 0, & 0\end{array}\right),\nonumber \\
v_{3} & = & \left(\begin{array}{cccccccccccccccc}
0, & -imp_{x}, & -ip_{y}p_{z}, & 0, & -ip^{0}p_{z}, & 0, & -p^{0}p_{x}, & -(p_{x}^{2}+p_{z}^{2}), & -p_{x}p_{y}, & 0, & 0, & mp_{z}, & 0, & 0, & 0, & 0\end{array}\right),\nonumber \\
v_{4} & = & \left(\begin{array}{cccccccccccccccc}
0, & imp_{y}, & -ip_{x}p_{z}, & -ip^{0}p_{z}, & 0, & 0, & p^{0}p_{y}, & p_{x}p_{y}, & p_{y}^{2}+p_{z}^{2}, & 0, & mp_{z}, & 0, & 0, & 0, & 0, & 0\end{array}\right),\nonumber \\
v_{5} & = & \left(\begin{array}{cccccccccccccccc}
0, & -imp^{0}, & 0, & ip_{y}p_{z}, & -ip_{x}p_{z}, & 0, & -(p^{0})^{2}+p_{z}^{2}, & -p^{0}p_{x}, & -p^{0}p_{y}, & 0, & 0, & 0, & 0, & 0, & 0, & mp_{z}\end{array}\right),\nonumber \\
v_{6} & = & \left(\begin{array}{cccccccccccccccc}
imp^{0}, & 0, & i(p^{0})^{2}-ip_{z}^{2}, & ip^{0}p_{x}, & ip^{0}p_{y}, & 0, & 0, & p_{y}p_{z}, & -p_{x}p_{z}, & 0, & 0, & 0, & mp_{z}, & 0, & 0, & 0\end{array}\right),\nonumber \\
v_{7} & = & \left(\begin{array}{cccccccccccccccc}
-imp_{x}, & 0, & -ip^{0}p_{x}, & -i(p_{x}^{2}+p_{z}^{2}), & -ip_{x}p_{y}, & 0, & p_{y}p_{z}, & 0, & p^{0}p_{z}, & 0, & 0, & 0, & 0, & 0, & mp_{z}, & 0\end{array}\right),\nonumber \\
v_{8} & = & \left(\begin{array}{cccccccccccccccc}
imp_{y}, & 0, & ip^{0}p_{y}, & ip_{x}p_{y}, & i(p_{y}^{2}+p_{z}^{2}), & 0, & p_{x}p_{z}, & p^{0}p_{z}, & 0, & 0, & 0, & 0, & 0, & mp_{z}, & 0, & 0\end{array}\right).
\end{eqnarray}
}Note that these vectors are neither properly normalized nor orthogonal
to each other. The general solution for $w$ can be written in terms
of the above eigenvectors,
\begin{equation}
w=\sum_{i=1}^{8}c_{i}v_{i}.
\end{equation}
The property of the Wigner function tells us that all components $\mathcal{F}$,
$\mathcal{P}$, $\mathcal{V}^{\mu}$, $\mathcal{A}^{\mu}$, and $\mathcal{S}^{\mu\nu}$
are real functions, but the eigenvectors $v_{i}$ with $i=2,3,\cdots,8$
are complex vectors. In order to make sure all the components of $w$
are real, we can obtain the following constraints for coefficients
$c_{i}$,
\begin{eqnarray}
p^{0}c_{6}-p_{x}c_{7}-p_{y}c_{8} & = & 0,\nonumber \\
c_{2}-p_{x}c_{3}+p_{y}c_{4}-p^{0}c_{5} & = & 0,\nonumber \\
-p_{y}p_{z}c_{3}-p_{x}p_{z}c_{4}+\left[(p^{0})^{2}-p_{z}^{2}\right]c_{6}-p^{0}p_{x}c_{7}+p^{0}p_{y}c_{8} & = & 0,\nonumber \\
-p^{0}p_{z}c_{4}+p_{y}p_{z}c_{5}+p^{0}p_{x}c_{6}-(p_{x}^{2}+p_{z}^{2})c_{7}+p_{x}p_{y}c_{8} & = & 0,\nonumber \\
-p^{0}p_{z}c_{3}-p_{x}p_{z}c_{5}+p^{0}p_{y}c_{6}-p_{x}p_{y}c_{7}+(p_{y}^{2}+p_{z}^{2})c_{8} & = & 0.
\end{eqnarray}
Using these constraints, the coefficients $c_{5}$, $c_{6}$, $c_{7}$,
and $c_{8}$ can be expressed in terms of $c_{2}$, $c_{3}$, and
$c_{4}$,
\begin{eqnarray}
c_{5}=\frac{c_{2}-p_{x}c_{3}+p_{y}c_{4}}{p^{0}}, &  & c_{6}=-\frac{p_{y}c_{3}+p_{z}c_{4}}{p_{z}},\nonumber \\
c_{7}=\frac{p_{y}c_{2}-p_{x}p_{y}c_{3}-(m^{2}+p_{x}^{2}+p_{z}^{2})c_{4}}{p^{0}p_{z}}, &  & c_{8}=\frac{p_{x}c_{2}+(m^{2}+p_{y}^{2}+p_{z}^{2})c_{3}+p_{x}p_{y}c_{4}}{p^{0}p_{z}}.
\end{eqnarray}
If the coefficients do not satisfy these relations, then the vector
$\sum_{i=1}^{8}c_{i}v_{i}$ may have an imaginary part, which cannot
be a correct solution for the Wigner function. So in order to construct
the general solution of $w(x,p)$, we only need four parameters $c_{i}$
with $i=1,2,3,4$. This means the general order-$\hbar$ solution
has only four independent degrees of freedom, which agrees with the
conclusion of the previous subsections \ref{subsec:Massive-case-1}
and \ref{subsec:Massive-case-2}.

\subsubsection{Shift of mass-shell}

In this section we will show how the energies of the particles are
shifted by the coupling between the electromagnetic field and the
dipole moment. First we find that the solutions (\ref{eq:order hbar solutions of FPS})
and (\ref{eq:VA from FPS-1}) are invariant under transformations
\begin{eqnarray}
\widehat{\Sigma}_{\mu\nu} & = & \Sigma_{\mu\nu}+(p^{2}-m^{2})\delta\Sigma_{\mu\nu},\nonumber \\
\widehat{V} & = & V-\frac{\hbar}{2}F^{\mu\nu}\delta\Sigma_{\mu\nu},\label{eq:off-shell-trans-1}
\end{eqnarray}
and
\begin{eqnarray}
\widehat{V} & = & V+(p^{2}-m^{2})\delta V,\nonumber \\
\widehat{\Sigma}_{\mu\nu} & = & \Sigma_{\mu\nu}-\hbar F_{\mu\nu}\delta V-\frac{\hbar}{m^{2}}p_{[\mu}F_{\nu]\alpha}p^{\alpha}\delta V.\label{eq:off-shell-trans-2}
\end{eqnarray}
Here $\delta\Sigma_{\mu\nu}$ and $\delta V$ are arbitrary functions
which should be non-singular on the mass-shell $p^{2}=m^{2}$. The
invariance can be easily proven by inserting the transformations into
Eqs. (\ref{eq:order hbar solutions of FPS}) and (\ref{eq:VA from FPS-1}),
and using the property of the Dirac delta-function $-x\delta^{\prime}(x)=\delta(x)$.
Note that the transformation (\ref{eq:off-shell-trans-1}) does not
affect the on-shell value of $\Sigma_{\mu\nu}$ because the factor
$p^{2}-m^{2}$ in front of the additional term vanishes on the mass-shell.
But the transformation (\ref{eq:off-shell-trans-1}) changes the on-shell
value of $V$ by a term $-\frac{\hbar}{2}F^{\mu\nu}\delta\Sigma_{\mu\nu}$.
Similarly, the transformation (\ref{eq:off-shell-trans-2}) does not
change the on-shell value of $V$ but changes the on-shell value of
$\Sigma_{\mu\nu}$.

Since $p^{2}\equiv(p^{0})^{2}-\mathbf{p}^{2}$, we have the following
relation
\begin{equation}
p^{0}=\pm\sqrt{(p^{2}-m^{2})+E_{\mathbf{p}}^{2}},
\end{equation}
where $E_{\mathbf{p}}\equiv\sqrt{m^{2}+\mathbf{p}^{2}}$ is the on-shell
energy. The sign of $p^{0}$ labels fermions ($p^{0}>0$) or anti-fermions
($p^{0}<0$). Now we define $\delta m^{2}\equiv p^{2}-m^{2}$ as a
new parameter, which describes the distance between the mass-shell
and a given $p^{\mu}$. Using the chain rule for computing the derivative
we obtain, for an arbitrary function $f(p^{0},\mathbf{p})$, the Taylor
expansion in $\delta m^{2}$ %
\begin{equation}
f(p^{0},\mathbf{p})=\left.f(p^{0},\mathbf{p})\right|_{p^{0}\rightarrow\pm E_{\mathbf{p}}}+\frac{1}{2}(p^{2}-m^{2})\left.\frac{\partial}{p^{0}\partial p^{0}}f(p^{0},\mathbf{p})\right|_{p^{0}\rightarrow\pm E_{\mathbf{p}}}+\mathcal{O}\left[(\delta m^{2})^{2}\right],\label{eq:near-shell expansion}
\end{equation}
where the first term is the on-shell value of $f(p^{0},\mathbf{p})$
while the second term is related to the on-shell value of $\frac{\partial}{p^{0}\partial p^{0}}f(p^{0},\mathbf{p})$.
Here we make the replacement $p^{0}\rightarrow E_{\mathbf{p}}$ if
we focus on fermions and $p^{0}\rightarrow-E_{\mathbf{p}}$ if we
focus on anti-fermions. Comparing the expansion (\ref{eq:near-shell expansion})
with the transformations (\ref{eq:off-shell-trans-1}) and (\ref{eq:off-shell-trans-2})
we immediately find that in these transformations, $\delta V$ and
$\delta\Sigma_{\mu\nu}$ change the on-shell values of $\frac{\partial}{p^{0}\partial p^{0}}f(p^{0},\mathbf{p})$.
If we take a specific choice as
\begin{equation}
\delta\Sigma_{\mu\nu}=-\left.\frac{\partial}{2p^{0}\partial p^{0}}\Sigma_{\mu\nu}\right|_{p^{0}\rightarrow\pm E_{\mathbf{p}}},\ \ \delta V=-\left.\frac{\partial}{2p^{0}\partial p^{0}}V\right|_{p^{0}\rightarrow\pm E_{\mathbf{p}}},
\end{equation}
then after the transformations (\ref{eq:off-shell-trans-1}) and (\ref{eq:off-shell-trans-2})
we have
\begin{eqnarray}
\widehat{\Sigma}_{\mu\nu} & = & \left.\Sigma_{\mu\nu}\right|_{p^{0}\rightarrow\pm E_{\mathbf{p}}}+\mathcal{O}\left[(p^{2}-m^{2})^{2}\right],\nonumber \\
\widehat{V} & = & \left.V\right|_{p^{0}\rightarrow\pm E_{\mathbf{p}}}+\mathcal{O}\left[(p^{2}-m^{2})^{2}\right].
\end{eqnarray}
All these functions take their on-shell values plus high order corrections
in $p^{2}-m^{2}$. %

If we take the energy integration for the covariant Wigner function
(this is how the equal-time Wigner function is obtained), we find
that the equal-time formula depends on the following terms,
\begin{equation}
\left.V\right|_{p^{0}\rightarrow\pm E_{\mathbf{p}}},\ \ \left.\bar{\Sigma}^{\mu\nu}\right|_{p^{0}\rightarrow\pm E_{\mathbf{p}}},\ \ \left.\frac{\partial}{\partial p^{0}}V\right|_{p^{0}\rightarrow\pm E_{\mathbf{p}}},\ \ \left.\frac{\partial}{\partial p^{0}}\bar{\Sigma}^{\mu\nu}\right|_{p^{0}\rightarrow\pm E_{\mathbf{p}}}.
\end{equation}
The transformations (\ref{eq:off-shell-trans-1}) and (\ref{eq:off-shell-trans-2})
indicate that the above four terms are not independent from each other.
For example, the transformation (\ref{eq:off-shell-trans-1}) changes
$\left.V\right|_{p^{0}\rightarrow\pm E_{\mathbf{p}}}$ and $\left.\frac{\partial}{\partial p^{0}}\bar{\Sigma}^{\mu\nu}\right|_{p^{0}\rightarrow\pm E_{\mathbf{p}}}$
at the same time. Since the covariant Wigner function is invariant
under the transformations (\ref{eq:off-shell-trans-1}) and (\ref{eq:off-shell-trans-2}),
the equal-time Wigner function should only depends on the following
invariant combinations,
\begin{eqnarray}
 &  & \left.V-\frac{\hbar}{2}F^{\mu\nu}\frac{\partial}{2p^{0}\partial p^{0}}\bar{\Sigma}^{\mu\nu}\right|_{p^{0}\rightarrow\pm E_{\mathbf{p}}},\nonumber \\
 &  & \left.\bar{\Sigma}^{\mu\nu}-\hbar\left(F_{\mu\nu}+\frac{1}{m^{2}}p_{[\mu}F_{\nu]\alpha}p^{\alpha}\right)\frac{\partial}{2p^{0}\partial p^{0}}V\right|_{p^{0}\rightarrow\pm E_{\mathbf{p}}}.
\end{eqnarray}
They are identified respectively as the fermion number distribution
and the dipole-moment tensor, which appear in the semi-classical solution
of the equal-time Wigner function in Ref. \cite{Wang:2019moi}.

The $\delta^{\prime}$ terms in the kinetic equations (\ref{eq:massive kinetic equations})
can be dropped if we properly choose a transformation. We take the
first equation in (\ref{eq:massive kinetic equations}) as an example.
For one on-shell $p^{\mu}$, we obtain $p\cdot\nabla^{(0)}V=\mathcal{O}(\hbar)$.
However, if there exists a transformation which satisfies
\begin{equation}
p\cdot\nabla^{(0)}\delta V=-\frac{p\cdot\nabla^{(0)}V-\left.p\cdot\nabla^{(0)}V\right|_{p^{0}\rightarrow\pm E_{\mathbf{p}}}}{p^{2}-m^{2}},
\end{equation}
then we obtain that $p\cdot\nabla^{(0)}\widehat{V}=\mathcal{O}(\hbar)$
holds for any $p^{\mu}$, either on-shell or off-shell. Similarly,
due to the ambiguity of $\delta\Sigma_{\alpha\beta}$ we can also
find a transformation which ensures $p\cdot\nabla^{(0)}\widehat{\bar{\Sigma}}_{\mu\nu}-F_{\ [\mu}^{\alpha}\widehat{\bar{\Sigma}}_{\nu]\alpha}=\mathcal{O}(\hbar)$
hold for any $p^{\mu}$. Note that after the transformations, the
terms with $\delta^{\prime}(p^{2}-m^{2})$ in Eq. (\ref{eq:massive kinetic equations})
are $\mathcal{O}(\hbar^{2})$ and we obtain
\begin{eqnarray}
0 & = & p\cdot\nabla^{(0)}\widehat{V}+\frac{\hbar}{4}(\partial_{x}^{\alpha}F^{\mu\nu})\partial_{p\alpha}\widehat{\Sigma}_{\mu\nu}+\mathcal{O}(\hbar^{2}),\nonumber \\
0 & = & p\cdot\nabla^{(0)}\widehat{\bar{\Sigma}}_{\mu\nu}-F_{\ [\mu}^{\alpha}\widehat{\bar{\Sigma}}_{\nu]\alpha}+\frac{\hbar}{2}(\partial_{x\alpha}F_{\mu\nu})\partial_{p}^{\alpha}\widehat{V}+\mathcal{O}(\hbar^{2}).\label{eq:kin-after-transformation-1}
\end{eqnarray}
These kinetic equations are used for deriving the thermal equilibrium
distribution in the presence of vorticity in Ref. \cite{Weickgenannt:2019dks}.

In practice, the transformations (\ref{eq:off-shell-trans-1}) and
(\ref{eq:off-shell-trans-2}) can be interpreted as a shift of the
mass-shell. We take the scalar component in Eq. (\ref{eq:order hbar solutions of FPS})
as an example. In the solution of the scalar component $\mathcal{F}$,
there is one term which is proportional to $\delta^{\prime}(p^{2}-m^{2})$
which contributes to the off-shell effect \cite{Gao:2012ix,Huang:2018wdl,Gao:2019znl,Weickgenannt:2019dks}.
We now focus on fermions and neglect anti-fermions. Assuming that
the average dipole-moment per particle is $\tilde{\Sigma}^{\mu\nu}$,
we have the following relation
\begin{equation}
\Sigma^{\mu\nu}=\tilde{\Sigma}^{\mu\nu}V,
\end{equation}
because $V$ is the fermion number distribution. Then the scalar component
can be written in terms of a modified on-shell condition
\begin{equation}
\mathcal{F}=m\theta(p^{0})\delta\left(p^{2}-m^{2}-\frac{\hbar}{2}F_{\mu\nu}\tilde{\Sigma}^{\mu\nu}\right)V(x,p)+\mathcal{O}(\hbar^{2}).\label{eq:modified on-shell condition}
\end{equation}
The modified on-shell delta-function is defined via a Taylor expansion,
\begin{equation}
\delta\left(p^{2}-m^{2}-\frac{\hbar}{2}F_{\mu\nu}\tilde{\Sigma}^{\mu\nu}\right)=\delta(p^{2}-m^{2})-\frac{\hbar}{2}F_{\mu\nu}\bar{\Sigma}^{\mu\nu}\delta^{\prime}(p^{2}-m^{2})+\mathcal{O}(\hbar^{2}).
\end{equation}
We find that the normal mass-shell is changed by a spin-magnetic coupling
term. This term can be expanded near $p^{0}=E_{\mathbf{p}}$, and
the term in the delta function can be written as
\begin{eqnarray}
p^{2}-m^{2}-\frac{\hbar}{2}F_{\mu\nu}\tilde{\Sigma}^{\mu\nu} & = & (p^{0})^{2}-E_{\mathbf{p}}^{2}-\frac{\hbar}{2}F_{\mu\nu}\left.\tilde{\Sigma}^{\mu\nu}\right|_{p^{0}\rightarrow E_{\mathbf{p}}}-(p^{0}-E_{\mathbf{p}})\frac{\hbar}{2}F_{\mu\nu}\left.\frac{\partial}{\partial p^{0}}\tilde{\Sigma}^{\mu\nu}\right|_{p^{0}\rightarrow E_{\mathbf{p}}}\nonumber \\
 & = & \left(p^{0}+\delta p^{0}\right)^{2}-\left(E_{\mathbf{p}}+\delta E_{\mathbf{p}}\right)^{2},
\end{eqnarray}
where we have dropped $\mathcal{O}(\hbar^{2})$ terms. The shift of
$p^{0}$ and the shift of $E_{\mathbf{p}}$ are defined as
\begin{eqnarray}
\delta p^{0} & = & -\frac{\hbar}{4}F_{\mu\nu}\left.\frac{\partial}{\partial p^{0}}\tilde{\Sigma}^{\mu\nu}\right|_{p^{0}\rightarrow E_{\mathbf{p}}},\nonumber \\
\delta E_{\mathbf{p}} & = & \frac{\hbar}{4E_{\mathbf{p}}}F_{\mu\nu}\left.\tilde{\Sigma}^{\mu\nu}\right|_{p^{0}\rightarrow E_{\mathbf{p}}}-\frac{\hbar}{4}F_{\mu\nu}\left.\frac{\partial}{\partial p^{0}}\tilde{\Sigma}^{\mu\nu}\right|_{p^{0}\rightarrow E_{\mathbf{p}}}.
\end{eqnarray}
Thus we obtain
\begin{equation}
\theta(p^{0})\delta\left(p^{2}-m^{2}-\frac{\hbar}{2}F_{\mu\nu}\bar{\Sigma}^{\mu\nu}\right)=\frac{1}{2\left(p^{0}+\delta p^{0}\right)}\theta\left(p^{0}-E_{\mathbf{p}}+\delta p^{0}-\delta E_{\mathbf{p}}\right).
\end{equation}
When integrating the scalar component $\mathcal{F}$ over $p^{0}$
we have
\begin{eqnarray}
\int dp^{0}\mathcal{F} & = & m\left.\frac{1}{2\left(p^{0}+\delta p^{0}\right)}V(x,p)\right|_{p^{0}\rightarrow E_{\mathbf{p}}+\delta E_{\mathbf{p}}-\delta p^{0}}\nonumber \\
 & = & m\left.\frac{1}{2p^{0}}V(x,p)\right|_{p^{0}\rightarrow E_{\mathbf{p}}}+m\frac{\hbar}{4E_{\mathbf{p}}}F_{\mu\nu}\left.\frac{\partial}{\partial p^{0}}\left[\frac{1}{2p^{0}}\tilde{\Sigma}^{\mu\nu}V\right]\right|_{p^{0}\rightarrow E_{\mathbf{p}}}.
\end{eqnarray}
Here in the last line, $p^{0}$ takes its on-shell value $E_{\mathbf{p}}$,
thus we can introduce a normal mass-shell delta function
\begin{equation}
\int dp^{0}\mathcal{F}=m\int dp^{0}\theta(p^{0})\delta(p^{2}-m^{2})\left[V+\frac{\hbar}{2}F_{\mu\nu}\frac{\partial}{\partial p^{0}}\left(\frac{1}{2p^{0}}\Sigma^{\mu\nu}\right)\right].\label{eq:int p0 F}
\end{equation}
Note that this result can also be derived by taking the $p^{0}$-integration
for the function $\mathcal{F}$ in Eq. (\ref{eq:order hbar solutions of FPS}),
and integrating by parts for the $\delta^{\prime}(p^{2}-m^{2})$ term.
Making a comparison with the transformation (\ref{eq:off-shell-trans-1}),
we find that if we take
\begin{equation}
\delta\Sigma^{\mu\nu}=-\frac{\partial}{\partial p^{0}}\left(\frac{1}{2p^{0}}\Sigma^{\mu\nu}\right),
\end{equation}
then Eq. (\ref{eq:int p0 F}) can be written in terms of the new distribution
$\widehat{V}$
\begin{equation}
\int dp^{0}\mathcal{F}=m\int dp^{0}\theta(p^{0})\delta(p^{2}-m^{2})\widehat{V}.
\end{equation}
In Eq. (\ref{eq:modified on-shell condition}), the distribution $V$
is on the modified mass-shell while in the above equation the new
distribution $\widehat{V}$ is on the normal mass-shell. Thus we conclude
that the transformations (\ref{eq:off-shell-trans-1}) and (\ref{eq:off-shell-trans-2})
change the mass-shell. We can always find some specific transformation,
after which we can put the coupling between the electromagnetic field
and the dipole moment into the distribution instead of into the mass-shell
delta function.

\subsubsection{Reference-frame dependence}

In the solutions of the massless Wigner function (\ref{sol:massless V and A}),
we have used a reference frame vector, which determines how to separate
the currents into a distribution part and a gradient part. Here we
will focus on the massive case and show how the reference vector can
be introduced. We will also show how to reproduce the massless results
from the massive ones.

First we focus on the tensor component in the solution (\ref{eq:order hbar solutions of FPS}).
Since any anti-symmetric tensor can be decomposed into an electric-like
part and a magnetic-like part, we have the following decomposition,
\begin{equation}
\mathcal{S}_{\mu\nu}=\mathcal{D}_{\mu}u_{\nu}-\mathcal{D}_{\nu}u_{\mu}-\epsilon_{\mu\nu\alpha\beta}u^{\alpha}\mathcal{M}^{\beta},\label{eq:decomposition of S}
\end{equation}
where the electric dipole moment and the magnetic dipole moment are
respectively given by,
\begin{eqnarray}
\mathcal{D}_{\mu} & = & \mathcal{S}_{\mu\nu}u^{\nu}\,=\,m\,\mathbb{P}_{\mu}\delta(p^{2}-m^{2})-m\hbar E_{\mu}V\delta^{\prime}(p^{2}-m^{2})+\mathcal{O}(\hbar^{2}),\nonumber \\
\mathcal{M}_{\mu} & = & -\frac{1}{2}\epsilon_{\mu\nu\alpha\beta}u^{\nu}\mathcal{S}^{\alpha\beta}\,=\,m\,\mathbb{M}_{\mu}\delta(p^{2}-m^{2})-m\hbar B_{\mu}V\delta^{\prime}(p^{2}-m^{2})+\mathcal{O}(\hbar^{2}).
\end{eqnarray}
Here $E_{\mu}\equiv F_{\mu\nu}u^{\nu}$ is the electric-field vector
and $B_{\mu}\equiv\tilde{F}_{\mu\nu}u^{\nu}$ is the magnetic-field
vector. We have defined the on-shell parts $\mathbb{P}_{\mu}$ and
$\mathbb{M}_{\mu}$,
\begin{eqnarray}
\mathbb{P}_{\mu} & = & \Sigma_{\mu\nu}u^{\nu}-\frac{\hbar}{2m}p_{\mu}u^{\nu}\nabla_{\nu}^{(0)}V+\frac{\hbar}{2m^{2}}(p\cdot u)\nabla_{\mu}^{(0)}V,\nonumber \\
\mathbb{M}_{\mu} & = & -\frac{1}{2}\epsilon_{\mu\nu\alpha\beta}u^{\nu}\Sigma^{\alpha\beta}+\frac{\hbar}{2m^{2}}\epsilon_{\mu\nu\alpha\beta}u^{\nu}p^{\alpha}\nabla^{(0)\beta}V,
\end{eqnarray}
The on-shell dipole-moment tensor $\Sigma_{\mu\nu}$ can be reproduced
via
\begin{equation}
\Sigma_{\mu\nu}=\mathbb{P}_{\mu}u_{\nu}-\mathbb{P}_{\nu}u_{\mu}+\epsilon_{\mu\nu\alpha\beta}u^{\alpha}\mathbb{M}^{\beta}+\frac{\hbar}{2m^{2}}p_{[\mu}\nabla_{\nu]}^{(0)}V.
\end{equation}
Due to the constraint equation (\ref{eq:constraint equation for Sigma}),
one obtains the relation between $\mathbb{P}_{\mu}$ and $\mathbb{M}_{\mu}$,
\begin{equation}
\delta(p^{2}-m^{2})\left[(p\cdot u)\mathbb{P}_{\mu}-\epsilon_{\mu\nu\alpha\beta}p^{\nu}u^{\alpha}\mathbb{M}^{\beta}-\frac{\hbar}{2}(g_{\mu\nu}-u_{\mu}u_{\nu})\nabla^{(0)\nu}V\right]=\mathcal{O}(\hbar^{2}),
\end{equation}
where we have used the dynamical equation $\delta(p^{2}-m^{2})p^{\nu}\nabla_{\nu}^{(0)}V=\mathcal{O}(\hbar)$.
From the above constraint, one can express $\mathbb{P}_{\mu}$ in
terms of $\mathbb{M}_{\mu}$,
\begin{equation}
\delta(p^{2}-m^{2})\mathbb{P}_{\mu}=\delta(p^{2}-m^{2})\left[\frac{1}{p\cdot u}\epsilon_{\mu\nu\alpha\beta}p^{\nu}u^{\alpha}\mathbb{M}^{\beta}+\frac{\hbar}{2(p\cdot u)}(g_{\mu\nu}-u_{\mu}u_{\nu})\nabla^{(0)\nu}V\right].
\end{equation}
Inserting back into the decomposition (\ref{eq:decomposition of S}),
we obtain
\begin{equation}
\mathcal{S}_{\mu\nu}=m\delta(p^{2}-m^{2})\left[-\frac{1}{p\cdot u}\epsilon_{\mu\nu\alpha\beta}p^{\alpha}\mathbb{M}^{\beta}-\frac{\hbar}{2(p\cdot u)}u_{[\mu}\nabla_{\nu]}^{(0)}V\right]-m\hbar F_{\mu\nu}V\delta^{\prime}(p^{2}-m^{2}).
\end{equation}
This indicates that $\mathcal{S}_{\mu\nu}$ can be written in terms
of the magnetic dipole-moment vector $\mathbb{M}_{\mu}$, which is
defined in frame $u^{\mu}$. Meanwhile, we can express the axial-vector
component of the Wigner function,
\begin{equation}
\mathcal{A}^{\mu}=\frac{1}{p\cdot u}\left[m^{2}\mathbb{M}^{\mu}-(p\cdot\mathbb{M})p^{\mu}\right]\delta(p^{2}-m^{2})+\frac{\hbar}{2(p\cdot u)}\epsilon^{\mu\nu\alpha\beta}p_{\nu}u_{\alpha}\nabla_{\beta}^{(0)}V+\hbar\tilde{F}_{\mu\nu}p^{\nu}V\delta^{\prime}(p^{2}-m^{2}).\label{eq:axial-vector as function of M}
\end{equation}
In the massless limit, we define
\begin{equation}
A\equiv-\lim_{m\rightarrow0}\frac{p\cdot\mathbb{M}}{p\cdot u}.
\end{equation}
Then $\mathcal{A}^{\mu}$ in Eq. (\ref{eq:axial-vector as function of M})
correctly reproduces the massless one in Eq. (\ref{sol:massless V and A}).

On the other hand, the polarization vector $n^{\mu}$ can be projected
into the direction of $p^{\mu}$ and an arbitrary frame vector $u^{\mu}$.
Since $p^{\mu}$ has the unit of energy, we write down the following
formula
\begin{equation}
m\,n^{\mu}=c_{\parallel}\left(p^{\mu}-\frac{m^{2}}{p\cdot u}u^{\mu}\right)+c_{\perp}m\,n_{\perp}^{\mu}.\label{eq:decomposition of polarization}
\end{equation}
Here the coefficient of $u^{\mu}$ is proportional to the coefficient
of $p^{\mu}$ in order to satisfy the constraint equation $p\cdot n\,\delta(p^{2}-m^{2})=0$.
The vector $n_{\perp}^{\mu}$ is assumed to be a normalized space-like
vector $n_{\perp}^{\mu}n_{\perp\mu}=-1$, which is perpendicular to
both $u^{\mu}$ and $p^{\mu}$. If we observe in the frame $u^{\mu}$,
the first term in Eq. (\ref{eq:decomposition of polarization}) would
be parallel to the momentum while the second term is perpendicular.
In this way the polarization vector is decomposed into one longitudinally
polarized part and one transversely polarized part in reference to
the particle's momentum direction. We now consider a special case
where
\begin{equation}
c_{\parallel}=A\neq0,\ c_{\perp}m\,n_{\perp}^{\mu}=\frac{\hbar}{2(p\cdot u)}\epsilon^{\mu\nu\alpha\beta}p_{\alpha}u_{\beta}\nabla_{\nu}^{(0)}V,\label{eq:specific choice of nmu}
\end{equation}
Inserting back into the Wigner function in (\ref{eq:massive solutions VA}),
we obtain the axial-vector current
\begin{equation}
\mathcal{A}^{\mu}=\delta(p^{2}-m^{2})\left[\left(p^{\mu}-\frac{m^{2}}{p\cdot u}u^{\mu}\right)A+\frac{\hbar}{2(p\cdot u)}\epsilon^{\mu\nu\alpha\beta}p_{\alpha}u_{\beta}\nabla_{\nu}^{(0)}V\right]+\hbar\tilde{F}^{\mu\nu}p_{\nu}V\delta^{\prime}(p^{2}-m^{2})+\mathcal{O}(\hbar^{2}),\label{eq:specific solution of A^=00005Cmu}
\end{equation}
When taking the massless limit $m\rightarrow0$, $\mathcal{A}^{\mu}$
in Eq. (\ref{eq:specific solution of A^=00005Cmu}) smoothly reproduces
the massless result in Eq. (\ref{sol:massless V and A}). Here we
can identify the transverse part $c_{\perp}m\,n_{\perp}^{\mu}$ as
the contribution from the side-jump effect \cite{Chen:2015gta,Hidaka:2016yjf,Huang:2018wdl}.

The dipole-moment tensor $\Sigma^{\mu\nu}$ can be expressed in terms
of $n^{\mu}$, as Eq. (\ref{eq:relation between Sigma and n}) shows.
Inserting $n^{\mu}$ and $\Sigma^{\mu\nu}$ into the vector $\mathcal{V}^{\mu}$
in Eq. (\ref{eq:massive solutions VA}), we obtain %
{}
\begin{eqnarray}
\mathcal{V}^{\mu} & = & \delta(p^{2}-m^{2})\left(p^{\mu}V+\frac{\hbar}{2m}\epsilon^{\mu\nu\alpha\beta}p_{\nu}\nabla_{\alpha}^{(0)}n_{\beta}\right)\nonumber \\
 &  & +\hbar\left(m\tilde{F}_{\mu\nu}n^{\nu}-\frac{p\cdot n}{m}\tilde{F}_{\mu\nu}p^{\nu}\right)\delta^{\prime}(p^{2}-m^{2})+\mathcal{O}(\hbar^{2}),
\end{eqnarray}
Note that here $p\cdot n\,\delta^{\prime}(p^{2}-m^{2})$ does not
have to vanish. In this formula we want to introduce a reference frame
for the second term. With the help of the Schouten identity (\ref{eq:Schouten identity-1})
and the kinetic equation for $n^{\mu}$ in Eq. (\ref{eq:kinetic equation for n_mu})
we can prove%
{}
\begin{eqnarray}
\delta(p^{2}-m^{2})\frac{\hbar}{2m}\epsilon_{\mu\nu\alpha\beta}p^{\nu}\nabla^{(0)\alpha}n^{\beta} & = & -\delta(p^{2}-m^{2})\frac{\hbar}{2m(p\cdot u)}p_{\mu}\epsilon_{\nu\alpha\beta\gamma}u^{\gamma}p^{\nu}\nabla^{(0)\alpha}n^{\beta}\nonumber \\
 &  & +\delta(p^{2}-m^{2})\frac{\hbar}{2(p\cdot u)}\epsilon_{\mu\nu\alpha\beta}u^{\nu}\nabla^{(0)\alpha}mn^{\beta}\nonumber \\
 &  & +\delta(p^{2}-m^{2})\frac{\hbar}{2m(p\cdot u)}\epsilon_{\mu\nu\alpha\beta}u^{\nu}p^{\alpha}\nabla^{(0)\beta}(p\cdot n).
\end{eqnarray}
We furthermore replace $n^{\mu}$ by Eqs. (\ref{eq:decomposition of polarization})
and (\ref{eq:specific choice of nmu}). Up to leading order in $\hbar$
we have $p\cdot n=\frac{(p^{2}-m^{2})}{m}A+\mathcal{O}(\hbar)$. Substituting
$n^{\mu}$ into the above equations, we obtain %
\begin{eqnarray}
\mathcal{V}^{\mu} & = & \delta(p^{2}-m^{2})p^{\mu}\left[V+\frac{\hbar}{2(p\cdot u)}\epsilon_{\nu\alpha\beta\gamma}p^{\nu}u^{\alpha}\nabla^{(0)\beta}\left(\frac{u^{\gamma}}{u\cdot p}A\right)\right]\nonumber \\
 &  & +\delta(p^{2}-m^{2})\frac{\hbar}{2(p\cdot u)}\epsilon^{\mu\nu\alpha\beta}p_{\nu}u_{\alpha}\nabla_{\beta}^{(0)}A+\hbar\tilde{F}^{\mu\nu}p_{\nu}A\delta^{\prime}(p^{2}-m^{2})\nonumber \\
 &  & -m^{2}\left[\delta(p^{2}-m^{2})\frac{\hbar}{2(p\cdot u)}\epsilon^{\mu\nu\alpha\beta}u_{\nu}\nabla_{\alpha}^{(0)}\left(\frac{u_{\beta}}{p\cdot u}A\right)+\delta^{\prime}(p^{2}-m^{2})\hbar\tilde{F}^{\mu\nu}\frac{u_{\nu}}{p\cdot u}A\right]+\mathcal{O}(\hbar^{2}),\nonumber \\
\label{eq:specific solution of V^=00005Cmu}
\end{eqnarray}
Taking the massless limit $m\rightarrow0$, and redefining the distribution
function $V$, Eq. (\ref{eq:specific solution of V^=00005Cmu}) agrees
with the massless results in (\ref{sol:massless V and A}).

In this subsection, we have found the relation between massive and
massless solutions. We emphasize that the polarization of massive
particles has three degrees of freedom. That is because in their rest
frames, their spins can point to any spatial directions. Thus in the
massive case we have four functions to describe the system, three
of which describe the polarization and the remaining one is the particle
distribution. However, in the massless case, particles are either
LH or RH and their spins are either parallel or anti-parallel to their
momenta. Thus for massless fermions the net fermion number distribution
and the axial-charge distribution are sufficient to describe the system.
In this section we decompose the polarization into a longitudinal
part and a transverse part. Since massless particles cannot be transversely
polarized, we presume that the transverse part only comes from the
side-jump effect. With this assumption, we find that the massive solutions
smoothly reproduce the massless ones. In heavy ion collisions, the
$u$, $d$ quarks can be treated as massless, because their current
masses are much smaller than the typical temperature of the QGP. However,
the current mass of the $s$ quark is about $95$ MeV, which is comparable
with the chemical freeze-out temperature ($\sim160$ MeV). Thus the
massless chiral kinetic theory is not sufficient for describing all
flavors. The kinetic theory for massive particles with spin was developed
many years ago \cite{Israel:1978up} and later reproduced via the
Wigner function approach. However, up to now there is no systematic
tool for the spin-evolution of massive fermions in heavy-ion collisions.
Our study in this section would be a starting point for future works
on the dynamical spin evolution. Comparing with the classical Boltzmann
equation and the BMT equation, we have obtained $\hbar$-order corrections,
which are attributed to the coupling between the spin and the electromagnetic
fields. In the future, we will use the method of moments to deal with
the kinetic equations and derive spin-hydrodynamics.

\newpage{}
$\ $
\newpage{}

\section{Physical quantities\label{sec:Physical quantities}}

In this section we will consider systems in thermal equilibrium. The
Wigner function for free fermions without chiral imbalance is computed
at the leading order in $\hbar$ in subsection \ref{subsec:Free-fermions}.
Higher-order terms in $\hbar$ have been obtained by the semi-classical
expansion in Eqs. (\ref{eq:order hbar solutions of FPS}) and (\ref{eq:VA from FPS-1})
in Sec. \ref{sec:Semi-classical-expansion}. On the other hand, the
case of fermions with chiral imbalance is analytically solved in subsection
\ref{subsec:Free-with-chiral}, which is called the chiral quantization
in this thesis. The chiral chemical potential is not well-defined
in the massive case, thus in the chiral quantization, $\mu_{5}$ is
treated as a variable for an additional self-energy correction term.
However, the chiral quantization can only deal with constant $\mu_{5}$.
In this section, through a comparison between semi-classical results
and chiral-quantization results, we will give a reasonable estimate
about the parameter region in which the semi-classical results are
applicable. In a constant magnetic field a thermal equilibrium can
also exist, which is then compared with the free-fermion case in this
section. The pair production in the presence of an electric field
is a dynamical problem, which is analytically solved in subsection
\ref{subsec:Fermions-in-electric}. In this section we will numerically
calculate the pair-production rates and display the thermal suppression
to the production rates.

\subsection{Physical quantities from quantum field theory\label{subsec:Physical-quantities}}

Throughout this thesis, we focus on spin-1/2 particles in electromagnetic
fields. Since the electromagnetic interaction is a long-range interaction,
short-range interactions such as the strong and weak interactions
will be neglected. In this subsection we will start from the QED Lagrangian
and derive some basic physical quantities. Then we will find a straightforward
relation between the Wigner function and these quantities. In subsections
\ref{subsec:Net-fermion-current} and \ref{subsec:Energy-momentum-tensor-and}
we will calculate these quantities using the Wigner function in thermal
equilibrium.

As an Abelian $U(1)$ gauge theory, the QED Lagrangian for a Dirac
spinor field in an electromagnetic field is given by
\begin{equation}
\hat{\mathcal{L}}=\frac{1}{2}\left[\hat{\bar{\psi}}\gamma^{\mu}i\partial_{x\mu}\hat{\psi}-\left(i\partial_{x\mu}\hat{\bar{\psi}}\right)\gamma^{\mu}\hat{\psi}\right]\bigg)-\hat{\bar{\psi}}(m+\gamma\cdot\mathbb{A})\hat{\psi}-\frac{1}{4}F^{\mu\nu}F_{\mu\nu},\label{eq:QED Lagrangian}
\end{equation}
where $\hat{\psi}$ is the spinor field operator, $\mathbb{A}^{\mu}$
is the gauge potential and $F^{\mu\nu}=\partial_{x}^{\mu}\mathbb{A}^{\nu}-\partial_{x}^{\nu}\mathbb{A}^{\mu}$
is the field-strength tensor, whose $0i$ components represent the
electric fields and $ij$ components represent the magnetic fields.
The Lagrangian is invariant under the following local gauge transformation
\begin{eqnarray}
\mathbb{A}^{\mu}(x) & \rightarrow & \mathbb{A}^{\mu}(x)-\partial_{x}^{\mu}\theta(x),\nonumber \\
\psi(x) & \rightarrow & e^{i\theta(x)}\psi(x),
\end{eqnarray}
where the gauge potential has an additional derivative term while
the Dirac field has a phase rotation. The gauge invariance indicates
an ambiguity of the gauge potential. In practice we can fix the gauge
by e.g. taking the Lorenz-gauge condition $\partial_{x}^{\mu}\mathbb{A}_{\mu}=0$,
or the temporal gauge $\mathbb{A}^{0}=0$, etc.. From the Lagrangian
we obtain the Euler-Lagrangian equations for $\hat{\psi}$, $\hat{\bar{\psi}}$
and $\mathbb{A}^{\mu}$, which are the Dirac equation, the conjugate
of the Dirac equation and the inhomogeneous Maxwell equation, respectively.
The Dirac equation and its conjugate are shown in Eq. (\ref{eq:Dirac equation and conjugate})
and the Maxwell equation reads
\begin{equation}
\partial_{x\mu}F^{\mu\nu}=\hat{\bar{\psi}}\gamma^{\nu}\hat{\psi}.
\end{equation}

According to the Noether's theorem, each continuous symmetry of the
Lagrangian corresponds to a conserved current. The gauge symmetry
is associated with an electric current,
\begin{equation}
\hat{\mathbb{N}}^{\mu}=\hat{\bar{\psi}}\gamma^{\mu}\hat{\psi}.
\end{equation}
Here we use hat to distinguish the operator from physical quantities.
The conserved currents corresponding to translations and Lorentz transformations
are the energy-momentum tensor and the angular-momentum tensor, respectively,
\begin{eqnarray}
\hat{\mathbb{T}}^{\mu\nu} & = & \frac{\partial\hat{\mathcal{L}}}{\partial(\partial_{x\mu}\hat{\psi})}\partial_{x}^{\nu}\hat{\psi}+\partial_{x}^{\nu}\hat{\bar{\psi}}\frac{\partial\hat{\mathcal{L}}}{\partial(\partial_{x\mu}\hat{\bar{\psi}})}+\frac{\partial\mathcal{\hat{\mathcal{L}}}}{\partial(\partial_{x\mu}\mathbb{A}_{\rho})}\partial_{x}^{\nu}\mathbb{A}_{\rho}-g^{\mu\nu}\hat{\mathcal{L}},\nonumber \\
\hat{\mathbb{M}}^{\rho\mu\nu} & = & x^{\mu}\hat{\mathbb{T}}^{\rho\nu}-x^{\nu}\hat{\mathbb{T}}^{\rho\mu}-i\frac{\partial\hat{\mathcal{L}}}{\partial(\partial_{x\rho}\hat{\psi})}\mathcal{S}^{\mu\nu}\hat{\psi}+i\hat{\bar{\psi}}\mathcal{S}^{\mu\nu}\frac{\partial\hat{\mathcal{L}}}{\partial(\partial_{x\rho}\hat{\bar{\psi}})}-i\frac{\partial\hat{\mathcal{L}}}{\partial(\partial_{x\rho}\mathbb{A}_{\alpha})}(\mathcal{J}^{\mu\nu})_{\alpha\beta}\mathbb{A}^{\beta},\label{eq:definition of Tmunu}
\end{eqnarray}
where generators of the spin are $\mathcal{S}^{\mu\nu}\equiv\frac{i}{4}[\gamma^{\mu},\gamma^{\nu}]=\frac{1}{2}\sigma^{\mu\nu}$
and $(\mathcal{J}^{\mu\nu})_{\alpha\beta}=i(\delta_{\alpha}^{\mu}\delta_{\beta}^{\nu}-\delta_{\beta}^{\mu}\delta_{\alpha}^{\nu})$.
Inserting the Lagrangian into the definition of these currents we
obtain %
\begin{eqnarray}
\hat{\mathbb{T}}^{\mu\nu} & = & \hat{\mathbb{T}}_{\text{mat}}^{\mu\nu}+\mathbb{A}^{\nu}\hat{\mathbb{N}}^{\mu}+\hat{\mathbb{T}}_{\text{field}}^{\mu\nu},\nonumber \\
\hat{\mathbb{M}}^{\rho,\mu\nu} & = & \hat{\mathbb{L}}^{\rho,\mu\nu}+\hat{\mathbb{S}}_{\text{mat}}^{\rho,\mu\nu}+\hat{\mathbb{S}}_{\text{field}}^{\rho,\mu\nu}.\label{eq:separation of Tmunu and Mrhomunu}
\end{eqnarray}
Here we have separated the total energy-momentum tensor and the total
angular-momentum tensor into several parts. The total orbital angular-momentum
tensor is given by
\begin{equation}
\hat{\mathbb{L}}^{\rho,\mu\nu}\equiv x^{\mu}\hat{\mathbb{T}}^{\rho\nu}-x^{\nu}\hat{\mathbb{T}}^{\rho\mu}.
\end{equation}
The matter parts of the energy-momentum tensor and the spin-angular-momentum
tensor are
\begin{eqnarray}
\hat{\mathbb{T}}_{\text{mat}}^{\mu\nu} & = & \frac{1}{2}\left[i\hat{\bar{\psi}}\gamma^{\mu}(\overrightarrow{\partial}_{x}^{\nu}+i\mathbb{A}^{\nu})\hat{\psi}-i\hat{\bar{\psi}}\gamma^{\mu}(\overleftarrow{\partial}_{x}^{\nu}-i\mathbb{A}^{\nu})\hat{\psi}\right],\nonumber \\
\hat{\mathbb{S}}_{\text{mat}}^{\rho,\mu\nu} & = & \frac{1}{4}\hat{\bar{\psi}}\{\gamma^{\rho},\sigma^{\mu\nu}\}\hat{\psi},\label{eq:cannonical T and S}
\end{eqnarray}
while the field parts are
\begin{eqnarray}
\hat{\mathbb{T}}_{\text{field}}^{\mu\nu} & = & \frac{1}{4}g^{\mu\nu}F^{\rho\sigma}F_{\rho\sigma}-F^{\mu\rho}\partial^{\nu}\mathbb{A}_{\rho},\nonumber \\
\hat{\mathbb{S}}_{\text{field}}^{\rho,\mu\nu} & = & -(F^{\rho\mu}\mathbb{A}^{\nu}-F^{\rho\nu}\mathbb{A}^{\mu}).
\end{eqnarray}
Note that after such a decomposition, the matter parts are gauge-invariant
while the remaining parts are not. The gauge dependence comes from
the definitions (\ref{eq:definition of Tmunu}), where derivative
operators are ordinary ones instead of covariant ones. Taking expectation
values of the above operators on one specific system $|\Omega\rangle$,
we can obtain the fermion number current $\mathbb{N}^{\mu}$, the
matter part of the energy-momentum tensor $\mathbb{T}_{\text{mat}}^{\mu\nu}$,
and the spin angular-momentum tensor $\mathbb{S}_{\text{mat}}^{\rho,\mu\nu}$,
\begin{eqnarray}
\mathbb{N}^{\mu}=\langle\Omega|\hat{\mathbb{N}}^{\mu}|\Omega\rangle, & \mathbb{T}_{\text{mat}}^{\mu\nu}=\langle\Omega|\hat{\mathbb{T}}_{\text{mat}}^{\mu\nu}|\Omega\rangle, & \mathbb{S}_{\text{mat}}^{\rho,\mu\nu}=\langle\Omega|\hat{\mathbb{S}}_{\text{mat}}^{\rho,\mu\nu}|\Omega\rangle.
\end{eqnarray}

The Noether currents are conserved, thus the fluid-dynamical quantities
satisfy the following conservation laws automatically,
\begin{eqnarray}
\partial_{x\mu}\mathbb{N}^{\mu}=0, & \partial_{x\mu}\mathbb{T}^{\mu\nu}=0, & \partial_{x\rho}\mathbb{M}^{\rho,\mu\nu}=0.\label{eq:conservation equations}
\end{eqnarray}
But note that $\mathbb{T}^{\mu\nu}$ and $\mathbb{M}^{\rho,\mu\nu}$
are separated into several parts as in Eq. (\ref{eq:separation of Tmunu and Mrhomunu}),
hence the matter parts of $\mathbb{T}_{\text{mat}}^{\mu\nu}$ and
$\mathbb{S}_{\text{mat}}^{\rho,\mu\nu}$ are not conserved themselves,
\begin{eqnarray}
\partial_{x\mu}\mathbb{T}_{\text{mat}}^{\mu\nu} & = & F^{\nu\alpha}\mathbb{N}_{\alpha},\nonumber \\
\partial_{x\rho}\mathbb{S}_{\text{mat}}^{\rho,\mu\nu} & = & -\mathbb{T}_{\text{mat}}^{\mu\nu}+\mathbb{T}_{\text{mat}}^{\nu\mu}.\label{eq:canonical-equations}
\end{eqnarray}
From the second line we observe that the anti-symmetric part of $\mathbb{T}_{\text{mat}}^{\mu\nu}$
is related to the derivative of the spin tensor. For a classical particle,
whose the spin degrees of freedom are ignored, the spin tensor vanishes,
which render a symmetric $\mathbb{T}_{\text{mat}}^{\mu\nu}$. But
in general the canonical energy-momentum tensor is not symmetric for
a system with spin.

The above tensors are not uniquely defined. The Lagrangian in Eq.
(\ref{eq:QED Lagrangian}) can have an additional term, like $\partial_{x\mu}\delta\mathcal{L}^{\mu}$
with an arbitrary $\delta\mathcal{L}^{\mu}$. When taking integration
over the whole space, this additional term gives a boundary term,
which in general is neglected. However, changing the definition of
the Lagrangian leads to a different energy-momentum tensor and a different
spin-angular-momentum tensor. All these different definitions are
exactly equivalent since they are related to the canonical form by
so-called pseudo-gauge transformations \cite{Hehl:1976vr,Becattini:2018duy}
\begin{eqnarray}
\mathbb{T}_{\text{mat}}^{\prime\mu\nu} & = & \mathbb{T}_{\text{mat}}^{\mu\nu}+\frac{1}{2}\partial_{\rho}(\mathbb{F}^{\rho,\mu\nu}+\mathbb{F}^{\mu,\nu\rho}+\mathbb{F}^{\nu,\mu\rho}),\nonumber \\
\mathbb{S}_{\text{mat}}^{\prime\rho,\mu\nu} & = & \mathbb{S}_{\text{mat}}^{\rho,\mu\nu}-\mathbb{F}^{\rho,\mu\nu},
\end{eqnarray}
where $\mathbb{F}^{\rho,\mu\nu}$ is an arbitrary tensor which is
anti-symmetric under $\mu\leftrightarrow\nu$. We can check that the
newly defined quantities still satisfy the conservation equations
(\ref{eq:canonical-equations}) %
. A specific choice for $\mathbb{F}^{\rho,\mu\nu}$ is $\mathbb{S}_{\text{mat}}^{\rho,\mu\nu}$,
which makes $\mathbb{S}_{\text{mat}}^{\prime\rho,\mu\nu}$ vanishes.
The new energy-momentum tensor is the Belinfante one,
\begin{equation}
\mathbb{T}_{\text{Bel}}^{\mu\nu}\equiv\mathbb{T}_{\text{mat}}^{\mu\nu}+\frac{1}{2}\partial_{\rho}(\mathbb{S}_{\text{mat}}^{\rho,\mu\nu}+\mathbb{S}_{\text{mat}}^{\mu,\nu\rho}+\mathbb{S}_{\text{mat}}^{\nu,\mu\rho}).
\end{equation}
It is easy to check that the Belinfante energy-momentum tensor is
the symmetric part of the canonical one,
\begin{equation}
\mathbb{T}_{\text{Bel}}^{\mu\nu}=\frac{1}{2}\left(\mathbb{T}_{\text{mat}}^{\mu\nu}+\mathbb{T}_{\text{mat}}^{\nu\mu}\right).
\end{equation}

Comparing the definition of the Wigner function in Eq. (\ref{def:Wigner function})
with the above fluid-dynamical quantities, we obtain the following
relations
\begin{eqnarray}
\mathbb{N}^{\mu}(x) & = & \int d^{4}p\ \mathcal{V}^{\mu}(x,p),\nonumber \\
\mathbb{T}_{\text{mat}}^{\mu\nu}(x) & = & \int d^{4}p\ p^{\nu}\mathcal{V}^{\mu}(x,p),\nonumber \\
\mathbb{S}_{\text{mat}}^{\rho,\mu\nu}(x) & = & -\frac{1}{2}\epsilon^{\rho\mu\nu\alpha}\int d^{4}p\ \mathcal{A}_{\alpha}(x,p),\label{eq:cannonical quantities}
\end{eqnarray}
where $\mathcal{V}^{\mu}(x,p)$ and $\mathcal{A}^{\mu}(x,p)$ are
the vector and axial-vector components of the Wigner function. On
the other hand, the Belinfante energy-momentum tensor is given by
\begin{equation}
\mathbb{T}_{\text{Bel}}^{\mu\nu}(x)=\frac{1}{2}\int d^{4}p\ \left[p^{\mu}\mathcal{V}^{\nu}(x,p)+p^{\nu}\mathcal{V}^{\mu}(x,p)\right].
\end{equation}
In the remaining part of this section, we will specify the Wigner
function and calculate the canonical quantities in Eq. (\ref{eq:cannonical quantities}).

\subsection{Thermal Equilibrium \label{subsec:Thermal-equilibrium}}

In Sec. \ref{sec:Analytically-solvable-cases} we have derived the
analytical solutions of the Wigner function. Note that in these solutions,
the distributions for fermions and anti-fermions are still undetermined.
In this subsection we will consider systems in thermal equilibrium
and give the equilibrium distributions. First we consider massless
particles at a given temperature $T$, chemical potential $\mu$,
and chiral chemical potential $\mu_{5}$. The helicity of a massless
particle is a conserved quantity, thus $\mu_{5}$ is well-defined.
The canonical partition function for such a system is given by
\begin{equation}
\hat{Z}=\exp[-\beta(\hat{H}_{0}-\mu\hat{N}-\mu_{5}\hat{N}_{5})],\label{eq:old partition function-1}
\end{equation}
where the Hamiltonian operator $\hat{H}_{0}$, the fermion number
operator $\hat{N}$, and the axial-charge number operator $\hat{N}_{5}$
are
\begin{eqnarray}
\hat{H}_{0} & = & \int d^{3}\mathbf{x}\hat{\psi}^{\dagger}(t,\mathbf{x})\left(-i\gamma^{0}\boldsymbol{\gamma}\cdot\boldsymbol{\partial}_{\mathbf{x}}+m\gamma^{0}\right)\hat{\psi}(t,\mathbf{x}),\nonumber \\
\hat{N} & = & \int d^{3}\mathbf{x}\hat{\psi}^{\dagger}(t,\mathbf{x})\hat{\psi}(t,\mathbf{x}),\nonumber \\
\hat{N}_{5} & = & \int d^{3}\mathbf{x}\hat{\psi}^{\dagger}(t,\mathbf{x})\gamma^{5}\hat{\psi}(t,\mathbf{x}).\label{eq:operators for equilibrium}
\end{eqnarray}
Using the quantized field operator in Eq. (\ref{def:quantized free field})
and the single-particle wavefunctions in Eq. (\ref{sol:free wave functions})
and (\ref{eq:product of Sqrt p=00005Csigma}), we derive %
\begin{eqnarray}
\hat{H}_{0} & = & \sum_{s}\int\frac{d^{3}\mathbf{p}}{(2\pi)^{3}}\left|\mathbf{p}\right|\left(\hat{a}_{\mathbf{p},s}^{\dagger}\hat{a}_{\mathbf{p},s}-\hat{b}_{\mathbf{p},s}\hat{b}_{\mathbf{p},s}^{\dagger}\right),\nonumber \\
\hat{N} & = & \sum_{s}\int\frac{d^{3}\mathbf{p}}{(2\pi)^{3}}\left(\hat{a}_{\mathbf{p},s}^{\dagger}\hat{a}_{\mathbf{p},s}+\hat{b}_{\mathbf{p},s}\hat{b}_{\mathbf{p},s}^{\dagger}\right),\nonumber \\
\hat{N}_{5} & = & \sum_{ss_{1}}\int\frac{d^{3}\mathbf{p}}{(2\pi)^{3}}\frac{\xi_{s}^{\dagger}(\boldsymbol{\sigma}\cdot\mathbf{p})\xi_{s_{1}}}{\left|\mathbf{p}\right|}\left(\hat{a}_{\mathbf{p},s}^{\dagger}\hat{a}_{\mathbf{p},s_{1}}+\hat{b}_{\mathbf{p},s}\hat{b}_{\mathbf{p},s_{1}}^{\dagger}\right),\label{eq:cannonical operators HNN_5}
\end{eqnarray}
where we have dropped terms $\hat{a}_{\mathbf{p},s}^{\dagger}\hat{b}_{-\mathbf{p},s}^{\dagger}$
and $\hat{b}_{\mathbf{p},s}\hat{a}_{-\mathbf{p},s}$. In general,
the term $\xi_{s}^{\dagger}(\boldsymbol{\sigma}\cdot\mathbf{p})\xi_{s_{1}}$
is not diagonalizable in spin space, which means that the physical
states are superposition of states with $s=+$ and states with $s=-$,
and then the thermal expectation value of $a_{\mathbf{p},+}^{\dagger}a_{\mathbf{p},-}$
will be non-zero. We can use the method in subsection \ref{subsec:Free-fermions}
to diagonalize the distribution. If we choose the spin quantization
direction as the direction of $\mathbf{p}$, which fulfills,
\begin{equation}
(\boldsymbol{\sigma}\cdot\mathbf{p})\xi_{s}=s\left|\mathbf{p}\right|\xi_{s},\label{eq:specific choice of spinors}
\end{equation}
then the operator $\hat{N}_{5}$ is diagonalized in spin space,
\begin{equation}
\hat{N}_{5}=\sum_{s}s\int\frac{d^{3}\mathbf{p}}{(2\pi)^{3}}\left(\hat{a}_{\mathbf{p},s}^{\dagger}\hat{a}_{\mathbf{p},s}+\hat{b}_{\mathbf{p},s}\hat{b}_{\mathbf{p},s}^{\dagger}\right).\label{eq:cannonical operator N_5}
\end{equation}
Inserting the quantized operators (\ref{eq:cannonical operators HNN_5}),
(\ref{eq:cannonical operator N_5}) into the canonical partition function,
we obtain
\begin{equation}
\hat{Z}=\exp\left\{ -\beta\sum_{s}\int\frac{d^{3}\mathbf{p}}{(2\pi)^{3}}\left[\left(\left|\mathbf{p}\right|-\mu-s\mu_{5}\right)\hat{a}_{\mathbf{p},s}^{\dagger}\hat{a}_{\mathbf{p},s}-\left(\left|\mathbf{p}\right|+\mu+s\mu_{5}\right)\hat{b}_{\mathbf{p},s}\hat{b}_{\mathbf{p},s}^{\dagger}\right]\right\} .
\end{equation}
This result coincides with our knowledge for the massless case: for
massless particles, the operators $\hat{N}$ and $\hat{N}_{5}$ commute
with the Hamiltonian $\hat{H}$, so the basis states can be chosen
as common eigenstates of the operators $\hat{H}$, $\hat{N}$, and
$\hat{N}_{5}$. The canonical partition function is then diagonalized.
The expectation value of any operator $\hat{O}$ is computed via
\begin{equation}
\left\langle \hat{O}\right\rangle =\frac{\mathrm{Tr\ }(\hat{O}\hat{Z})}{\mathrm{Tr\ }\hat{Z}}.\label{eq:expectation value of O-1}
\end{equation}
Here $\mathrm{Tr}$ runs over all possible quantum states. Taking
the expectation values of the fermion number operator $\hat{a}_{\mathbf{p},s}^{\dagger}\hat{a}_{\mathbf{p},s^{\prime}}$
and anti-fermion number operator $\hat{b}_{\mathbf{p},s}^{\dagger}\hat{b}_{\mathbf{p},s^{\prime}}$
we obtain the Fermi-Dirac distributions,
\begin{equation}
f_{ss^{\prime}}^{(\pm)}(x,\mathbf{p})=\frac{1}{1+\exp\left[\beta\left(\left|\mathbf{p}\right|\mp\mu\mp s\mu_{5}\right)\right]}\delta_{ss^{\prime}},\label{eq:Fermi distribution for fermion}
\end{equation}
which coincide with the equilibrium distributions for chiral particles.
Since the spinors satisfy Eq. (\ref{eq:specific choice of spinors}),
we can calculate
\begin{equation}
\xi_{s}^{\dagger}\hat{n}^{\mu}(\mathbf{p})\xi_{s}=sp^{\mu}.
\end{equation}
Then the Wigner function in Eq. (\ref{eq:Wigner function in free case})
reproduces the massless results of Refs. \cite{Gao:2012ix,Hidaka:2016yjf,Huang:2018wdl}.

For the massive case, one possible choice for the thermal equilibrium
distributions is the naive extension from the massless ones in Eq.
(\ref{eq:Fermi distribution for fermion}) by substituting $\left|\mathbf{p}\right|\rightarrow E_{\mathbf{p}}$
,
\begin{equation}
f_{ss^{\prime}}^{(\pm)}(x,\mathbf{p})=\frac{1}{1+\exp\left[\beta\left(E_{\mathbf{p}}\mp\mu\mp s\mu_{5}\right)\right]}\delta_{ss^{\prime}}.\label{eq:naively extension}
\end{equation}
However, the axial-charge number $\hat{N}_{5}$ is not conserved in
the massive case because it does not commute with the Hamiltonian,
so $\mu_{5}$ is not well-defined. The correct way is to include an
additional self-energy term $\mu\psi^{\dagger}\psi+\mu_{5}\psi^{\dagger}\gamma^{5}\psi$
in the Hamiltonian, where $\mu_{5}$ is the conjugate variable of
the axial-charge. Here $\mu_{5}$ controls the chiral imbalance and
is the counterpart of the chiral chemical potential in the massless
case. Meanwhile, $\mu$ is interpreted as the vector chemical potential.
The single-particle wavefunction, as well as the Wigner function have
been computed in subsection \ref{subsec:Free-with-chiral}, see Eq.
(\ref{eq:Wigner function with chiral}). The Hamiltonian is quantized
in Eq. (\ref{eq:quantized Hamiltonian}). Since the chemical potentials
are already included in the Hamiltonian, the canonical partition function
is given by \cite{Dolan:1973qd}
\begin{equation}
\hat{Z}=\exp(-\beta\hat{H}),\label{eq:new partition function}
\end{equation}
with the total Hamiltonian
\begin{equation}
\hat{H}=\hat{H}_{0}-\mu\hat{N}-\mu_{5}\hat{N}_{5},
\end{equation}
where the free Hamiltonian $\hat{H}_{0}$, the net fermion number
$\hat{N}$, and the axial-charge number $\hat{N}_{5}$ are defined
in Eq. (\ref{eq:operators for equilibrium}). The equilibrium distributions
for fermions and anti-fermions are derived using this partition function,
which agree with the Fermi-Dirac distributions,
\begin{eqnarray}
f_{s}^{(+)}(\mathbf{p}) & = & \frac{1}{1+\exp\left[\beta(E_{\mathbf{p},s}-\mu)\right]},\nonumber \\
f_{s}^{(-)}(\mathbf{p}) & = & \frac{1}{1+\exp\left[\beta(E_{-\mathbf{p},s}+\mu)\right]},\label{eq:explicity distributions}
\end{eqnarray}
where $f_{s}^{(+)}(\mathbf{p})\equiv\left\langle a_{\mathbf{p},s}^{\dagger}a_{\mathbf{p},s}\right\rangle $
and $f_{s}^{(-)}(\mathbf{p})\equiv\left\langle b_{\mathbf{p},s}^{\dagger}b_{\mathbf{p},s}\right\rangle $
are expectation values of the fermion number and anti-fermion number
at given $\mathbf{p}$ and $s$. Note that here the energy $E_{\mathbf{p},s}=\sqrt{m^{2}+(\left|\mathbf{p}\right|-s\mu_{5})^{2}}$
is reduced to $E_{\mathbf{p},s}=\left|\left|\mathbf{p}\right|-s\mu_{5}\right|$
in the massless limit and these distributions agree with previous
results in Eq. (\ref{eq:Fermi distribution for fermion}) when $\left|\mathbf{p}\right|>\mu_{5}$.
In subsections \ref{subsec:Net-fermion-current} and \ref{subsec:Energy-momentum-tensor-and}
we will use both the naive distributions in (\ref{eq:naively extension})
and the explicit ones in (\ref{eq:explicity distributions}) to compute
dynamical quantities. They will show a coincidence with each other
in some parameter region.

In a constant magnetic field, the Dirac equation is solved in subsection
\ref{subsec:Fermions-in-const-B} and the Hamiltonian is quantized
as shown in Eq. (\ref{eq:quantized Hamiltonian in magnetic}). The
corresponding canonical partition function is again defined by Eq.
(\ref{eq:new partition function}). Taking the expectation values
of $\hat{a}_{s}^{(n)\dagger}(p^{x},p^{z})\hat{a}_{s}^{(n)}(p^{x},p^{z})$
and $\hat{b}_{s}^{(n)\dagger}(p^{x},p^{z})\hat{b}_{s}^{(n)}(p^{x},p^{z})$,
we obtain the following distributions respectively,
\begin{eqnarray}
f_{s}^{(+)(n)}(p^{z}) & = & \frac{1}{1+\exp\left[\beta\left(E_{p^{z}s}^{(n)}-\mu\right)\right]},\nonumber \\
f_{s}^{(-)(n)}(p^{z}) & = & \frac{1}{1+\exp\left[\beta\left(E_{-p^{z},s}^{(n)}+\mu\right)\right]}.
\end{eqnarray}
Note that the equilibrium distributions are independent of the parameter
$p^{x}$ because the energy states, $E_{p^{z}}^{(0)}=\sqrt{m^{2}+(p^{z}-\mu_{5})^{2}}$
and $E_{p^{z}s}^{(n)}=\sqrt{m^{2}+\left[\sqrt{(p^{z})^{2}+2nB_{0}}-s\mu_{5}\right]^{2}}$
for $n>0$, are independent of $p^{x}$. This agrees with our knowledge
about the Landau levels: the transverse momentum is quantized and
described by the quantum number $n$, and $p^{x}$ is now a parameter
for the center position of the wavefunction in the $y$ direction.

The coupling between the spin and the magnetic field is already considered
when computing the Wigner function because electromagnetic fields
are included in the definition of the Wigner function (\ref{def:Wigner function}).
However in the presence of vorticity, additional spin-vorticity coupling
terms are necessary, otherwise the vortical effects cannot be derived.
The vorticity of charged particles generates an effective magnetic
field, which couples with the magnetic dipole moment of the particles,
thus the additional coupling term is
\begin{equation}
\Delta\hat{H}=\frac{\hbar}{4}\omega^{\mu\nu}\hat{\psi}^{\dagger}\gamma^{0}\sigma_{\mu\nu}\hat{\psi},
\end{equation}
where $\omega^{\mu\nu}\equiv\partial_{x}^{\mu}(\beta u^{\nu})-\partial_{x}^{\nu}(\beta u^{\mu})$
is the thermal vorticity and $\frac{\hbar}{2}\hat{\psi}^{\dagger}\gamma^{0}\sigma_{\mu\nu}\hat{\psi}$
is the operator of the dipole-moment tensor. In general $\Delta\hat{H}$
is diagonal for the eigenstates of Hamiltonian $\hat{H}$ if and only
if they commute $\left[\hat{H},\Delta\hat{H}\right]=0$. Assuming
that the dipole-moment tensor of particles points along the direction
$m_{\mu\nu}$, then we obtain
\begin{equation}
\Delta\hat{H}=\frac{\hbar}{4}\omega^{\mu\nu}m_{\mu\nu}\int\frac{d^{3}\mathbf{p}}{(2\pi)^{3}}\sum_{s}s\left(a_{\mathbf{p},s}^{\dagger}a_{\mathbf{p},s}-b_{\mathbf{p},s}b_{\mathbf{p},s}^{\dagger}\right).
\end{equation}
This can be achieved in the case of free fermions without chiral imbalance
because the spin quantization direction is not specified, as discussed
in subsection \ref{subsec:Free-fermions}. Then the distribution functions
would have an order-$\hbar$ correction,
\begin{equation}
f_{ss^{\prime}}^{(\pm)}(x,\mathbf{p})=\frac{1}{1+\exp\left[\beta\left(E_{\mathbf{p}}\mp\mu\pm s\frac{\hbar}{4}\omega^{\mu\nu}m_{\mu\nu}\right)\right]}\delta_{ss^{\prime}},\label{eq:Fermi distribution for fermion-1}
\end{equation}
which agrees with the suggestion of Ref. \cite{Becattini:2013fla}.
Note that the validity of this distribution needs more careful discussion.
That is because in the presence of a magnetic field, there are two
specific directions: the direction of the magnetic field and the direction
of the vorticity. If they differ from each other, the spin-magnetic
coupling term and the spin-vorticity coupling term cannot be diagonal
simultaneously. Even if the vorticity and the magnetic field are along
the same direction, we find it difficult to introduce the chiral chemical
potential because the axial-charge $\hat{N}_{5}$ is diagonal only
if the spin is quantized along $p^{\mu}$. More precisely, the vorticity
in general depends on spatial coordinates while the dipole moment
depends on the particle's momentum. If we consider these dependences,
$\Delta\hat{H}$ cannot commute with $\hat{H}_{0}$, which indicates
that it is impossible to find the common eigenstates of $\Delta\hat{H}$
and $\hat{H}_{0}$. Thus the total Hamiltonian is in general not diagonal
and the equilibrium in the presence of vorticity cannot be as simple
as shown in Eq. (\ref{eq:Fermi distribution for fermion-1}). The
Wigner function contains the vortical effect at order $\hbar$, but
in this thesis we will not discuss the vorticity effect because the
correct way to define the thermal equilibrium distributions with spin-vorticity
coupling is still under discussion.

\subsection{Fermion number current and polarization\label{subsec:Net-fermion-current}}

As discussed in subsection \ref{subsec:Physical-quantities}, the
fermion number current can be derived from the vector component of
the Wigner function. The chiral imbalance for massive particles should
be included in the Dirac equation, as we did in subsection \ref{subsec:Free-with-chiral}.
But if the particle masses is small enough, they can be treated as
massless particles. In this case, the chiral chemical potential $\mu_{5}$
is included in the thermal equilibrium distribution (\ref{eq:Fermi distribution for fermion}).
The Wigner function is solved up to $\mathcal{O}(\hbar)$ using the
semi-classical expansion of Eqs. (\ref{eq:order hbar solutions of FPS})
and (\ref{eq:VA from FPS-1}). Note that this method is a straightforward
extension for the massless case. Another more exact approach is chiral
quantization. In this subsection we will compare these two approaches.
Meanwhile, we will discuss the magnetic-field dependence of physical
quantities.

\subsubsection{Semi-classical results}

If we use the semi-classical results in Eq. (\ref{eq:VA from FPS-1}),
the fermion number current is given by
\begin{equation}
\mathbb{N}^{\mu}=\int d^{4}p\ p^{\mu}\left[V\delta(p^{2}-m^{2})-\frac{\hbar}{2}F_{\alpha\beta}\Sigma^{\alpha\beta}\delta^{\prime}(p^{2}-m^{2})\right]+\frac{\hbar}{2}\partial_{x\nu}\int d^{4}p\ \Sigma^{\mu\nu}\delta(p^{2}-m^{2}).
\end{equation}
Since $\int d^{4}p\ \Sigma^{\mu\nu}\delta(p^{2}-m^{2})$ is the leading-order
dipole-moment tensor, the second term is identified as the magnetization
current. Meanwhile, the axial vector current is given by $\mathcal{A}^{\mu}$,
\begin{equation}
\mathbb{N}_{5}^{\mu}=-\frac{1}{2}\epsilon_{\mu\nu\alpha\beta}\int d^{4}p\ p^{\nu}\Sigma^{\alpha\beta}\delta(p^{2}-m^{2})+\hbar\tilde{F}_{\mu\nu}\int d^{4}p\ p^{\nu}V\delta^{\prime}(p^{2}-m^{2}).
\end{equation}
Since we have no idea how to determine the dipole-moment tensor, we
presume that all particles are longitudinally polarized. That is,
at the leading order in $\hbar$, the spins of the particles are either
parallel or anti-parallel to their momenta in the observer's rest
frame. The frame dependence of spin polarization is discussed in subsection
\ref{subsec:Ambiguity-of-functions}, see Eqs. (\ref{eq:decomposition of polarization})
and (\ref{eq:specific choice of nmu}). In this case the vector and
axial-vector components of the Wigner function are given by Eqs. (\ref{eq:specific solution of A^=00005Cmu}),
(\ref{eq:specific solution of V^=00005Cmu}). Using the equilibrium
distributions in (\ref{eq:naively extension}), we obtain the fermion
number distribution $V$ and the axial-charge distribution $A$,
\begin{eqnarray}
V & = & \frac{2}{(2\pi)^{3}}\sum_{s}\left\{ \theta(p\cdot u)\frac{1}{1+\exp\left[\beta(p\cdot u-\mu-s\mu_{5})\right]}\right.\nonumber \\
 &  & \qquad\left.+\theta(-p\cdot u)\frac{1}{1+\exp\left[\beta(-p\cdot u+\mu+s\mu_{5})\right]}-\theta(-p\cdot u)\right\} .\nonumber \\
A & = & \frac{2}{(2\pi)^{3}}\sum_{s}s\left\{ \theta(p\cdot u)\frac{1}{1+\exp\left[\beta(p\cdot u-\mu-s\mu_{5})\right]}\right.\nonumber \\
 &  & \qquad\left.+\theta(-p\cdot u)\frac{1}{1+\exp\left[\beta(-p\cdot u+\mu+s\mu_{5})\right]}-\theta(-p\cdot u)\right\} .\label{eq:equilibrium distributions VA}
\end{eqnarray}
Note that in Eq. (\ref{eq:equilibrium distributions VA}) there are
terms from vacuum contributions. When calculating the fermion number
density $n$ and the CME conductivity $\sigma_{\chi}$, these vacuum
terms lead to divergence and should be dropped. Here we have replaced
$p^{0}$, or the energy $E_{\mathbf{p}}$, by $p\cdot u$, which is
the energy in the frame $u^{\mu}$. From Eq. (\ref{eq:specific solution of V^=00005Cmu})
we obtain the fermion number current by integrating over $d^{4}p$.
\begin{equation}
\mathbb{N}^{\mu}=\int d^{4}p\ p^{\mu}\left[V\delta(p^{2}-m^{2})-\hbar\tilde{F}^{\nu\alpha}\frac{p_{\nu}u_{\alpha}}{p\cdot u}A\delta^{\prime}(p^{2}-m^{2})\right]+\frac{\hbar}{2}\epsilon^{\mu\nu\alpha\beta}\partial_{x\nu}\int d^{4}p\frac{p_{\alpha}u_{\beta}}{p\cdot u}A\delta(p^{2}-m^{2}).
\end{equation}
We find that the first term agrees with the classical fermion current
with an energy shift from the spin-magnetic coupling, while the second
term gives the analogue of the CVE. Note that if we compare with Maxwell's
equation we immediately find that the second term is nothing but the
magnetization current. In the integration over four-momentum, we need
to deal with $\delta^{\prime}(p^{2}-m^{2})$, which can be achieved
by integrating by parts,
\begin{equation}
\mathbb{N}^{\mu}=\int d^{4}p\ \left[p^{\mu}V+\frac{\hbar}{2}\tilde{F}^{\mu\nu}u_{\nu}\frac{1}{p\cdot u}A+\frac{\hbar}{2}u^{\mu}\tilde{F}^{\nu\alpha}\frac{p_{\nu}u_{\alpha}}{p\cdot u}\frac{\partial}{\partial(p\cdot u)}A+\frac{\hbar}{2}u^{\mu}\tilde{F}^{\nu\alpha}\frac{p_{\nu}u_{\alpha}}{(p\cdot u)^{2}}A\right]\delta(p^{2}-m^{2}).
\end{equation}
Note that the functions $V$ and $A$ only depends on $(p\cdot u)$.
Thus the above current can be parametrized as
\begin{equation}
\mathbb{N}^{\mu}=u^{\mu}n+\sigma_{\chi}\hbar\tilde{F}^{\mu\nu}u_{\nu},\label{eq:parametrize n=00005Cmu}
\end{equation}
where
\begin{eqnarray}
n & = & \int d^{4}p\ (p\cdot u)V\delta(p^{2}-m^{2}),\nonumber \\
\sigma_{\chi} & = & \frac{1}{2}\int d^{4}p\ \frac{1}{p\cdot u}A\delta(p^{2}-m^{2})\label{eq:massive n and sigma-chi}
\end{eqnarray}
are the fermion number density and the CME conductivity respectively.
Using the distributions in Eq. (\ref{eq:equilibrium distributions VA}),
these quantities can be numerically calculated.

On the other hand, the axial-vector current, or the spin-polarization
density is calculated from the axial-vector current by taking an integration
over $d^{4}p$%
\begin{eqnarray}
\mathbb{N}_{5}^{\mu} & = & \int d^{4}p\ \left[\left(p^{\mu}-\frac{m^{2}}{p\cdot u}u^{\mu}\right)A-\hbar\tilde{F}^{\mu\nu}u_{\nu}\frac{\partial}{\partial(p\cdot u)}V+\frac{\hbar}{2}\tilde{F}^{\mu\nu}p_{\nu}\frac{1}{p\cdot u}\frac{\partial}{\partial(p\cdot u)}V\right.\nonumber \\
 &  & \left.\qquad+\frac{\hbar}{2}u^{\mu}\tilde{F}^{\nu\alpha}\frac{p_{\nu}u_{\alpha}}{p\cdot u}\frac{\partial}{\partial(p\cdot u)}V\right]\delta(p^{2}-m^{2}),
\end{eqnarray}
where we have neglected the spatial derivative of $u^{\mu}$. Analogous
to the fermion number current, the spin polarization can be parametrized
as
\begin{equation}
\mathbb{N}_{5}^{\mu}=u^{\mu}n_{5}+\sigma_{5}\hbar\tilde{F}^{\mu\nu}u_{\nu},\label{eq:parametrize n_5=00005Cmu}
\end{equation}
where $n_{5}$ is the axial-charge density and $\sigma_{5}$ is the
coefficient for the CSE,
\begin{eqnarray}
n_{5} & = & \int d^{4}p\ \left(u\cdot p-\frac{m^{2}}{u\cdot p}\right)A\delta(p^{2}-m^{2}),\nonumber \\
\sigma_{5} & = & -\frac{1}{2}\int d^{4}p\ \delta(p^{2}-m^{2})\frac{\partial}{\partial(u\cdot p)}V.\label{eq:massive n5 and sigma5}
\end{eqnarray}
Using the equilibrium distributions in Eq. (\ref{eq:equilibrium distributions VA}),
$n_{5}$ and $\sigma_{5}$ can be numerically calculated.

\begin{figure}
\includegraphics[width=8cm]{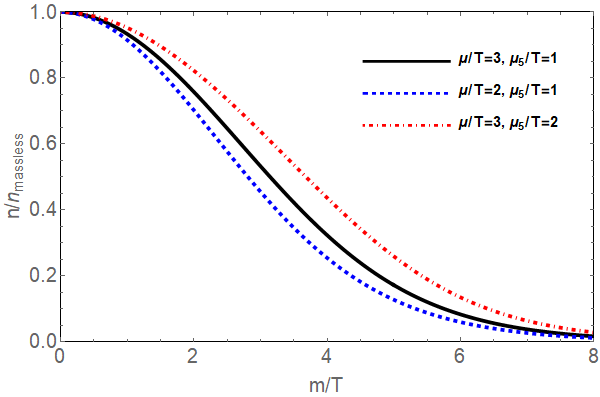}

\caption{\label{fig:Mass-dependence-n}Mass dependence of the fermion number
density $n$ computed using Eq. (\ref{eq:massive n and sigma-chi})
and normalized by the massless value in Eq. (\ref{eq:massless n}).
The particles' mass $m$ and chemical potentials $\mu$, $\mu_{5}$
are normalized by the temperature $T$.}
\end{figure}

\begin{figure}
\includegraphics[width=8cm]{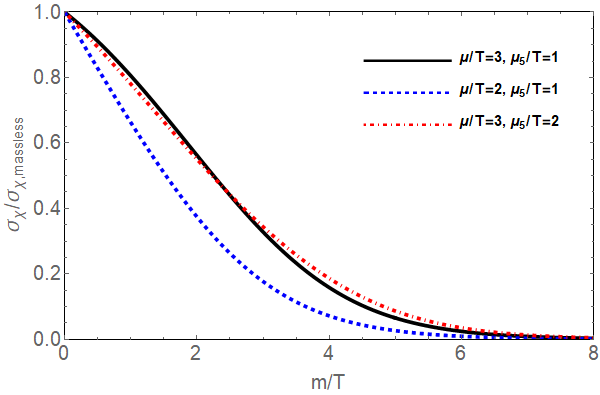}

\caption{\label{fig:Mass-dependence-sigma-chi}Mass dependence of the CME conductivity
$\sigma_{\chi}$ calculated using Eq. (\ref{eq:massive n and sigma-chi})
and normalized by the massless value $\sigma_{\chi,\text{massless}}=\mu_{5}/(2\pi^{2})$.
The particles' mass $m$ and chemical potentials $\mu$, $\mu_{5}$
are normalized by the temperature $T$.}
\end{figure}

\begin{figure}
\includegraphics[width=10cm]{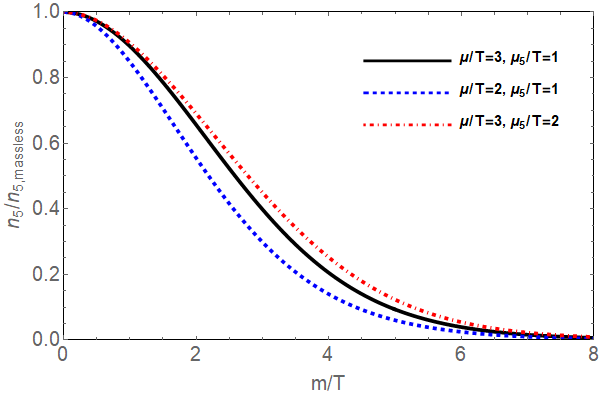}

\caption{\label{fig:Mass-dependence-n5}Mass dependence of the axial-charge
density $n_{5}$ calculated using Eq. (\ref{eq:massive n5 and sigma5})
and normalized by the massless value in Eq. (\ref{eq:massless n5}).
The particles' mass $m$ and chemical potentials $\mu$, $\mu_{5}$
are normalized by the temperature $T$.}
\end{figure}

\begin{figure}
\includegraphics[width=8cm]{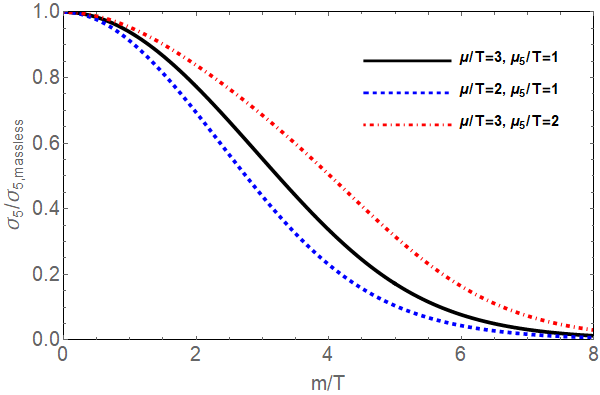}

\caption{\label{fig:Mass-dependence-sigma5}Mass dependence of the CSE coefficient
$\sigma_{5}$ calculated using Eq. (\ref{eq:massive n5 and sigma5})
and normalized by the massless value $\sigma_{5,\text{massless}}=\mu/(2\pi)^{2}$
. The particles' mass $m$ and chemical potentials $\mu$, $\mu_{5}$
are normalized by the temperature $T$.}
\end{figure}

In the massless case, the fermion number density is given by
\begin{equation}
n_{\text{massless}}=\int\frac{d^{3}\mathbf{p}}{(2\pi)^{3}}\sum_{rs}r\frac{1}{1+\exp\left[\beta(\left|\mathbf{p}\right|-r\mu-rs\mu_{5})\right]},\label{eq:massless n}
\end{equation}
while the CME conductivity is a constant $\sigma_{\chi,\text{massless}}=\mu_{5}/(2\pi^{2})$
\cite{Vilenkin:1980fu,Kharzeev:2007jp,Fukushima:2008xe,Landsteiner:2012kd}.
The axial-charge density is
\begin{equation}
n_{5,\text{massless}}=\int\frac{d^{3}\mathbf{p}}{(2\pi)^{3}}\sum_{rs}rs\frac{1}{1+\exp\left[\beta(\left|\mathbf{p}\right|-r\mu-rs\mu_{5})\right]},\label{eq:massless n5}
\end{equation}
while the coefficient for the chiral separation effect is $\sigma_{5,\text{massless}}=\mu/(2\pi)^{2}$
\cite{Son:2004tq,Metlitski:2005pr,Landsteiner:2012kd}. In Figs. \ref{fig:Mass-dependence-n}-\ref{fig:Mass-dependence-sigma5}
we computed the ratio between the massive results (\ref{eq:massive n and sigma-chi})
and (\ref{eq:massive n5 and sigma5}) and the massless ones (\ref{eq:massless n})
and (\ref{eq:massless n5}) for the net fermion number density $n$,
the CME conductivity $\sigma_{\chi}$, the axial-charge density $n_{5}$,
and the CSE coefficient $\sigma_{5}$, respectively. From these figures
we observe that the quantities in the massive case smoothly reproduce
the massless ones by the fact that the ratios become $1$ in the massless
limit. All of these quantities decrease with increasing mass. This
is because, when the particles' momentum is fixed, heavier masses
lead to larger energy and thus the states are less likely to be occupied.

\subsubsection{Results from chiral quantization}

\begin{figure}
\includegraphics[width=8cm]{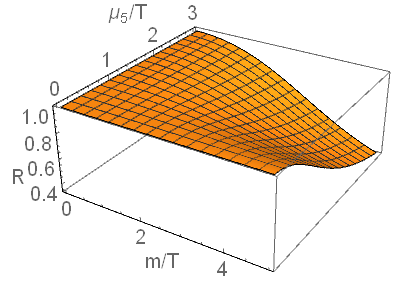}\caption{\label{fig:Ratio-of-n}Ratio of the net fermion number density $n$
of the result in (\ref{eq:chiral as a self energy}) from chiral quantization
and the semi-classical result in (\ref{eq:massive n and sigma-chi}).
Here we fix the chemical potential $\mu/T=3$ and plotted the dependence
on $m$ and $\mu_{5}$.}
\end{figure}

The coincidence of the massive and massless results is beyond our
expectation because the equilibrium distributions are taken to be
the ones in Eq. (\ref{eq:naively extension}), which are naive extensions
of the massless ones. However, as we have discussed in subsection
\ref{subsec:Thermal-equilibrium}, the chiral chemical potential in
the massive case should be considered as a self-energy term, which
appears in the Dirac equation. The corresponding Wigner function has
been given in Eq. (\ref{eq:Wigner function with chiral}). Here we
look at $\mathcal{V}^{0}$ and $\mathcal{A}^{0}$, which give the
fermion number density and the axial-charge density respectively.
In thermal equilibrium, these quantities are given by
\begin{eqnarray}
n & = & \int\frac{d^{3}\mathbf{p}}{(2\pi)^{3}}\sum_{s}\left\{ \frac{1}{1+\exp\left[\beta\left(E_{\mathbf{p},s}-\mu\right)\right]}-\frac{1}{1+\exp\left[\beta\left(E_{\mathbf{p},s}+\mu\right)\right]}+1\right\} ,\nonumber \\
n_{5} & = & \int\frac{d^{3}\mathbf{p}}{(2\pi)^{3}}\sum_{s}\frac{s\left|\mathbf{p}\right|-\mu_{5}}{E_{\mathbf{p},s}}\left\{ \frac{1}{1+\exp\left[\beta\left(E_{\mathbf{p},s}-\mu\right)\right]}+\frac{1}{1+\exp\left[\beta\left(E_{\mathbf{p},s}+\mu\right)\right]}-1\right\} ,\label{eq:chiral as a self energy}
\end{eqnarray}
where $E_{\mathbf{p},s}=\sqrt{m^{2}+(\left|\mathbf{p}\right|-s\mu_{5})^{2}}$
are eigenenergies of the Hamiltonian with chiral modification. When
computing the net fermion number density, we will simply drop the
vacuum contribution, i.e., the last term in the first line of Eq.
(\ref{eq:chiral as a self energy}). In Fig. \ref{fig:Ratio-of-n}
we compare $n$ in Eq. (\ref{eq:chiral as a self energy}) with the
semi-classical result in Eq. (\ref{eq:massive n and sigma-chi}).
We find that they coincide with each other when the mass $m$ or the
chiral chemical potential $\mu_{5}$ is not too large compared with
the temperature. When we have a large $m$ or $\mu_{5}$, the semi-classical
result overestimates the number density because the ratio is smaller
than $1$.

\begin{figure}
\includegraphics[width=8cm]{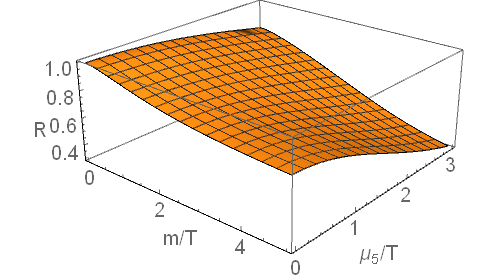}\caption{\label{fig:Ratio-of-n5}The ratio of the axial-charge density $n_{5}$
of the result (\ref{eq:chiral as a self energy}) and the semi-classical
result in (\ref{eq:massive n5 and sigma5}). Here we fix the chemical
potential $\mu/T=3$ and plotted the dependence on $m$ and $\mu_{5}$.}
\end{figure}

Meanwhile, we compare in Fig. \ref{fig:Ratio-of-n5} the axial-charge
density $n_{5}$ calculated using Eq. (\ref{eq:chiral as a self energy})
with the one in Eq. (\ref{eq:massive n5 and sigma5}). Here in the
calculation we have dropped the vacuum contribution. We find that
in the massless limit they do not agree with each other. Fortunately,
we can attribute the difference to the vacuum contribution. The vacuum
part for $n_{5}$ in Eq. (\ref{eq:chiral as a self energy}) will
be divergent for non-zero mass, but in the the massless case it has
a finite value %
\begin{equation}
\Delta n_{5,vac}=\frac{\mu_{5}^{3}}{3\pi^{2}}.
\end{equation}
Taking this into account, the results in Eq. (\ref{eq:chiral as a self energy})
agrees with (\ref{eq:massless n5}) in the massless limit. But in
general the semi-classical results over-estimates the axial-charge
density.

Note that the effects of electromagnetic fields are not included in
the chiral-quantization description. In order to derive the CME or
CSE, one needs to use the semi-classical method to derive the first-order
contribution in $\hbar$, while the results in subsection \ref{subsec:Free-with-chiral}
only serve as the zeroth-order solutions. However, note that when
the chemical potentials $\mu$ and $\mu_{5}$ appear in the Dirac
equation, the kinetic equation (\ref{eq:Dirac equation for Wigner})
of the Wigner function will also have additional terms which are related
to $\mu$ and $\mu_{5}$. So the present semi-classical discussions
in Sec. \ref{sec:Semi-classical-expansion} need to be repeated for
finite $\mu$ and $\mu_{5}$. In this thesis, the method of chiral
quantization only serves for comparing with the semi-classical method
in Sec. \ref{sec:Semi-classical-expansion}, and thus the chiral effects
are not discussed using this method.

\subsubsection{Results in magnetic field}

\begin{figure}
\includegraphics[width=8cm]{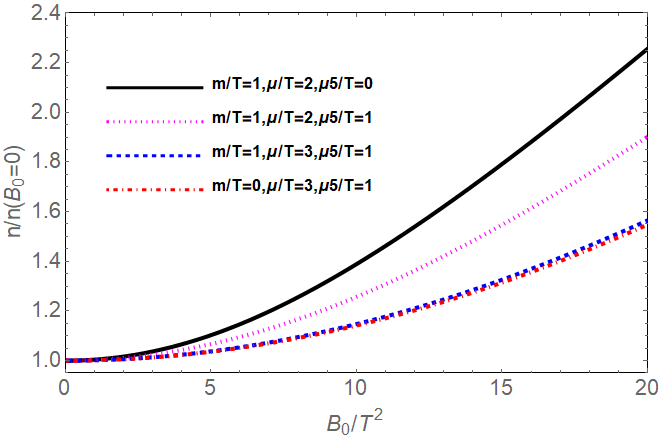}

\caption{\label{fig:Magnetic-dependence-n}The magnetic-field dependence of
the fermion number density $n$. Here we compute the ratio of the
one in Eq. (\ref{eq:magnetic dependent n}), which is derived via
Landau quantization, to the one in Eq. (\ref{eq:chiral as a self energy}),
which is derived via chiral quantization. Four configurations of $m$,
$\mu$, and $\mu_{5}$ are considered.}

\end{figure}

In the presence of a constant magnetic field, fermions are quantized
according to the Landau levels. We have derived the Wigner function
in subsection \ref{subsec:Fermions-in-const-B}, where the vector
components read
\begin{eqnarray}
\mathcal{V}^{0} & = & (p^{0}+\mu)\sum_{n=0}\Lambda_{+}^{(n)}(p_{T})V_{n}+(p^{0}+\mu)\sum_{n>0}\frac{p^{z}}{\sqrt{(p^{z})^{2}+2nB_{0}}}\Lambda_{-}^{(n)}(p_{T})A_{n},\nonumber \\
\mathcal{V}^{x} & = & p^{x}\sum_{n>0}\frac{2nB_{0}}{p_{T}^{2}}\left[V_{n}-\frac{\mu_{5}}{\sqrt{(p^{z})^{2}+2nB_{0}}}A_{n}\right]\Lambda_{+}^{(n)}(p_{T}),\nonumber \\
\mathcal{V}^{y} & = & p^{y}\sum_{n>0}\frac{2nB_{0}}{p_{T}^{2}}\left[V_{n}-\frac{\mu_{5}}{\sqrt{(p^{z})^{2}+2nB_{0}}}A_{n}\right]\Lambda_{+}^{(n)}(p_{T}),\nonumber \\
\mathcal{V}^{z} & = & (p^{z}-\mu_{5})\Lambda^{(0)}(p_{T})V_{0}+p^{z}\sum_{n>0}\left[V_{n}-\frac{\mu_{5}}{\sqrt{(p^{z})^{2}+2nB_{0}}}A_{n}\right]\Lambda_{+}^{(n)}(p_{T})\nonumber \\
 &  & \qquad+\sum_{n>0}\left[\sqrt{(p^{z})^{2}+2nB_{0}}A_{n}-\mu_{5}V_{n}\right]\Lambda_{-}^{(n)}(p_{T}),
\end{eqnarray}
The distributions are assumed to take their thermal equilibrium forms,
\begin{eqnarray}
V_{n} & \equiv & \frac{2}{(2\pi)^{3}}\sum_{s}\delta\left\{ (p^{0}+\mu)^{2}-[E_{p^{z}s}^{(n)}]^{2}\right\} \nonumber \\
 &  & \qquad\times\left\{ \theta(p^{0}+\mu)\frac{1}{1+\exp(\beta p^{0})}+\theta(-p^{0}-\mu)\left[\frac{1}{1+\exp(-\beta p^{0})}-1\right]\right\} ,\\
A_{n} & \equiv & \frac{2}{(2\pi)^{3}}\sum_{s}s\delta\left\{ (p^{0}+\mu)^{2}-[E_{p^{z}s}^{(n)}]^{2}\right\} \\
 &  & \qquad\times\left\{ \theta(p^{0}+\mu)\frac{1}{1+\exp(\beta p^{0})}+\theta(-p^{0}-\mu)\left[\frac{1}{1+\exp(-\beta p^{0})}-1\right]\right\} ,\label{eq:def-VnAn-1}
\end{eqnarray}
and
\begin{eqnarray}
V_{0} & = & \frac{2}{(2\pi)^{3}}\delta\left\{ (p^{0}+\mu)^{2}-[E_{p^{z}}^{(0)}]^{2}\right\} \nonumber \\
 &  & \qquad\times\left\{ \theta(p^{0}+\mu)\frac{1}{1+\exp(\beta p^{0})}+\theta(-p^{0}-\mu)\left[\frac{1}{1+\exp(-\beta p^{0})}-1\right]\right\} .\label{eq:def-V0-1}
\end{eqnarray}
Here the eigenenergies are given by $E_{p_{z}}^{(0)}=\sqrt{m^{2}+(p^{z})^{2}}$
for the lowest Landau level and $E_{p^{z}s}^{(n)}=\sqrt{m^{2}+\left[\sqrt{(p^{z})^{2}+2nB_{0}}-s\mu_{5}\right]^{2}}$
for the higher Landau levels with $n>0$. These eigenenergies are
analytically derived from the Dirac equation in subsection \ref{subsec:Fermions-in-const-B}.
Note that these distributions are independent of the transverse momentum
$p^{x}$ and $p^{y}$, thus we observe that $\mathcal{V}^{x}$ is
odd in $p^{x}$. When integrating over $d^{4}p$, the component $\mathcal{V}^{x}$
gives zero, which means there is no current along the $x$ direction.
Meanwhile, the current along the $y$ direction vanishes for a similar
reason: $\mathcal{V}^{y}$ is odd in $p^{y}$. The fermion number
density and current along the magnetic field are non-vanishing, thus
the current can be parametrized as shown in Eq. (\ref{eq:parametrize n=00005Cmu}),
with
\begin{eqnarray}
n & = & \sum_{n=0}\int d^{4}p\ (p^{0}+\mu)\Lambda_{+}^{(n)}(p_{T})V_{n},\nonumber \\
\sigma_{\chi} & = & \frac{1}{B_{0}}\int d^{4}p(p^{z}-\mu_{5})\Lambda^{(0)}(p_{T})V_{0}+\frac{1}{B_{0}}\sum_{n>0}\int d^{4}p\ p^{z}\left[V_{n}-\frac{\mu_{5}}{\sqrt{(p^{z})^{2}+2nB_{0}}}A_{n}\right]\Lambda_{+}^{(n)}(p_{T}).\nonumber \\
\label{eq:calculation n sigma_chi}
\end{eqnarray}
Here we have dropped the terms of $\Lambda_{-}^{(n)}(p_{T})$ because
according to Eq. (\ref{eq:pt integration of Lambda_pm}), these terms
vanish when taking an integration over $\mathbf{p}_{T}$. With the
help of Eq. (\ref{eq:pt integration of Lambda_pm}), $\Lambda_{+}^{(n)}(p_{T})$
can be integrated out, which gives the density of states for the Landau
levels. Furthermore, we find that the higher Landau levels $n>0$
do not contribute to $\sigma_{\chi}$ because they are odd in $p^{z}$.
Inserting the distributions (\ref{eq:def-VnAn-1}) and (\ref{eq:def-V0-1})
into Eq. (\ref{eq:calculation n sigma_chi}), we finally obtain
\begin{eqnarray}
n & = & \frac{B_{0}}{(2\pi)^{2}}\int dp^{z}\sum_{n,s}\left\{ \frac{1}{1+\exp\left[\beta\left(E_{p^{z}s}^{(n)}-\mu\right)\right]}-\frac{1}{1+\exp\left[\beta\left(E_{p^{z}s}^{(n)}+\mu\right)\right]}+1\right\} ,\nonumber \\
\sigma_{\chi} & = & \frac{1}{(2\pi)^{2}}\int dp^{z}\frac{p^{z}-\mu_{5}}{E_{p^{z}}^{(0)}}\left\{ \frac{1}{1+\exp\left[\beta\left(E_{p^{z}}^{(0)}-\mu\right)\right]}+\frac{1}{1+\exp\left[\beta\left(E_{p^{z}}^{(0)}+\mu\right)\right]}-1\right\} .\label{eq:magnetic dependent n}
\end{eqnarray}
Here $\sigma_{\chi}$ can be analytically computed,
\begin{eqnarray}
\sigma_{\chi} & = & -\frac{1}{(2\pi)^{2}\beta}\left.\left\{ \ln\left[1+\exp\left(-\beta E_{p^{z}}^{(0)}+\beta\mu\right)\right]+\ln\left[1+\exp\left(-\beta E_{p^{z}}^{(0)}-\beta\mu\right)\right]\right\} \right|_{-\Lambda}^{\Lambda}\nonumber \\
 &  & \qquad-\frac{1}{(2\pi)^{2}}\left.\sqrt{m^{2}+(p^{z}-\mu_{5})^{2}}\right|_{-\Lambda}^{\Lambda}\nonumber \\
 & = & \frac{\mu_{5}}{2\pi^{2}},
\end{eqnarray}
which agrees with the massless results. The result shows that the
CME is independent of mass \cite{Fukushima:2008xe}. In Fig. \ref{fig:Magnetic-dependence-n}
we plot the dependence of the net fermion number $n$ on the magnetic
field. Here the $x$-axis is the field strength, for which we considered
a large range from $0$ to $20\,T^{2}$. Note that if we take $T=100$
MeV, then $20\,T^{2}\sim10m_{\pi}^{2}$ is of the order of the maximum
field strength in Pb+Pb collisions at the LHC energy. In Fig. \ref{fig:Magnetic-dependence-n}
we compute the ratio of $n$ in Eq. (\ref{eq:magnetic dependent n})
to that in Eq. (\ref{eq:chiral as a self energy}). The ratio reaches
$1$ in the weak-field limit, as expected because the result in Eq.
(\ref{eq:chiral as a self energy}) is obtained without the magnetic
field. Four parameter configurations are considered: 1) $m/T=1$,
$\mu/T=2$, and $\mu_{5}/T=0$, which represents a chirally symmetric
system, 2) $m/T=1$, $\mu/T=2$, and $\mu_{5}/T=1$, which represents
a system with chiral imbalance, 3) $m/T=1$, $\mu/T=3$, and $\mu_{5}/T=1$,
which can show the effect of the chemical potential by comparing with
case 2), and 4) $m/T=0$, $\mu/T=3$, and $\mu_{5}/T=1$, which represents
a system of massless fermions. From Fig. \ref{fig:Magnetic-dependence-n}
we observe that the ratio is sensitive to the chemical potentials
while insensitive to the mass.

\begin{figure}
\includegraphics[width=8cm]{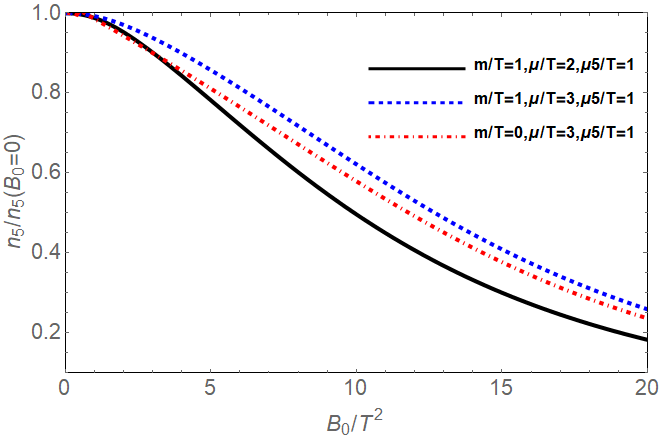}

\caption{\label{fig:Magnetic-field-n5}The axial-charge density $n_{5}$ in
Eq. (\ref{eq:n5 and sigma5}) normalized by Eq. (\ref{eq:chiral as a self energy})
as functions of the magentic field normalized by the temperature square.
Three parameter configurations are considered.}
\end{figure}

Analogous to the vector component, for the axial-vector part, we firstly
list all the relevant components of the Wigner function,
\begin{eqnarray}
\mathcal{A}^{0} & = & (p^{z}-\mu_{5})V_{0}\Lambda^{(0)}(p_{T})+\sum_{n>0}\left[\sqrt{(p^{z})^{2}+2nB_{0}}A_{n}-\mu_{5}V_{n}\right]\Lambda_{+}^{(n)}(p_{T})\nonumber \\
 &  & \qquad+p^{z}\sum_{n>0}\left[V_{n}-\frac{\mu_{5}}{\sqrt{(p^{z})^{2}+2nB_{0}}}A_{n}\right]\Lambda_{-}^{(n)}(p_{T}),\nonumber \\
\mathcal{A}^{z} & = & (p^{0}+\mu)V_{0}\Lambda^{(0)}(p_{T})+(p^{0}+\mu)\sum_{n>0}V_{n}\Lambda_{-}^{(n)}(p_{T})\nonumber \\
 &  & \qquad+(p^{0}+\mu)p^{z}\sum_{n>0}A_{n}\frac{1}{\sqrt{(p^{z})^{2}+2nB_{0}}}\Lambda_{+}^{(n)}(p_{T}).
\end{eqnarray}
Here the $x$- and $y$-components are not listed because they do
not contribute to the axial-vector current for reasons of symmetry.
The current $\mathbb{N}_{5}^{\mu}$ is then parametrized as Eq. (\ref{eq:parametrize n_5=00005Cmu}),
where the axial-charge density $n_{5}$ and coefficient $\sigma_{5}$
for the CSE are given by
\begin{eqnarray}
n_{5} & = & \int d^{4}p\ (p^{z}-\mu_{5})V_{0}\Lambda^{(0)}(p_{T})+\sum_{n>0}\int d^{4}p\ \left[\sqrt{(p^{z})^{2}+2nB_{0}}A_{n}-\mu_{5}V_{n}\right]\Lambda_{+}^{(n)}(p_{T}),\nonumber \\
\sigma_{5} & = & \frac{1}{B_{0}}\int d^{4}p(p_{0}+\mu)V_{0}\Lambda^{(0)}(p_{T}).
\end{eqnarray}
Using the distributions in Eqs. (\ref{eq:def-VnAn-1}), (\ref{eq:def-V0-1})
and the property of $\Lambda_{+}^{(n)}(p_{T})$ in Eq. (\ref{eq:pt integration of Lambda_pm}),
the integration over $\mathbf{p}_{T}$ can be performed and we obtain
\begin{eqnarray}
n_{5} & = & \frac{B_{0}}{(2\pi)^{2}}\int dp^{z}\frac{p^{z}-\mu_{5}}{E_{p^{z}}^{(0)}}\left\{ \frac{1}{1+\exp\left[\beta\left(E_{p^{z}}^{(0)}-\mu\right)\right]}+\frac{1}{1+\exp\left[\beta\left(E_{p^{z}}^{(0)}+\mu\right)\right]}-1\right\} \nonumber \\
 &  & +\frac{B_{0}}{(2\pi)^{2}}\sum_{n>0}\sum_{s}\int dp^{z}\frac{s\sqrt{(p^{z})^{2}+2nB_{0}}-\mu_{5}}{E_{p^{z}s}^{(n)}}\nonumber \\
 &  & \qquad\times\left[\frac{1}{1+\exp\left[\beta\left(E_{p^{z}s}^{(n)}-\mu\right)\right]}+\frac{1}{1+\exp\left[\beta\left(E_{p^{z}s}^{(n)}+\mu\right)\right]}-1\right],\nonumber \\
\sigma_{5} & = & \frac{1}{(2\pi)^{2}}\int dp^{z}\left\{ \frac{1}{1+\exp\left[\beta\left(E_{p^{z}}^{(0)}-\mu\right)\right]}-\frac{1}{1+\exp\left[\beta\left(E_{p^{z}}^{(0)}+\mu\right)\right]}+1\right\} .\label{eq:n5 and sigma5}
\end{eqnarray}
In general these quantities cannot be analytically done. In Fig. \ref{fig:Magnetic-field-n5}
we compare the axial-charge density $n_{5}$ in a magnetic field,
i.e., Eq. (\ref{eq:n5 and sigma5}), with that without the magnetic
field, i.e., Eq. (\ref{eq:chiral as a self energy}). We consider
three cases: 1) $m/T=1$, $\mu/T=2$, $\mu_{5}=1$, 2) $m/T=1$, $\mu/T=3$,
$\mu_{5}=1$, and 3) $m/T=0$, $\mu/T=3$, $\mu_{5}=1$. Comparing
these cases we find that the ratio varies with both the mass $m$
and the chemical potential $\mu$. In the weak-field limit, the ratio
reaches $1$, which indicates that Eq. (\ref{eq:n5 and sigma5}) agrees
with Eq. (\ref{eq:chiral as a self energy}) in this limit. But when
the magnetic field increases, we find that the axial-charge density
decreases for fixed $m$, $\mu$, and $\mu_{5}$.

\begin{figure}
\includegraphics[width=8cm]{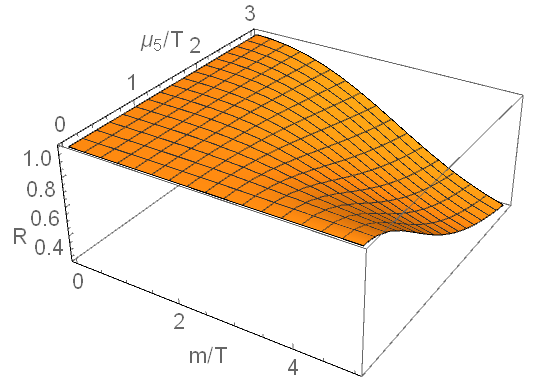}

\caption{\label{fig:Ratio-between-sigma5}The ratio of the CSE coefficient
$\sigma_{5}$ calculated via Eq. (\ref{eq:n5 and sigma5}) to the
semi-classical result in (\ref{eq:massive n5 and sigma5}), as functions
of $m$ and $\mu_{5}$ at fixed $\mu/T=3$. }
\end{figure}

\begin{figure}
\includegraphics[width=8cm]{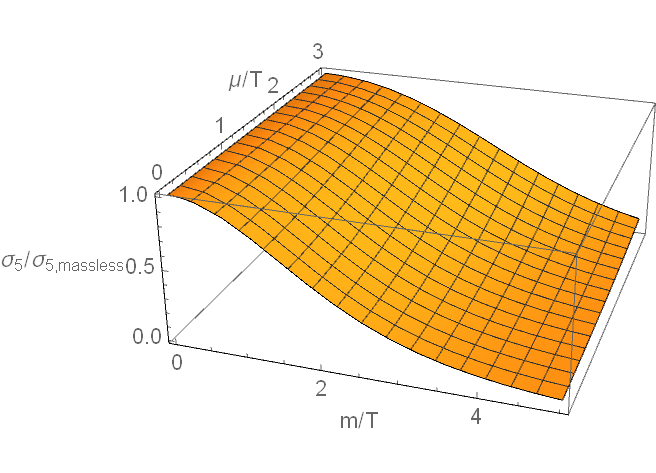}

\caption{\label{fig:Mass-and-chemical-sigma5}The ratio of $\sigma_{5}$ calculated
from Landau quantization in Eq. (\ref{eq:n5 and sigma5}) to the massless
one $\sigma_{5,\text{massless}}=\mu/(2\pi)^{2}$, as a function of
mass and chemical-potential (it is independent of $\mu_{5}/T$).}
\end{figure}

On the other hand, $\sigma_{5}$ is independent of the chiral chemical
potential $\mu_{5}$, which can be proven by a shift of the integration
variable $p^{z}\rightarrow p^{z}+\mu_{5}$. However, in the semi-classical
result (\ref{eq:massive n5 and sigma5}), $\sigma_{5}$ depends on
$\mu_{5}$, which is conflict with the one in Eq. (\ref{eq:n5 and sigma5}).
In Fig. \ref{fig:Ratio-between-sigma5} we compute the ration of $\sigma_{5}$
calculated via Eq. (\ref{eq:n5 and sigma5}) to the semi-classical
result (\ref{eq:massive n5 and sigma5}). The figure shows that these
two agree with each other if $\mu_{5}$ and $m$ are not too large.
The ratio is smaller than $1$, which means that the semi-classical
result overestimates the chiral separation effect. In the limit $m\ll T$,
we can expand $\sigma_{5}$ into a series of $\beta m$. The leading
two terms read
\begin{equation}
\sigma_{5}=\frac{\mu}{2\pi^{2}}-\frac{(\beta m)^{2}}{(2\pi)^{2}\beta}\int_{0}^{\infty}dp\frac{e^{\beta(p-\mu)}(e^{2\beta\mu}-1)(e^{2\beta p}-1)}{p\left[1+e^{\beta(p+\mu)}\right]^{2}\left[1+e^{\beta(p-\mu)}\right]^{2}}+\mathcal{O}\left[(\beta m)^{4}\right].
\end{equation}
The leading-order term agrees with the massless result. In Fig. \ref{fig:Mass-and-chemical-sigma5}
we plotted the mass and chemical-potential dependence of $\sigma_{5}$.
This shows that a finite mass suppresses the chiral separation effect.

\begin{figure}
\includegraphics[width=8cm]{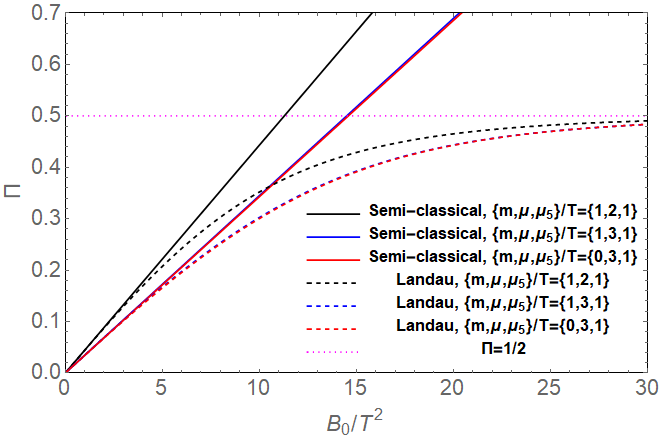}

\caption{\label{fig:Average-spin-polarization}The average spin polarization
as functions of the magnetic-field strength. The solid lines are the
results from the semi-classical expansion, where the fermion number
density $n$ and coefficient $\sigma_{5}$ are given in Eqs. (\ref{eq:massive n and sigma-chi})
and (\ref{eq:massive n5 and sigma5}). The dashed lines are the results
calculated via Landau quantization, where $n$ and $\sigma_{5}$ are
given in Eqs. (\ref{eq:magnetic dependent n}) and (\ref{eq:n5 and sigma5}),
respectively. The dotted line shows the fully polarized case, where
$\Pi=\frac{1}{2}$.}

\end{figure}

Since we have already derived the vector and the axial-vector currents,
now we compute the average spin polarization in a magnetic field.
Using the fermion number density $n$ and the coefficient $\sigma_{5}$,
the average polarization can be expressed as
\begin{equation}
\Pi=\frac{\hbar}{2}\frac{\sigma_{5}B_{0}}{n},
\end{equation}
where the factor $\frac{\hbar}{2}$ is the unit of spin and $\frac{\hbar}{2}\sigma_{5}B_{0}$
is the total spin-polarization density in a constant magnetic field.
In Fig. \ref{fig:Average-spin-polarization} we plot the average polarization
$\Pi$ as a function of the magnetic-field strength $B_{0}$, where
both the semi-classical result and Landau-quantized result are presented.
We find that the average polarization of the semi-classical result
grows to infinity when $B_{0}$ increases. This is because in the
semi-classical result $n$ and $\sigma_{5}$ are independent of $B_{0}$
and thus the average polarization is linear in $B_{0}$. Meanwhile,
the result from Landau quantization has the upper limit $1/2$. This
is because in sufficiently strong magnetic fields, fermions will stay
in the lowest Landau level. As we discussed in subsection \ref{subsec:Fermions-in-const-B},
the spins in the lowest Landau level are fixed. So the system reaches
a fully-polarized state if the field strength is large enough and
the average spin polarization approaches $\hbar/2$.

\subsection{Energy-momentum tensor and spin tensor\label{subsec:Energy-momentum-tensor-and}}

\subsubsection{Semi-classical results}

The canonical energy-momentum tensor and the spin-angular-momentum
tensor in the quantum field theory are given in Eq. (\ref{eq:cannonical quantities}).
In the semi-classical description, $\mathcal{V}^{\mu}$ and $\mathcal{A}^{\mu}$
are given by Eq. (\ref{eq:VA from FPS-1}). Inserting them into Eq.
(\ref{eq:cannonical quantities}) we obtain
\begin{eqnarray}
\mathbb{T}_{\text{mat}}^{\mu\nu} & = & \int d^{4}p\,p^{\mu}p^{\nu}\left[V\delta(p^{2}-m^{2})-\frac{\hbar}{2}F_{\alpha\beta}\Sigma^{\alpha\beta}\delta^{\prime}(p^{2}-m^{2})\right]\nonumber \\
 &  & +\frac{\hbar}{2}\partial_{x\alpha}\int d^{4}p\ p^{\nu}\Sigma^{\mu\alpha}\delta(p^{2}-m^{2})+\frac{\hbar}{2}\int d^{4}p\ \Sigma^{\mu\alpha}F_{\alpha}^{\ \nu}\delta(p^{2}-m^{2}),\nonumber \\
\mathbb{S}_{\text{mat}}^{\rho,\mu\nu} & = & \frac{1}{2}\int d^{4}p\,\left(p^{\rho}\Sigma^{\mu\nu}+\rho^{\mu}\Sigma^{\nu\rho}-\rho^{\nu}\Sigma^{\mu\rho}\right)\delta(p^{2}-m^{2})\nonumber \\
 &  & -\frac{\hbar}{2}\int d^{4}p\,\left(p^{\rho}F^{\mu\nu}+\rho^{\mu}F^{\nu\rho}-\rho^{\nu}F^{\mu\rho}\right)V\delta^{\prime}(p^{2}-m^{2}).
\end{eqnarray}
Note that in the classical limit $\hbar\rightarrow0$ the energy-momentum
tensor is symmetric with respect to its indices and agrees with the
classical result. But the leading-order term can be non-symmetric.
The spin-angular-momentum tensor $\mathbb{S}_{\text{mat}}^{\rho,\mu\nu}$
has a straightforward connection to the axial-vector current $\mathbb{S}_{\text{mat}}^{\rho,\mu\nu}=-\frac{1}{2}\epsilon^{\rho\mu\nu\lambda}\mathbb{N}_{5,\lambda}$.
Since the axial-vector current has been discussed in the previous
subsection, in this subsection we only focus on the energy-momentum
tensor.

\begin{figure}
\includegraphics[width=8cm]{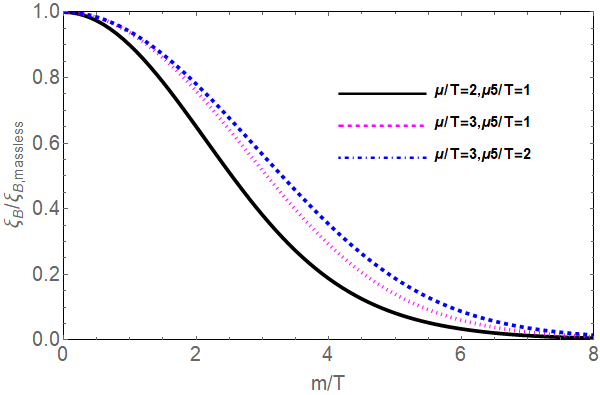}

\caption{\label{fig:Mass-dependence-xiB}The mass dependence of the ratio between
the semi-classical $\xi_{B}$ in Eq. (\ref{eq:semi-classical xiB})
and the massless one in Eq. (\ref{eq:massless xiB}). We take several
chemical-potential configurations, 1) $\mu/T=2$, $\mu_{5}/T=1$ (solid
line), 2) $\mu/T=3$, $\mu_{5}/T=1$ (dashed line), 3) $\mu/T=3$,
$\mu_{5}/T=2$ (dash-dotted line).}

\end{figure}

Similar to the previous subsection, we have no idea about what the
equilibrium dipole-moment tensor looks like. Thus we adopt the specific
solution in Eqs. (\ref{eq:specific solution of A^=00005Cmu}) and
(\ref{eq:specific solution of V^=00005Cmu}), which smoothly recovers
the massless limit. %
{} Inserting the dipole-moment tensor into $\mathbb{T}_{\text{mat}}^{\mu\nu}$,
we obtain
\begin{eqnarray}
\mathbb{T}_{\text{mat}}^{\mu\nu} & = & \int d^{4}p\ p^{\mu}p^{\nu}\left[V\delta(p^{2}-m^{2})-\hbar\tilde{F}^{\alpha\beta}\frac{p_{\alpha}u_{\beta}}{u\cdot p}A\delta^{\prime}(p^{2}-m^{2})\right]\nonumber \\
 &  & +\frac{\hbar}{2}\epsilon^{\mu\alpha\beta\gamma}\partial_{x\alpha}u_{\gamma}\int d^{4}p\ p^{\nu}p_{\beta}\frac{1}{u\cdot p}A\delta(p^{2}-m^{2})\nonumber \\
 &  & -\frac{\hbar}{2}\epsilon^{\mu\alpha\beta\gamma}F_{\beta}^{\ \nu}u_{\gamma}\int d^{4}p\ \frac{p_{\alpha}}{u\cdot p}A\delta(p^{2}-m^{2}).
\end{eqnarray}
Here we assume that the distributions take their equilibrium form
in (\ref{eq:equilibrium distributions VA}), which depends on $u\cdot p$
in the fluid's comoving frame $u^{\mu}$. %
{} The energy-momentum tensor can be parametrized as
\begin{equation}
\mathbb{T}_{\text{mat}}^{\mu\nu}=u^{\mu}u^{\nu}\epsilon-(g^{\mu\nu}-u^{\mu}u^{\nu})P+\hbar\left(u^{\mu}\tilde{F}^{\nu\beta}u_{\beta}+u^{\nu}\tilde{F}^{\mu\beta}u_{\beta}\right)\xi_{B},
\end{equation}
where
\begin{eqnarray}
\epsilon & = & \int d^{4}p\ (u\cdot p)^{2}V\delta(p^{2}-m^{2}),\nonumber \\
P & = & \frac{1}{3}\int d^{4}p\ \left[(u\cdot p)^{2}-m^{2}\right]V\delta(p^{2}-m^{2}),\nonumber \\
\xi_{B} & = & \frac{1}{2}\int d^{4}p\ A\delta(p^{2}-m^{2}),\label{eq:semi-classical xiB}
\end{eqnarray}
are the energy density, the pressure, and the coefficient for the
energy flux along the magnetic field. In the massless limit, the coefficient
$\xi_{B}$ takes the following form \cite{Landsteiner:2012kd}
\begin{equation}
\xi_{B,\text{massless}}=\frac{\mu\mu_{5}}{2\pi^{2}}.\label{eq:massless xiB}
\end{equation}
In Fig. \ref{fig:Mass-dependence-xiB} we compare the semi-classical
results (\ref{eq:semi-classical xiB}) for massive fermions with $\xi_{B}$
for massless fermions (\ref{eq:massless xiB}). We consider a wide
range for the value of mass and find that the energy flux decreases
for a larger mass. Several configurations of chemical potentials are
considered. In the massless limit, the semi-classical results coincide
with the massless ones for all the cases considered.

\subsubsection{Results from chiral quantization}

\begin{figure}
\includegraphics[width=8cm]{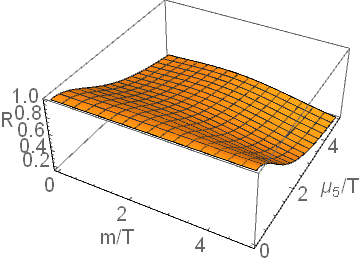}

\caption{\label{fig:Ratio-of-energy}The ratio of energy density calculated
using Eq. (\ref{eq:energy-density with chiral}) to the semi-classical
result in Eq. (\ref{eq:semi-classical xiB}), as a function of the
mass $m$ and the chiral chemical potential $\mu_{5}$ in the unit
of $T$. Here the vector chemical potential is set to $\mu/T=3$ . }

\end{figure}

\begin{figure}
\includegraphics[width=8cm]{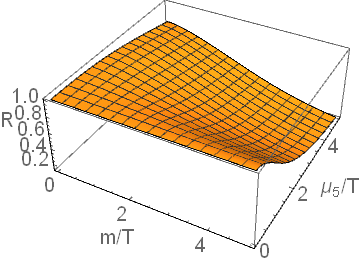}

\caption{\label{fig:Ratio-of-pressure}The ratio of pressure calculated using
Eq. (\ref{eq:pressure with chiral}) to the semi-classical result
in Eq. (\ref{eq:semi-classical xiB}), as a function of the mass $m$
and the chiral chemical potential $\mu_{5}$. Here the vector chemical
potential is set to $\mu/T=3$. }

\end{figure}

If we adopt the chiral quantization description, the energy-momentum
tensor can be calculated at zeroth order in $\hbar$ using the results
in Eq. (\ref{eq:Wigner function with chiral}). Since the formula
is not Lorentz-covariant, we take the fluid velocity $u^{\mu}=(1,0,0,0)^{T}$.
Then the energy density is given by
\begin{equation}
\epsilon=\int\frac{d^{3}\mathbf{p}}{(2\pi)^{3}}\sum_{s}E_{\mathbf{p},s}\left\{ \frac{1}{1+\exp\left[\beta\left(E_{\mathbf{p},s}-\mu\right)\right]}+\frac{1}{1+\exp\left[\beta\left(E_{\mathbf{p},s}+\mu\right)\right]}\right\} ,\label{eq:energy-density with chiral}
\end{equation}
and the pressure is
\begin{equation}
P=\int\frac{d^{3}\mathbf{p}}{(2\pi)^{3}}\sum_{s}\frac{\left|\mathbf{p}\right|^{2}}{3E_{\mathbf{p},s}}\left(1-s\frac{\mu_{5}}{\left|\mathbf{p}\right|}\right)\left\{ \frac{1}{1+\exp\left[\beta\left(E_{\mathbf{p},s}-\mu\right)\right]}+\frac{1}{1+\exp\left[\beta\left(E_{\mathbf{p},s}+\mu\right)\right]}\right\} ,\label{eq:pressure with chiral}
\end{equation}
while other components of the energy-momentum tensor vanish. In Figs.
\ref{fig:Ratio-of-energy} and \ref{fig:Ratio-of-pressure} we compare
the semi-classical results with the ones from the chiral quantization.
We find that in the limit $\mu_{5}\rightarrow0$ these results agree
with each other, while for large $\mu_{5}$, the semi-classical ones
over-estimate both the energy density and the pressure.

\subsubsection{Results in magnetic field}

\begin{figure}
\includegraphics[width=8cm]{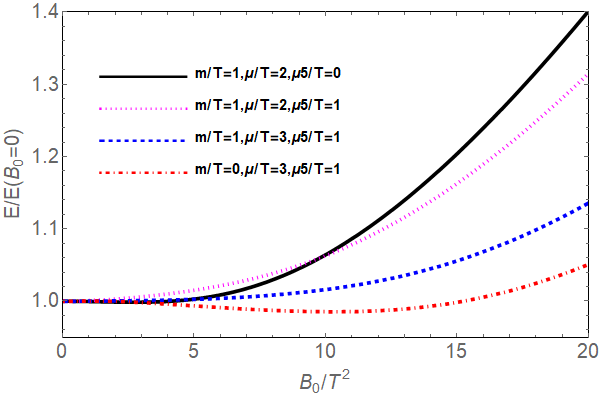}\caption{\label{fig:Energy-density-B} The energy density ratio of the result
by Eq. (\ref{eq:quantities in Landau levels}) to that by Eq. (\ref{eq:energy-density with chiral}),
as functions of the magnetic field. }

\end{figure}

\begin{figure}
\includegraphics[width=8cm]{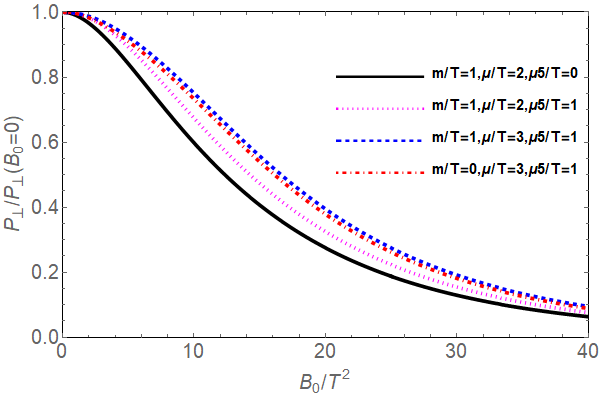}\caption{\label{fig:Transverse-pressure}The transverse pressure ratio of the
result by Eq. (\ref{eq:quantities in Landau levels}) to that by Eq.
(\ref{eq:energy-density with chiral}), as functions of the magnetic
field. }
\end{figure}

\begin{figure}
\includegraphics[width=8cm]{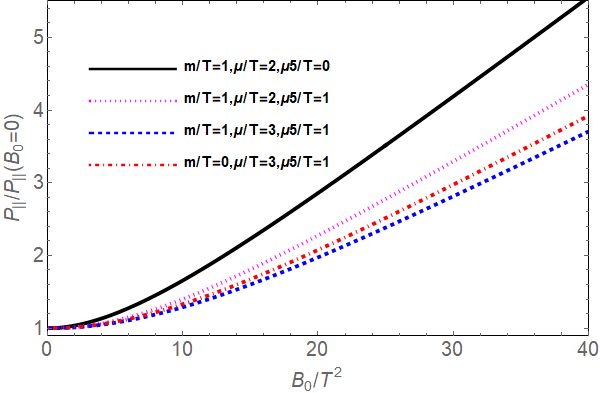}\caption{\label{fig:Longitudinal-pressure-as}The longitudinal pressure ratio
of the result by Eq. (\ref{eq:quantities in Landau levels}) to that
by Eq. (\ref{eq:energy-density with chiral}) as functions of the
magnetic field. }
\end{figure}

In the presence of a magnetic field, the Wigner function has been
computed in Eq. (\ref{eq:sol-Wigner-function}). The constant-magnetic
field assumption breaks the Lorentz symmetry because an electric field
will appear if we perform a Lorentz boost along a direction which
is not parallel to the magnetic field. Here we take the observer's
frame as $u^{\mu}=(1,0,0,0)^{T}$. In this frame, the energy density
is $\mathbb{T}_{\text{mat}}^{00}$ and the pressure is $\mathbb{T}_{\text{mat}}^{ii}$,
$i=1,2,3$. We first compute the components of $\mathbb{T}_{\text{mat}}^{\mu\nu}$
with the help of Eq. (\ref{eq:cannonical quantities}) and the solutions
(\ref{eq:sol-Wigner-function}) in a magnetic field, %
\begin{eqnarray}
\mathbb{T}_{\text{mat}}^{00} & = & \int d^{4}p\ (p_{0}+\mu)^{2}\sum_{n=0}V_{n}\Lambda_{+}^{(n)}(p_{T}),\nonumber \\
\mathbb{T}_{\text{mat}}^{11}=\mathbb{T}_{\text{mat}}^{22} & = & \int d^{4}p\ \sum_{n>0}nB_{0}\left[V_{n}-\frac{\mu_{5}}{\sqrt{(p^{z})^{2}+2nB_{0}}}A_{n}\right]\Lambda_{+}^{(n)}(p_{T}),\nonumber \\
\mathbb{T}_{\text{mat}}^{33} & = & \int d^{4}p\ p^{z}(p^{z}-\mu_{5})V_{0}\Lambda^{(0)}(p_{T})\nonumber \\
 &  & \qquad+\int d^{4}p\ (p^{z})^{2}\sum_{n>0}\left[V_{n}-\frac{\mu_{5}}{\sqrt{(p^{z})^{2}+2nB_{0}}}A_{n}\right]\Lambda_{+}^{(n)}(p_{T}),\nonumber \\
\mathbb{T}_{\text{mat}}^{03} & = & \int d^{4}p\ p^{z}(p_{0}+\mu)V_{0}\Lambda^{(0)}(p_{T}),
\end{eqnarray}
where we have dropped terms which vanish when integrating over four-momentum.
All the unlisted components are zero. Here the distributions $V_{n}$
and $A_{n}$ are assumed to take their equilibrium forms in Eqs. (\ref{eq:def-VnAn-1})
and (\ref{eq:def-V0-1}). In a Lorentz-covariant form, the energy-momentum
tensor can be generalized as follows
\begin{equation}
\mathbb{T}_{\text{mat}}^{\mu\nu}=\epsilon u^{\mu}u^{\nu}-P_{\perp}(g^{\mu\nu}-u^{\mu}u^{\nu}+b^{\mu}b^{\nu})+P_{\parallel}b^{\mu}b^{\nu}+\hbar B_{0}u^{\mu}b^{\nu}\xi_{B},
\end{equation}
where $b^{\mu}$ is the direction of the magnetic field. The energy
density $\epsilon$, the transverse pressure $P_{\perp}$, the longitudinal
pressure $P_{\parallel}$, and the coefficients $\xi_{B}$ are respectively
given by
\begin{eqnarray}
\epsilon & = & \frac{B_{0}}{(2\pi)^{2}}\int dp^{z}\sum_{n,s}E_{p^{z}s}^{(n)}\left\{ \frac{1}{1+\exp\left[\beta\left(E_{p^{z}s}^{(n)}-\mu\right)\right]}+\frac{1}{1+\exp\left[\beta\left(E_{p^{z}s}^{(n)}+\mu\right)\right]}-1\right\} ,\nonumber \\
P_{\perp} & = & \frac{(B_{0})^{2}}{(2\pi)^{2}}\int dp^{z}\sum_{n>0,s}\frac{n}{E_{p^{z}s}^{(n)}}\left[1-\frac{s\mu_{5}}{\sqrt{(p^{z})^{2}+2nB_{0}}}\right]\nonumber \\
 &  & \qquad\times\left\{ \frac{1}{1+\exp\left[\beta\left(E_{p^{z}s}^{(n)}-\mu\right)\right]}+\frac{1}{1+\exp\left[\beta\left(E_{p^{z}s}^{(n)}+\mu\right)\right]}-1\right\} ,\nonumber \\
P_{\parallel} & = & \frac{B_{0}}{(2\pi)^{2}}\int dp^{z}\frac{p^{z}(p^{z}-\mu_{5})}{E_{p^{z}}^{(0)}}\left\{ \frac{1}{1+\exp\left[\beta\left(E_{p^{z}}^{(0)}-\mu\right)\right]}+\frac{1}{1+\exp\left[\beta\left(E_{p^{z}}^{(0)}+\mu\right)\right]}-1\right\} \nonumber \\
 &  & +\frac{B_{0}}{(2\pi)^{2}}\int dp^{z}\sum_{n>0,s}\frac{(p^{z})^{2}}{E_{p^{z}s}^{(n)}}\left[1-\frac{s\mu_{5}}{\sqrt{(p^{z})^{2}+2nB_{0}}}\right]\nonumber \\
 &  & \qquad\times\left\{ \frac{1}{1+\exp\left[\beta\left(E_{p^{z}s}^{(n)}-\mu\right)\right]}+\frac{1}{1+\exp\left[\beta\left(E_{p^{z}s}^{(n)}+\mu\right)\right]}-1\right\} ,\nonumber \\
\xi_{B} & = & \frac{1}{4\pi^{2}}\int dp^{z}\ p^{z}\left\{ \frac{1}{1+\exp\left[\beta\left(E_{p^{z}}^{(0)}-\mu\right)\right]}-\frac{1}{1+\exp\left[\beta\left(E_{p^{z}}^{(0)}+\mu\right)\right]}+1\right\} .\label{eq:quantities in Landau levels}
\end{eqnarray}
In these formula we have kept vacuum contributions, e.g. the last
term ``$1$'' in curly braces. But in practice one should neglect
the vacuum part otherwise the results will diverge. In Figs. \ref{fig:Energy-density-B},
\ref{fig:Transverse-pressure}, and \ref{fig:Longitudinal-pressure-as}
we compare the energy density and pressure from Landau quantization
with those from chiral quantization in Eqs. (\ref{eq:energy-density with chiral})
and (\ref{eq:pressure with chiral}). We find that in the weak-field
limit, these two approaches coincide with each other. When the field
strength increases, the transverse pressure decreases while the longitudinal
pressure increases. The decrease of the transverse pressure is attributed
to the lowest Landau level: in a strong field, the fermions are more
likely to stay in the lowest Landau level, which does not contribute
to the transverse pressure. The field-strength dependence of the energy
density is a little complicated, for some parameter configurations,
the ratio first decreases and then increases with growing field strength.

\begin{figure}
\includegraphics[width=8cm]{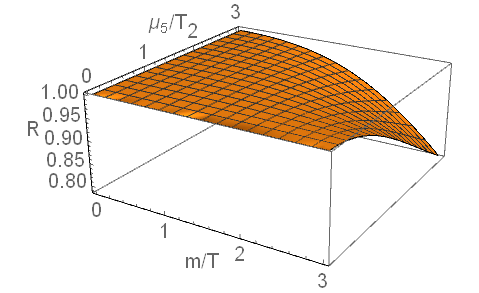}\caption{\label{fig:Coefficient-xiB}The ratio between the coefficient $\xi_{B}$
calculated using Eq. (\ref{eq:quantities in Landau levels}) and the
one in Eq. (\ref{eq:semi-classical xiB}), as a function of the mass
$m/T$ and $\mu_{5}/T$. The chemical potential is set to $\mu/T=3$.}

\end{figure}

On the other hand, the chiral chemical potential also induces an energy
flux along the magnetic field direction. Note that if we adopt the
Landau quantization, the energy flux $\mathbb{T}_{\text{mat}}^{03}$
exists but the momentum density $\mathbb{T}_{\text{mat}}^{30}$ vanishes,
which results in a non-symmetric $\mathbb{T}_{\text{mat}}^{\mu\nu}$.
But in the semi-classical results, $\mathbb{T}_{\text{mat}}^{03}$
and $\mathbb{T}_{\text{mat}}^{30}$ take the same value and $\mathbb{T}_{\text{mat}}^{\mu\nu}$
is symmetric. In Fig. \ref{fig:Coefficient-xiB} we compute the ratio
between the coefficient $\xi_{B}$ in the Landau-quantization calculation
in Eq. (\ref{eq:quantities in Landau levels}) and the one from the
semi-classical method in Eq. (\ref{eq:semi-classical xiB}). From
this figure we observe that these two results coincide in a wide parameter
range. When $m/T$ and $\mu_{5}/T$ are significantly large, the semi-classical
method again overestimates $\xi_{B}$.

\subsection{Pair production\label{subsec:Pair-production}}

\subsubsection{In Sauter-type field}

In this subsection we will focus on pair-production processes in the
presence of an electric field. As shown in Eq. (\ref{eq:density of pairs}),
the number of pairs can be expressed in terms of the Wigner function.
First we focus on the Sauter-type field, using Eqs. (\ref{eq:solution of Wigner function in Electric}),
(\ref{eq:basis functions 123}), and (\ref{eq:density of pairs}),
we obtain
\begin{equation}
n_{\text{pair}}(t,\mathbf{p})=\frac{m_{T}\chi_{2}(t,\mathbf{p})+p^{z}\chi_{1}(t,\mathbf{p})}{2E_{\mathbf{p}}}C_{1}\left(\mathbf{p}-\int_{t_{0}}^{t}dt^{\prime}E(t^{\prime})\mathbf{e}^{z}\right)+\text{const}.,\label{eq:pair spectrum Sauter}
\end{equation}
where the transverse mass $m_{T}=\sqrt{m^{2}+\mathbf{p}_{T}^{2}}$
and energy $E_{\mathbf{p}}=\sqrt{m^{2}+\mathbf{p}^{2}}$. Here we
have dropped all the spatial dependence. We further assumed that the
distribution function takes the equilibrium form,
\begin{equation}
C_{1}(\mathbf{p})=\frac{2}{(2\pi)^{3}}\left[\frac{1}{1+\exp[\beta(E_{\mathbf{p}}-\mu)]}+\frac{1}{1+\exp[\beta(E_{\mathbf{p}}+\mu)]}-1\right],
\end{equation}
In the pair spectrum (\ref{eq:pair spectrum Sauter}), the constant
term is expected to cancel with the vacuum contribution, thus here
we take the constant term to be $2/(2\pi)^{3}$.

\begin{figure}
\includegraphics[width=8cm]{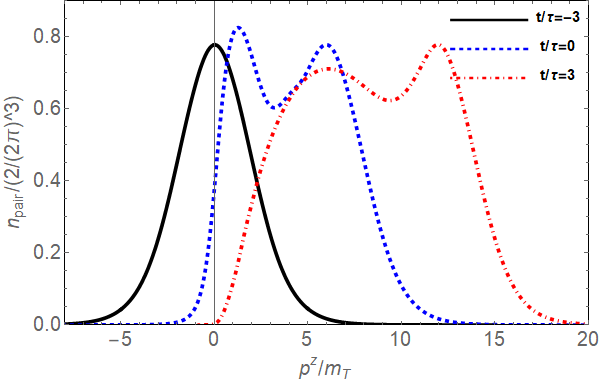}

\caption{\label{fig:Spectrums-of-pairs}Pair spectra at $t=-3\tau$ (solid
line), $t=0$ (dashed line), and $t=3\tau$ (dot-dashed line) for
a thermal system with $T/m_{T}=1$ and $\mu/m_{T}=2$ in a Sauter-type
field $E(t)=E_{0}\cosh^{-2}(t/\tau)$ with the peak value $E_{0}/m_{T}^{2}=3$
and the width $\tau=2/m_{T}$. The transverse mass $m_{T}=\sqrt{m^{2}+\mathbf{p}_{T}^{2}}$
is taken to be the energy unit. The $x$-axis is the dimensionless
longitudinal kinetic momentum, while the $y$-axis is the pair density
in phase space. }
\end{figure}

For the Sauter-type field, the coefficients $\chi_{2}(t,\mathbf{p})$
and $\chi_{1}(t,\mathbf{p})$ can be numerically computed from Eq.
(\ref{eq:equilibrium C1C2}). The solution is proven to depend on
the transverse mass $m_{T}$ and the longitudinal kinetic momentum
$p^{z}$. For simplicity we take $m_{T}$ as the energy unit and all
quantities are described by dimensionless variables, such as the temperature
$\tilde{T}=T/m_{T}$, the chemical potential $\tilde{\mu}=\mu/m_{T}$,
and the longitudinal kinetic momentum $\tilde{p}^{z}=p^{z}/m_{T}$.
As an example, we numerically calculate the evolution of the pair
spectrum for a Sauter-type field with the peak value $E_{0}/m_{T}^{2}=3$
and width $\tau=2/m_{T}$ in a thermal system with $T/m_{T}=1$ and
$\mu/m_{T}=2$. We take three typical moments in time $t=-3\tau$,
$0$, and $3\tau$. The pair spectra are given in Fig. \ref{fig:Spectrums-of-pairs}.
Since the electric field will be less than $1\%$ for $t<-3\tau$,
the initial condition for this system at $t=-\infty$ is similar to
the one for $t=-3\tau$ because before this moment the electric field
is not strong enough to generate any effect. After the time $t=3\tau$,
the functions $\chi_{1}(t,\mathbf{p})$ and $\chi_{2}(t,\mathbf{p})$
still evolve with time, but the pair spectrum stays unchanged, which
means that there is no more pair production after $t=3\tau$. In Fig.
(\ref{fig:Spectrums-of-pairs}), we observe that the spectra at $t=0$
and $t=3\tau$ have two peaks. We identify the peaks on the right-hand-side
as contribution from initially existing particles which are accelerated
by the electric field. Since the shift of the longitudinal momentum
is given by $\int_{-\infty}^{t}dt^{\prime}E(t^{\prime})$, we can
obtain that this shift is $E_{0}\tau$ at $t=0$, and $2E_{0}\tau$
at $t=3\tau$, which agrees with the locations of peaks. The peaks
on the left-hand-side in the spectra at $t=0$ and $t=3\tau$ can
be identified as the contribution from the pair production.

\begin{figure}
\includegraphics[width=8cm]{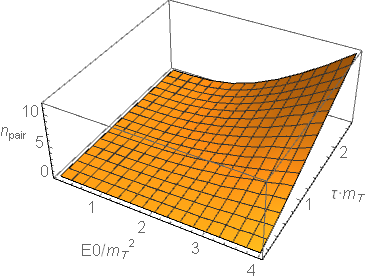}

\caption{\label{fig:Pair-production-Sauter}The total number of pairs produced
in a Sauter-type field in a thermal system with $T/m_{T}=1$ and $\mu/m_{T}=3$
as a function of $E_{0}/m_{T}^{2}$ and $\tau m_{T}$.}

\end{figure}

Now we compute the total number of pairs generated in the Sauter-type
field. Here we have four variables in this case: the peak value $E_{0}$
and the width $\tau$ for the Sauter-type field, and the temperature
$T$ and the chemical potential $\mu$ for the initial thermal equilibrium
state. First we set the thermodynamical quantities to $T/m_{T}=1$
and $\mu/m_{T}=3$ and study the dependence with respect to $E_{0}$
and $\tau$. In Fig. \ref{fig:Pair-production-Sauter} we plot the
total number of produced pairs as a function of $E_{0}$ and $\tau$.
We find that more pairs are generated for a larger peak value $E_{0}$
or a longer lifetime $\tau$, as expected.

\begin{figure}
\includegraphics[width=8cm]{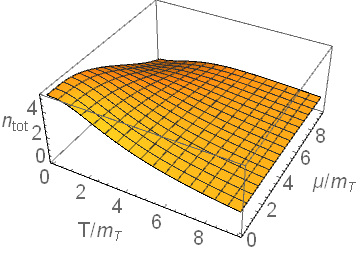}

\caption{\label{fig:Total pair number Tmu dependence}The total number of pairs
generated in a Saute-type field $E(t)=E_{0}\cosh^{-2}(t/\tau)$ with
the peak value $E_{0}/m_{T}^{2}=3$ and the width $\tau=2/m_{T}$
as a function of the thermodynamical quantities $T/m_{T}$ and $\mu/m_{T}$.}

\end{figure}

Then we take $E_{0}/m_{T}^{2}=3$ and $\tau=2/m_{T}$ and study the
dependence with respect to $T$ and $\mu$. The results are shown
in Fig. \ref{fig:Total pair number Tmu dependence}. We find that
the total number of produced pairs reaches a maximum value at $\mu=T=0$.
When the temperature increases, or the chemical potential increases,
the pair-production is suppressed. This agrees with our expectation
because in high-$T$ or high-$\mu$ system the quantum states are
more likely to be occupied and the production of new pairs is suppressed
due to the Pauli exclusion principle.

\subsubsection{In parallel electromagnetic fields}

In subsection \ref{subsec:Fermions-in-electric} we have analytically
computed the Wigner function in the case of a constant electric field.
Meanwhile we also analytically computed the Wigner function in the
case of constant parallel electromagnetic fields in subsection \ref{subsec:Fermions-in-parallel-EB}.
Since the results in parallel electromagnetic fields reduce to the
ones in a pure electric field, we will skip the pair production in
a constant electric field and directly focus on the process in the
presence of both electric and magnetic fields. The pair spectrum is
related to the Wigner function as shown in Eq. (\ref{eq:density of pairs}).
Analogous to this equation, for a system in a constant magnetic field,
the eigenenergies are replaced by the Landau energy levels $E_{p^{z}}^{(n)}=\sqrt{m^{2}+(p^{z})^{2}+2nB_{0}}$,
and the number of pairs in the $n$-th Landau level is
\begin{equation}
n^{(n)}(t,\mathbf{p})=\frac{m\mathcal{F}^{(n)}(t,\mathbf{p})+\mathbf{p}\cdot\boldsymbol{\mathcal{V}}^{(n)}(t,\mathbf{p})}{2E_{p^{z}}^{(n)}}+\text{const.}.
\end{equation}
Here $\mathcal{F}^{(n)}$ and $\boldsymbol{\mathcal{V}}^{(n)}$ are
 components of the Wigner function where the superscript $(n)$ labels
the contribution from the $n$-th Landau level. Then the pair-production
rate in the $n$-th Landau level is calculated via
\begin{equation}
\frac{d}{dt}n^{(n)}(t)=\frac{1}{2}\frac{d}{dt}\int d^{3}\mathbf{p}\frac{m\mathcal{F}^{(n)}(t,\mathbf{p})+\mathbf{p}\cdot\boldsymbol{\mathcal{V}}^{(n)}(t,\mathbf{p})}{E_{p^{z}}^{(n)}}.
\end{equation}
Employing the results in Eq. (\ref{eq:Wigner function in parallel EM fields})
into the pair-production rate, we obtain
\begin{eqnarray}
\frac{d}{dt}n^{(n)} & = & \frac{1}{2}\frac{d}{dt}\int d^{3}\mathbf{p}\left[\frac{\eta^{(n)}}{E_{p^{z}}^{(n)}}\sqrt{\frac{E_{0}}{2}}d_{2}\left(\eta^{(n)},\sqrt{\frac{2}{E_{0}}}p^{z}\right)+\frac{p^{z}}{E_{p^{z}}^{(n)}}d_{1}\left(\eta^{(n)},\sqrt{\frac{2}{E_{0}}}p^{z}\right)\right]\nonumber \\
 &  & \qquad\times C_{1}^{(n)}(p^{z}-E_{0}t)\Lambda_{+}^{(n)}(p_{T}),
\end{eqnarray}
where $C_{1}^{(n)}$ is given in (\ref{eq:equilibrium C1C2}). The
integration over $\mathbf{p}_{T}$ can be performed using relation
(\ref{eq:pt integration of Lambda_pm}). We also replace the kinetic
momentum $p^{z}$ by the canonical one $q^{z}=p^{z}-E_{0}t$. Then
we obtain the pair-production rate in a multi-particle system,
\begin{equation}
\frac{d}{dt}n^{(n)}=\int dq^{z}\left[1-f^{(+)(n)}(q^{z})-f^{(-)(n)}(q^{z})\right]\frac{d}{dt}n_{vac}^{(n)}(t,q^{z}),\label{eq:pair production rate}
\end{equation}
where $\frac{d}{dt}n_{\text{vac}}^{(n)}(t,q^{z})$ is the pair-production
rate in vacuum for given quantum numbers $n$ and $p^{z}$,
\begin{eqnarray}
\frac{d}{dt}n_{\text{vac}}^{(n)}(t,q^{z}) & = & -\left(1-\frac{\delta_{n0}}{2}\right)\frac{B_{0}E_{0}}{(2\pi)^{2}}\frac{d}{dq^{z}}\left\{ \frac{\eta^{(n)}}{E_{q^{z}+E_{0}t}^{(n)}}\sqrt{\frac{E_{0}}{2}}d_{2}\left[\eta^{(n)},\sqrt{\frac{2}{E_{0}}}\left(q^{z}+E_{0}t\right)\right]\right.\nonumber \\
 &  & \qquad\left.+\frac{q^{z}+E_{0}t}{E_{q^{z}+E_{0}t}^{(n)}}d_{1}\left[\eta^{(n)},\sqrt{\frac{2}{E_{0}}}\left(q^{z}+E_{0}t\right)\right]\right\} .
\end{eqnarray}
Summing Eq. (\ref{eq:pair production rate}) over all Landau levels
yields the total pair-production rate. Here we notice that the pair
production in the lowest Landau level is suppressed by the factor
$1-\frac{\delta_{n0}}{2}$. This is because the spin is not degenerate
for the lowest Landau level while is two-fold degenerate for the higher
Landau levels. In Eq. (\ref{eq:pair production rate}), the distribution
of fermions and anti-fermions appears in the square bracket, which
suppresses the pair-production due to the Pauli exclusion principle.
Moreover, if $f^{(+)(n)}(q^{z})+f^{(-)(n)}(q^{z})>1$, the pair-production
rate will have the opposite sign with the one in the vacuum. This
case corresponds to a system where almost fermion and anti-fermion
states are already occupied. Thus the pair annihilation is more likely
to happen than the pair creation. In a thermal equilibrium system
with zero chemical potential and non-zero temperature, the suppression
factor is $\tanh(\beta E_{q^{z}}^{(n)}/2)$, which suppresses the
production of pairs with small energies. This factor agrees with Ref.
\cite{Kim:2008em}. Later on we will discuss the pair-production in
finite chemical potential and temperature.

First we consider the pair-production in vacuum. The distribution
function $f^{(\pm)(n)}(q^{z})$ is set to zero and the pair-production
rate in Eq. (\ref{eq:pair production rate}) can be calculated using
the method of integrating by parts. The asymptotic behavior of $d_{1,2}$
is given in Eqs. (\ref{eq:asympotic behaviour 1}), (\ref{eq:asympotic behaviour 2}),
thus the results read
\begin{equation}
\frac{d}{dt}n_{\text{vac}}^{(n)}=\left(1-\frac{\delta_{n0}}{2}\right)\frac{B_{0}E_{0}}{2\pi^{2}}\exp\left(-\pi\frac{m^{2}+2nB_{0}}{E_{0}}\right),
\end{equation}
and the total pair-production rate is
\begin{equation}
\frac{d}{dt}\sum_{n=0}^{\infty}n_{\text{vac}}^{(n)}=\frac{B_{0}E_{0}}{2\pi^{2}}\exp\left(-\pi\frac{m^{2}}{E_{0}}\right)\coth\left(\pi\frac{B_{0}}{E_{0}}\right),
\end{equation}
which agrees with previous results of Refs. \cite{Nikishov:1969tt,Bunkin:1970iz,Popov:1971iga}.
We find that this rate is enhanced for a large magnetic field comparing
to the one in a pure electric field.

\begin{figure}[h]
\includegraphics[width=8cm]{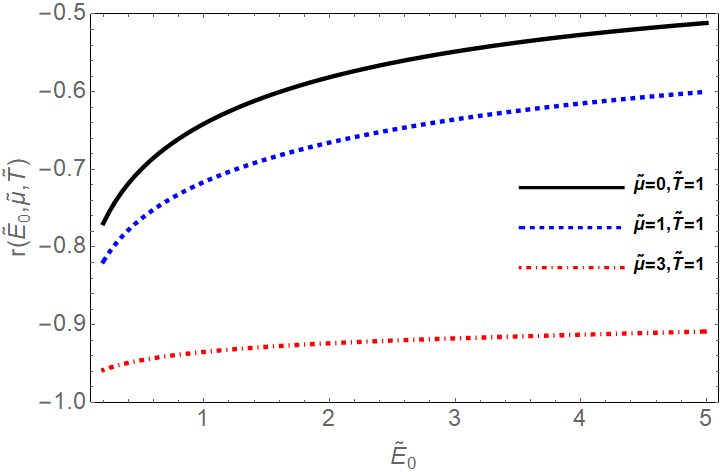}\caption{\label{fig:Ratio-between-thermal}The ratio of the thermal contribution
to the vacuum one. We fix the temperature $\tilde{T}=1$ and take
three typical values for the chemical potential, 1) $\tilde{\mu}=0$
(solid line) for a system without net fermion number, 2) $\tilde{\mu}=1$
(dashed line) for a system with a medium chemical potential, and 3)
$\tilde{\mu}=3$ (dash-dotted line) for a system with significantly
large chemical potential.}
\end{figure}

\begin{figure}
\includegraphics[width=8cm]{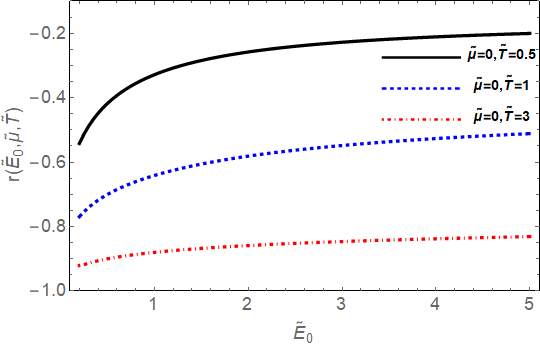}\caption{\label{fig:Ratio-between-thermal-1}The ratio of the thermal contribution
to the vacuum one in a system without net fermion number $\tilde{\mu}=0$.
We choose a low temperature $\tilde{T}=0.5$ (solid line), a medium
temperature $\tilde{T}=1$ (dashed line) and a high temperature $\tilde{T}=3$
(dash-dotted line).}
\end{figure}

For more general distribution functions $f^{(\pm)(n)}(q^{z})$, the
pair-production rate (\ref{eq:pair production rate}) requires a numerical
calculation. However we find that the function $d_{2}$ oscillate
strongly for large $q^{z}$, which makes the numerical integration
slowly converging. In order to solve this problem, we separate the
production rate into a vacuum part and a thermal part, where the thermal
part is given by
\begin{equation}
\frac{d}{dt}n_{\text{thermal}}^{(n)}=-\int dq^{z}\left[f^{(+)(n)}(q^{z})+f^{(-)(n)}(q^{z})\right]\frac{d}{dt}n_{\text{vac}}^{(n)}(t,q^{z}).
\end{equation}
In general the distributions converge quickly at large $q^{z}$. So
the numerical calculation for this thermal part is easier than directly
calculating Eq. (\ref{eq:pair production rate}). In order to show
the thermal effect we consider a thermal equilibrium system where
the distribution functions are given by
\begin{equation}
f^{(\pm)(n)}(q^{z})=\frac{1}{1+\exp\left[\beta\left(E_{q^{z}}^{(n)}\mp\mu\right)\right]},
\end{equation}
We introduce a function to describe the ratio of the thermal contribution
to the vacuum contribution. The ratio depends on several dimensionless
parameters, the time $\tilde{t}=m^{(n)}t$, the electric field strength
$\tilde{E}_{0}=E_{0}/\left(m^{(n)}\right)^{2}$, the temperature $\tilde{T}=T/m^{(n)}$
and the chemical potential $\tilde{\mu}=\mu/T$. The pair-production
rate in the $n$-th Landau level is given by
\begin{equation}
\frac{d}{dt}n^{(n)}=\left[1+r\left(\tilde{t},\tilde{E}_{0},\tilde{T},\tilde{\mu}\right)\right]\frac{d}{dt}n_{\text{vac}}^{(n)}.
\end{equation}
Now we choose the moment $\tilde{t}=0$ when the kinetic momentum
equals the canonical one. This reflects the pair-production around
thermal equilibrium. In Figs. \ref{fig:Ratio-between-thermal} and
\ref{fig:Ratio-between-thermal-1} we plot $r\left(0,\tilde{E}_{0},\tilde{T},\tilde{\mu}\right)$
as a function of $\tilde{E}_{0}$ of for various temperature and chemical-potential
configurations. We observe that the ratio lies between $\left[-1,\,0\right]$,
which means that the pair production is suppressed in the thermal
system. We also observe that the suppression, which was described
by the absolute value of $r\left(\tilde{t},\tilde{E}_{0},\tilde{T},\tilde{\mu}\right)$,
is larger for larger chemical potentials or larger temperatures. This
agrees with our understanding about the Pauli exclusion principle:
energy states are more likely to be occupied in a system with high
$T$ or high $\mu$, and the existing particles will prohibit the
generation of pairs with the same quantum number.

\newpage{}

$\,$

\newpage{}

\section{Summary\label{sec:Summary}}

We study the Wigner-function for spin-$1/2$
particles and then chiral effects and
pair production in electromagnetic fields. The Wigner
function is defined as a semi-classical distribution function in phase space, which is a complex valued $4\times4$ matrix. It
can be expanded in terms of the generators of the Clifford algebra
$\Gamma_{i}$. The expansion coefficients are scalar,
pseudoscalar, vector, axial-vector, and tensor according to their
transformation properties under the Lorentz and parity transformations.
An integration of the Wigner function over the momentum can give various kinds
of macroscopic physical quantities such as the fermion current, the
spin polarization and the magnetic dipole moment.

Since the Wigner function is constructed from the Dirac field, we
can obtain the kinetic equations for the Wigner function from the
Dirac equation. In this thesis, we derive the equation in a Dirac form,
which is linear in differential operators. Meanwhile, we also obtain
the equation in a Klein-Gordon form, which is second order in differential operators.
These equations are then decomposed in terms of $\Gamma_{i}$, as
we do for the Wigner function, which provide several partial
differential equations. Fortunately, the equations for Wigner-function components are
not independent from each other. Eliminating the redundant equations,
we obtain two possible ways of computing the Wigner function in
the massive case. The redundancy is based on the fact that the vector
and axial-vector components $\mathcal{V}^{\mu}$ and $\mathcal{A}^{\mu}$
can be expressed in terms of the scalar, pseudo-scalar,
and tensor components $\mathcal{F}$, $\mathcal{P}$, and $\mathcal{S}^{\mu\nu}$,
or vice versa. Thus one approach to solve the system is to choose $\mathcal{V}^{\mu}$
and $\mathcal{A}^{\mu}$ as basis functions and focus on their on-shell
conditions. Meanwhile, the other approach is to choose $\mathcal{F}$,
$\mathcal{P}$, and $\mathcal{S}^{\mu\nu}$ as basis functions. Carrying
out an expansion in $\hbar$, known as the semi-classical
expansion, we obtain a general solution of the Wigner function up
to first order in $\hbar$. The two approaches mentioned above are
proven to be equivalent to each
other. The final solution only has four independent degrees of freedom,
which is proven through an eigenvalue analysis. At the linear order in $\hbar$,
the normal mass-shell $p^{2}-m^{2}=0$ is shifted by the spin-magnetic
coupling.

We also reproduce the Wigner function in the massless case through
the semi-classical expansion. In the massless case, fermions can
be separated into two groups according to their chirality. Using $\mathcal{V}^{\mu}$
and $\mathcal{A}^{\mu}$, we can construct the LH and RH currents, which
are then solved up to the linear order in $\hbar$. The remaining components $\mathcal{F}$,
$\mathcal{P}$, and $\mathcal{S}^{\mu\nu}$ are proportional to the
particle mass and thus vanish in the massless limit. We find a direct relation between the massless Wigner
function and the massive one. This indicates that our massive results
are more general than the Chiral Kinetic Theory.

In this thesis we have discussed several analytically solvable cases.
In the following three cases, single-particle wavefunctions
can be derived from the Dirac equation analytically, which are then used to compute
the Wigner function. We only list the leading-order contributions
in spatial gradients for the Wigner functions, but deriving higher order contributions
is straightforward in our approach.
\begin{enumerate}
\item Plane-wave quantization: In this case the Dirac equation does not
contain any external interaction and thus has free plane-wave solutions.
The results obtained in this approach are the cornerstone for the
method of the semi-classical expansion: they serve as the solutions
to the zeroth order in $\hbar$, while higher-order ones can be obtained order by order.
\item Chiral quantization: In this case we introduce $\mu$ and $\mu_{5}$
as constant variables. Electromagnetic fields
are still not included. Corresponding to these variables is a contribution $\mu\hat{\mathbb{N}}+\mu_{5}\hat{\mathbb{N}}_{5}$
to the total Hamiltonian, where $\hat{\mathbb{N}}$ and $\hat{\mathbb{N}}_{5}$
are operators for the fermion number and the axial-charge.
In the massless limit, we identify $\mu$ as the vector chemical potential
and $\mu_{5}$ as the chiral chemical potential. We emphasize that
the chiral chemical potential is not well-defined in the massive case
because its conjugate quantity, the axial-charge, is not conserved.
Thus, $\mu_{5}$ is just a variable which describes
the spin imbalance. The modified Hamiltonian leads to a new Dirac
equation, which can be solved when $\mu$ and $\mu_{5}$ are constants.
The Wigner function is then constructed from the single-particle wavefunctions.
However, since the presence of $\mu$ and $\mu_{5}$ changes the Dirac
equation, the kinetic equations for the Wigner function are modified. Moreover, we cannot obtain the single-particle
wavefunction for a general space-/time-dependent $\mu$ or $\mu_{5}$.
Hence the chiral-quantization method only serve as a cross-check
for the method of semi-classical expansion.
\item Landau quantization: Based on case 2, we further introduce a constant
magnetic field. The energy levels are described by the Landau
levels with modifications from $\mu$ and $\mu_{5}$. This allows
us to explicitly study phenomena in a magnetic field, such as the
CME, the CSE, and the anomalous energy flux. Since the
field changes the energy spectrum, the fermion number
density, the energy density, and the pressure depend on the strength
of the magnetic field.
\end{enumerate}
Based on the plane-wave quantization, we carry out
a semi-classical expansion in $\hbar$. The Wigner function is then
solved up to $\mathcal{O}(\hbar)$. Note that the method of the semi-classical
expansion can be used for an arbitrary space-time dependent electromagnetic
field. In this method, we put $\mu$ and $\mu_{5}$ into thermal
equilibrium distributions instead of into the Hamiltonian and make
the specific assumption that all fermions are longitudinally polarized.
These treatments are naive extensions of the massless case.
Numerical calculations show that the fermion number density and
the axial-charge density coincide with the ones
from the chiral quantization if $\mu_{5}$ and $m$ are comparable
or smaller than the temperature, and so do the energy density and pressures.

Besides the above three analytically solvable cases, we also discuss
the Wigner function in an electric field. Based on the results from
the plane-wave quantization and those from the Landau quantization,
we obtained the Wigner function in a constant electric
field via a dynamical treatment. Pair-production rate is then computed,
which proves to be enhanced by a parallel magnetic field, and suppressed
by the temperature and chemical potential. The suppression of pair
production in a thermal system is attributed to the Pauli exclusion
principle.

The method of the semi-classical expansion provides a general way
to compute spin corrections. At the zeroth order in $\hbar$, we reproduce
the classical spinless Boltzmann equation. At the linear order in $\hbar$, spin
corrections, such as the energy shift by the spin-magnetic coupling,
arise naturally. We have obtain a general Boltzmann
equation and a general BMT equation, which govern the evolution
of the particle distribution and the spin polarization density, respectively.
However, particle collisions are not yet included. Following
the method of moments, we can extend the semi-classical results to
a hydrodynamical description, which is topic of future work.
In the semi-classical expansion method, electromagnetic
fields appear at the linear order in $\hbar$, which works in the weak-field limit.
However in the initial stage of heavy-ion collisions, the magnetic
field strength is comparable with $m_{\pi}^{2}$.
In strong-laser physics, the electromagnetic fields are significantly
strong but there are nearly no particles. Whether the semi-classical
expansion can be used in these cases needs more careful consideration.
The study of a constant magnetic field in this thesis may serve as
a starting point for the kinetic theory in a strong background field.

Another possible extension of this thesis is axial-charge production.
In the presence of parallel electric and magnetic fields, the electric
field can excite fermion pairs from vacuum and the produced
pairs are polarized by the magnetic field. As a consequence, the pair-production
in the lowest Landau level contributes to the axial-charge density.
The real-time axial-charge production of massive particles in a thermal
background has not yet been computed. The Wigner function approach
in this thesis may provide a possible approach towards this problem.

\newpage{}
$\ $
\newpage{}

\begin{acknowledgments}
First and foremost, I want to thank my supervisors Qun Wang and Dirk
H. Rischke. This project, and all the published works are under their
guidance and support. I learned a lot about physics through many discussions
with them in the past few years.

I also want to thank my collaborators, Prof. David Vasak, Ren-Hong
Fang, Nora Weickgenannt, and Dr. Enrico Speranza. We finished some
works together and both of them are expert in some field.

I want to thank the members of Qun Wang's group of University of Science
and Technology of China. I have learned a lot about programming and
numerical simulations from Hao-jie Xu and Jun-jie Zhang, and a lot
of techniques of Mathematica from Xiao-liang Xia. I specially thank
Ren-hong Fang and Yu-kun Song, who brought me to the way of Wigner
function. In life we are friends and we spent an enjoyable time at
USTC.

I want to thank members of the magneto-hydrodynamics group of Frankfurt.
I admire Dr. Leonardo Tinti for his wide range of knowledge. I also
want to thank Nora Weickgenannt and David Wagner for helpful discussions
about the future of the Wigner-function approach.

I want to thank Prof. Igor Shovkovy, Prof. Shijun Mao, Prof. Shi Pu,
Dr. Xingyu Guo, Dr. Kai Zhou, Dr. Ziyue Wang, and Yu-chen Liu for
enlightening discussions.

I especially want to thank Yuyu Zhang, without her company I could
not finish this work.

This project was supported by the China Scholarship Council and the
Deutsche Forschungsgemeinschaft (DFG, German Research Foundation)
through the CRC-TR 211 ``Strong-interaction matter under extreme
conditions'' - project number 315477589 - TRR 211.
\end{acknowledgments}

\newpage{}

\appendix

\section{Gamma matrices\label{sec:Gamma-matrices}}

In this section we list the gamma matrices used throughout this paper
and discuss their properties. The gamma matrices $\gamma^{\mu}$ should
satisfy the following anti-commutation relation
\begin{equation}
\left\{ \gamma^{\mu},\gamma^{\nu}\right\} =2g^{\mu\nu}\mathbb{I}_{4},\label{eq:anticommutaion}
\end{equation}
where $g^{\mu\nu}$ is the Minkowski metric. In principle there are
many ways to construct the gamma matrices and the above anti-commutation
relation is the only constraint. We can find one $4\times4$ representation,
in which the gamma matrices are given by
\begin{equation}
\gamma^{\mu}=\left(\begin{array}{cc}
0 & \sigma^{\mu}\\
\bar{\sigma}^{\mu} & 0
\end{array}\right),\label{eq:gamma matrices-1}
\end{equation}
with $\sigma^{\mu}=\left(\mathbb{I}_{2},\,\sigma^{1},\,\sigma^{2},\,\sigma^{3}\right)$
and $\bar{\sigma}^{\mu}=\left(\mathbb{I}_{2},\,-\sigma^{1},\,-\sigma^{2},\,-\sigma^{3}\right)$.
Here $\mathbb{I}_{2}$ is a $2\times2$ unit matrix and $\sigma^{\{1,2,3\}}$
are the Pauli matrices. This representation is known as the Weyl or
chiral representation, which is convenient to deal with massless particles.
The Weyl representation is used through out this thesis.

The Hermitian conjugate of the gamma matrices $\gamma^{\mu}$ satisfy
the following relation,
\begin{equation}
(\gamma^{\mu})^{\dagger}=\gamma^{0}\gamma^{\mu}\gamma^{0}.
\end{equation}
The anti-commutation relation (\ref{eq:anticommutaion}) indicates
that any product of several gamma matrices can be expressed in terms
of the anti-symmetric combinations $\gamma^{[\mu_{1}}\gamma^{\mu_{2}}\cdots\gamma^{\mu_{n}]}$
with $n=0,1,2,3,4$. The maximum $n$ is $4$, which equals the number
of $\gamma^{\mu}$ according to the Pigeonhole principle \cite{Pigeonhole:website}.
When taking the Hermitian conjugate, these anti-symmetric combinations
satisfy
\begin{equation}
(\gamma^{[\mu_{1}}\gamma^{\mu_{2}}\cdots\gamma^{\mu_{n}]})^{\dagger}=\begin{cases}
\gamma^{0}\gamma^{[\mu_{1}}\gamma^{\mu_{2}}\cdots\gamma^{\mu_{n}]}\gamma^{0}, & n=0,1,4,\\
-\gamma^{0}\gamma^{[\mu_{1}}\gamma^{\mu_{2}}\cdots\gamma^{\mu_{n}]}\gamma^{0}, & n=2,3.
\end{cases}
\end{equation}
Thus we define

\begin{eqnarray}
\gamma^{5} & \equiv & i\gamma^{0}\gamma^{1}\gamma^{2}\gamma^{3},\nonumber \\
\sigma^{\mu\nu} & \equiv & \frac{i}{2}\left[\gamma^{\mu},\gamma^{\nu}\right],
\end{eqnarray}
and then any combination of $\gamma^{\mu}$ can be written in terms
of $\Gamma_{i}=\{\mathbb{I}_{4},\ i\gamma^{5},\ \gamma^{\mu},\ \gamma^{5}\gamma^{\mu},\ \frac{1}{2}\sigma^{\mu\nu}\}$,
which are 16 matrices in total. These matrices are also known as the
independent generators of the Clifford algebra, which automatically
satisfy
\begin{equation}
(\Gamma_{i})^{\dagger}=\gamma^{0}\Gamma_{i}\gamma^{0},\label{eq:Cliddord algebra}
\end{equation}
and will be used to expand our Wigner function.

In Sec. \ref{sec:Overview-of-Wigner} we find it useful to calculate
the commutators and anti-commutators between $\sigma^{\mu\nu}$ and
the generators of the Clifford algebra $\Gamma_{i}$. %
{} Here we list all results
\begin{eqnarray}
[\sigma^{\mu\nu},\mathbb{I}_{4}] & = & 0,\nonumber \\{}
[\sigma^{\mu\nu},-i\gamma^{5}] & = & 0,\nonumber \\{}
[\sigma^{\mu\nu},\gamma^{\sigma}] & = & 2i(g^{\nu\sigma}\gamma^{\mu}-g^{\mu\sigma}\gamma^{\nu}),\nonumber \\{}
[\sigma^{\mu\nu},\gamma^{5}\gamma^{\sigma}] & = & 2i(g^{\nu\sigma}\gamma^{5}\gamma^{\mu}-g^{\mu\sigma}\gamma^{5}\gamma^{\nu}),\nonumber \\{}
[\sigma^{\mu\nu},\sigma^{\sigma\rho}] & = & 2i\left(g^{\mu\rho}\sigma^{\nu\sigma}+g^{\nu\sigma}\sigma^{\mu\rho}-g^{\mu\sigma}\sigma^{\nu\rho}-g^{\nu\rho}\sigma^{\mu\sigma}\right),\nonumber \\
\{\sigma^{\mu\nu},\mathbb{I}_{4}\} & = & 2\sigma^{\mu\nu},\nonumber \\
\{\sigma^{\mu\nu},-i\gamma^{5}\} & = & \epsilon^{\mu\nu\alpha\beta}\sigma_{\alpha\beta},\nonumber \\
\{\sigma^{\mu\nu},\gamma^{\alpha}\} & = & 2\epsilon^{\mu\nu\alpha\beta}\gamma^{5}\gamma_{\beta},\nonumber \\
\{\sigma^{\mu\nu},\gamma^{5}\gamma^{\alpha}\} & = & 2\epsilon^{\mu\nu\alpha\beta}\gamma_{\beta},\nonumber \\
\{\sigma^{\mu\nu},\sigma^{\sigma\rho}\} & = & 2g^{\mu[\sigma}g^{\rho]\nu}+2i\epsilon^{\mu\nu\sigma\rho}\gamma^{5}.\label{eq:Commutators and anticommutators}
\end{eqnarray}
On the other hand, the matrices $\Gamma_{i}$ can be constructed from
the Pauli matrices by taking tensor products. The tensor-product form
would be useful when calculating the Wigner function from the quantized
field operator,
\begin{eqnarray}
\mathbb{I}_{4} & = & \mathbb{I}_{2}\otimes\mathbb{I}_{2},\nonumber \\
\gamma^{5} & = & -\sigma^{3}\otimes\mathbb{I}_{2},\nonumber \\
\gamma^{0} & = & \sigma^{1}\otimes\mathbb{I}_{2},\nonumber \\
\boldsymbol{\gamma} & = & i\sigma^{2}\otimes\boldsymbol{\sigma},\nonumber \\
\gamma^{5}\gamma^{0} & = & -i\sigma^{2}\otimes\mathbb{I}_{2},\nonumber \\
\gamma^{5}\boldsymbol{\gamma} & = & -\sigma^{1}\otimes\boldsymbol{\sigma},\nonumber \\
\sigma^{0j} & = & i\gamma^{0}\gamma^{j}=-i\sigma^{3}\otimes\sigma^{j},\nonumber \\
\sigma^{jk} & = & i\gamma^{j}\gamma^{k}=\epsilon^{jkl}\mathbb{I}_{2}\otimes\sigma^{l}.\label{eq:relation between gamma matrices and Pauli}
\end{eqnarray}
The tensor product of two matrices, also known as the Kronecker product
and denoted by $\otimes$, is a generalization of the outer product
for two vectors. For example, considering two matrices $A$ and $B$,
the tensor product $A\otimes B$ is given by
\begin{equation}
A\otimes B=\left(\begin{array}{cccc}
a_{11}B & a_{12}B & \cdots & a_{1n}B\\
a_{21}B & a_{22}B & \cdots & a_{2n}B\\
\vdots & \vdots & \ddots & \vdots\\
a_{m1}B & a_{m2}B & \cdots & a_{mn}B
\end{array}\right),
\end{equation}
where $a_{ij}$ is the element of $A$ in the $i$-th row and $j$-th
column. We find it useful to emphasize the mixed-product property,
\begin{equation}
(A\otimes B)(C\otimes D)=(AC)\otimes(BD),\label{eq:mixed-product}
\end{equation}
where $A,\ B,\ C,$ and $D$ are matrices with proper size such that
the matrix products make sense. When taking the Hermitian conjugate,
we have the following property
\begin{equation}
(A\otimes B)^{\dagger}=A^{\dagger}\otimes B^{\dagger}.
\end{equation}
These properties are used when analytically deriving the Wigner function
in subsections \ref{subsec:Free-fermions}, \ref{subsec:Free-with-chiral},
and \ref{subsec:Fermions-in-const-B}.

\section{Auxiliary functions\label{sec:Auxiliary-functions}}

When calculating the Wigner function in electromagnetic fields, we
define some useful auxiliary functions. In this appendix we will list
these functions and briefly discuss their properties.

When calculating the Wigner function in a magnetic field, we need
to calculate the following integral
\begin{equation}
I_{ij}(p^{x},p^{y})\equiv\int dy^{\prime}\exp(ip^{y}y^{\prime})\phi_{i}\left(p^{x},\frac{y^{\prime}}{2}\right)\phi_{j}\left(p^{x},-\frac{y^{\prime}}{2}\right),\label{eq:definition of I_ij}
\end{equation}
with $\phi_{n}$ being the eigenstates of the harmonic oscillator
defined in Eq. (\ref{eq:def-phi}). Using the explicit form of $\phi_{n}$
one can calculate the integral $I_{ij}$ and obtains %
\begin{eqnarray}
I_{mn}(p^{x},p^{y}) & = & \sqrt{\frac{B_{0}}{\pi}}\frac{1}{\sqrt{2^{m+n}m!n!}}\exp\left(-\frac{p_{T}^{2}}{B_{0}}\right)\int dy^{\prime}\exp\left[-B_{0}\left(\frac{y^{\prime}}{2}-\frac{ip^{y}}{B_{0}}\right)^{2}\right]\nonumber \\
 &  & \times H_{m}\left[\sqrt{B_{0}}\left(\frac{1}{2}y^{\prime}-\frac{ip^{y}}{B_{0}}\right)+\frac{p^{x}+ip^{y}}{\sqrt{B_{0}}}\right]H_{n}\left[-\sqrt{B_{0}}\left(\frac{1}{2}y^{\prime}-\frac{ip^{y}}{B_{0}}\right)+\frac{p^{x}-ip^{y}}{\sqrt{B_{0}}}\right],\nonumber \\
\end{eqnarray}
where $p_{T}^{2}\equiv(p^{x})^{2}+(p^{y})^{2}$ is the transverse
momentum squared. Here $H_{n}(x)$ are the Hermite polynomials, whose
Taylor expansion reads
\begin{equation}
H_{n}(x+y)=\sum_{i=0}^{n}\frac{2^{i}y^{i}n!}{i!(n-i)!}H_{n-i}(x).
\end{equation}
Then $I_{mn}(p^{x},p^{y})$ can be calculated by firstly expanding
the Hermite polynomials around $\pm\sqrt{B_{0}}\left(\frac{1}{2}y^{\prime}-\frac{ip^{y}}{B_{0}}\right)$,
then using the symmetric property $H_{n}(-x)=(-1)^{n}H_{n}(x)$ and
the following orthonormality condition
\begin{eqnarray}
 &  & \sqrt{\frac{B_{0}}{\pi}}\int dy'\exp\left[-B_{0}(\frac{y'}{2}-\frac{ip^{y}}{B_{0}})^{2}\right]H_{m-i}\left[\sqrt{B_{0}}(\frac{1}{2}y'-\frac{ip^{y}}{B_{0}})\right]H_{n-j}\left[\sqrt{B_{0}}(\frac{1}{2}y'-\frac{ip^{y}}{B_{0}})\right]\nonumber \\
 &  & \qquad\qquad\qquad\qquad\qquad\qquad\qquad\qquad\qquad\qquad\qquad\qquad=2^{m-i+1}(m-i)!\delta_{m-i,n-j}.
\end{eqnarray}
Finally we obtain the result
\begin{eqnarray}
I_{mn}(p^{x},p^{y}) & = & \frac{1}{\sqrt{2^{m+n}m!n!}}\exp\left(-\frac{p_{T}^{2}}{B_{0}}\right)\nonumber \\
 &  & \qquad\times\sum_{i=0}^{m}\sum_{j=0}^{n}\frac{2^{m+1+j}m!n!}{i!j!(n-j)!}\left(\frac{p^{x}+ip^{y}}{\sqrt{B_{0}}}\right)^{i}\left(\frac{p^{x}-ip^{y}}{\sqrt{B_{0}}}\right)^{j}(-1)^{m-i}\delta_{m-i,n-j}.\label{eq:result of I_mn}
\end{eqnarray}
If we take $m=n$, then it can be written in terms of the Laguerre
polynomials
\begin{equation}
I_{nn}(p^{x},p^{y})=2(-1)^{n}\exp\left(-\frac{p_{T}^{2}}{B_{0}}\right)L_{n}\left(\frac{2p_{T}^{2}}{B_{0}}\right),
\end{equation}
where the Laguerre polynomials are defined as
\begin{equation}
L_{n}(x)=\sum_{i=0}^{n}\frac{(-1)^{i}n!}{i!i!(n-i)!}x^{i}.
\end{equation}
For simplicity we define a set of new functions $\Lambda_{\pm}^{(n)}(p_{T})$,
which depend on the magnitude of transverse momentum $p_{T}$,
\begin{equation}
\Lambda_{\pm}^{(n)}(p_{T})\equiv(-1)^{n}\left[L_{n}\left(\frac{2p_{T}^{2}}{B_{0}}\right)\mp L_{n-1}\left(\frac{2p_{T}^{2}}{B_{0}}\right)\right]\exp\left(-\frac{p_{T}^{2}}{B_{0}}\right),\label{eq:def-Lambda-n}
\end{equation}
where $n>0$ because $L_{n-1}(x)$ is not well-defined. Then $I_{mn}(p^{x},p^{y})$
can be related to these $\Lambda_{\pm}^{(n)}(p_{T})$,
\begin{eqnarray}
\frac{I_{nn}\pm I_{n-1,n-1}}{2} & = & \Lambda_{\pm}^{(n)}(p_{T}),\nonumber \\
\frac{I_{n,n-1}+I_{n-1,n}}{2} & = & \frac{p^{x}\sqrt{2nB_{0}}}{p_{T}^{2}}\Lambda_{+}^{(n)}(p_{T}),\nonumber \\
\frac{I_{n,n-1}-I_{n-1,n}}{2} & = & \frac{ip^{y}\sqrt{2nB_{0}}}{p_{T}^{2}}\Lambda_{+}^{(n)}(p_{T}),\label{eq:relation between I_mn and Lambda_pm}
\end{eqnarray}
where the last two lines can be checked by expanding $\Lambda_{+}^{(n)}(p_{T})$
into a polynomial and comparing with the left-hand-side terms, which
are calculated by Eq. (\ref{eq:result of I_mn}). For the case $n=0$
we specially define
\begin{equation}
I_{00}(p_{x},p_{y})=\Lambda_{\pm}^{(0)}(p_{T})\equiv2\exp\left(-\frac{p_{T}^{2}}{B_{0}}\right),\label{eq:def-Lambda-0}
\end{equation}
which is independent of the subscript $\pm$.

\begin{figure}
\includegraphics[width=8cm]{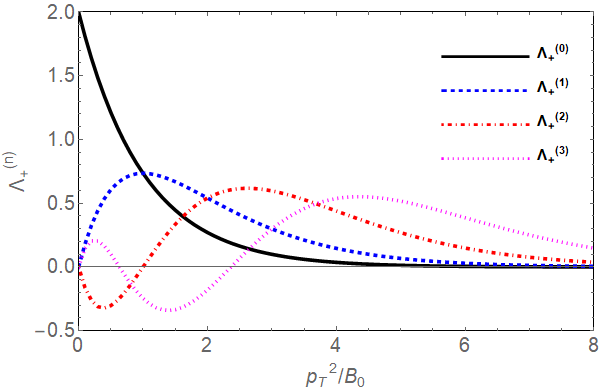}\caption{\label{fig:The-first-four-Lambda+}The first four functions of $\Lambda_{+}^{(n)}$.}
\end{figure}

\begin{figure}
\includegraphics[width=8cm]{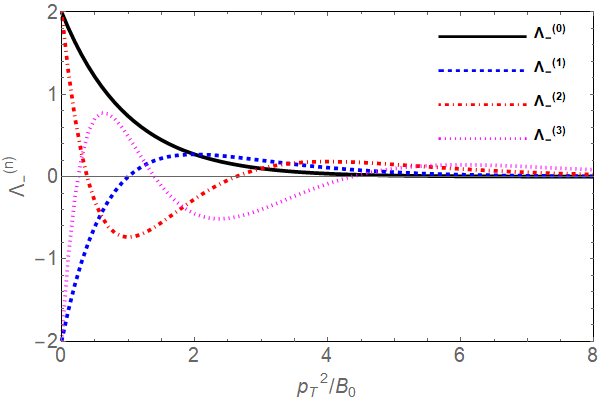}\caption{\label{fig:The-first-four-Lambda-}The first four functions of $\Lambda_{-}^{(n)}$.}
\end{figure}

\begin{figure}
\includegraphics[width=8cm]{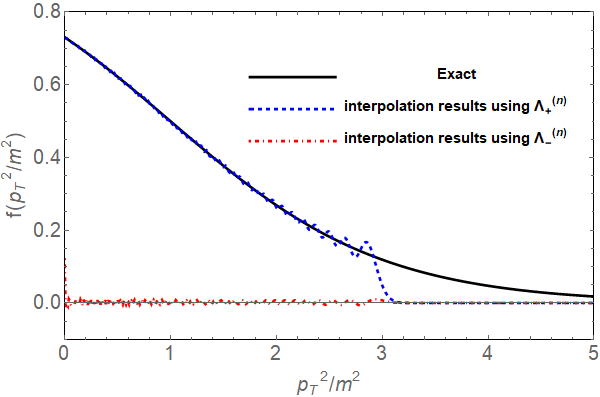}

\caption{\label{fig:Numerical-proof-in}Numerical proof in the weak-field limit.
For numerical convenience we take the particles rest mass as natural
unit of energy or momentum, and set $B_{0}=0.01\ m^{2}$. The sum
over $n$ is truncated at $n=150$. The test function (solid line)
coincides with the interpolation function, which is constructed from
$\Lambda_{+}^{(n)}$ (dashed line), while the one constructed from
$\Lambda_{-}^{(n)}$ (dot-dashed line) agrees with zero. }

\end{figure}

In the Wigner function, the functions $\Lambda_{\pm}^{(n)}(p_{T})$
in Eqs. (\ref{eq:def-Lambda-n}), (\ref{eq:def-Lambda-0}) play roles
of distribution with respect to transverse momentum. When integrating
over transverse momentum $\mathbf{p}_{T}$, $\Lambda_{+}^{(n)}(p_{T})$
give the density of states in the $n$-th Landau level, while $\Lambda_{-}^{(n)}(p_{T})$
give zero for any $n>0$,%
\begin{eqnarray}
\int\frac{d^{2}\mathbf{p}_{T}}{(2\pi)^{2}}\Lambda_{+}^{(n)}(p_{T}) & = & \frac{B_{0}}{2\pi},\nonumber \\
\int\frac{d^{2}\mathbf{p}_{T}}{(2\pi)^{2}}\Lambda_{-}^{(n)}(p_{T}) & = & 0,\ \ \ \ (n\neq0).\label{eq:pt integration of Lambda_pm}
\end{eqnarray}
Both of $\Lambda_{\pm}^{(n)}$ are symmetric with respect to $p^{x}$
and $p^{y}$ because they only depend on the magnitude $p_{T}$. In
Figs. \ref{fig:The-first-four-Lambda+} and \ref{fig:The-first-four-Lambda-}
we plot the first four $\Lambda_{\pm}^{(n)}$. We find that all these
functions converge to zero in the limit $p_{T}\rightarrow\infty$,
which is ensured by the exponential term in their definitions (\ref{eq:def-Lambda-n})
and (\ref{eq:def-Lambda-0}). In the point $p_{T}=0$, the functions
$\Lambda_{+}^{(n)}$ have zero values for all $n>0$, while $\Lambda_{-}^{(n)}(0)=2(-1)^{n}$
oscillate between $\{-2,\,2\}$. The oscillation of $\Lambda_{-}^{(n)}(0)$
is similar to the Runge's phenomenon, which occurs when using polynomial
interpolation. In fact, $\Lambda_{+}^{(n)}$ plays the role of an
interpolation function because numerically we can prove
\begin{equation}
f(p_{T}^{2})=\lim_{B_{0}\rightarrow0}\left[\frac{1}{2}f^{(0)}\Lambda_{+}^{(0)}(p_{T})+\sum_{n>0}f^{(n)}\Lambda_{+}^{(n)}(p_{T})\right],\label{eq:weak field limit-1}
\end{equation}
where $f(x)$ is an arbitrary function and $f^{(n)}$ are the values
of the function $f$ at the points $2nB_{0}$. In the weak-magnetic
field limit $B_{0}\rightarrow0$, the interpolation form on the right-hand-side
reproduces the function $f(p_{T}^{2})$. On the other hand, when $B_{0}\rightarrow0$
we also have
\begin{equation}
\lim_{B_{0}\rightarrow0}\left[\frac{1}{2}f^{(0)}\Lambda_{-}^{(0)}(p_{T})+\sum_{n>0}f^{(n)}\Lambda_{-}^{(n)}(p_{T})\right]=0,\label{eq:weak field limit-2}
\end{equation}
which is numerically proven. In Fig. \ref{fig:Numerical-proof-in}
we take an example $f(p_{T}^{2})=1/\left[1+\exp(p_{T}^{2}/m^{2}-1)\right]$,
where $m$ is the particles rest mass. Here $m$ plays the role of
the energy unit. For convenience we take $B_{0}=0.01\ m^{2}$ and
truncate the sum at $n=150$. We find that the interpolation result
formed from $\Lambda_{+}^{(n)}(p_{T})$ coincides with the original
function, while the one formed from $\Lambda_{-}^{(n)}(p_{T})$ coincides
with zero. There is some disagreement in the large $p_{T}^{2}$ region,
which is caused by the truncation at $n=150$. If the sum is calculated
without an upper limit of $n$, the results would agree with Eq. (\ref{eq:weak field limit-1}).

When taking the derivative with respect to $p^{x}$, we have the following
relations,
\begin{eqnarray}
\hbar B_{0}\partial_{p^{x}}\Lambda_{+}^{(n)}(p_{T}) & = & -2p^{x}\Lambda_{-}^{(n)}(p_{T}),\nonumber \\
\hbar B_{0}\partial_{p^{x}}\Lambda_{-}^{(n)}(p_{T}) & = & -2p^{x}\left(1-\frac{2nB_{0}}{p_{T}^{2}}\right)\Lambda_{+}^{(n)}(p_{T}),
\end{eqnarray}
where the $p^{y}$-derivative can be derived by replacingt $p^{x}\leftrightarrow p^{y}$.
These relations help when calculating the Wigner function in parallel
electromagnetic fields in subsection \ref{subsec:Fermions-in-parallel-EB}.

We define four set of basis vectors, which are 4-dimensional column
vectors,
\begin{eqnarray}
e_{1}^{(n)}(p_{T})=\left(\begin{array}{c}
\Lambda_{+}^{(n)}(p_{T})\\
0\\
0\\
\Lambda_{-}^{(n)}(p_{T})
\end{array}\right), &  & e_{2}^{(n)}(p_{T})=\left(\begin{array}{c}
\Lambda_{-}^{(n)}(p_{T})\\
0\\
0\\
\Lambda_{+}^{(n)}(p_{T})
\end{array}\right),\nonumber \\
e_{3}^{(n)}(\mathbf{p}_{T})=\left(\begin{array}{c}
0\\
p^{x}\\
p^{y}\\
0
\end{array}\right)\frac{\sqrt{2nB_{0}}}{p_{T}^{2}}\Lambda_{+}^{(n)}(p_{T}), &  & e_{4}^{(n)}(\mathbf{p}_{T})\equiv\left(\begin{array}{c}
0\\
-p^{y}\\
p^{x}\\
0
\end{array}\right)\frac{\sqrt{2nB_{0}}}{p_{T}^{2}}\Lambda_{+}^{(n)}(p_{T}).\label{eq:def-basis}
\end{eqnarray}
The first two basis vectors only depend on the magnitude $p_{T}$
of the transverse momentum, while $e_{3}^{(n)}(\mathbf{p}_{T})$ and
$e_{4}^{(n)}(\mathbf{p}_{T})$ also depend on the direction of $\mathbf{p}_{T}$.
Here the functions $\Lambda_{\pm}^{(n)}$ are defined in Eqs. (\ref{eq:def-Lambda-n})
and (\ref{eq:def-Lambda-0}). Note that when $n=0$, the last two,
$e_{3}^{(0)}(\mathbf{p}_{T})$ and $e_{4}^{(0)}(\mathbf{p}_{T})$
are not well-defined because they are zero vectors. At the same time,
due to the fact that $\Lambda_{+}^{(0)}=\Lambda_{-}^{(0)}$, we have
$e_{1}^{(0)}(p_{T})=e_{2}^{(0)}(p_{T})$. The basis vector $e_{1}^{(0)}(p_{T})$
is properly normalized with respect to an inner product of transverse
momentum $\mathbf{p}_{T}$,
\begin{equation}
\int d^{2}\mathbf{p}_{T}\,e_{1}^{(0)T}(p_{T})e_{1}^{(0)}(p_{T})=4\pi B_{0}.\label{eq:normality e_1^0}
\end{equation}
At the same time, the basis vectors with $n>0$ are orthogonal to
$e_{1}^{(0)}(p_{T})$ for the lowest Landau level,
\begin{equation}
\int d^{2}\mathbf{p}_{T}\,e_{1}^{(0)T}(p_{T})e_{i}^{(n)}(\mathbf{p}_{T})=0,\ \ \ \ (n>0,\,i=1,2,3,4).\label{eq:orthogonal e_1^0 e_i^n}
\end{equation}
Meanwhile, we have the following orthonormality conditions,
\begin{equation}
\int d^{2}\mathbf{p}_{T}\,e_{i}^{(m)T}(\mathbf{p}_{T})e_{j}^{(n)}(\mathbf{p}_{T})=2\pi B_{0}\delta_{ij}\delta_{mn},\label{eq:orthonormality e_i^m e_j^n}
\end{equation}
for any $m,n>0$ and $i,j=1,2,3,4$. The basis vectors defined in
Eq. (\ref{eq:def-basis}) will be used for the Wigner function in
the presence of a constant magnetic field in subsections \ref{subsec:Fermions-in-const-B}
and \ref{subsec:Fermions-in-parallel-EB}.

When a constant electric field exists, the following functions $d_{1}$,
$d_{2}$, $d_{3}$ are used in subsections \ref{subsec:Fermions-in-electric}
and \ref{subsec:Fermions-in-parallel-EB},
\begin{eqnarray}
d_{1}(\eta,u) & = & -1+e^{-\frac{\pi\eta}{4}}\eta\left|D_{-1-i\eta/2}(-ue^{i\frac{\pi}{4}})\right|^{2},\nonumber \\
d_{2}(\eta,u) & = & e^{-\frac{\pi\eta}{4}}e^{i\frac{\pi}{4}}D_{-1-i\eta/2}(-ue^{i\frac{\pi}{4}})D_{i\eta/2}(-ue^{-i\frac{\pi}{4}})+c.c.,\nonumber \\
d_{3}(\eta,u) & = & e^{-\frac{\pi\eta}{4}}e^{-i\frac{\pi}{4}}D_{-1-i\eta/2}(-ue^{i\frac{\pi}{4}})D_{i\eta/2}(-ue^{-i\frac{\pi}{4}})+c.c.,\label{eq:auxiliary functions-1}
\end{eqnarray}
where $D_{\nu}(z)$ is the parabolic cylinder function and ``$c.c.$''
is short for ``complex conjugate''. The complex conjugate of $D_{\nu}(z)$
is $\left[D_{\nu}(z)\right]^{\ast}=D_{\nu^{\ast}}(z^{\ast})$. Note
that the parabolic cylinder functions satisfy the recurrence relations
\begin{eqnarray}
D_{\nu+1}(z)-zD_{\nu}(z)+\nu D_{\nu-1}(z) & = & 0,\nonumber \\
\frac{\partial}{\partial z}D_{\nu}(z)+\frac{1}{2}zD_{\nu}(z)-\nu D_{\nu-1}(z) & = & 0.
\end{eqnarray}
Combining them we obtain a relation between $D_{\nu}(z)$ and $D_{\nu+1}(z)$,
\begin{equation}
\frac{\partial}{\partial z}D_{\nu}(z)-\frac{1}{2}zD_{\nu}(z)+D_{\nu+1}(z)=0.
\end{equation}
Using this relation we can obtain the differential equations for $d_{1}$,
$d_{2}$, $d_{3}$ %
\begin{eqnarray}
\frac{\partial}{\partial u}d_{1}(\eta,u) & = & \eta d_{3}(\eta,u),\nonumber \\
\frac{\partial}{\partial u}d_{2}(\eta,u) & = & -ud_{3}(\eta,u),\nonumber \\
\frac{\partial}{\partial u}d_{3}(\eta,u) & = & -2d_{1}(\eta,u)+ud_{2}(\eta,u),
\end{eqnarray}
where we have used
\begin{equation}
\left|D_{i\eta/2}(-ue^{-i\frac{\pi}{4}})\right|^{2}=e^{\frac{\pi\eta}{4}}-\frac{1}{2}\eta\left|D_{-1-i\eta/2}(-ue^{i\frac{\pi}{4}})\right|^{2}.\label{eq:constant d4}
\end{equation}
In order to prove relation (\ref{eq:constant d4}) we first construct
another function whose variables are $\eta$ and $u$,
\begin{equation}
d_{4}(\eta,u)=\left|D_{i\eta/2}(-ue^{-i\frac{\pi}{4}})\right|^{2}+\frac{1}{2}\eta\left|D_{-1-i\eta/2}(-ue^{i\frac{\pi}{4}})\right|^{2}.
\end{equation}
Then we can prove that this function does not depend on $u$ because
$\frac{\partial}{\partial u}d_{4}(\eta,u)=0$. Furthermore, the value
at $u=0$ can be calculated using $D_{\nu}(0)=2^{\nu/2}\sqrt{\pi}/\Gamma\left(\frac{1-\nu}{2}\right)$,
where $\Gamma(z)$ is the Gamma function.
\begin{eqnarray}
d_{4}(\eta,0) & = & \frac{\pi}{\left|\Gamma\left(\frac{1}{2}-\frac{i\eta}{4}\right)\right|^{2}}+\frac{\pi\eta}{4\left|\Gamma\left(1+\frac{i\eta}{4}\right)\right|^{2}}\nonumber \\
 & = & \cosh\left(-i\frac{\pi\eta}{4}\right)+\sinh\left(i\frac{\pi\eta}{4}\right)\nonumber \\
 & = & e^{\frac{\pi\eta}{4}},
\end{eqnarray}
where we have used the property of the Gamma function $\Gamma(1+z)=z\Gamma(z)$
and a special case of the multiplication theorem,%
\begin{eqnarray}
\left|\Gamma(bi)\right|^{2} & = & \frac{\pi}{b\sinh(\pi b)},\nonumber \\
\left|\Gamma\left(\frac{1}{2}+bi\right)\right|^{2} & = & \frac{\pi}{\cosh(\pi b)},
\end{eqnarray}
where $b$ is a real constant.

\begin{figure}
\includegraphics[width=8cm]{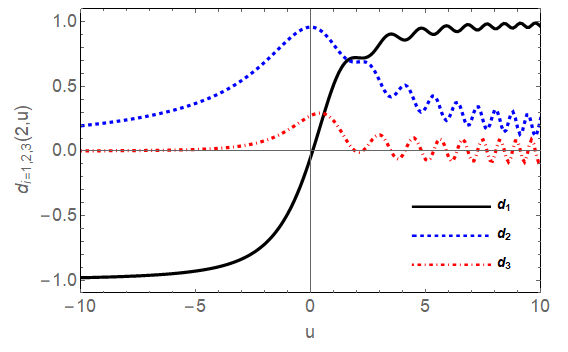}

\caption{\label{fig:functions di}Dependence of the functions $d_{i}(\eta,u)$
for $i=1,2,3$ with respect to $u$. Here $\eta=2$ corresponds to
$E_{0}=\frac{1}{2}m_{T}^{2}$.}
\end{figure}

\begin{figure}
\includegraphics[width=8cm]{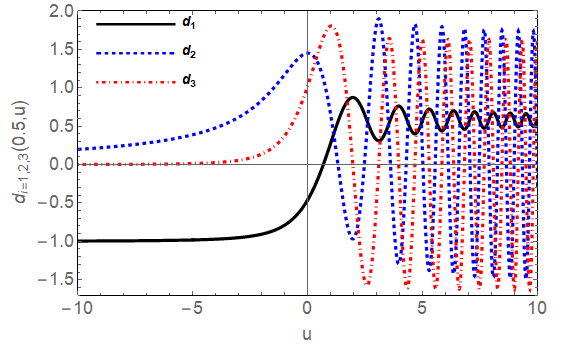}

\caption{\label{fig:functions di-2}Dependence of the functions $d_{i}(\eta,u)$
for $i=1,2,3$ with respect to $u$. Here $\eta=0.5$ corresponds
to $E_{0}=2m_{T}^{2}$.}
\end{figure}

In Figs. \ref{fig:functions di} and \ref{fig:functions di-2} we
plot the $u$-dependence of $d_{i}(\eta,u)$ for two typical values
$\eta=2$ and $\eta=0.5$. All these functions are convergent when
$u\rightarrow-\infty$,
\begin{equation}
\lim_{u\rightarrow-\infty}d_{1}(\eta,u)=-1,\ \lim_{u\rightarrow-\infty}d_{2}(\eta,u)=0,\ \lim_{u\rightarrow-\infty}d_{3}(\eta,u)=0,\label{eq:asympotic behaviour 1}
\end{equation}
Meanwhile, in the limit $u\rightarrow+\infty$, the functions $d_{2}(\eta,u)$
and $d_{3}(\eta,u)$ are highly oscillatory and are not convergent.
The function $d_{1}(\eta,u)$ is also oscillatory but the oscillation
amplitude turns to zero when $u\rightarrow+\infty$, thus $d_{1}(\eta,u)$
converges to a finite value. Explicit analysis of the parabolic cylinder
functions give their asymptotic behavior,
\begin{equation}
\lim_{u\rightarrow+\infty}D_{-1-i\eta/2}(-ue^{i\frac{\pi}{4}})=\frac{\sqrt{2\pi}}{\Gamma(1+i\eta/2)}\exp\left\{ i\left[\frac{u^{2}}{4}+\frac{\eta}{2}\log(u)\right]-\frac{\pi\eta}{8}\right\} ,
\end{equation}
where the Gamma function is given by
\begin{equation}
\Gamma(1+i\eta/2)=\int_{0}^{\infty}x^{i\eta/2}e^{-x}dx,
\end{equation}
Thus we obtain the asymptotic behavior of function $d_{1}(\eta,u)$,
\begin{equation}
\lim_{u\rightarrow+\infty}d_{1}(\eta,u)=-1+e^{-\frac{\pi\eta}{2}}\frac{2\pi\eta}{\left|\Gamma(1+i\eta/2)\right|^{2}}=1-2e^{-\pi\eta}.\label{eq:asympotic behaviour 2}
\end{equation}

\section{Wave-packet description \label{sec:Wave-packet-description}}

Due to the uncertainty principle, in quantum mechanics the momentum
and position of one particle cannot be determined at the same time.
If we adopt the plane-wave description, then the momentum is fixed,
which indicates that the uncertainty of position is infinity. This
agrees with the spatial homogeneity of the plane wave. However, we
want to have a more realistic description. Thus, in this appendix
we will introduce the wave-packet description of a particle. In quantum
mechanics, the wave packet is interpreted as probability amplitude,
whose square describes the probability of detecting a particle with
given position and momentum.

\begin{figure}
\includegraphics[width=8cm]{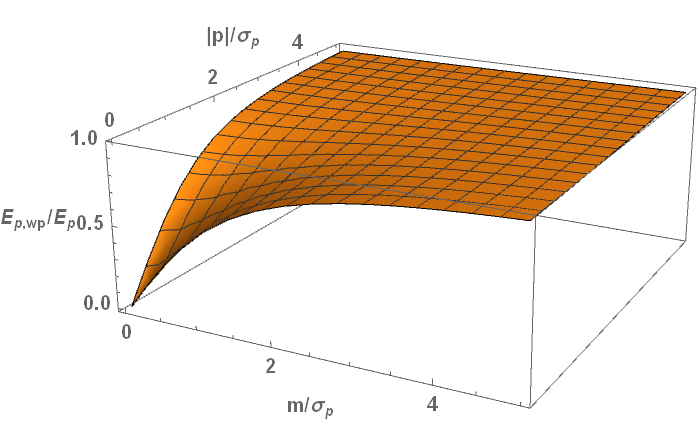}

\caption{\label{fig:Energy-of-wave-packet}The ratio between the energy of
a wave packet with the most probable momentum $\mathbf{p}$ and that
of a plane wave with the same momentum $\mathbf{p}$. The ratio depends
on dimensionless variables, $m/\sigma_{p}$ and $\left|\mathbf{p}\right|/\sigma_{p}$,
where $\sigma_{p}$ is the uncertainty of momentum.}
\end{figure}

The single-particle state and anti-particle state for the plane-wave
case are given by acting with the creation operators onto the vacuum
state,
\begin{equation}
\left|\mathbf{p},s,+\right\rangle =a_{\mathbf{p},s}^{\dagger}\left|0\right\rangle ,\ \ \left|\mathbf{p},s,-\right\rangle =b_{\mathbf{p},s}^{\dagger}\left|0\right\rangle ,
\end{equation}
where $\left|0\right\rangle $ is the vacuum state. Using the single-particle/anti-particle
states, we can calculate the expectation values of energy, momentum,
and polarization, respectively. Note that these states have fixed
momentum $\mathbf{p}$, e.g. the uncertainty of momentum is zero.
On the other hand, the wave packet for one particle is defined as
a superposition of plane waves with different wave numbers,
\begin{equation}
\left|\mathbf{p},s,+\right\rangle \,_{\text{wp}}=\frac{1}{N}\int\frac{d^{3}\mathbf{p^{\prime}}}{(2\pi)^{3}}\exp\left[-\frac{(\mathbf{p}-\mathbf{p}^{\prime})^{2}}{4\sigma_{p}^{2}}\right]a_{\mathbf{p}^{\prime},s}^{\dagger}\left|0\right\rangle ,\label{eq:wave packet state}
\end{equation}
where the most probable momentum is $\mathbf{p}$ and the uncertainty
of momentum is described by the parameter $\sigma_{p}$. The normalization
factor $N$ is determined by $\,_{\text{wp}}\left\langle \mathbf{p},s,+|\mathbf{p},s,+\right\rangle \,_{\text{wp}}=1$,
\begin{equation}
N=\sqrt{\frac{\sigma_{p}^{3}}{2\sqrt{2\pi^{3}}}}.
\end{equation}
Now we calculate the energy of the wave packet. The total energy is
given by %
\begin{equation}
E_{\mathbf{p},\text{wp}}=\,_{\text{wp}}\left\langle \mathbf{p},s,+\left|\hat{\mathcal{H}}\right|\mathbf{p},s,+\right\rangle \,_{\text{wp}}=\frac{1}{N^{2}}\int\frac{d^{3}\mathbf{p^{\prime}}}{(2\pi)^{3}}\exp\left[-\frac{(\mathbf{p}-\mathbf{p}^{\prime})^{2}}{2\sigma_{p}^{2}}\right]\sqrt{m^{2}+(\mathbf{p}^{\prime})^{2}}.
\end{equation}
We can compare the energy with the one $E_{\mathbf{p}}=\sqrt{m^{2}+(\mathbf{p})^{2}}$
for the plane wave with the same momentum. We find that the ratio
depends on the dimensionless parameters $m/\sigma_{p}$ and $\left|\mathbf{p}\right|/\sigma_{p}$.
The ratio is plot in Fig. \ref{fig:Energy-of-wave-packet}. %
{} We can observe from this figure that if the mass and center momentum
are much larger than the uncertainty $\sigma_{p}$, the ratio becomes
$1$. This indicates that a wave packet with the most probable momentum
$\mathbf{p}$ has the same energy as a plane wave with the same momentum
$\mathbf{p}$ when $E_{\mathbf{p}}\gg\sigma_{p}$.

On the other hand, the wavefunction of the wave packet in coordinate
space can be obtained by superposition of the single-particle wavefunction,
with the superposition coefficients equal to the ones for the state
in Eq. (\ref{eq:wave packet state}). For example, the particle's
wavefunction in the wave-packet description is
\begin{equation}
\psi_{s,\text{wp}}^{(+)}(x)=\frac{1}{N}\int\frac{d^{3}\mathbf{p^{\prime}}}{(2\pi)^{3}}\exp\left[-\frac{(\mathbf{p}-\mathbf{p}^{\prime})^{2}}{4\sigma_{p}^{2}}-\frac{i}{\hbar}E_{\mathbf{p},wp}t+\frac{i}{\hbar}\mathbf{p}^{\prime}\cdot\mathbf{x}\right]\left(\begin{array}{c}
\sqrt{p_{\mu}^{\prime}\sigma^{\mu}}\xi_{s}\\
\sqrt{p_{\mu}^{\prime}\bar{\sigma}^{\mu}}\xi_{s}
\end{array}\right).
\end{equation}
In the limit $m,\,\left|\mathbf{p}\right|\gg\sigma_{p}$, this wavefunction
can be expressed using the plane wave in Eq. (\ref{sol:free wave functions})
\begin{equation}
\psi_{s,\text{wp}}^{(+)}(x)\simeq\exp\left(-\sigma_{p}^{2}\frac{\mathbf{x}^{2}}{\hbar^{2}}\right)\psi_{s}^{(+)}(x).
\end{equation}
The overall factor suppresses the probability of detecting the particle
in one point which is far from the original point. Thus the most probable
position of the above wavefunction is the original point, while the
uncertainty in spatial position is
\begin{equation}
\sigma_{x}=\frac{\hbar}{2\sigma_{p}},
\end{equation}
which agrees with the uncertainty principle (\ref{eq:Heisenberg uncertainty principle}).
Thus we conclude that the wave-packet description can be used for
quantum particles with given center positions and average momentums.

\section{Pair production in Wigner-function formalism\label{sec:Pair-production-in}}

In this appendix we will show the relation between the Schwinger pair-production
process in a strong electric field and the Wigner function. This is
helpful for the calculation of pair-production rate in subsections
\ref{subsec:Fermions-in-electric} and \ref{subsec:Fermions-in-parallel-EB}.
In Quantum Kinetic Theory, the field operator is quantized in Heisenberg
picture as
\begin{equation}
\hat{\psi}(t,\mathbf{x})=\sum_{s}\int\frac{d^{3}\mathbf{q}}{(2\pi)^{3}}e^{i\mathbf{q}\cdot\mathbf{x}}\left[u_{s}(t,\mathbf{q})\hat{a}_{s}(\mathbf{q})+v_{s}(t,-\mathbf{q})\hat{b}_{s}^{\dagger}(-\mathbf{q})\right],\label{eq:Heisenberg quantization}
\end{equation}
where $E_{\mathbf{q}}$ is the canonical energy, $\mathbf{q}$ is
the canonical momentum, and $u_{s}(t,\mathbf{q})$ and $v_{s}(t,-\mathbf{q})$
are normalized single-particle wavefunctions. On the other hand, we
have
\begin{equation}
\hat{\psi}(t,\mathbf{x})=\sum_{s}\int\frac{d^{3}\mathbf{q}}{(2\pi)^{3}}e^{i\mathbf{q}\cdot\mathbf{x}}\left[\tilde{u}_{s}(t,\mathbf{q})\hat{\tilde{a}}_{s}(t,\mathbf{q})+\tilde{v}_{s}(t,-\mathbf{q})\hat{\tilde{b}}_{s}^{\dagger}(t,-\mathbf{q})\right],\label{eq:adiabatic quantization}
\end{equation}
where $\tilde{u}_{s}$ and $\tilde{v}_{s}$ are adiabatic wavefunctions,
which are chosen as $\tilde{u}_{s}(t,\mathbf{q})=\tilde{u}_{s}(\mathbf{p})$
with the kinetic momentum $\mathbf{p}=\mathbf{q}-e\mathbf{A}(t)$,
while for anti-fermions $\tilde{v}_{s}(t,-\mathbf{q})=\tilde{u}_{s}(-\mathbf{p})$.
Note that the wavefunctions should be normalized as
\begin{eqnarray}
u_{r}^{\dagger}(t,\mathbf{q})u_{s}(t,\mathbf{q})=\delta_{rs}, &  & v_{r}^{\dagger}(t,-\mathbf{q})v_{s}(t,-\mathbf{q})=\delta_{rs},\nonumber \\
\tilde{u}_{r}^{\dagger}(t,\mathbf{q})\tilde{u}_{s}(t,\mathbf{q})=\delta_{rs}, &  & \tilde{v}_{r}^{\dagger}(t,-\mathbf{q})\tilde{v}_{s}(t,-\mathbf{q})=\delta_{rs}.
\end{eqnarray}
Thus we can solve these adiabatic operators from the quantized field
in Eq. (\ref{eq:adiabatic quantization}) using the normalization
properties
\begin{eqnarray}
\hat{\tilde{a}}_{s}(t,\mathbf{q}) & = & \int d^{3}\mathbf{x}\,e^{-i\mathbf{q}\cdot\mathbf{x}}\tilde{u}_{s}^{\dagger}(t,\mathbf{q})\hat{\psi}(t,\mathbf{x}),\nonumber \\
\hat{\tilde{b}}_{s}^{\dagger}(t,-\mathbf{q}) & = & \int d^{3}\mathbf{x}\,e^{-i\mathbf{q}\cdot\mathbf{x}}\tilde{v}_{s}^{\dagger}(t,-\mathbf{q})\hat{\psi}(t,\mathbf{x}).
\end{eqnarray}
Inserting the quantized field operator in Eq. (\ref{eq:Heisenberg quantization})
into the above we obtain %
\begin{eqnarray}
\tilde{a}_{s}(t,\mathbf{q}) & = & \tilde{u}_{s}^{\dagger}(t,\mathbf{q})\sum_{r}u_{r}(t,\mathbf{q})a_{r}(\mathbf{q})+\tilde{u}_{s}^{\dagger}(t,\mathbf{q})\sum_{r}v_{r}(t,-\mathbf{q})b_{r}^{\dagger}(-\mathbf{q}),\nonumber \\
\tilde{b}_{s}^{\dagger}(t,-\mathbf{q}) & = & \tilde{v}_{s}^{\dagger}(t,-\mathbf{q})\sum_{r}u_{r}(t,\mathbf{q})a_{r}(\mathbf{q})+\tilde{v}_{s}^{\dagger}(t,-\mathbf{q})\sum_{r}v_{r}(t,-\mathbf{q})b_{r}^{\dagger}(-\mathbf{q}).
\end{eqnarray}
They give the relation between adiabatic operators and the ones in
the Heisenberg picture. This relation is also known as Bogoliubov
transformation. The particle number and anti-particle number for a
system are defined as the expectation values
\begin{eqnarray}
f_{s}^{(+)}(t,\mathbf{q}) & = & \left\langle \Omega\left|\hat{\tilde{a}}_{s}^{\dagger}(t,\mathbf{q})\hat{\tilde{a}}_{s}(t,\mathbf{q})\right|\Omega\right\rangle ,\nonumber \\
f_{s}^{(-)}(t,\mathbf{q}) & = & \left\langle \Omega\left|\hat{\tilde{b}}_{s}^{\dagger}(t,\mathbf{q})\hat{\tilde{b}}_{s}(t,\mathbf{q})\right|\Omega\right\rangle .
\end{eqnarray}
Here we use $\left|\Omega\right\rangle $ to represent the quantum
state for the considered system. Then the average pair number is defined
as
\begin{equation}
n_{\text{pair}}(t)\equiv\frac{1}{2}\int\frac{d^{3}\mathbf{q}}{(2\pi)^{3}}\sum_{s}\left[f_{s}^{(+)}(t,\mathbf{q})+f_{s}^{(-)}(t,-\mathbf{q})\right].
\end{equation}
Inserting the distribution functions into the average pair number
we finally obtain %
\begin{eqnarray}
n_{\text{pair}}(t) & = & \frac{1}{4}\int\frac{d^{3}\mathbf{q}}{(2\pi)^{3}}\text{Tr}\left\{ \frac{\boldsymbol{\gamma}\cdot\mathbf{p}+m}{2E_{\mathbf{p}}}\sum_{r,r^{\prime}}\bar{u}_{r^{\prime}}(\mathbf{q},t)\otimes u_{r}(\mathbf{q},t)\left\langle \Omega\left|\hat{a}_{r^{\prime}}^{\dagger}(\mathbf{q})\hat{a}_{r}(\mathbf{q})\right|\Omega\right\rangle \right\} \nonumber \\
 &  & +\frac{1}{4}\int\frac{d^{3}\mathbf{q}}{(2\pi)^{3}}\text{Tr}\left\{ \frac{\boldsymbol{\gamma}\cdot\mathbf{p}+m}{2E_{\mathbf{p}}}\sum_{r,r^{\prime}}\bar{v}_{r^{\prime}}(-\mathbf{q},t)\otimes v_{r}(-\mathbf{q},t)\left\langle \Omega\left|\hat{b}_{r^{\prime}}(-\mathbf{q})\hat{b}_{r}^{\dagger}(-\mathbf{q})\right|\Omega\right\rangle \right\} \nonumber \\
 &  & -\frac{1}{4}\int\frac{d^{3}\mathbf{q}}{(2\pi)^{3}}\sum_{r}\left\langle \Omega\left|a_{r}^{\dagger}(\mathbf{q})a_{r}(\mathbf{q})+b_{r^{\prime}}(-\mathbf{q})b_{r}^{\dagger}(-\mathbf{q})\right|\Omega\right\rangle .
\end{eqnarray}
Note that we are working in the Heisenberg picture, where the quantum
state $\left|\Omega\right\rangle $ is independent of time and thus
the last term is also independent of $t$. Using the equal-time Wigner
function, we have
\begin{equation}
n_{\text{pair}}(t)=\frac{1}{4}\int d^{3}\mathbf{x}\int\frac{d^{3}\mathbf{p}}{(2\pi)^{3}}\text{Tr}\left\{ \frac{\boldsymbol{\gamma}\cdot\mathbf{p}+m}{2E_{\mathbf{p}}}W(t,\mathbf{x},\mathbf{p})\right\} +\text{const.},
\end{equation}
where we have replaced the integration over canonical momentum by
the one over kinetic momentum. From this formula we can derive the
density of pairs,
\begin{equation}
n_{\text{pair}}(t,\mathbf{x},\mathbf{p})=\frac{\mathbf{p}\cdot\boldsymbol{\mathcal{V}}+m\mathcal{F}}{2E_{\mathbf{p}}}+\text{const.},\label{eq:density of pairs}
\end{equation}
where $\mathcal{F}$ and $\boldsymbol{\mathcal{V}}$ are the scalar
and vector components of the Wigner function, respectively, as shown
in Eq. (\ref{def:Wigner function decomposition}).

\bibliographystyle{apsrev4-1}
\bibliography{References}

\end{document}